%% file: start.tex
\documentclass[sigconf,10pt]{acmart}
\renewcommand\footnotetextcopyrightpermission[1]{} 
\input{helper}
\DeclareMathAlphabet{\mathcal}{OMS}{cmsy}{m}{n}

\begin{document}
\title{Facilitating SQL Query Composition and Analysis}

\author{Zainab Zolaktaf}
\affiliation{%
  \institution{University of British Columbia}
  \city{Vancouver}
  \state{B.C}
}
\email{zolaktaf@cs.ubc.ca}

\author{Mostafa Milani}
\affiliation{%
  \institution{University of British Columbia}
  \city{Vancouver}
  \state{B.C}
}
\email{mkmilani@cs.ubc.ca}

\author{Rachel Pottinger}
\affiliation{%
  \institution{University of British Columbia}
  \city{Vancouver}
  \state{B.C}
}
\email{rap@cs.ubc.ca}

\begin{abstract}
\input{abstract}
\end{abstract}


\maketitle

\input{main}

\bibliographystyle{ACM-Reference-Format}
\small

\bibliography{cr-sigmod}

\iffullpaperComposition
\appendix

\input{appendix}

\fi
\end{document}

%% file: helper.tex
\usepackage{booktabs}
\usepackage{import}

\usepackage{hyperref}
\usepackage{graphicx}
\usepackage{subfig}

\usepackage[inline]{enumitem}   
\usepackage{ltablex}
\usepackage{color}	
\usepackage{amsmath,scalerel}
\usepackage{amsfonts}
\usepackage{url}
\usepackage{balance}
\usepackage{bm}
\usepackage{bbm} 
\usepackage{multirow}
\usepackage{listings}
\usepackage[labelfont=bf,textfont=bf]{caption}
\usepackage{dblfloatfix}
\usepackage[export]{adjustbox}
\usepackage{enumitem}
\usepackage{rotating}
\usepackage{soul}
\usepackage{supertabular}
\usepackage{tikz}
\usetikzlibrary{fit,positioning,arrows.meta}
\tikzset{neuron/.style={shape=circle, minimum size=1.25cm,
  inner sep=0, draw, font=\normalsize}, io/.style={neuron, fill=gray!20}}

\usepackage{colortbl}
\newtoggle{fond}
\providecommand{\chfd}%
    {\iftoggle{fond}%
        {\rowcolor[rgb]{1,.1,1}}%
        {}%
    }%

\togglefalse{fond}

\usepackage{xspace}
\usepackage{todonotes}
\usepackage{tabularx}
\usepackage{multirow}
\usepackage[ruled, linesnumbered, titlenumbered]{algorithm2e}

\usepackage{amsthm}

\makeatletter
\newcommand\footnoteref[1]{\protected@xdef\@thefnmark{\ref{#1}}\@footnotemark}
\makeatother

\newcommand{\blue}[1]{\textcolor{blue}{#1}}

\newcommand{\nit}[1]{\textit{#1}\xspace}
\newcommand{\HL}[1]{\framebox[1.1\width]{\textbf{#1}}}
\newcommand{\mc}[1]{\mathcal{#1}}

\newcommand{\settingOne}{Homogeneous Instance\xspace}
\newcommand{\settingTwo}{Homogeneous Schema\xspace}
\newcommand{\settingThree}{Heterogeneous Schema\xspace}
\newcommand{\ignore}[1]{}
\newcommand{\Label}[1]{{\tt #1}}
\newcommand{\Sqlkeyword}[1]{\textcolor{blue}{\tt #1}}
\newcommand{\nowebh}{\Label{no\_web\_hit}\xspace}
\newcommand{\unknown}{\Label{unknown}\xspace}
\newcommand{\admin}{\Label{admin}\xspace}
\newcommand{\program}{\Label{program}\xspace}
\newcommand{\anonym}{\Label{anonymous}\xspace}
\newcommand{\browser}{\Label{browser}\xspace}
\newcommand{\bott}{\Label{bot}\xspace}
\newcommand{\scss}{\Label{success}\xspace}
\newcommand{\severe}{\Label{severe}\xspace}
\newcommand{\nsevere}{\Label{non\_severe}\xspace}

\newcommand{\clevel}{number of characters\xspace}
\newcommand{\wlevel}{number of words\xspace}
\newcommand{\fcount}{number of functions\xspace}
\newcommand{\jcount}{number of joins\xspace}
\newcommand{\pccount}{number of  predicate table names\xspace}
\newcommand{\pcount}{number of  predicates\xspace}
\newcommand{\tcount}{number of  unique table names\xspace}
\newcommand{\sccount}{number of selected columns\xspace}
\newcommand{\ncount}{nestedness level\xspace}
\newcommand{\nacount}{nested aggregation\xspace}

\newcommand{\ctfidf}{\Label{ctfidf}\xspace}
\newcommand{\wtfidf}{\Label{wtfidf}\xspace}
\newcommand{\clstm}{\Label{clstm}\xspace}
\newcommand{\wlstm}{\Label{wlstm}\xspace}
\newcommand{\ccnn}{\Label{ccnn}\xspace}
\newcommand{\wcnn}{\Label{wcnn}\xspace}
\newcommand{\median}{\Label{median}\xspace}
\newcommand{\opt}{\Label{opt}\xspace}
\newcommand{\mfreq}{\Label{mfreq}\xspace}
\newcommand{\prm}{$p$\xspace}

\numberwithin{equation}{section}

\newif\ifproposal
\proposalfalse

\newif\ifPhDThesis
\PhDThesisfalse

\newif\iffullpaperComposition
\fullpaperCompositiontrue

\newif\iffiltereddata
\filtereddatafalse

\newif\ifextraparts 
\extrapartsfalse

\newif\ifbootstrapexp
\bootstrapexptrue
\newcommand{\eat}[1]{} 
\newenvironment{changemargin} [2]{\begin{list}{}{
         \setlength{\topsep}{0pt}\setlength{\leftmargin}{0pt}
         \setlength{\rightmargin}{0pt}
         \setlength{\listparindent}{\parindent}
         \setlength{\itemindent}{\parindent}
         \setlength{\parsep}{0pt plus 1pt}
         \addtolength{\leftmargin}{#1}\addtolength{\rightmargin}{#2}
         }\item }{\end{list}}

  {
    \begin{changemargin}{-8pt}{-0cm}
    \vspace{-13pt}
    \hspace{-8pt}
    \begin{itemize}
    \setlength{\itemsep}{-1pt}
  }
  {
    \end{itemize}
    \end{changemargin}
  }

  {
    \begin{changemargin}{-8pt}{-0cm}
    \vspace{-13pt}
    \hspace{-8pt}
    \begin{enumerate}
    \setlength{\itemsep}{-1pt}
  }
  {
    \end{enumerate}
    \end{changemargin}
  }

\newcommand{\ssh}[1]{SQLShare\xspace}
\newcommand{\sdss}[1]{SDSS}
\newcommand{\vl}[1]{``#1''}
\newcommand{\ch}[1]{{\tt `#1'}}

\newlength\mystoreparindent

\newtheorem{DEFINE}{Definition}
\newenvironment{defin}{\begin{DEFINE}\begin{rm}} {{\hfill$\Box$}\end{rm}\end{DEFINE}}

\newtheorem{EXAMPLE}{Example}
\newenvironment{example}{\begin{EXAMPLE}\begin{rm}} {{\hfill$\Box$}\end{rm}\end{EXAMPLE}}

\numberwithin{equation}{section}

\usepackage{xcolor,listings}
\usepackage{textcomp}
\usepackage{color}

\definecolor{codegreen}{rgb}{0,0.6,0}
\definecolor{codeblue}{rgb}{0,0,0.9}
\definecolor{codegray}{rgb}{0.5,0.5,0.5}
\definecolor{codepurple}{HTML}{C42043}
\definecolor{backcolour}{HTML}{FFFFFF}
\definecolor{bookColor}{cmyk}{0,0,0,0.90}
\color{bookColor}

\lstdefinestyle{mystyle}{
    backgroundcolor=\color{backcolour},
    commentstyle=\color{codegreen},
    keywordstyle=\color{codeblue},
    stringstyle=\color{codeblue},
    basicstyle=\tiny\ttfamily,
    breakatwhitespace=false,
    breaklines=true,
    captionpos=b,
    keepspaces=true,
    showspaces=false,
    showstringspaces=false,
    showtabs=false,
    frame=single,
}

\lstdefinestyle{qstyle}{
    backgroundcolor=\color{backcolour},
    commentstyle=\color{codegreen},
    keywordstyle=\color{codeblue},
    stringstyle=\color{codeblue},
    basicstyle=\small\ttfamily,
    breakatwhitespace=false,
    breaklines=true,
    captionpos=b,
    keepspaces=true,
    showspaces=false,
    showstringspaces=false,
    showtabs=false,
    frame=single,
}
\newcommand\numberstyle[1]{%
    \footnotesize
    \color{codegray}%
    \ttfamily
    \ifnum#1<10 0\fi#1 |%
}
\newsavebox{\mylistingbox}

\hypersetup{
  colorlinks   = true, 
  urlcolor     = blue, 
  linkcolor    = blue, 
  citecolor   = blue 
}

%% file: abstract.tex
\iffullpaperComposition
Formulating efficient SQL queries  requires several  cycles of  tuning and execution, particularly for inexperienced users.  We examine methods that can accelerate and improve this interaction by providing insights about SQL queries \textit{prior to execution}. 
 We achieve this by predicting properties such as the query answer size, its run-time, and error class.   
Unlike existing approaches, our approach does not rely on any statistics from the database instance or query execution plans. This is particularly important in settings with limited access to the database instance.

Our approach is based on  using data-driven machine learning techniques that rely on large query workloads  to model  SQL queries and their properties.   We evaluate the utility of neural network models and traditional machine learning models.  
We use two real-world query workloads:  the Sloan Digital Sky Survey (SDSS) and the SQLShare query workload. 
Empirical results show that the neural network models are more accurate in predicting the query error class, achieving a higher F-measure on classes with fewer samples as well as performing better on other problems such as run-time and answer size prediction.
These results are encouraging and confirm that SQL query workloads and data-driven machine learning methods can be leveraged to  facilitate query composition and analysis. 
\else 
Formulating efficient SQL queries  requires several  cycles of  tuning and execution.  We examine methods that can accelerate and improve this interaction by providing insights about SQL queries \textit{prior to execution}. 
 We achieve this by predicting properties such as the query answer size, its run-time, and error class.   
Unlike existing approaches, our approach does not rely on any statistics from the database instance or query execution plans. 
Our approach is based on  using data-driven machine learning techniques that rely on large query workloads  to model  SQL queries and their properties.   
 Empirical results show that the neural network models are more accurate in predicting several query properties.
 \fi  


%% file: main.tex
\input{Introduction}

\input{Motivation}
\input{ProblemDefinitions}

\input{DatasetAndAnalysis}

\input{Methods}

\input{ExperimentalSetup}

\input{Results}

\input{RelatedWork}

\input{Conclusion}


%% file: Introduction.tex
\section{Introduction}
\label{sec:dbrec-introduction}
Formulating effective SQL queries is one of the main challenges of interacting with large relational databases. Particularly for inexperienced users,  writing good SQL queries may require several cycles of tuning and execution. This diminishes the user experience, prevents them from easily accessing information~\cite{jagadish2007making}, and   can also be costly. For example, cloud-based services, such as Google BigQuery, charge their users for running queries~\cite{snowflake,bigquery,redshift}. Moreover, inefficient SQL queries can pose a burden on the database's resources. Our focus in this work is on facilitating user interaction with the database by providing additional information about SQL queries prior to their execution.

We focus on two groups of users:  {\em end users} and  {\em database administrators (DBAs)}. To  help end users compose SQL queries, we  study three problems: {\em query answer size prediction}, {\em query run-time prediction}, and {\em query error prediction}.  We can save end users time and effort by pointing out when their queries are inefficient, unlikely to work at all, or are likely to take radically different time than they are expecting (and thus are likely to be not the queries that they are trying to write).

We also improve user interaction for DBAs. To characterize how end users and programs use the service, DBAs need to analyze incoming requests and queries and categorize them into \textit{classes of clients}~\cite{singh2007skyserver}. This in turn allows them to improve service quality for end users.  To help DBAs with this  analysis,  we  study the problem of \textit{session type classification}, which is the automatic identification of the  class of clients that generated the queries in a session.

While we use estimates of SQL query properties to improve usability, these estimates have typically been used to improve tasks like admission control, access control, scheduling, and costing during query optimization~\cite{li2012robust,liu2015cardinality,ganapathi2009predicting,akdere2012learning}. Most of these studies, however, are based on manually constructed cost models in the query optimizer and require access to the database instance. But the analytical cost models in the  query optimizer can be imprecise due to simplifying assumptions, e.g.,~uniform data distributions~\cite{leis2015good,ganapathi2009predicting,ai-meets-ai}. Moreover, access to the database instance can be infeasible in an increasingly large number of settings, e.g., cloud-based data warehouses like Google BigQuery~\cite{bigquery}, databases on the hidden web, sources located behind wrappers in data integration systems~\cite{bergamaschi2011keyword}, and instances with limited access due to cost or privacy issues~\cite{bigquery,snowflake,redshift,hacigumucs2002executing}. Due to these restrictions, there is growing need for work that does not assume direct access to database instances.

Our approach for modeling the queries and their properties relies on using ~\emph{SQL query workloads}, which contain logs of past queries submitted to the database. Query workloads are an alternative resource  in settings  with limited database instance access. They have been used to improve query performance estimates for tasks like query optimization and scheduling~\cite{akdere2012learning,li2012robust,ganapathi2009predicting,liu2015cardinality}.  In addition to  requiring access to the database  instance and schema, these  works examine synthetic or small-scale query workloads, e.g.,~TPC-H~\cite{TPC-H1}.  They extract hand-engineered   features from query execution plans, and apply a prediction model.  However, synthetic and small-scale query workloads do not represent the full capacity and challenges of potential queries, as shown empirically in~\cite{leis2015good}.

We use SQL query workloads that are {\em large-scale and real-world} and present an abundance of realistic usage patterns from a variety of different users. These workloads are broadly used and publicly available in scientific and academic research domains. Examples are the Sloan Digital Sky Survey (SDSS) query workload~\cite{szalay2002sdss,raddick2014ten1,raddick2014ten2} and SQLShare~\cite{jain2016sqlshare}. Companies and organizations, such as Snowflake~\cite{jain2018query2vec,jain2018database}, often maintain their own large-scale private workloads. DBMSs support logging features that make it easy to generate and maintain these workloads.

When large-scale workloads do not exist for a database, knowledge learned from  workloads over other databases may be used for query property prediction. Consequently, we define different {\em query facilitation problem settings}  that vary in their data  heterogeneity. This makes the prediction problem more challenging due to different   underlying data distributions for the SQL query and the workload. These settings require models that generalize well and can transfer knowledge learned from workloads to predict query properties over a different database. We empirically show that some models generalize better, which allows reusing large-scale workloads for query performance property prediction.  We hope in light of research like ours that shows the benefits of large-scale workloads, more companies and organizations develop, maintain, and share their query workloads, which can ultimately improve transparency and customer engagement.

We use data-driven machine learning techniques that require large-scale workloads to build effective models and predict different query properties. Since there are no standard models for our problems, we start by establishing appropriate baseline models. Our query facilitating problems are essentially query labeling tasks. A closely related area is text categorization, where the goal is to predict categories for documents written in natural language~\cite{kim2014convolutional,zhang2015character,conneau2016very,johnson2014effective,johnson2015semi,johnson2016supervised,johnson2017deep}.
We chose a broad set of applicable models from this domain;
on large-scale datasets, the dominant approaches use
Long-Term Short-Term (LSTM) and  Convolutional Neural Network (CNN) models~\cite{lecun1989backpropagation,conneau2016very,bai2018empirical,yin2017comparative}.  LSTMs treat texts as sequential inputs, while CNNs can automatically identify n-grams. \ignore{Compared to more complex models, CNNs have competitive results, and are easier to train and interpret. One part of our work is to examine whether LSTMs and CNNs are equally suitable  for predicting properties of SQL queries.} SQL queries have significant differences with natural language sentences, e.g., they include mathematical expressions that are important to retain in the query representation. We therefore applied all models at both character and word level.

Our work is closest to~\cite{jain2018query2vec,jain2018database}, which assume no direct access to database instances and apply data-driven models to large-scale query workloads.  They address workload management tasks, such as index recommendation and security audits. We introduce additional problems related to facilitating query composition and analysis,~e.g., session classification. Moreover, we formalise and study different problem settings and conduct an extensive workload analysis that results in encouraging empirical  results.

In summary we contribute the following:
\begin{itemize}
\item We motivate (Section~\ref{sec:dbrec_motivation}) and formally define (Section~\ref{sec:dbrec_problem_definition}) four problems that help end users and DBAs with composition and analysis by providing query insights prior to execution. 

\item Our approach relies on exploiting large  query workloads.
We use  two  real-world query workloads that are publicly available,  SDSS and SQLShare. In Section~\ref{sec:dbrec_dataset_and_analysis}, we describe these workloads. Moreover, to enable  better model selection and evaluation, we conduct a comprehensive workload analysis that covers structural and syntactic features of the queries and their labels.

\item   We examine a broad set of models to establish the baselines and  assess the feasibility of our problems in Section~\ref{sec:dbrec_methods}.  We adapted  two classes of  neural network models to our problems and compared them with simple baselines and traditional NLP  models.

\item Our  empirical evaluation (Section~\ref{sec:dbrec_emp_eval}) shows  the neural networks are more accurate in predicting the query error class, achieving a higher F-measure on classes with fewer samples. For run-time and answer size prediction, the neural networks obtain better results,  particularly  on complex queries. Additionally, we found character level CNNs are able to generalize better under various problem settings.

\end{itemize}
Section~\ref{sec:dbrec-related-work} describes related work and Section~\ref{sec:dbrec-conclusion} concludes.

%% file: Motivation.tex
\section{Motivating Example}
\label{sec:dbrec_motivation}

\iffullpaperComposition

\begin{figure}
\begin{lrbox}{\mylistingbox}%
\begin{minipage}{1\linewidth}%
\begin{lstlisting}
SELECT COUNT(*) FROM Galaxy WHERE ...

\end{lstlisting}
\end{minipage}%
\end{lrbox}%
\subfloat[Users are advised to submit a count query before executing the actual query.]
{ \usebox{\mylistingbox}\label{dbrec:sdss-ex-0}}

\begin{lrbox}{\mylistingbox}%
\begin{minipage}{1\linewidth}%
\begin{lstlisting}
SELECT ... FROM PhotoObj 
WHERE flags & dbo.fPhotoFlags('BLENDED') > 0

\end{lstlisting}
\end{minipage}%
\end{lrbox}%
\subfloat[An inefficient query. It  retrieves a large number of objects,  the function call in the WHERE clause is called once per matching row.]
{ \usebox{\mylistingbox}\label{dbrec:sdss-ex-1}}

\caption{Users are advised to optimize their queries on  SDSS to prevent long wait times.} 
\label{dbrec:sdss-ex}
\end{figure}

\else

\begin{figure}
\begin{lrbox}{\mylistingbox}%
\begin{minipage}{1\linewidth}%
\begin{lstlisting}
SELECT ... FROM PhotoObj 
WHERE flags & dbo.fPhotoFlags('BLENDED') > 0

\end{lstlisting}
\end{minipage}%
\end{lrbox}%
{ \usebox{\mylistingbox}\label{dbrec:sdss-ex-1}}
\caption{An inefficient query in which the function in the WHERE clause is called once per matching row.}
\label{dbrec:sdss-ex}
\end{figure}

\fi

We use the Sloan Digital Sky Survey (SDSS) as a motivating example.
SDSS is an astronomy project that provides a digital map of the sky~\cite{raddick2014ten1,raddick2014ten2,szalay2002sdss,szalay2018skyserver}. \iffullpaperComposition
SDSS data includes raw images of the sky along with numerical estimates for the  physical attributes of objects in the images. These numerical estimates, known as \textit{scientific attributes}, are stored in the Catalog Archive Server (CAS) databases, which can be queried via SQL.
Users  access SDSS data  through several access interfaces, including  an asynchronous job interface called CasJobs and a web interface.
~\footnote{\url{http://skyserver.sdss.org/dr15/en/tools/search/sql.aspx}}

\else
The data in SDSS  can be queried through SQL.
\fi
A diverse set of end users, ranging from high school students to  astronomers,  with varied levels of astronomy  and SQL knowledge, use SDSS~\cite{szalay2018skyserver}. To help users compose queries, SDSS provides some resources. These include a tutorial  to help users learn SQL basics  and  a  set of sample SQL queries that can be used as templates. \iffullpaperComposition
~\footnote{\url{http://skyserver.sdss.org/dr8/en/help/docs/realquery.asp}}
\fi
There are also descriptions of costly queries with hints on how to rewrite them to ensure they run faster. \iffullpaperComposition~\footnote{\url{http://skyserver.sdss.org/dr8/en/help/docs/pdf/sql_help.pdf}}\fi
 For example,  users are advised to always start with a ``count''
 \iffullpaperComposition
 query (Figure~\ref{dbrec:sdss-ex-0})
 \else
 query 
 \fi
 to estimate of the query answer size and  prevent long wait times. 
 Figure~\ref{dbrec:sdss-ex} shows  an inefficient query example, given on the SDSS website,  which runs the function {\tt dbo.fPhotoFlags('BLENDED')} for several records in {\tt PhotoObj}. The users are advised to rewrite it to a query that invokes the function only once. \iffullpaperComposition~\footnote{\label{footnote-sdss-ex}\url{http://skyserver.sdss.org/dr8/en/help/docs/sql_help.asp\#optquery}}\fi

While these resources are helpful, 
they are not sufficient. In particular, the step-by-step SQL tutorial is generic and  helps inexperienced users learn SQL syntax rather than write meaningful queries.
The sample set of queries is small and static  compared to the size of the database and the complexity of potential queries. The  SDSS schema   has 87 tables, 46 views, 467 functions, 21 procedures, and 82 indices. \iffullpaperComposition~\footnote{\url{http://skyserver.sdss.org/dr12/en/help/browser/browser.aspx}}\fi The schema size and complexity makes it hard for users to become familiar enough with the SDSS database to effectively optimize and tune their queries. 
Ad-hoc hints do not cover all possible optimization  opportunities. In this context, real-time query performance estimates  like  answer size or execution time  can  increase user productivity and efficiency.

DBAs are another group  who interact with SDSS. One of their tasks is to analyze the incoming requests and queries during a session and decide the class of the end user (e.g. human, bot, program) that generated the queries. This is called the session classification problem, where a session is defined as a sequence of interactions between an end user and the system. Session classification allows DBAs to  improve the services offered based on usage patterns~\cite{szalay2002sdss}.

Session classification is challenging. First,  identifying individual sessions is difficult. This is because users are anonymous and do not necessarily login to the system, their IP addresses can change, and the same IP address can access different SDSS interfaces. In fact, there has been research on automatic session identification as a separate task~\cite{khoussainova2010snipsuggest}. We follow~\cite{szalay2002sdss} and assume that a session is characterized by an ordered sequence of \textit{hits} (i.e., SQL query or web request) from a single IP address, such that the gaps between hits in the sequence is no longer than 30 minutes~\cite{szalay2002sdss, raddick2014ten1}.

The second step in session classification is specifying the correct label for an identified session. These labels can be used to enforce certain policies and optimize system services  (e.g.,~for resource allocation, or to design different interaction modalities  based on the  usage patterns of different types of sessions~\cite{szalay2002sdss}).  Although the web requests contain a string that describes the browser or program that generated the request, this ``agent string'' is not reliable. Consequently, SDSS user sessions are labeled using a combination of agent string, IP address, and behavior during session. This procedure does not consider the content or syntactic properties of queries. Therefore, a question that arises is whether the raw query itself can be used for performing the label assignment.  This functionality would provide a complementary resource for assisting DBAs. Additionally, it helps automate  identification of human traffic, which is needed for  downstream  usability problems, like query recommendation for end users~\cite{khoussainova2010snipsuggest}.




%% file: ProblemDefinitions.tex
\section{Problem Definition}
\label{sec:dbrec_problem_definition}
We use small letters for scalars, capital letters for  sets or sequences, bold small letters for vectors, and  bold capital letters for matrices.

\begin{defin}
A query $Q = (t_1, ..., t_n)$ is a sequence of tokens from a vocabulary $V$. We consider two sets of vocabularies: a vocabulary that contains characters and a vocabulary that contains words. We define $v$ to denote the size of $V$ and $\mathcal{Q}$ denotes the collection of all queries over $V$.\end{defin}

\iffullpaperComposition
\begin{example}
For the query in Figure~\ref{dbrec:fig_Q1},  $Q_1=$(\vl{\Sqlkeyword{SELECT}}, \vl{*}, \vl{\Sqlkeyword{FROM}}, ...) is a query with a vocabulary of words, and $Q_2=$ (\ch{S}, \ch{E}, \ch{L}, \ch{E}, ...) is a query with a vocabulary of characters.
\end{example}
\fi

\begin{defin} For a token $t_i$ in a query $Q$, we define $\bm{e}_i \in \{0,1\}^v$ as the one-hot encoding of $t_i$, i.e. a vector of bits for tokens in $V$ where the bit that corresponds to $t_i$ is $1$ and all the other bits are $0$. We use $\bm{x}_i \in \mathbb{R}^d$ to refer to the distributed representation of $t_i$ in latent space obtained using an embedding matrix $\bm{X} \in \mathbb{R}^{d \times v}$ ($\bm{x}_i = \bm{X}\bm{e}_i$). We define an $n$-gram in $Q$ as a sequence of $n$ tokens that appear in $Q$.\end{defin}

\iffullpaperComposition
\begin{example}
The one-hot encoding of $t_1=$\vl{\Sqlkeyword{SELECT}} and $t_2=$\vl{*} w.r.t. a vocabulary $V=\{t_1,t_2,t_3,t_4\}$ is $\bm{e}_1=[1\;0\;0\;0]$ and $\bm{e}_2=[0\;1\;0\;0]$. Given an embedding matrix $\bm{X}=[[3\;6\;4]\;[9\;5\;8]\;[4\;3\;$ $0]\;[6\;0\;4]]$, $\bm{x}_1=[3\;6\;4]$ and $\bm{x}_2=[9\;5\;8]$ are the distributed representation of $t_1$ and $t_2$, respectively. The sequence (\vl{\Sqlkeyword{SELECT}}, \vl{*}, \vl{\Sqlkeyword{FROM}}) is a $3$-gram of words and (\ch{F},\ch{R},\ch{O},\ch{M}) is a $4$-gram of characters.
\end{example}
\else
\begin{example}
For the query in Figure~\ref{dbrec:fig_Q1},  $Q_1=$(\vl{\Sqlkeyword{SELECT}}, \vl{*}, \vl{\Sqlkeyword{FROM}}, ...) is a query with a vocabulary of words, and $Q_2=$ (\ch{S}, \ch{E}, \ch{L}, \ch{E}, ...) is a query with a vocabulary of characters.
The one-hot encoding of $t_1=$\vl{\Sqlkeyword{SELECT}} and $t_2=$\vl{*} w.r.t. a vocabulary $V=\{t_1,t_2,t_3,t_4\}$ is $\bm{e}_1=[1\;0\;0\;0]$ and $\bm{e}_2=[0\;1\;0\;0]$. Given an embedding matrix $\bm{X}=[[3\;6\;4]\;[9\;5\;8]\;[4\;3\;$ $0]\;[6\;0\;4]]$, $\bm{x}_1=[3\;6\;4]$ and $\bm{x}_2=[9\;5\;8]$ are the distributed representation of $t_1$ and $t_2$, respectively. The sequence (\vl{\Sqlkeyword{SELECT}}, \vl{*}, \vl{\Sqlkeyword{FROM}}) is a $3$-gram of words and (\ch{F},\ch{R},\ch{O},\ch{M}) is a $4$-gram of characters.
\end{example}
\fi

\input{sql_example_table_short}

We want to model queries and their properties   to generate feedback for end users and DBAs. Similar to~\cite{bergamaschi2011keyword}, we assume a-priori access  to the database  instance, i.e., the tuples and their statistics,  is not available. This is commonly the case for users of \iffullpaperComposition systems like SDSS, SQLShare, or \fi web services like Google BigQuery. Instead, our approach exploits the rich content of large  query workloads:

\begin{defin} [Query workload] Let $\mc{W}=\{(Q_i,y_i)\}_{i=1}^n$ denote the input query workload, where $Q_i$ is a query statement and $y_i$ is a query label. The label  is a  query property that is obtained by submitting  $Q_i$ to the database.
\end{defin}

The  query statement $Q_i$ is typically a SQL query and  can  contain clauses such as \Sqlkeyword{SELECT}, \Sqlkeyword{EXECUTE}, \Sqlkeyword{CREATE},  \Sqlkeyword{ALTER},  or combinations such as \Sqlkeyword{DELETE} $|$ \Sqlkeyword{UPDATE} $|$ \Sqlkeyword{INSERT} clauses.
However, in realistic  workloads such as SDSS, the end user can  submit \emph{any} query to the system, including  a random natural language sentence. So,  the query type is not restricted.

The query  label $y_i$  can correspond to different query properties, e.g.,~answer size  or CPU time. We focus on four types of query  labels.   Our goal is to develop models that  can predict these labels --- prior to execution. For each query  $Q_i$ the error class $y^e_i$ is a numeric  indicator of whether the query successfully executed or not. The  query  total CPU time $y^c_i$ is  a real number and represents the query execution time. The query answer size  $y^a_i$ is an integer and represents the number of rows retrieved for the query. The session class  $y^s_i$  is  the class of client that  generated the  query (and its session). 
 Figure~\ref{tab:example_sql_queries} shows sample  queries from  our SDSS query workload, along with their properties.

\ignore{
\begin{defin} [Query Facilitation Problems] \label{df:problems} We address the following  four problems:
\begin{itemize}[leftmargin=*]

\item {\bf Query error classification.} Given a  query workload $\mc{W}=\{(Q_i,$ $y_i^e)\}_{i=1}^n$ comprised of the query statements and their  corresponding error classes, and a  new query   $Q_*$, predict its error class  $y^e_*$. 

\item {\bf Query CPU time prediction.} Given a  query workload $\mc{W}=\{(Q_i,y_i^c)\}_{i=1}^n$ comprised of the query statements and their  corresponding CPU times,  and a  new query   $Q_*$, predict its  total CPU time  $y^c_*$. 

\item {\bf Query answer size prediction.}  Given a  query workload $\mc{W}=\{(Q_i,y_i^a)\}_{i=1}^n$ comprised of queries  and answer sizes, and a  new query   $Q_*$, predict its answer size  $y^a_*$ .

\item {\bf Query session  classification.} Given a  query workload $\mc{W}=\{(Q_i,y_i^s)\}_{i=1}^n$ comprised of  queries  and the class of clients that generated each query, and a  new query   $Q_*$,  predict the  class of client $y^s_*$  that generated $Q_*$.
\end{itemize}\vspace{-4.5mm}
\end{defin}
}

\begin{defin} [Query Facilitation Problems] \label{df:problems} Given a query workload $\mc{W}=\{(Q_i,$ $y_i)\}_{i=1}^n$ and a query $Q_*$, {\em a query facilitating problem} is to predict the label $y_*$ of $Q_*$. We define four query facilitating problems depending on the label: \textit{error classification problem}, \textit{CPU time prediction problem}, \textit{answer size prediction problem}, and \textit{session classification problem}, where the label corresponds to either the error classes, CPU times, answer sizes, or the session classes.\end{defin}

The underlying assumption in Definition~\ref{df:problems} is that  $Q_*$  and $\mc{W}$ have similar execution conditions, e.g., run over the same database instance. However, this restriction does not hold in many real applications. For example, cloud-based, multi-tenant, and multi-database platforms  receive millions of queries from end users based on hundreds of schemas~\cite{jain2018database}. It is vital to consider such settings and  develop  models  that  generalize well to unseen queries.
\iffullpaperComposition Therefore, we relax this assumption and define the following settings.
\else Formally, we define:
\fi

\begin{defin}[Query Facilitation Problem Settings]
\label{df:settings}
 Given a query workload $\mc{W}$ and a new query $Q_*$, the problems in Definition~\ref{df:problems} can be studied under the following settings:
\begin{enumerate}[leftmargin=*]
\item {\em \settingOne:} $Q_*$ and  $\mc{W}$ are posed to the same database instance.

\item {\em \settingTwo:} $Q_*$ and $\mc{W}$ are posed to different database instances with the same schema.

\item {\em \settingThree:} $Q_*$ and  $\mc{W}$ are posed to different databases with different schemas.
\end{enumerate}\vspace{-4.99mm}\end{defin}

In this definition, we assume $\mc{W}$ and $Q_*$ are executed in the same DBMS.  However, the definition can be extended to include settings that vary  w.r.t. other execution conditions for $\mc{W}$ and $Q_*$, e.g.,  their  SQL version or  their DBMSs. Moreover, as the  problem setting heterogeneity increases, the prediction problem becomes more challenging. Our empirical study in Section~\ref{sec:results} confirms that while the prediction error of all models increases with increasing  problem heterogeneity,  some models can generalize better across settings.

%% file: sql_example_table_short.tex
\begin{figure}

\begin{lrbox}{\mylistingbox}%
\begin{minipage}{1\linewidth}%
\iffullpaperComposition
\begin{lstlisting}
SELECT * FROM PhotoTag WHERE objId=0x112d075f80360018

\end{lstlisting}
\else
\begin{lstlisting}
SELECT * FROM PhotoTag WHERE objId=0x112d075f80360018

\end{lstlisting}
\fi
\end{minipage}%
\end{lrbox}%
\subfloat[Error class: \scss,  session class: \bott,  answer size: 1,  and CPU time: 0.015] 
{ \usebox{\mylistingbox}\label{dbrec:fig_Q1}}

\begin{lrbox}{\mylistingbox}%
\begin{minipage}{1\linewidth}%
\iffullpaperComposition
\begin{lstlisting}
SELECT p.objid,p.ra,p.dec,p.u,p.g,p.r,p.i,p.z 
FROM PhotoObj AS p 
WHERE type=6 
  AND p.ra BETWEEN (156.519031-0.200000) AND (156.519031+0.200000) 
  AND p.dec BETWEEN (62.835405-0.200000) AND (62.835405+0.200000) 
  ORDER BY p.objid
\end{lstlisting}
\else
\begin{lstlisting}
SELECT p.objid,p.ra,p.dec,p.u,p.g,p.r,p.i,p.z 
FROM PhotoObj AS p 
WHERE type=6 AND p.ra BETWEEN (156.519-0.2) AND (156.519+0.2) AND
      p.dec BETWEEN (62.835-0.2) AND (62.835+0.2) 
ORDER BY p.objid
\end{lstlisting}
\fi
\end{minipage}%
\end{lrbox}%
\subfloat[Error class:  \scss,  session class: \browser,  answer size: 98877, and CPU time: 41.342999]
{ \usebox{\mylistingbox}\label{dbrec:fig_Q20}}
\vspace{-2mm}
\caption{Example SQL queries in SDSS workload.}
\label{tab:example_sql_queries}
\end{figure}

%% file: DatasetAndAnalysis.tex
\section{Workloads and Analysis}
\label{sec:dbrec_dataset_and_analysis}

We   describe  our workloads in Sections~\ref{subsec:dbrec_sdss_dataset} and~\ref{subsec:dbrec_sqlshare_dataset}, analyze them  in Section~\ref{subsec:dbrec_analysis}, and   summarize the implications of our analysis on  model selection and  evaluation, in Section~\ref{subsec:dbrec_analysis_implications}.


\input{Datasets}
\input{Analysis}

%% file: Datasets.tex
\subsection{SDSS Workload}
\label{subsec:dbrec_sdss_dataset}

The SDSS dataset contains logs of queries and requests submitted to SDSS servers. It is  described in~\cite{raddick2014ten1}, which we briefly summarize here.
\iffullpaperComposition
A \textit{hit} is defined as a SQL query or web request.  A \textit{session} is defined as an ordered sequence of hits from a single IP address, such that the gaps between hits in the sequence is no longer than 30 minutes~\cite{szalay2002sdss, raddick2014ten1}.
\fi
For hits, logged data includes the submitted query statement, the version of the database that was queried,  the IP address of the computer that generated the hit, the web agent string which  specifies the software system that generated the hit, and a time stamp for the hit~\cite{raddick2014ten1}.
In the SDSS schema, hits are recorded in  the ``SqlLog''  and  ``Weblog'' tables, while   session information is recorded in the ``Session'' and ``SessionLog'' tables.  Additional tables  record auxiliary information about the hits. The ``SqlLog'' table  contains around 194 million SQL \textit{query log entries} that are grouped into approximately 1.6 million sessions. We extracted the following information from the SDSS dataset:

\begin{itemize}[leftmargin=*]
\item  The raw  \textbf{query statement},   extracted from  the ``SqlStatement.statement'' column.  This statement can range from a correct SQL statement to random text. 

\item The \textbf{query CPU time} label,    extracted from the ``SqlLog.busy'' column. This value is a real number and represents the query CPU time in seconds~\cite{o2005batch}.

\item The \textbf{query answer size} label, extracted from the  ``SqlLog.rows'' column. This value is an integer and represents the number of rows retrieved for the query.

\item The  \textbf{query error class} label,  extracted from  the ``SqlLog.error''  column.  \iffullpaperComposition The error class indicates  whether the query successfully executed, had a severe error, or a non-severe error.
The schema on the SDSS website (also in Appendix~\ref{appendix:sdss_schema}), assigns  type ``int'' for this attribute,  with the definition ``0 if ok, otherwise the sql error \#; negative numbers are generated by the procedure''. In the workload, however, we found  three values, or classes,  for the ``SqlLog.error'' attribute.
\fi
The three error classes include  \scss  (the numeric value $0$ means that the query successfully executed),  \nsevere error (the numeric value $1$),  and   \severe  error (the numeric value $-1$,   indicates an invalid query  that was rejected by the web portal and was not submitted to the database server).

\item The  \textbf{query session class} label is extracted through a series of joins on the following tables in the SDSS schema:  WebAgentString, AgentStringID, WebAgent, WebLog, SessionLog, Session, and SqlLog \iffullpaperComposition(details in Appendix~\ref{appendix:sdss_data_preprocessing}). \else(details in~\cite{t-report}). \fi  
The seven session classes are \nowebh (the session is not established through the Web), \unknown (the session is established through the Web but no agent string is reported), \bott (e.g.,~search engine crawler), \admin (administrative service, e.g.,~performance monitor), \program (a user program, e.g.,~data downloader), \browser (a web browser).
\end{itemize}

The large size of the SDSS dataset (including 194 million query logs in the SqlLog table) poses a computational challenge in developing machine learning models. In addition, the SDSS dataset has data redundancy~\cite{singh2007skyserver,jain2016sqlshare}. The first type of redundancy is because many sessions can contain thousands of  query logs with the same template for their query statements, e.g., bot sessions or administrative sessions typically submit the same query template but with different constants. The second type of redundancy is  caused  when the same query statement  appears in different  query logs,  with varying values for properties like session class, error class, and answer size. This is because the same statement can be submitted in different sessions, via different access interfaces, and against different versions of the database. 

To resolve the redundancy and  size issues, we extract a workload  by sampling a subset of the SDSS dataset. For the first redundancy issue we randomly sample a SQL query log from each session to ensure a large and diverse subset (the input of our  problems  is a raw   query statement  and is independent of other queries in the same session). The result contains 1,563,386 query logs. For the second redundancy issue, we group query logs with the same query statement. We found 18.5\% of the query statements
appear in more than one query log \iffullpaperComposition
(see Figure~\ref{fig:statement_rep_info} in Appendix~\ref{appendix:repetition_query_statement}).
\else
(see~\cite{t-report}).
\fi
Therefore, we aggregate their meta-data labels. In particular, for answer size and CPU time we use the average of these values as the label. For session class, and error class, we use the majority class as the label (with ties broken randomly). Our final query workload contains 618,053 unique query statements.
\iffullpaperComposition
 Details are in Appendix~\ref{appendix:sdss_schema}.
 \else
Details are in~\cite{t-report}.
\fi

\subsection{SQLShare Workload}
\label{subsec:dbrec_sqlshare_dataset}
The SQLShare  query workload~\cite{jain2016sqlshare} is the result of a multi-year deployment of a database-as-a-service platform, where users upload their data, write queries, and share their results. This workload  represents short-term, ad-hoc analytics over user-uploaded datasets. We use the SQLShare workload in our work  and extracted the following information:
\begin{itemize}[leftmargin=*]

\item The raw  \textbf{query statement},   extracted from  the ``Query'' column. This may be a syntactically incorrect SQL query.  

\item The \textbf{query CPU time} label,    extracted from the ``QExecTime'' column. This value is an integer and represents the query CPU time in  seconds.
\end{itemize} 

%% file: Analysis.tex
\subsection{Workload Analysis}
\label{subsec:dbrec_analysis}


\ifPhDThesis
\input{AnalysisFiguresPhDThesis}

\else
\input{AnalysisFigures}
\fi

\subsubsection{Query Statement Analysis}
\label{subsec:dbrec_statement_analysis}
We analyze the query statement properties to understand the type of queries posed and their syntactic properties and statistics. 
Regarding the query statement types, \Sqlkeyword{SELECT} statements comprise  approximately 96.5\%  and  approximately 98\% of  statements on SDSS and SQLShare, respectively. The remaining 3.36\%  (21540)  and  2.02\% (544) of statements on SDSS and SQLShare, correspond to types   such as \Sqlkeyword{EXECUTE}, \Sqlkeyword{CREATE}, \Sqlkeyword{DROP}, \Sqlkeyword{UPDATE}, \Sqlkeyword{ALTER}, and various combinations like~\Sqlkeyword{DELETE}\;|\;\Sqlkeyword{UPDATE}\;|\;\Sqlkeyword{INSERT}.



We used the ANTLR parser~\cite{parr2013definitive} to generate the Abstract Syntax Trees (AST) of query statements   and extract 10  syntactic properties:
\iffullpaperComposition
\begin{enumerate}
\item \clevel: the number of characters in a query.
\item \wlevel: the number of words in a query (digits are replaced with the $<$DIGIT$>$ token).
\item \fcount: the number of function calls.
\item \jcount: the number of join operators.
\item \tcount: the number of unique table names in the query.
\item \sccount: the number of selected columns in the query.
\item \pcount: the number of predicates (logical conditions, e.g.,~\Label{s.flags\_s=0}) used in a query.
\item \pccount: the number of table names  in the predicates.
\item \ncount: the level of nestedness. 
\item \nacount: it is true if nested queries involve aggregation and false otherwise.
\end{enumerate}

\else
\begin{enumerate*}
\item the number of characters in a query,
\item the number of words in a query (digits are replaced with the $<$DIGIT$>$ token),
\item the number of function calls,
\item the number of join operators,
\item the number of unique table names in the query,
\item the number of selected columns in the query,
\item the number of predicates (logical conditions, e.g.,~\Label{s.flags\_s=0}) used in a query,
\item the number of table names  in the predicates,
\item the query nestedness level, and 
\item  a nested aggregation indicator that is true when nested queries involve aggregation.
\end{enumerate*}
\fi

\iffullpaperComposition
\begin{example} The query in Figure~\ref{fig:syntax} has the following syntactic properties:
\begin{enumerate}
    \item \fcount\!=$2$ (\Label{dbo.fGetURLExpid} and \Sqlkeyword{min}),
    \item \tcount\!=$2$ (\Label{SpecPhoto} and \Label{PhotoObj}),
    \item \sccount\!=$3$ (\Label{objid}, \Label{modelmag\_u}, \Label{modelmag\_g}),
    \item \pcount\!=$5$ (1 in the main query and $4$ in the sub-query including the predicate for the inner join operator),
    \item \pccount\!=$7$ ($7$ logical conditions),
    \item \ncount\!=$1$,
    \item \nacount\!=true (the nestedness involves \Sqlkeyword{min}).
\end{enumerate}
\end{example}

\begin{figure}
    \centering
    \begin{lstlisting}
SELECT dbo.fGetURLExpid(objid)
FROM SpecPhoto
WHERE modelmag_u-modelmag_g =
    (SELECT min(modelmag_u-modelmag_g)
     FROM SpecPhoto AS s INNER JOIN PhotoObj AS p
     ON s.objid=p.objid
     WHERE (s.flags_g=0 OR p.psfmagerr_g<=0.2 AND
     p.psfmagerr_u<=0.2)
\end{lstlisting}
\vspace{-4mm}
    \caption{A sample query from SDSS}
    \label{fig:syntax}
\end{figure}
\fi

Statistics of the syntactic properties of SDSS statements are shown in Figure~\ref{fig:sdss_query_statement_analysis}
\iffullpaperComposition
(cf. Figure~\ref{fig:sqlshare_query_statement_analysis} for SQLShare).
\else
(see~\cite{t-report} for SQLShare).
\fi
Figures~\ref{fig:sdss_analysis_char_level} and ~\ref{fig:sdss_analysis_word_level} plot the distribution of characters and words  for SDSS.
 \iffullpaperComposition The maximum number of characters and words  is 7,795 and 2,975  (5164 and 28227 for SQLShare),  respectively.
 \else
 The maximum number of characters and words  is 7,795 and 2,975, respectively.
 \fi
  Around $30\%$ of the queries have more than 62 characters, and more than 224 words (whcih are the corresponding distribution means). Figures~\ref{fig:sdss_function_count}-~\ref{fig:sdss_nested_count} report key structural metrics such as the \jcount and \pcount  \iffullpaperComposition
  for SDSS (cf. Figures~\ref{fig:sqlshare_function_count}-~\ref{fig:sqlshare_nested_count} for SQLShare)
\else
for SDSS.
  \fi
 Approximately  $5.91\%$ ($1.68\%$) of the queries in SDSS (SQLShare) have at least one join operator,
$14.01\%$ ($29.74\%$) of the queries in SDSS (SQLShare) access more than one table,
$0.34\%$ ($7.88\%$) of the queries in SDSS (SQLShare) are nested queries, and $0.03\%$ ($0.71\%$) are nested queries with aggregation. Note that the small percentage of nested queries still corresponds to a considerable number of queries ($2,112$ for SDSS and $2,107$ for SQLShare).

Our analysis of the syntax of queries in SDSS and SQLShare shows that these workloads have queries of various complexity w.r.t. the syntactic properties that we studied. Comparing the syntactic properties of the statements in SDSS with those in SQLShare, we observe that  while queries are typically longer in SQLShare, the mean number of predicates in the where clause  for SDSS is approximately four times that of
 \iffullpaperComposition SQLShare (Figure~\ref{fig:sqlshare_predicate_count} vs.~Figure~\ref{fig:sqlshare_predicate_count}).
\else
 SQLShare.
 \fi  Although SQLShare queries  access more tables on
\iffullpaperComposition average (as depicted by a higher mean value and maximum in Figure~\ref{fig:sdss_table_count} vs.~Figure~\ref{fig:sqlshare_table_count}),
\else
average,
\fi
 SDSS queries perform  more  joins on
\iffullpaperComposition average (as depicted by a higher mean value and maximum  Figure~\ref{fig:sdss_join_count} vs.~Figure~\ref{fig:sqlshare_join_count}).
\else
average.
\fi Finally, SQLShare's queries are more complex in both nestedness and aggregation.  




\ifPhDThesis
\input{LabelDistribtionFiguresPhDThesis}
\else
\input{LabelDistribtionFigures}
\fi


\subsubsection{Label Analysis}
\label{subsec:dbrec_label_analysis}
  Figures~\ref{fig:analysis_error} and~\ref{fig:analysis_session_class} show the label distributions of the classification problems for SDSS. As shown in Figure~\ref{fig:analysis_error}, the error classes  are imbalanced; 97.22\% of the queries ran without an error (\scss), while  1.93\% had
  \nsevere\ errors, and  0.85\% had \severe\ errors. Figure~\ref{fig:analysis_session_class} shows that session classes are also imbalanced, e.g., \program 
  and \bott 
  comprise 7.93\% and 25.98\% of the workload, respectively. 
  Note, a simple model that only predicts the majority class  (e.g. \scss in error classification) will achieve a high accuracy. We address this issue in our evaluations by separately calculating the per-class F-measure.




Figures~\ref{fig:analysis_rows}-\ref{fig:sqlshare_analysis_busy} show the label distributions for regression problems.
Figure~\ref{fig:analysis_rows} shows SDSS answer size distribution, which ranges from a minimum of -1 (the query did not run due to an error) to a maximum value of 966,278,220 tuples. Despite the wide range of values, the data is concentrated around smaller values with a median of 1, i.e., half of the queries either do not run, return no answer, or return only one answer. Figure~\ref{fig:analysis_busy} shows the CPU time distribution on SDSS. The time ranges between $0$ and $10^8$ seconds with the majority of queries taking little CPU time. Figure~\ref{fig:sqlshare_analysis_busy} shows the CPU time distribution in SQLShare ranges approximately between $0$ and $4*10^6$ seconds. 

 \subsection{Workload Analysis Implications}
\label{subsec:dbrec_analysis_implications}
\subsubsection{Model Selection and Train Loss Functions}
\label{subsec:dbrec_analysis_implications_model_selection}
Our query facilitation problems  in Definition~\ref{df:problems} can broadly be categorized as supervised classification and regression problems.   Text classification in NLP is a closely related area. Traditional NLP models, work in two stages: a feature extraction phase, where  input features are  hand-engineered, and a prediction phase.
 As shown in Figure~\ref{fig:sdss_query_statement_analysis}, queries range in complexity and extracting  an adequate set of features can be challenging.
Neural network  architectures  can learn features automatically. They combine the feature extraction and prediction stages in a joint training task, which  allows them to develop features and representations for the task~\cite{conneau2016very}. LSTMs are a type of recurrent neural network (RNNs),  and are one of the dominant models for text classification. They  treat text as sequential inputs and try to preserve long-term dependencies between tokens.  However, query statements are long (Figures~\ref{fig:sdss_analysis_char_level}  and~\ref{fig:sdss_analysis_word_level}), and this property can negatively affect the performance of LSTMs. As an alternative, we assess CNNs, which are feed-forward networks. Rather than preserving long-term dependencies,  CNNs   automatically identify   local patterns (i.e., n-grams) in the input and preserve them in their feature representations. For  NLP tasks, CNNs are known to be  competitive with several more sophisticated architectures (e.g., LSTMs) and are easier to train and interpret~\cite{bai2018empirical,yin2017comparative}.


Moreover, we observe that SQL queries
\iffullpaperComposition
(and code in general)
\fi
often contain mathematical expressions consisting of numbers and operators. These expressions   significantly affect  query properties like query answer size or CPU time~\cite{ganapathi2009predicting}. It  is beneficial to  retain relevant information in the representation of queries.  However, the set of variable names and digits used in code snippets  is unbounded, and there are many rare words.  For word-levels models, this leads to the unbounded or open vocabulary problem, which creates practical issues when learning representations in machine learning~\cite{kim2016character}.  We apply the models  in  Section~\ref{sec:dbrec_methods},  at  both  character and word level. For the latter, we replace the digits with a $<$DIGIT$>$ token to control for the vocabulary size.

Regarding the train loss functions, we observe that the error class and session class labels are imbalanced (Figures~\ref{fig:analysis_error} and \ref{fig:analysis_session_class}). For some applications,  like bot detection,  models that perform accurately on certain classes may be required.  Typically,  application-dependant assumptions are enforced by either re-sampling the data, or using weighted loss functions  during training of models. Because our work does not focus on specific applications (e.g.,~bot detection),  we treat all classes equally and use an unweighted cross entropy loss function for training the classification models in Section~\ref{sec:dbrec_methods}. Our evaluation (Section~\ref{sec:dbrec_emp_eval}), however,  considers this class imbalance where we analyze performance w.r.t.~each class.  

The regression labels have a wide range of values and are highly skewed, with the majority of queries concentrated around small values  (Figures~\ref{fig:analysis_rows}-~\ref{fig:sqlshare_analysis_busy}). To prevent the models from being too sensitive to  queries with a large label value (outliers), we perform two steps.
We  apply a logarithmic transformation  to the values  of these labels $y'_{i} = \ln(y_{i} + \epsilon- \min(\bm{y}))\ignore{\label{dbrec:log_tranform}}$, where $y_i$ is the label of query $i$, and $\bm{y}$ is a vector representing the labels (answer size or CPU time) of all queries, and $y'_i$ is the log-transformed value. When $y_{i}=\min(\bm{y})$, $\epsilon >0$ prevents the input of the $\ln(.)$ function from being zero.    We set $\epsilon=1$ to make the transformation non-negative.   We use the log-transformed values of CPU time and answer size as the labels of queries  in regression problems. Moreover,  to ensure that the regression models are robust to outliers in the data, during training  we use the well-known Huber loss~\cite{huber1964robust}, which is a hybrid between $l_2$-norm for small residuals and $l_1$-norm for large residuals.

\subsubsection{Model Evaluation}
\label{subsec:dbrec_statement_complexity}

\iffullpaperComposition
\begin{figure}
\centering

        \subfloat[SDSS]
        {
		\includegraphics[width=1\linewidth]{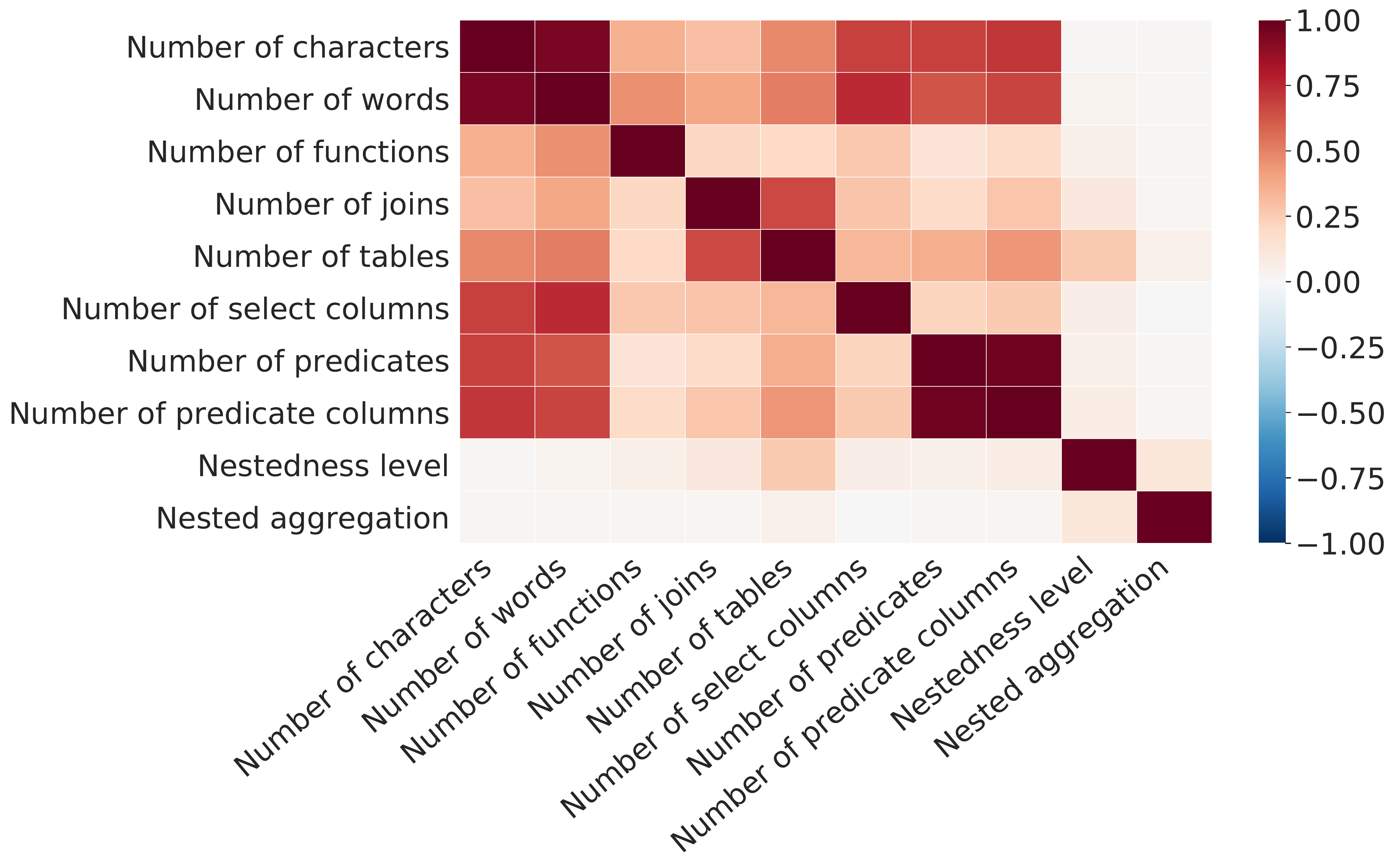}
		\label{fig:sdss_heatmap}
        }

         \subfloat[SQLShare]
        {
		\includegraphics[width=1\linewidth]{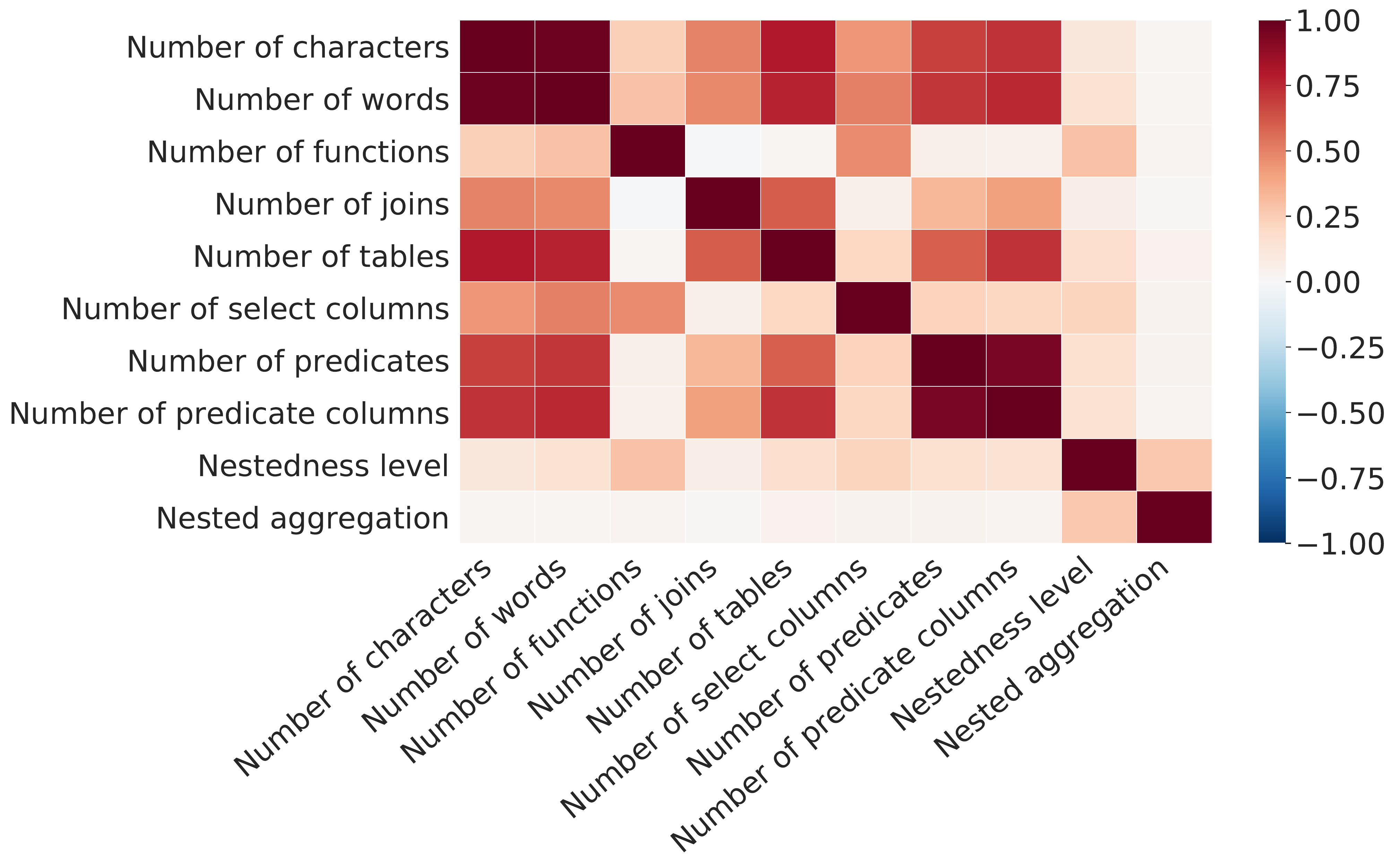}
		\label{fig:sqlshare_heatmap}
        }

\caption{Correlation matrix of strutural properties in Figures~\ref{fig:sdss_query_statement_analysis} and Figures~\ref{fig:sqlshare_query_statement_analysis}.
}
\label{fig:heatmap_fig}
\end{figure}
\else
\begin{figure}
\centering
\includegraphics[width=1\linewidth]{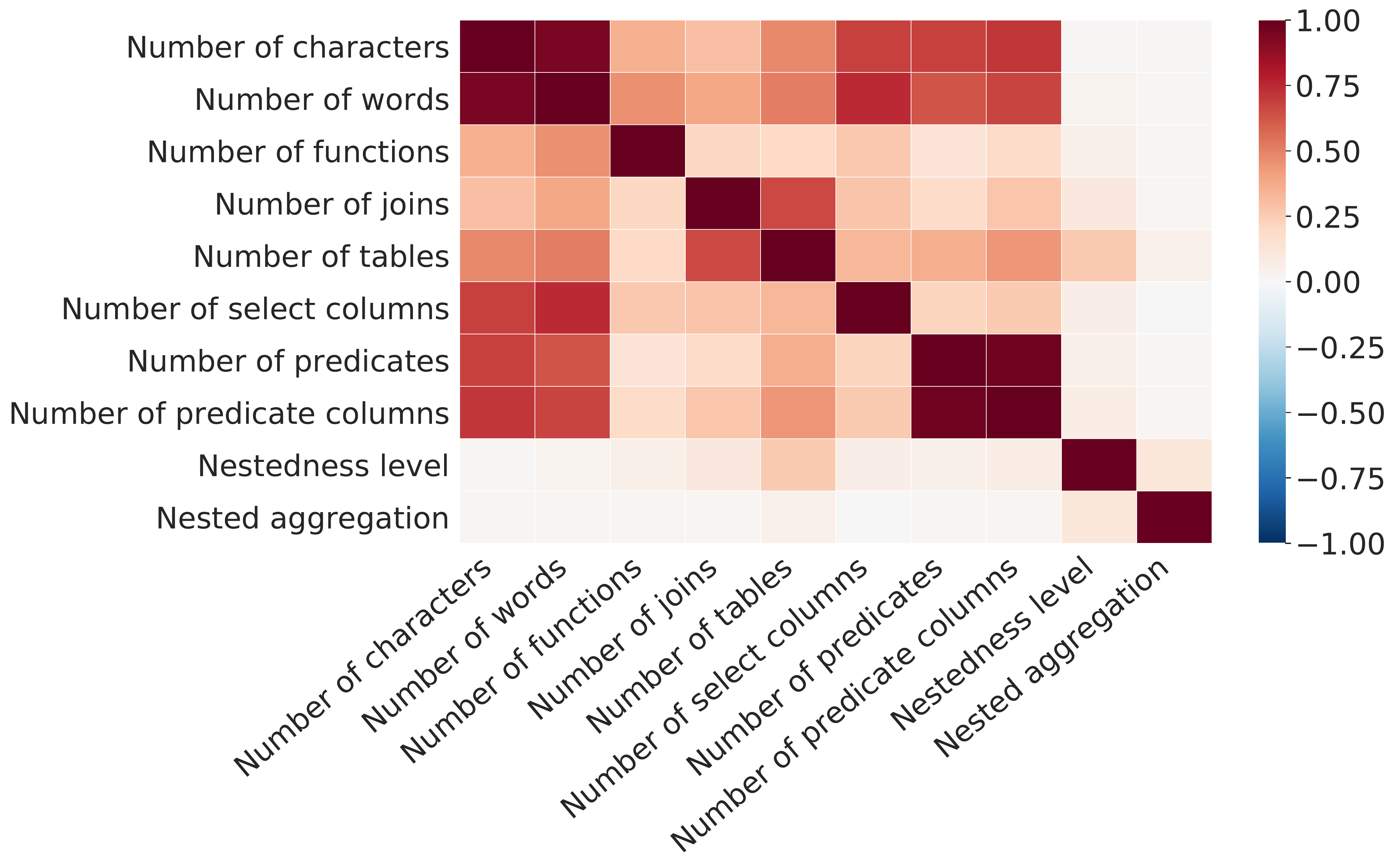}
\label{fig:sdss_heatmap}
\caption{SDSS structural property correlations.}  
\label{fig:heatmap_fig}
\end{figure}
\fi

In our work, we want to help end users write queries. This  is particularly important in settings where  query statements are complex and for users who have little experience.  Therefore, we need to assess model  performance on complex query statements. However, statement complexity information is not included explicitly in the data.  To effectively assess feasibility of complex queries, we must define both a  notion of query statement complexity and a proxy measure that captures it.

Similar to~\cite{jain2016sqlshare}, we want a proxy  metric for complexity that reflects the  cognitive effort  required to write  the query  statement.  
Metrics based on  query run time~\cite{ganapathi2009predicting}  are  not   adequate in this context. The   run  time depends on factors like the load of the database or the size of data selected, which are not relevant to the cognitive effort of the user, e.g., a simple query that selects all rows of a large table can have a long running time.
 In~\cite{jain2016sqlshare}, query complexity is defined  in terms of the query's ASCII length and  the number of distinct operators in the query execution plan. However,  we do not assume access to the database instance or execution plan. The ASCII length, on the other hand,  might not be a sufficient proxy for complexity, e.g., when a query has a simple structure with similar operations repeated many times. Additional syntactic properties may be required.

However, it is not clear  which  set of syntactic properties capture a meaningful notion of statement complexity.
Figure~\ref{fig:heatmap_fig} shows the correlation matrix for the syntactic properties in Figure~\ref{fig:sdss_query_statement_analysis} for SDSS. We observe some  properties are positively and linearly correlated with other types of properties, and hence indicative of them.  For example, the \clevel   is linearly correlated with the \wlevel, the \pccount,
\iffullpaperComposition the \pcount, and the \sccount
\else and the \pcount.
\fi
 So the  latter properties are redundant since the \clevel is indicative of them.  But  \clevel is not   positively  correlated with properties like  \nacount and \ncount and may not capture those complexities.  As another example, the \jcount  is only linearly correlated with the \tcount. We observed similar patterns for SQLShare
\iffullpaperComposition
in Figure~\ref{fig:sqlshare_heatmap}.
\else
(see~\cite{t-report}).
\fi


Overall, these observations suggest  that a subset of syntactic properties might be required to capture the full range of potential query complexities. Based on the query statement feature  correlations shown in Figure~\ref{fig:heatmap_fig}, we chose a subset of five syntactic properties for the qualitative analysis in Section~\ref{sec:dbrec_qualatative_analysis}. These include  the \clevel, the \fcount, the \jcount, the \ncount, and the \nacount indicator.

\input{SessionClassBreakDownFigure}

One viable assumption may be that different classes of  users  write queries with different  complexity levels. For example, \bott queries  may use linear  predicates in the where clause, while queries via  \browser may be more complex.   Thus,  session class, if available, can be an indirect proxy  for query complexity.
 \iffullpaperComposition
 We examined the SDSS queries and broke down several of their properties by session class.\fi
Figure~\ref{fig:bsa_row} plots the distribution of the answer size for each session class. The \nowebh and \browser classes have similar distributions, with the latter having slightly smaller values (likely due to the limitations  for queries posed via the web-based interface).
In  Figure~\ref{fig:bsa_busy} the distribution of the CPU time for each session class is shown. Queries in the \nowebh class have a wider range of values.  
Figures~\ref{fig:full_bsa_length_word} and \ref{fig:full_bsa_length_char} show the query size distributions  by session class. Overall, queries from \nowebh and \browser classes have similar distributions at both the character and word level.
These figures suggest that queries in the \nowebh and \browser  class are more  complex. \iffullpaperComposition
The drawback is that   session class information may not always be available, e.g.,  SQLShare workload  does not include it.\fi

\iffiltereddata
\begin{figure}[t]
\centering
        \subfloat[Character level, original statements]
        {
		\includegraphics[width=0.25\textwidth]{results/figures-filtered/original_char_lst.pdf}
		\label{fig:analysis-session_class}
        }
		\subfloat[Word level, space inserted and digits replaced with placeholders]
        {
		\includegraphics[width=0.25\textwidth]{results/figures-filtered/digit_replaced_word_lst.pdf}
		\label{fig:analysis-session_class}
        }

\caption{Query size distributions filtered}
\label{fig:query_size_dist_filtered}
\end{figure}
\fi

%% file: AnalysisFiguresPhDThesis.tex
\begin{figure*}
\centering
        \subfloat[]
        {
		\includegraphics[width=0.32\linewidth]{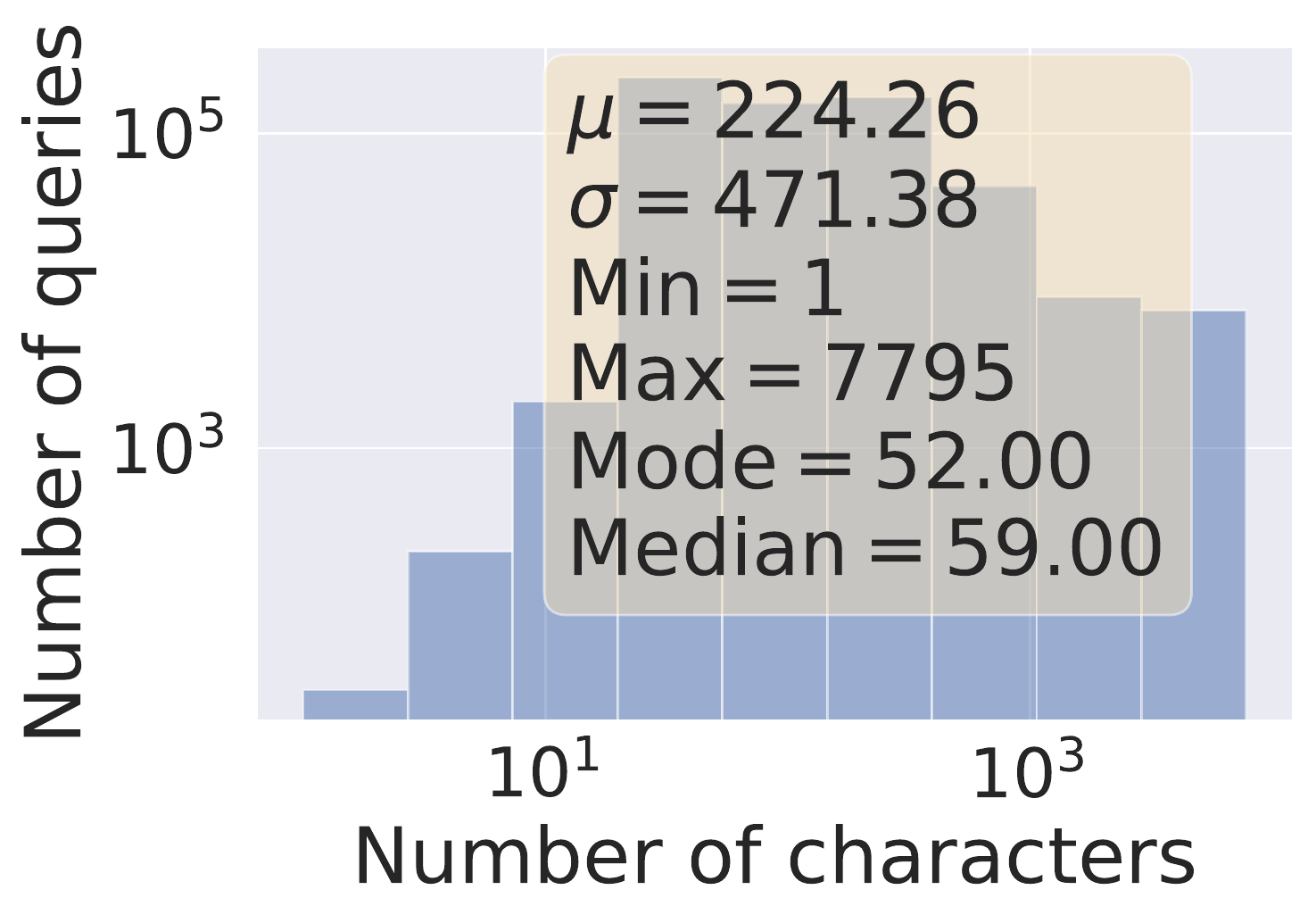} 
		\label{fig:sdss_analysis_char_level}
        }
		\subfloat[]
        {
		\includegraphics[width=0.32\linewidth]{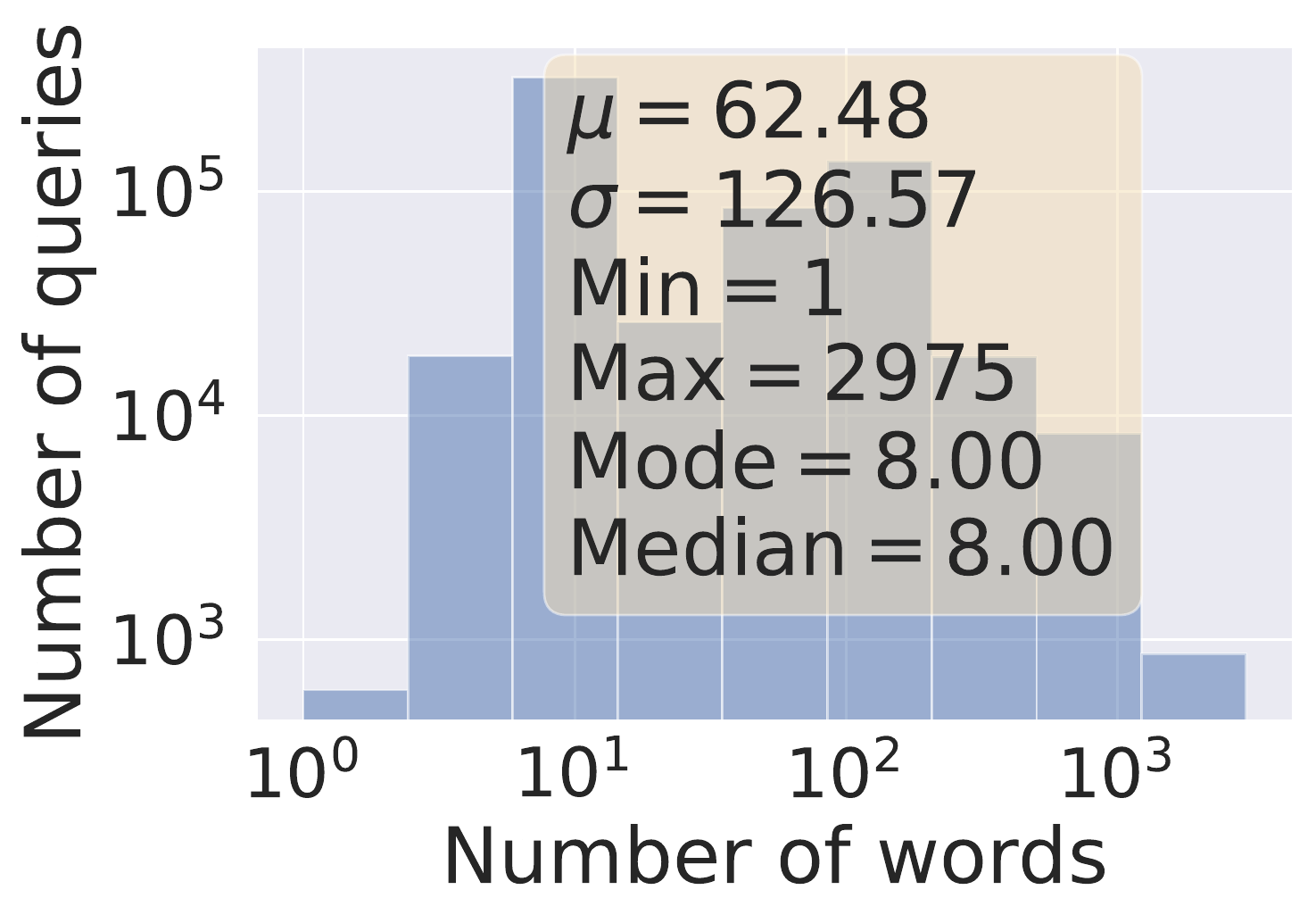} 
		\label{fig:sdss_analysis_word_level}
        } 
        \subfloat[]
        {
		\includegraphics[width=0.32\linewidth]{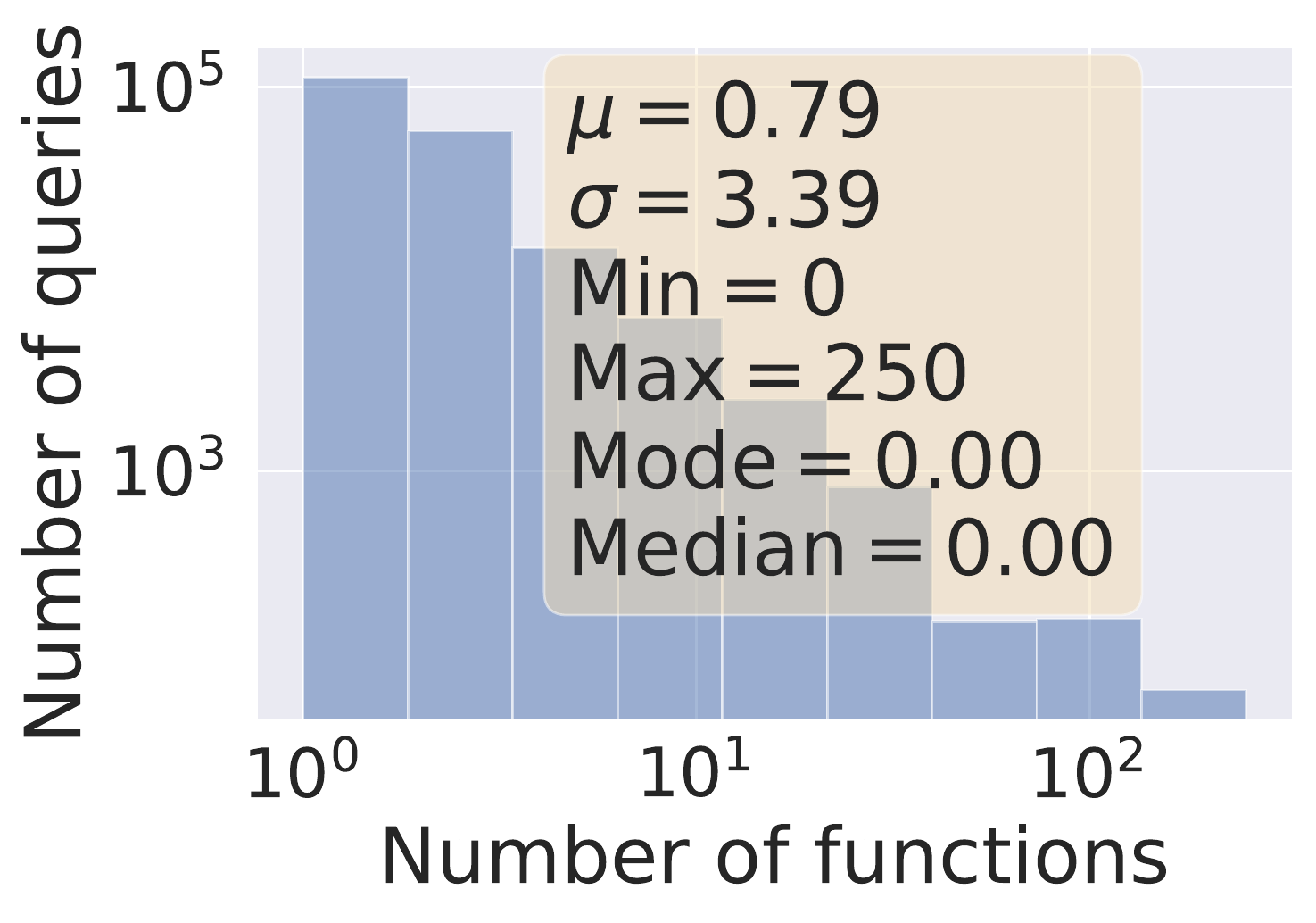} 
		\label{fig:sdss_function_count}
        }   
         
        \subfloat[]
        {
		\includegraphics[width=0.32\linewidth]{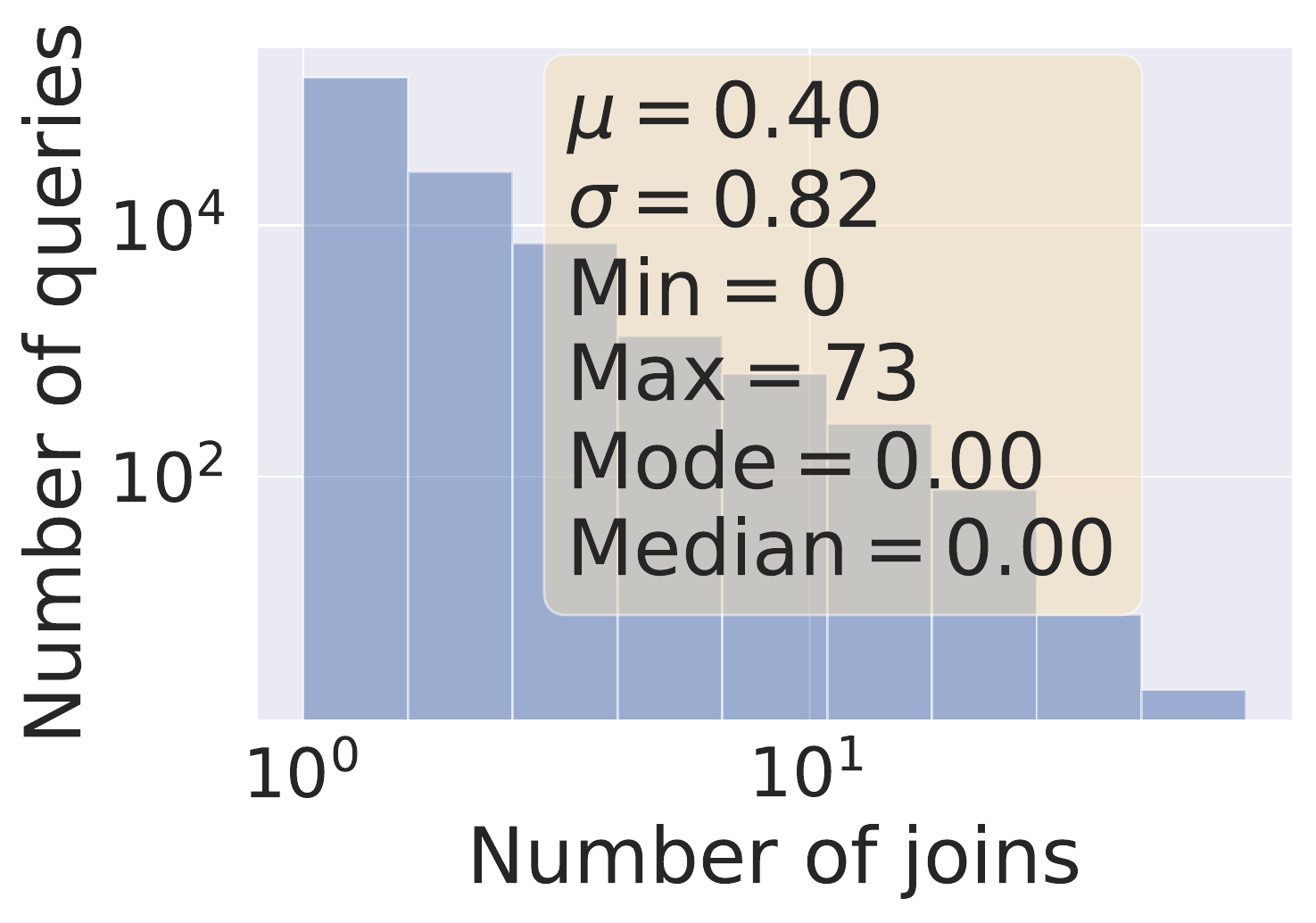} 
		\label{fig:sdss_join_count}
        }     
        \subfloat[]
        {
		\includegraphics[width=0.32\linewidth]{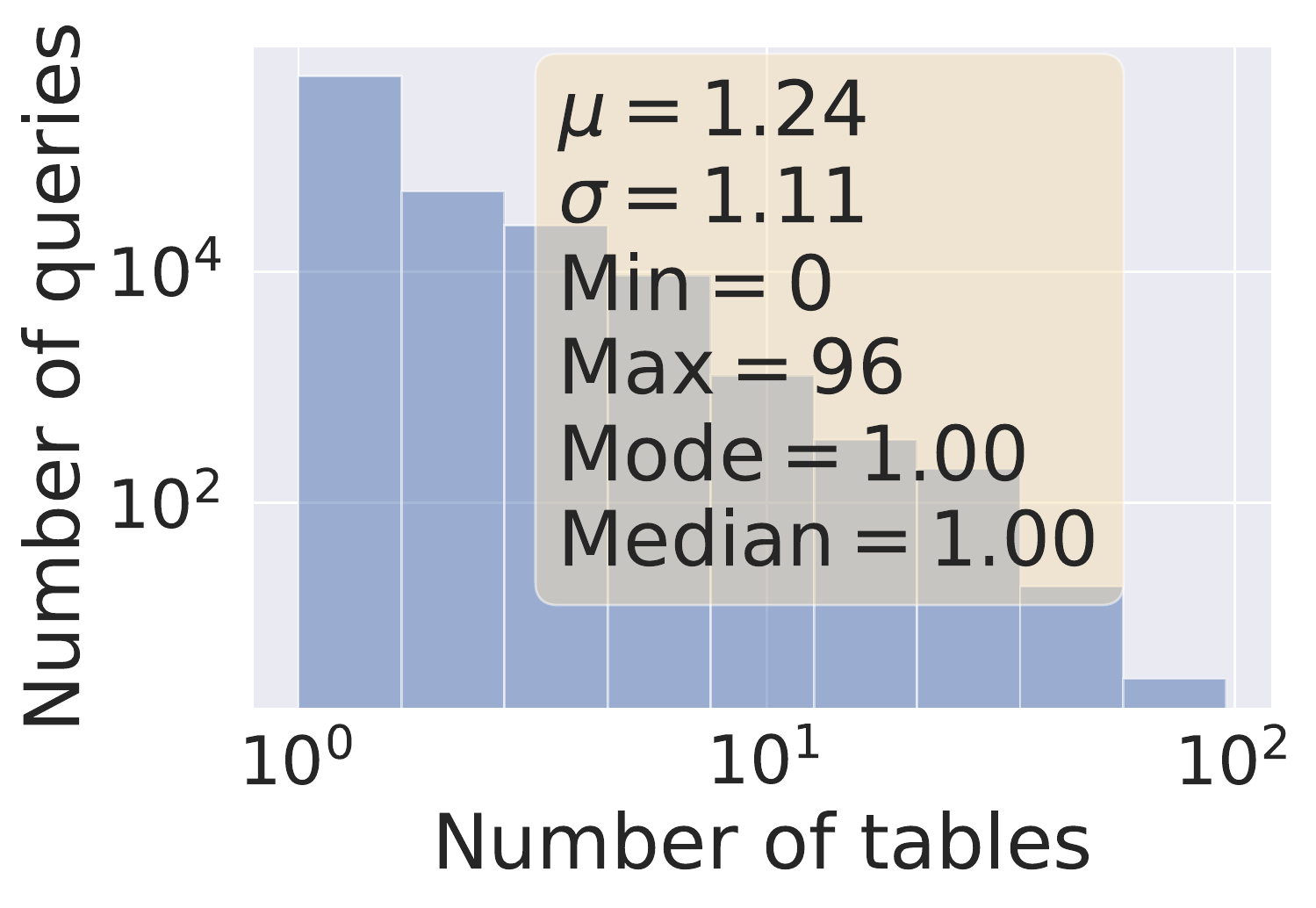} 
		\label{fig:sdss_table_count}
        }
        \subfloat[]
        {
		\includegraphics[width=0.32\linewidth]{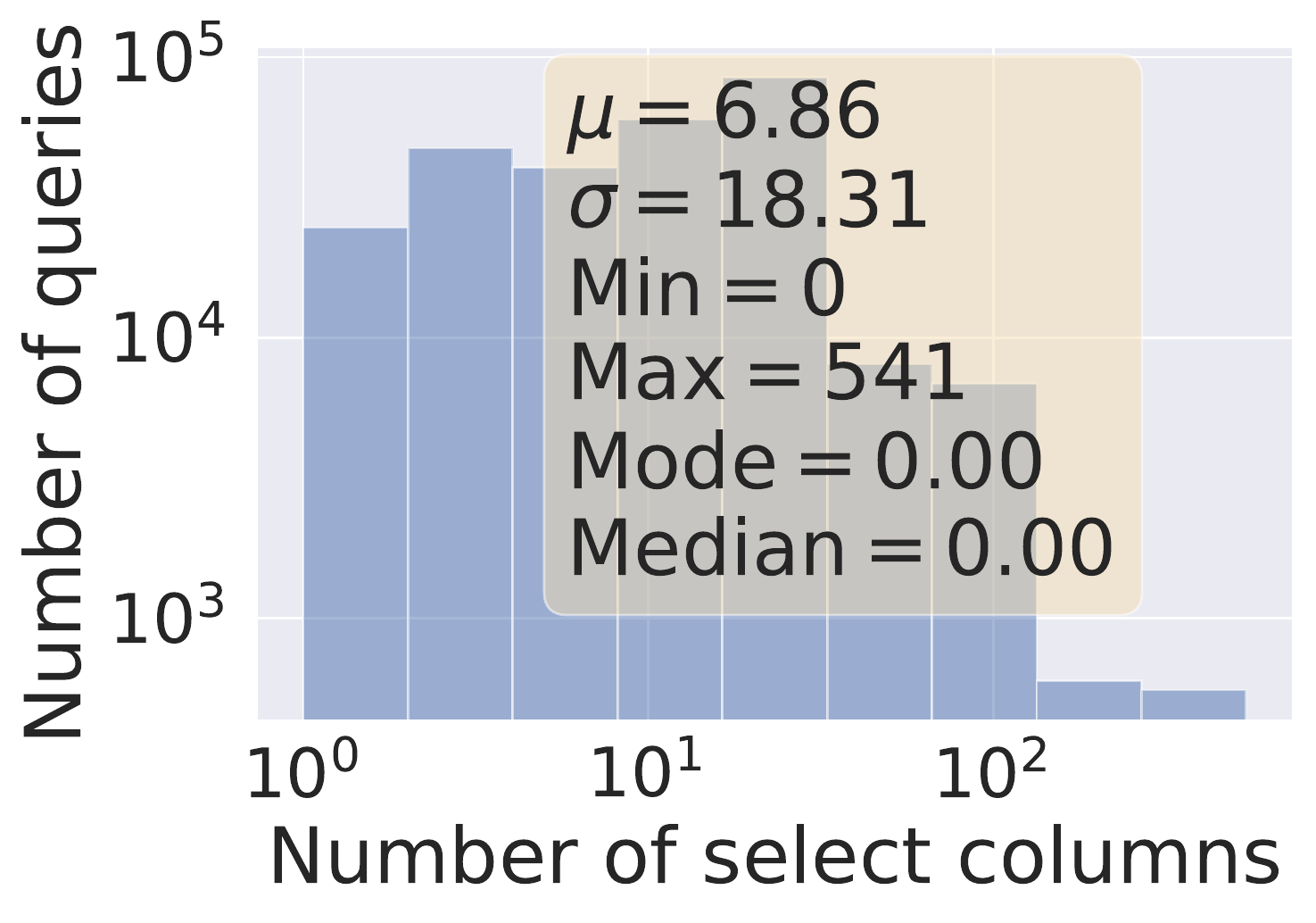} 
		\label{fig:sdss_select_columns}
        }  
        
        \subfloat[]
        {
		\includegraphics[width=0.32\linewidth]{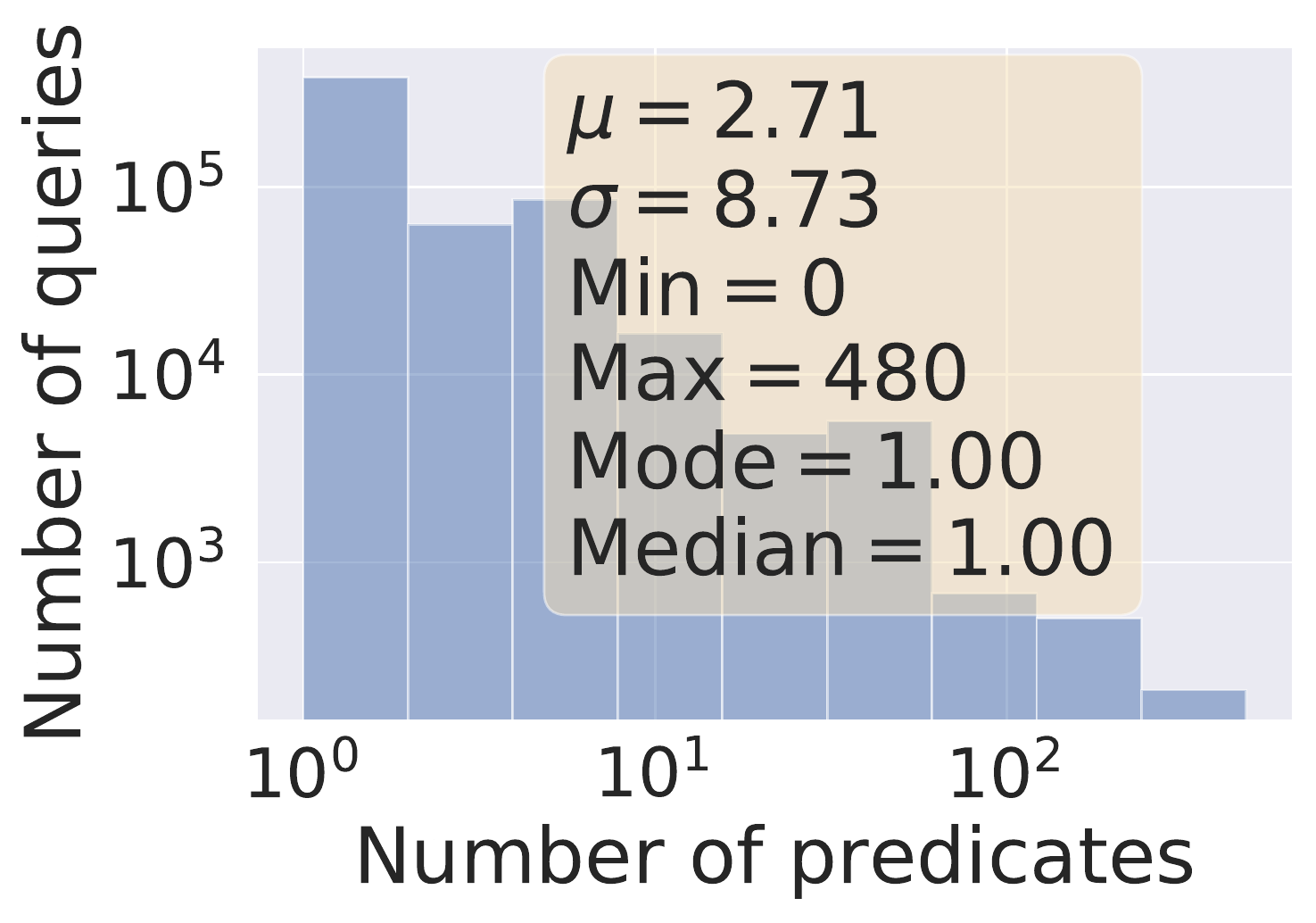} 
		\label{fig:sdss_predicate_count}
        }  
		 \subfloat[]
        {
		\includegraphics[width=0.32\linewidth]{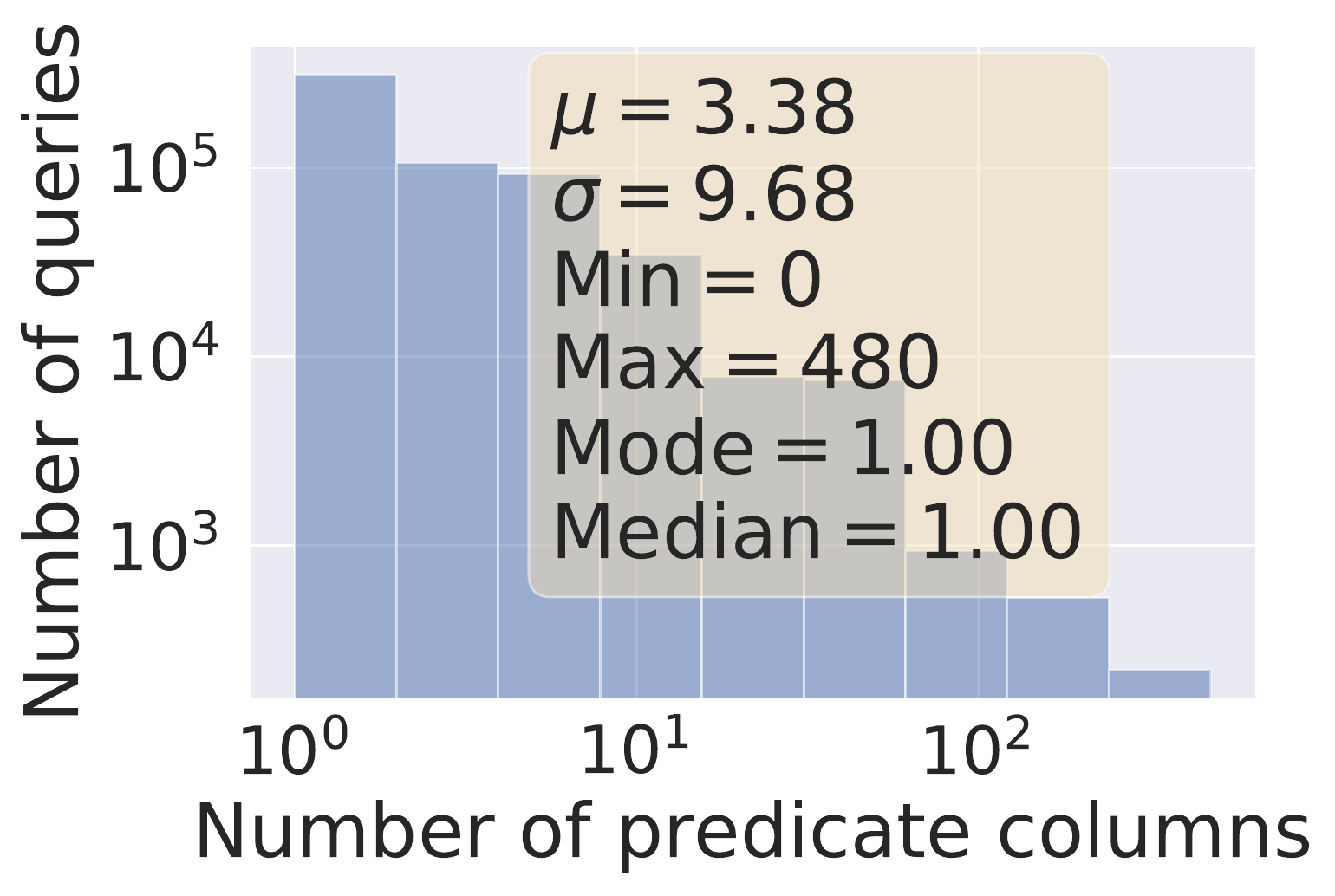} 
		\label{fig:sdss_predicate_column_count}
        }    
        \subfloat[]
        {
		\includegraphics[width=0.32\linewidth]{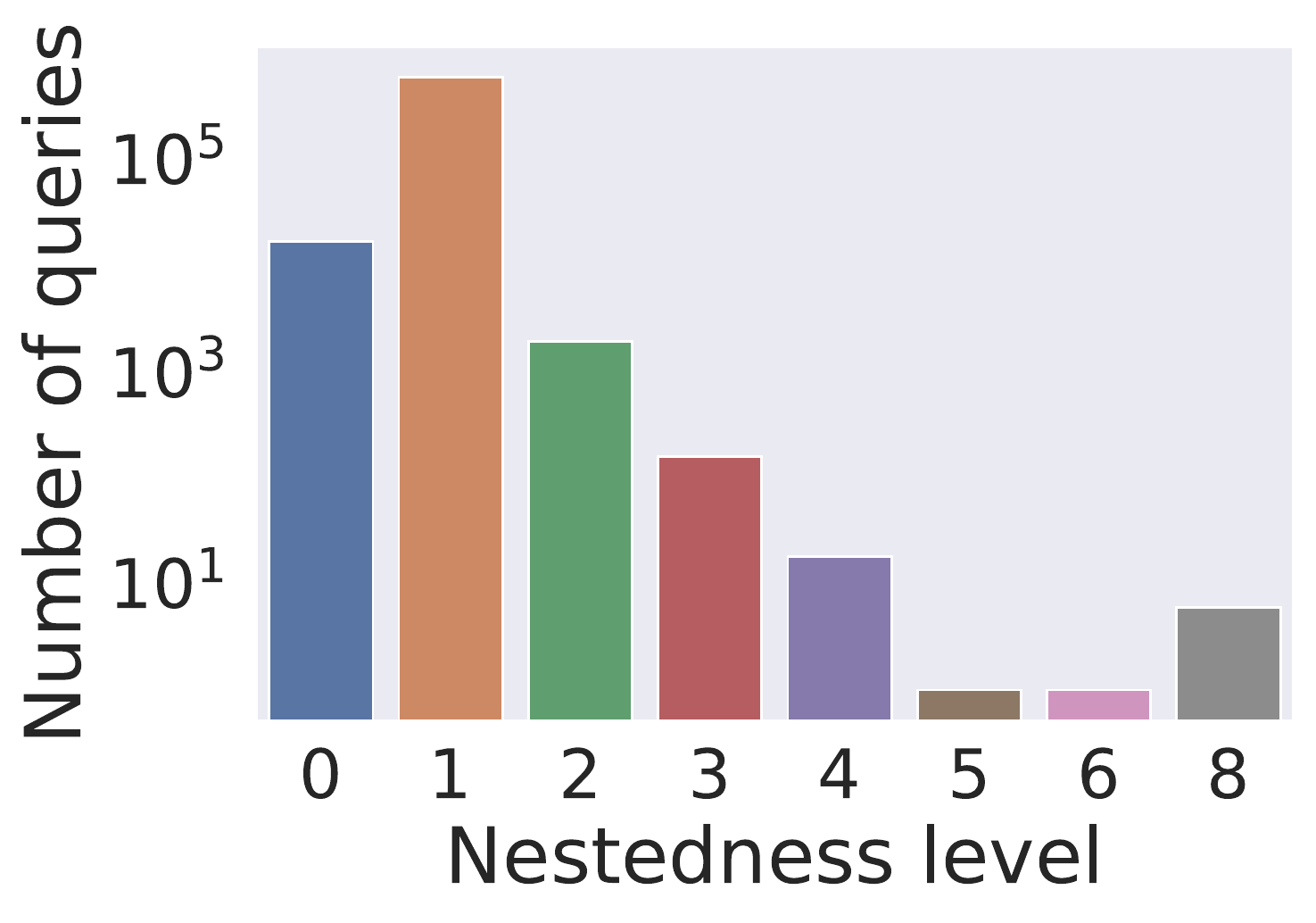} 
		\label{fig:sdss_nested_count}
        }
        
        \subfloat[]
        {
		\includegraphics[width=0.32\linewidth]{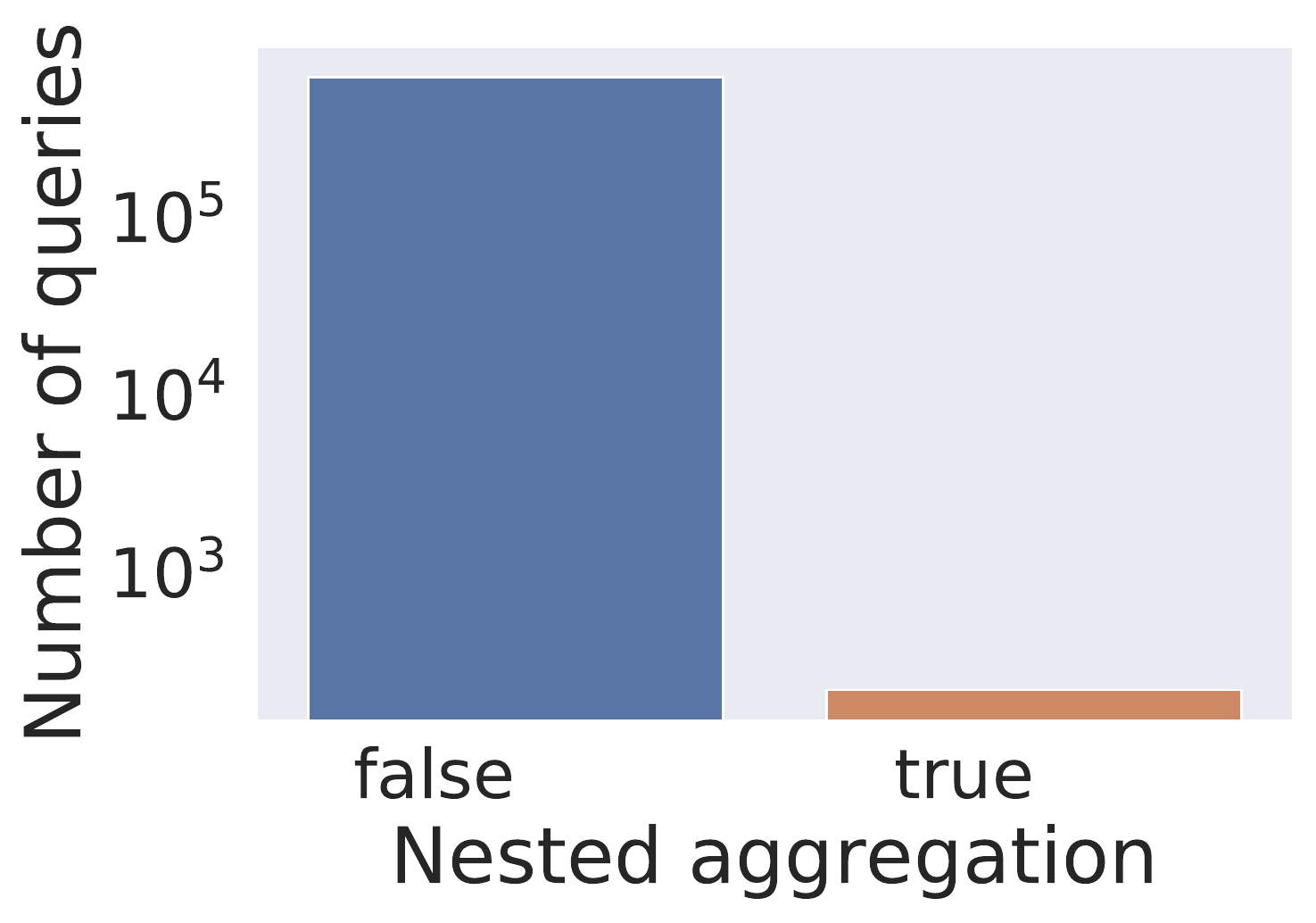} 
		\label{fig:sdss_nested_aggregate}
        }

\caption{Structural properties of SDSS query statements. Most plots are on log-log scale, due to the wide data range. 
}
\label{fig:sdss_query_statement_analysis}
\end{figure*}

\begin{figure*}
\centering
        \subfloat[]
        {
		\includegraphics[width=0.32\linewidth]{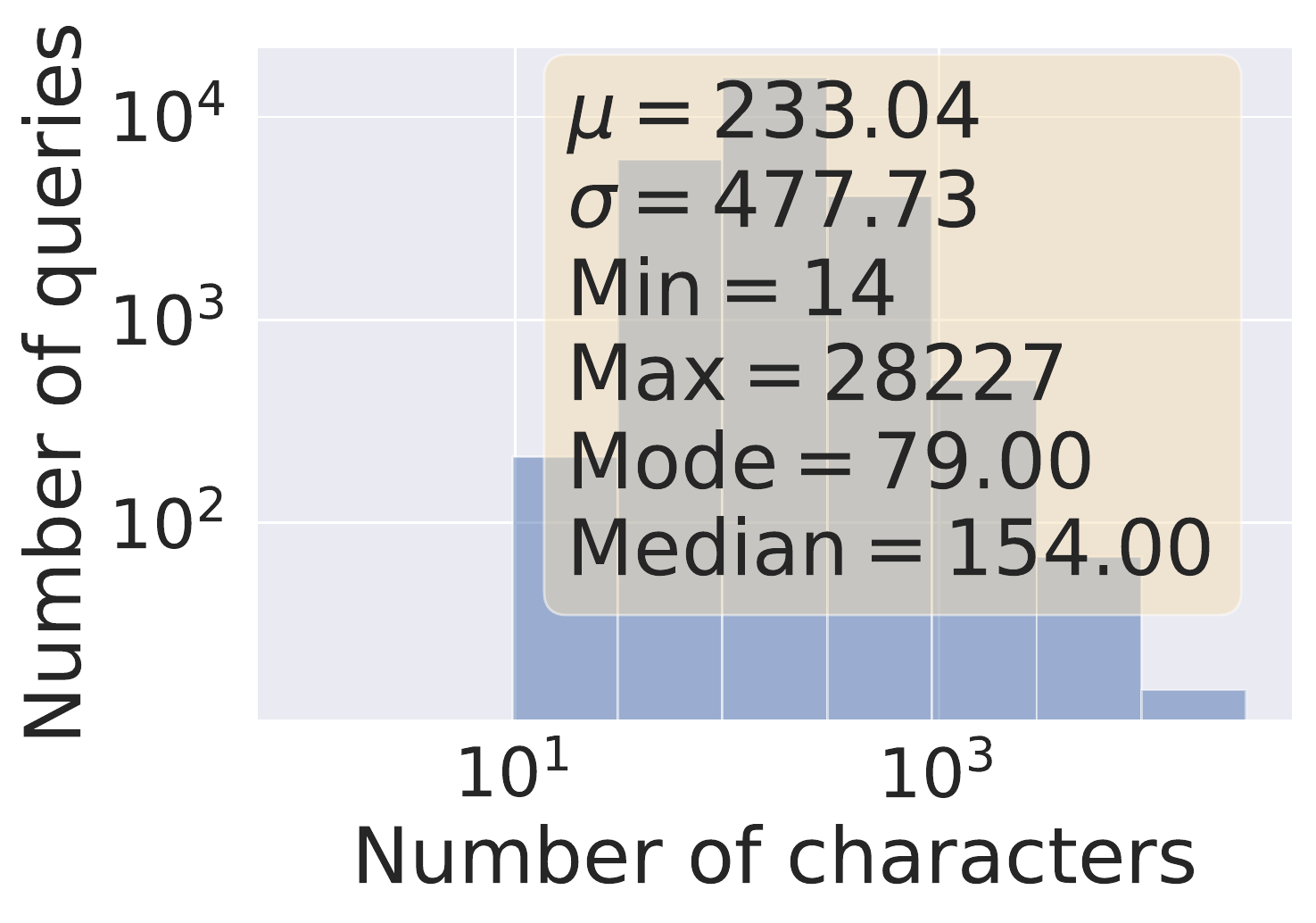} 
		\label{fig:sqlshare_analysis_char_level}
        }
		\subfloat[]
        {
		\includegraphics[width=0.32\linewidth]{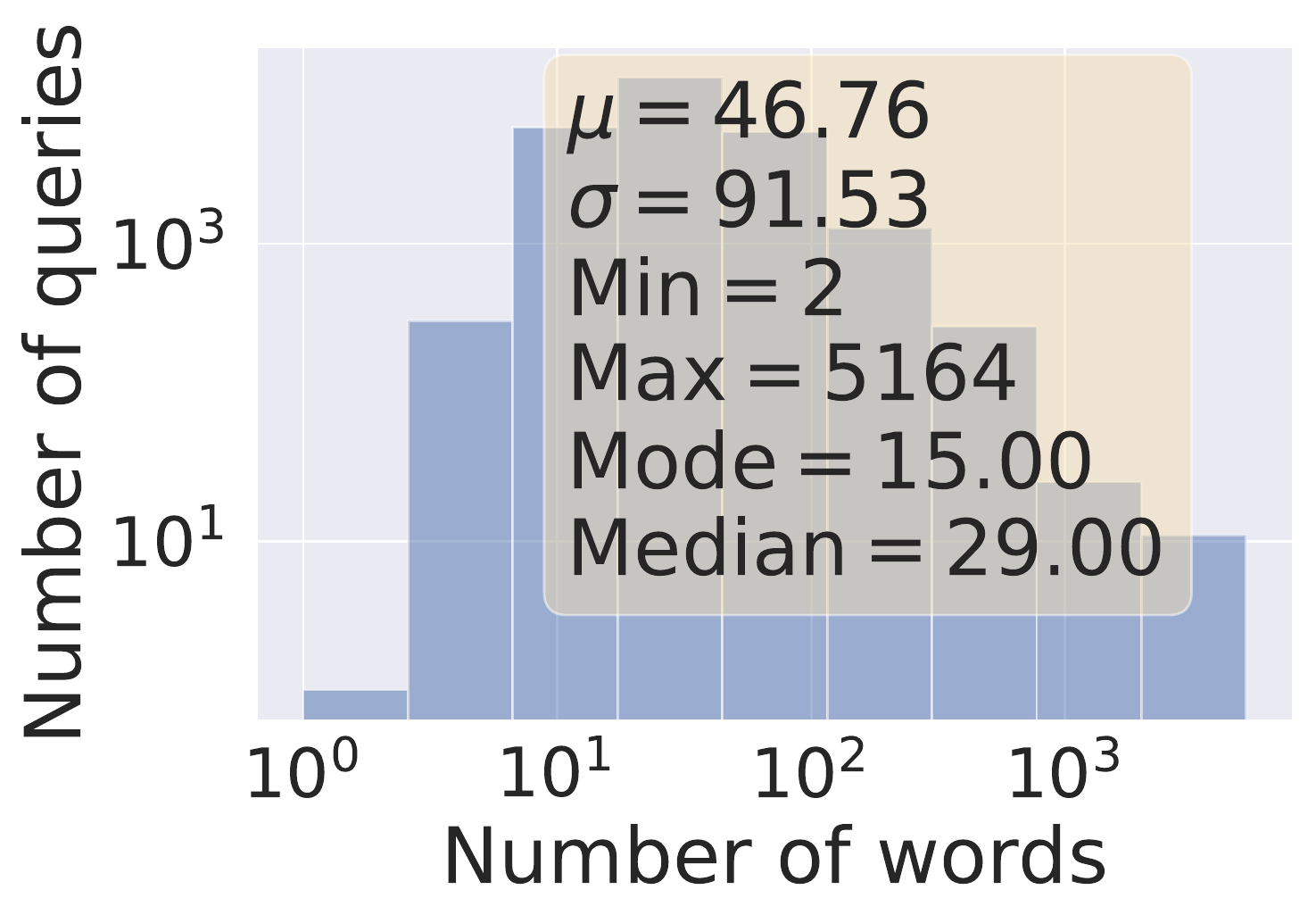}
		\label{fig:sqlshare_analysis_word_level}
        }
        \subfloat[]
        {
		\includegraphics[width=0.32\linewidth]{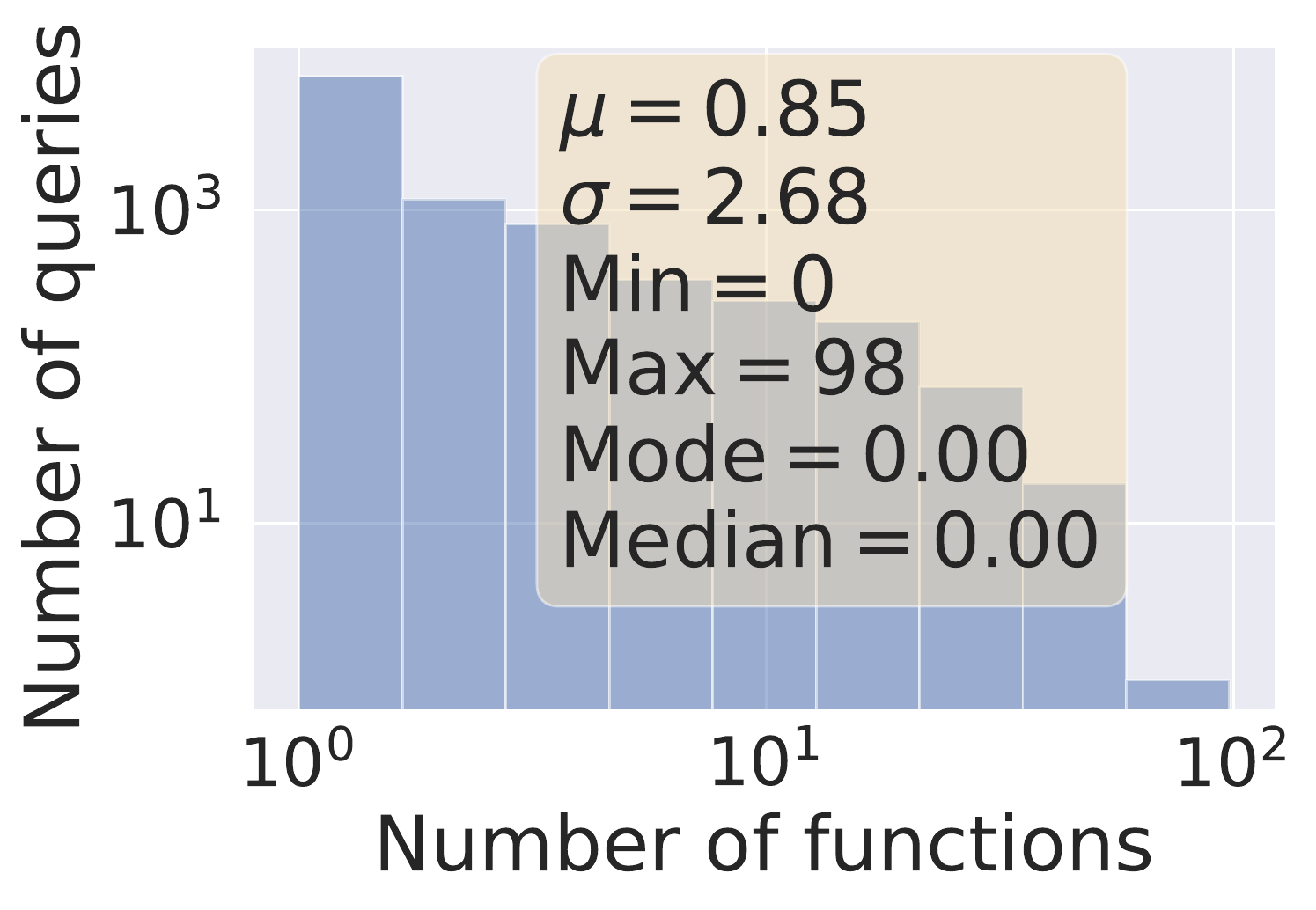} 
		\label{fig:sqlshare_function_count}
        }  
          
        \subfloat[]
        {
		\includegraphics[width=0.32\linewidth]{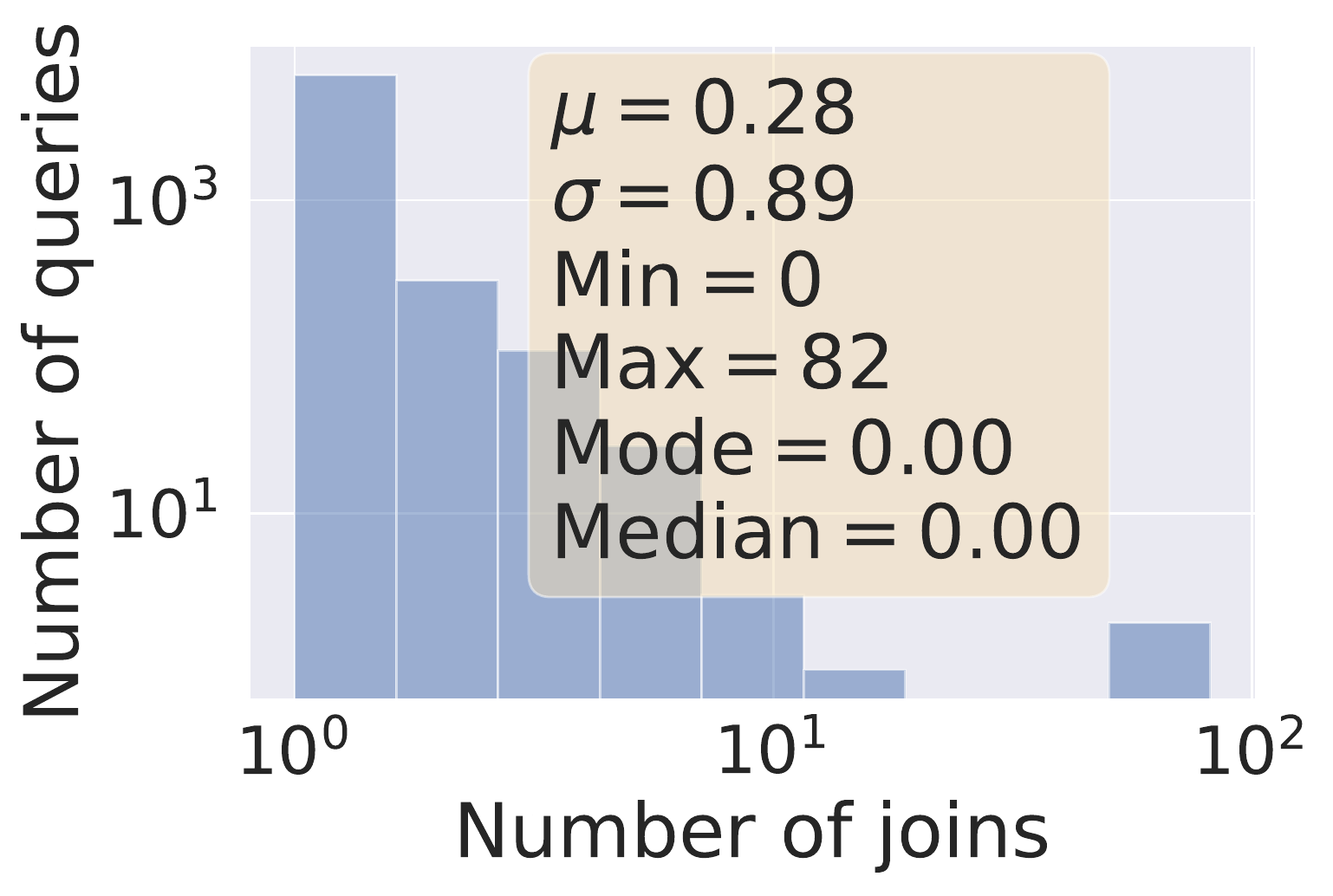} 
		\label{fig:sqlshare_join_count}
        } 
        \subfloat[]
        {
		\includegraphics[width=0.32\linewidth]{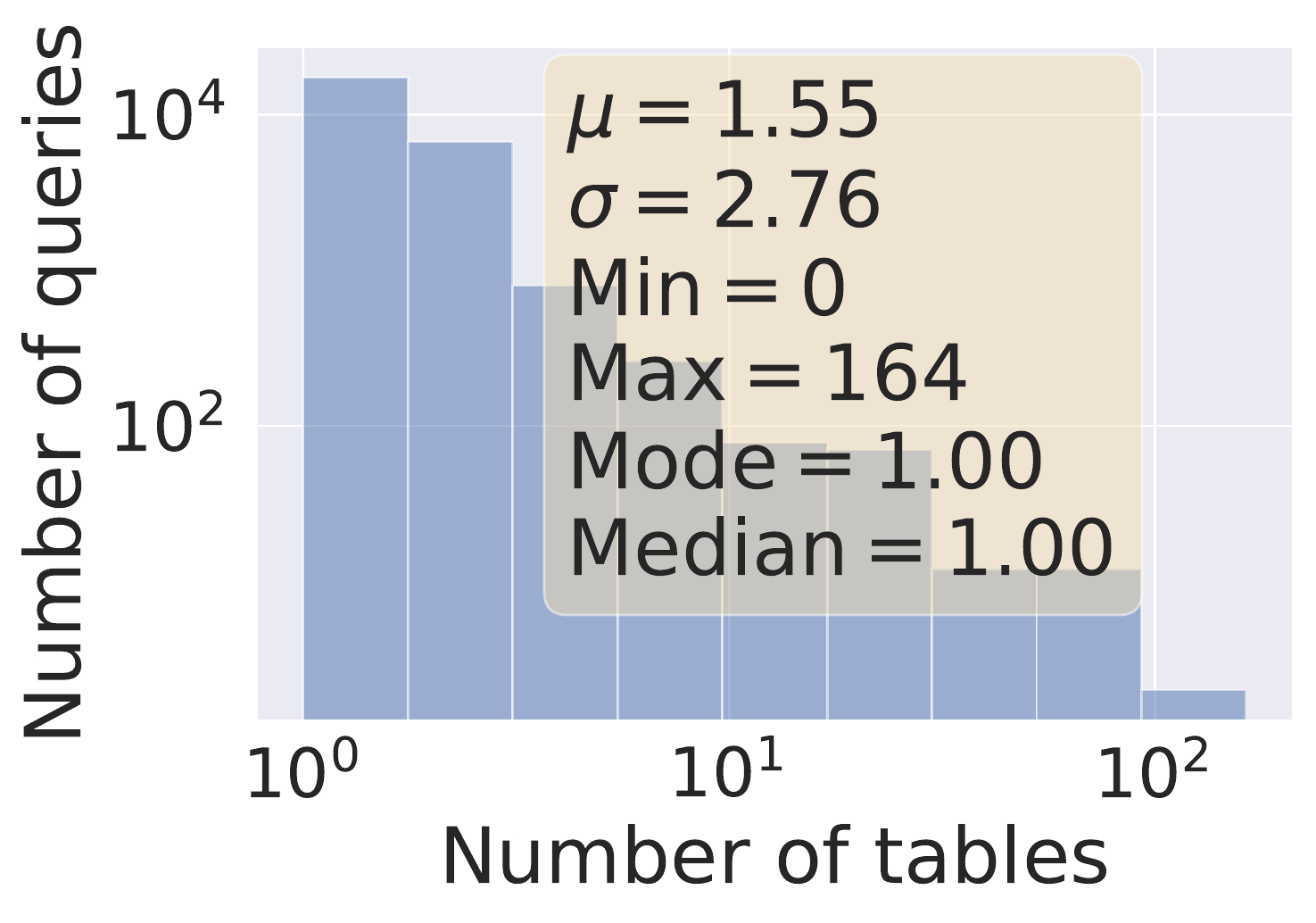} 
		\label{fig:sqlshare_table_count}
        } 
        \subfloat[]
        {
		\includegraphics[width=0.32\linewidth]{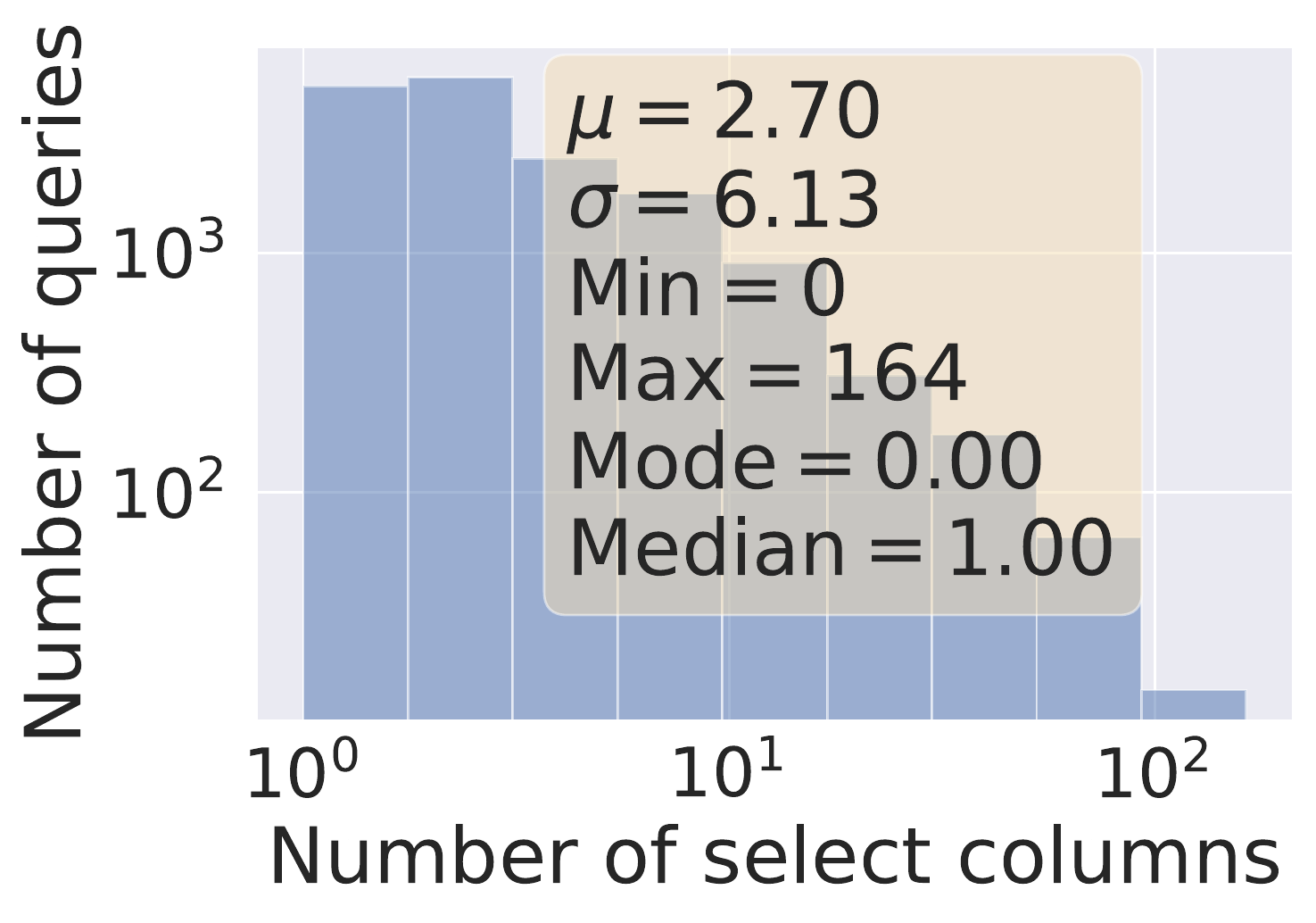} 
		\label{fig:sqlshare_select_columns}
        } 
         
         \subfloat[]
        {
		\includegraphics[width=0.32\linewidth]{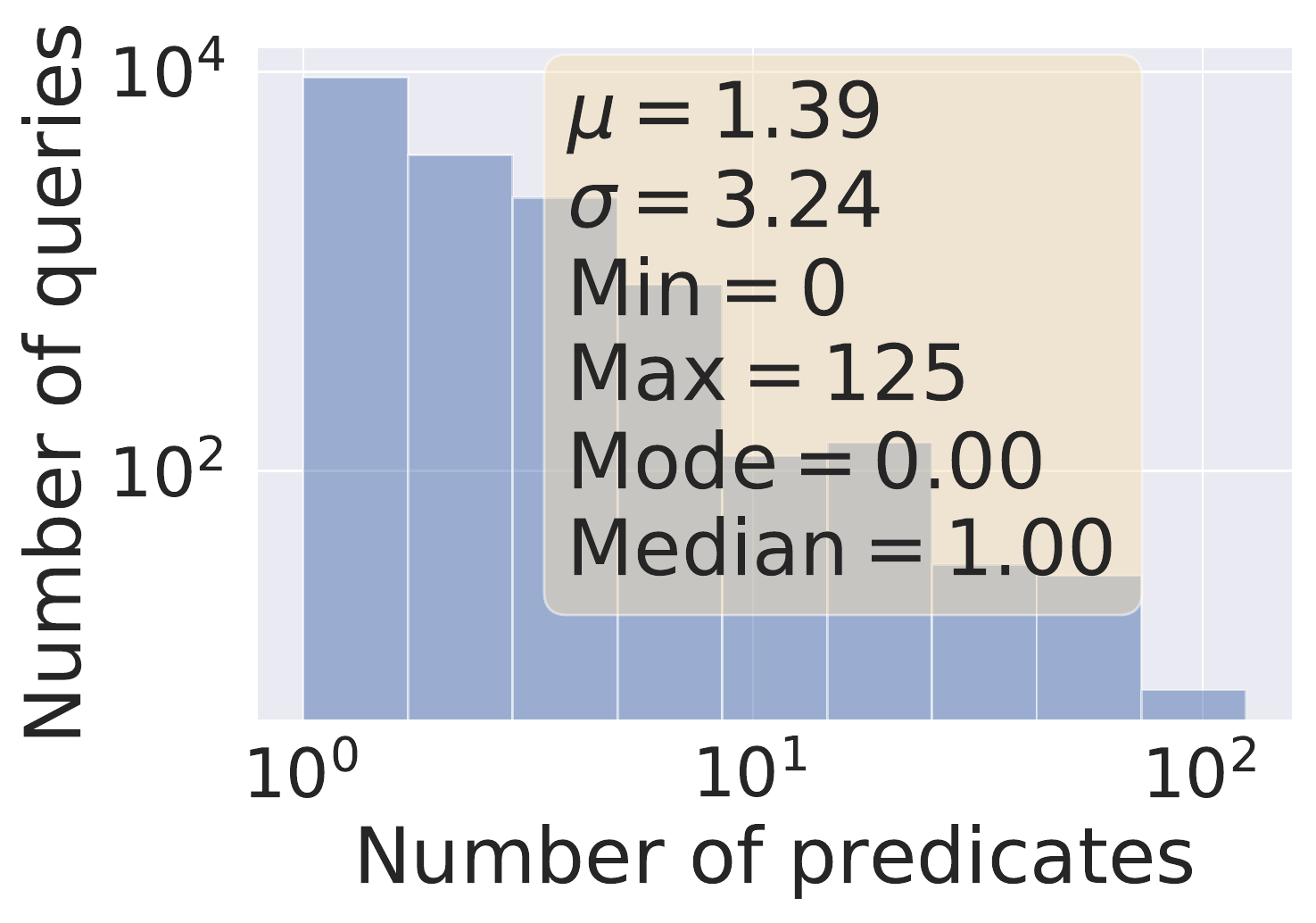} 
		\label{fig:sqlshare_predicate_count}
        } 
        \subfloat[]
        {
		\includegraphics[width=0.32\linewidth]{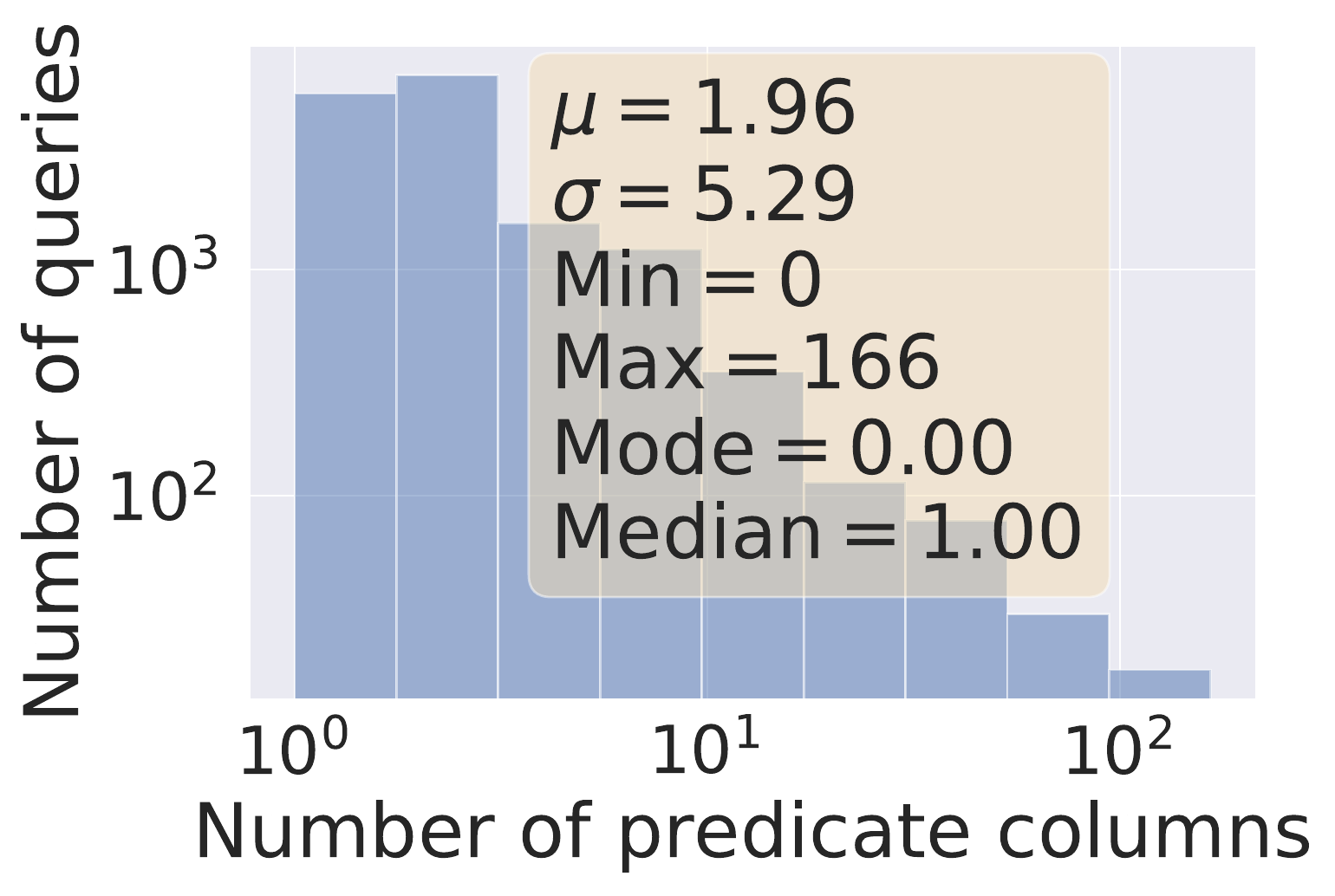} 
		\label{fig:sqlshare_predicate_column_count}
        }
        \subfloat[]
        {
		\includegraphics[width=0.32\linewidth]{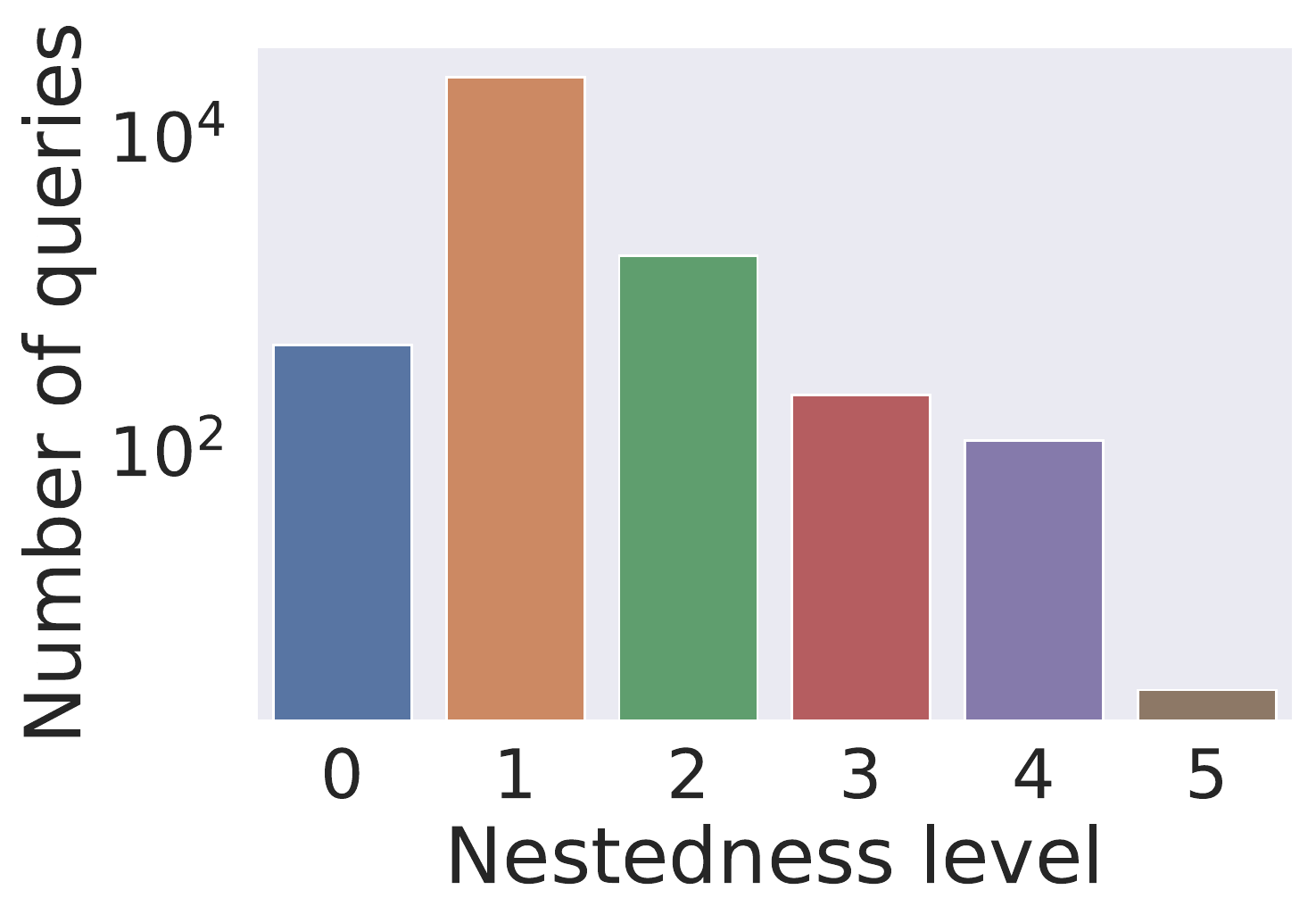} 
		\label{fig:sqlshare_nested_count}
        }
        
        \subfloat[]
        {
		\includegraphics[width=0.32\linewidth]{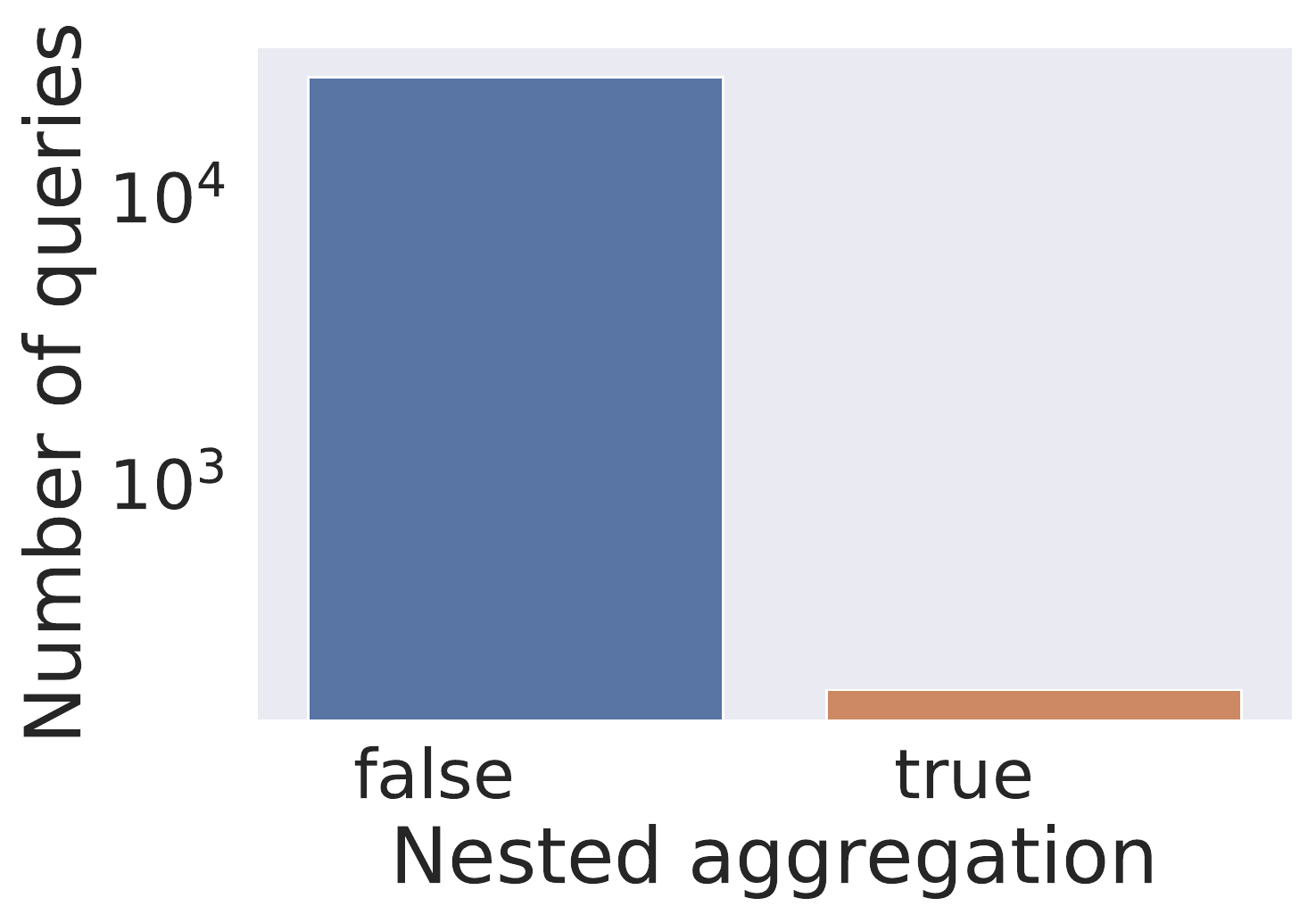} 
		\label{fig:sqlshare_nested_aggregate}
        }

\caption{Structural properties of SQLShare query statements. Most plots are on log-log scale, due to the wide data range. 
}
\label{fig:sqlshare_query_statement_analysis}
\end{figure*}

%% file: AnalysisFigures.tex
\begin{figure*}
\centering
        \subfloat[]
        {
		\includegraphics[width=0.19\linewidth]{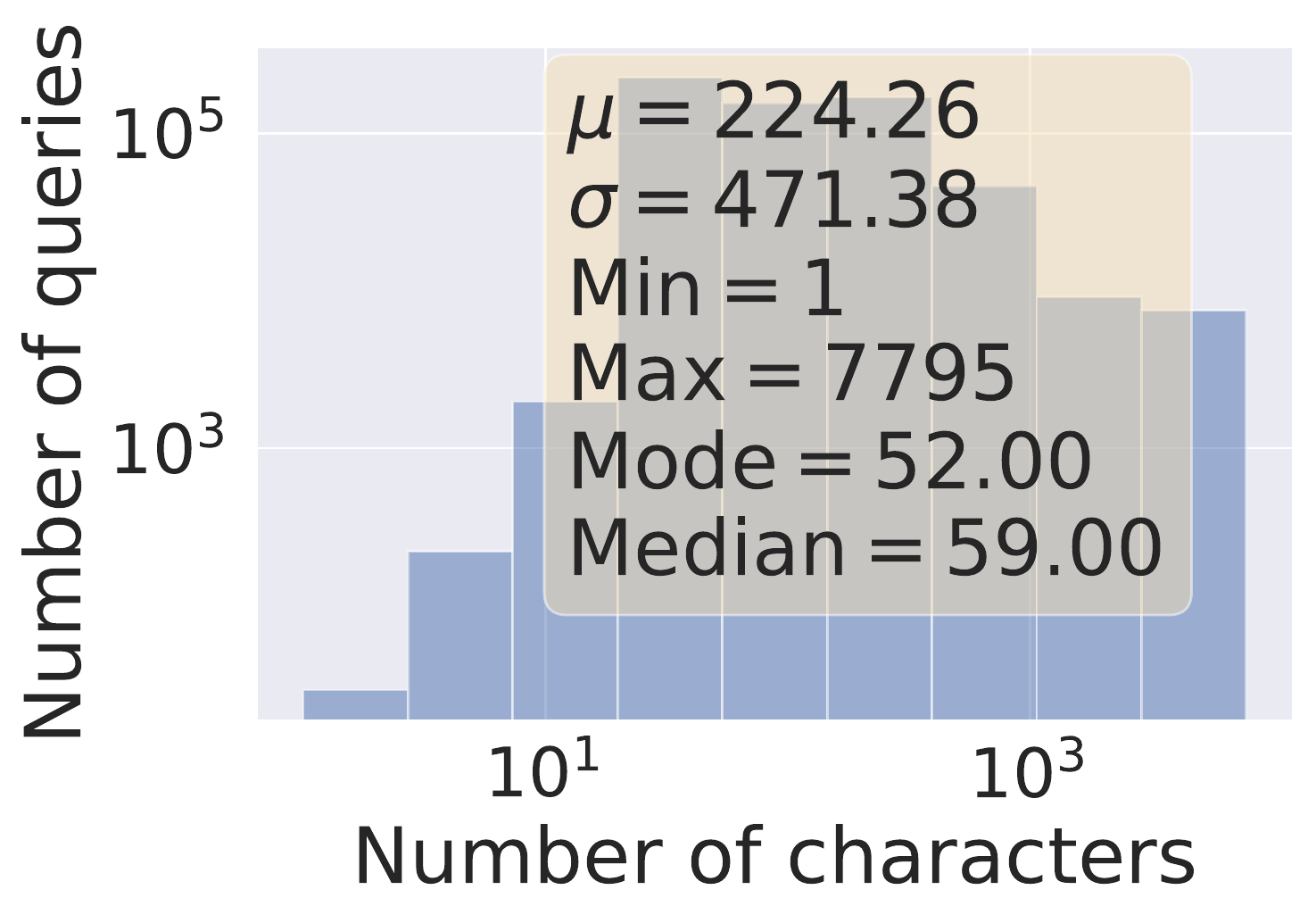} 
		\label{fig:sdss_analysis_char_level}
        }
		\subfloat[]
        {
		\includegraphics[width=0.19\linewidth]{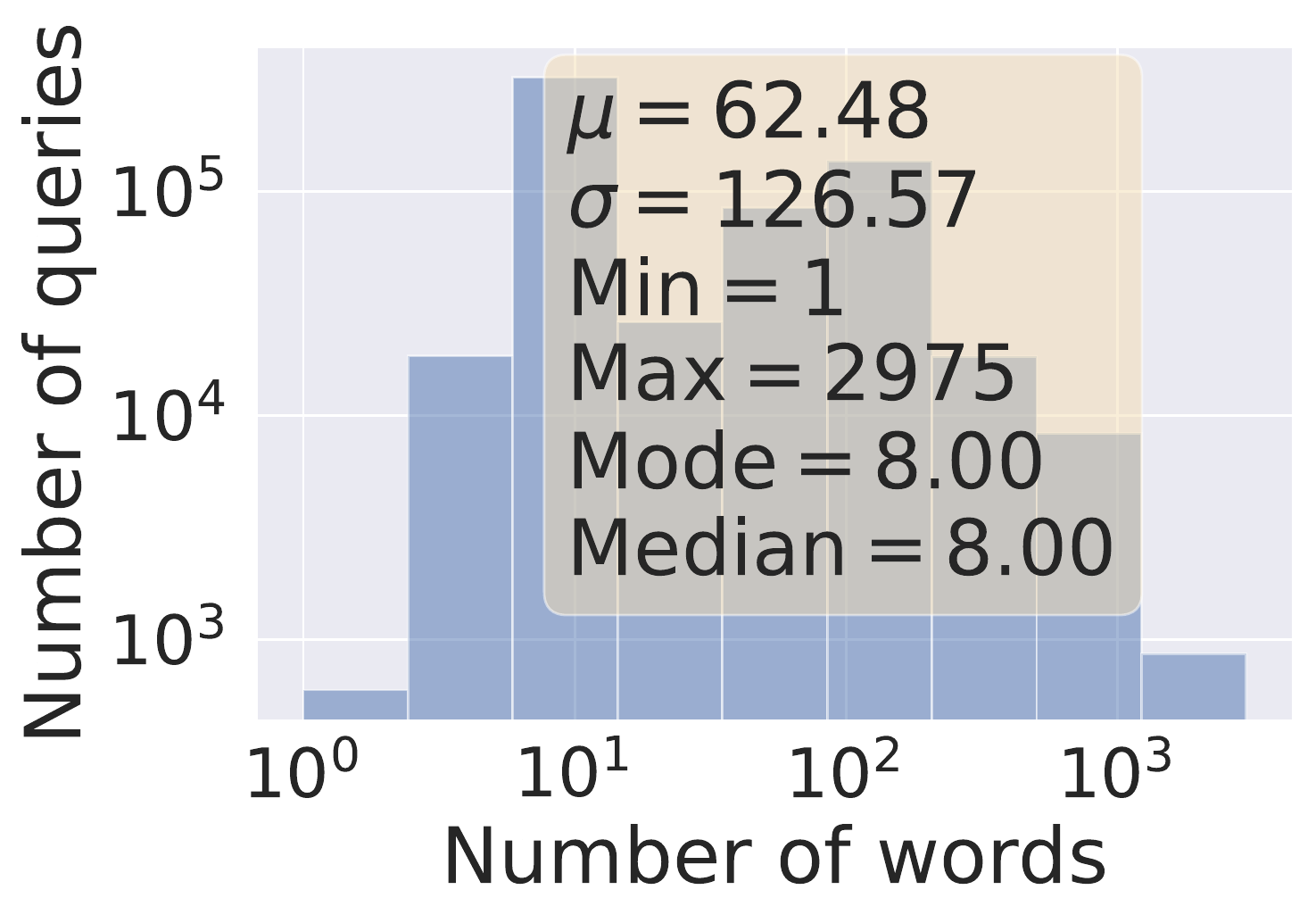} 
		\label{fig:sdss_analysis_word_level}
        } 
        \subfloat[]
        {
		\includegraphics[width=0.19\linewidth]{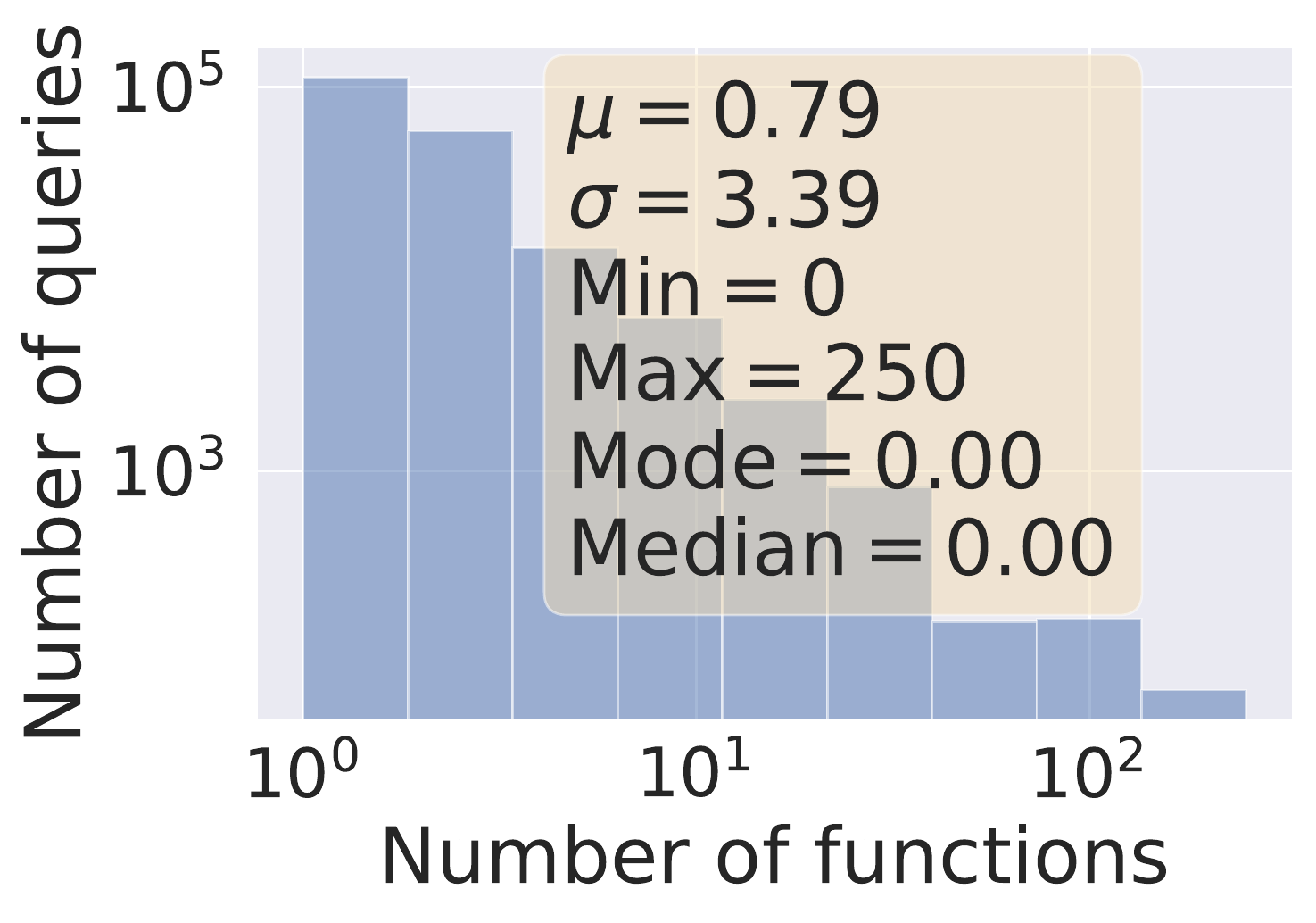} 
		\label{fig:sdss_function_count}
        }   
        \subfloat[]
        {
		\includegraphics[width=0.19\linewidth]{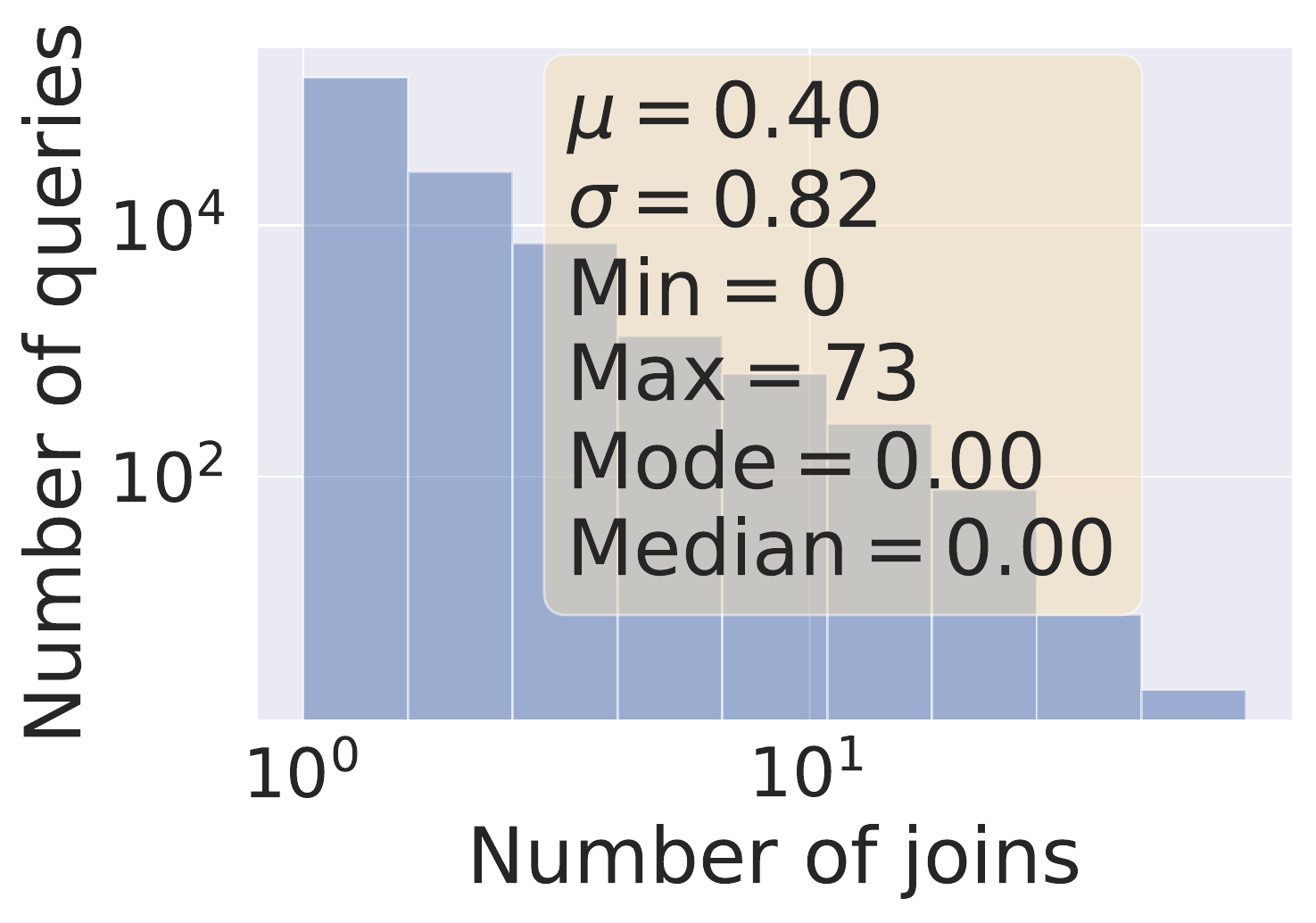} 
		\label{fig:sdss_join_count}
        }     
        \subfloat[]
        {
		\includegraphics[width=0.19\linewidth]{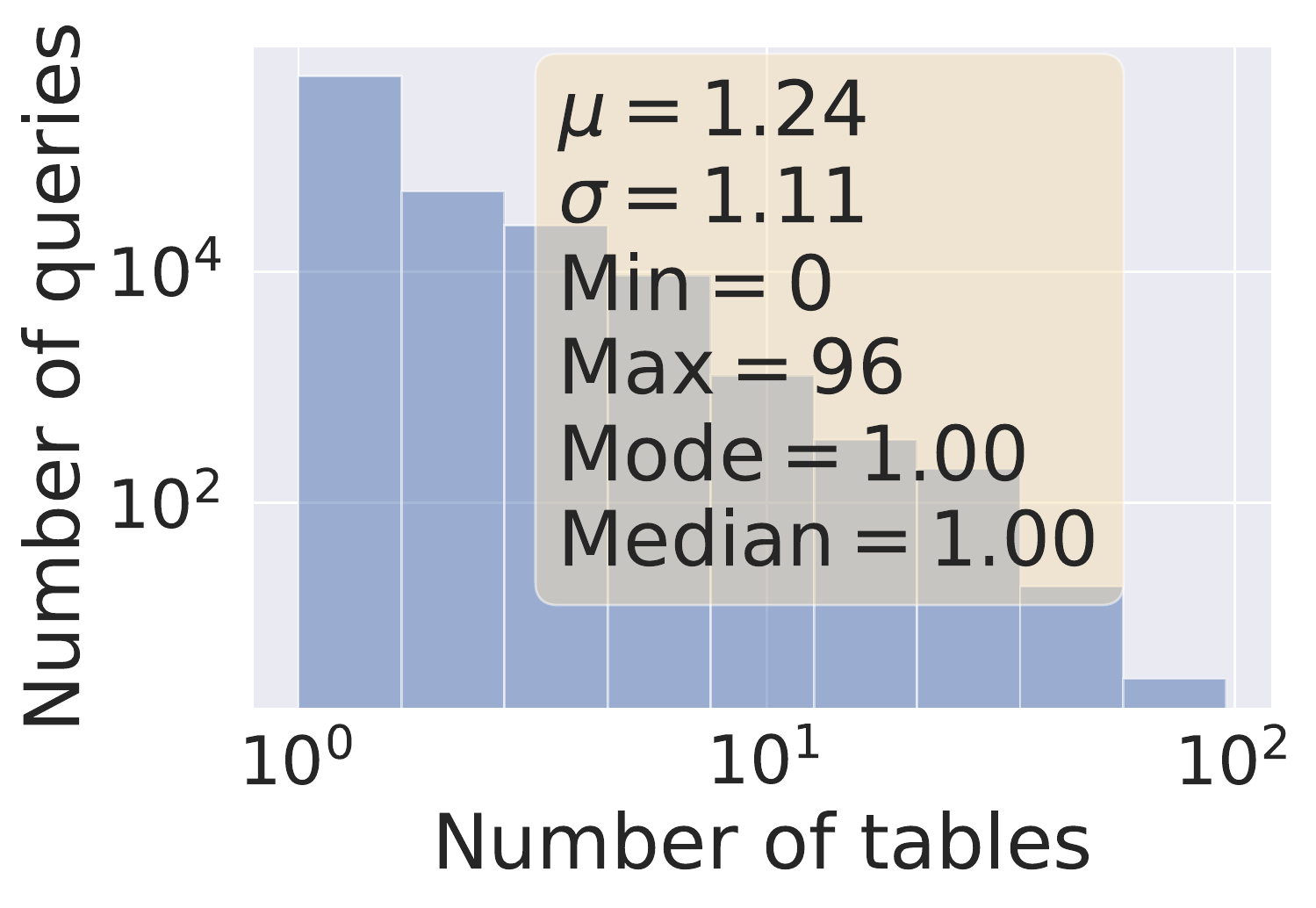} 
		\label{fig:sdss_table_count}
        }
        
        \subfloat[]
        {
		\includegraphics[width=0.19\linewidth]{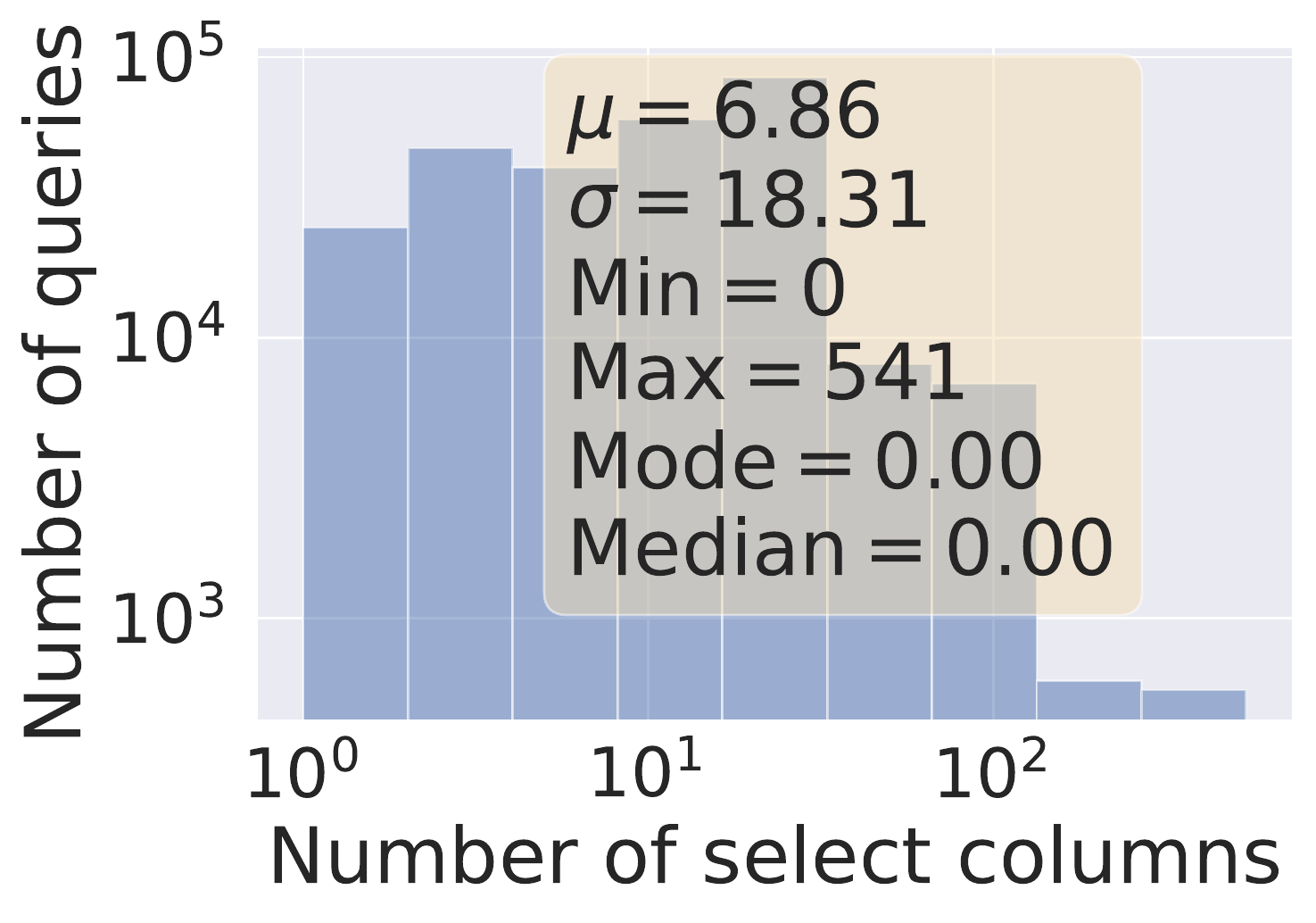} 
		\label{fig:sdss_select_columns}
        }  
        \subfloat[]
        {
		\includegraphics[width=0.19\linewidth]{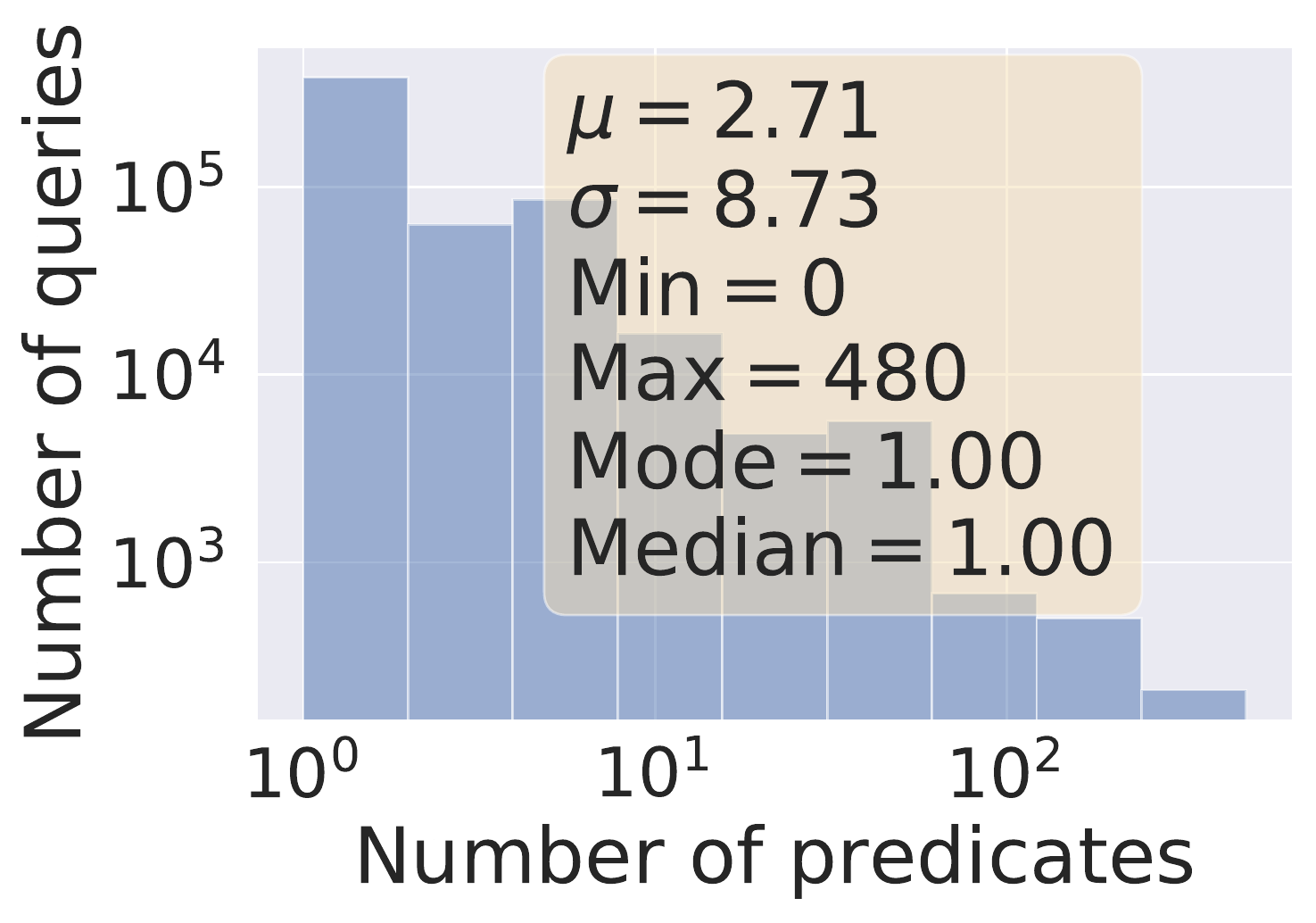} 
		\label{fig:sdss_predicate_count}
        }  
		 \subfloat[]
        {
		\includegraphics[width=0.19\linewidth]{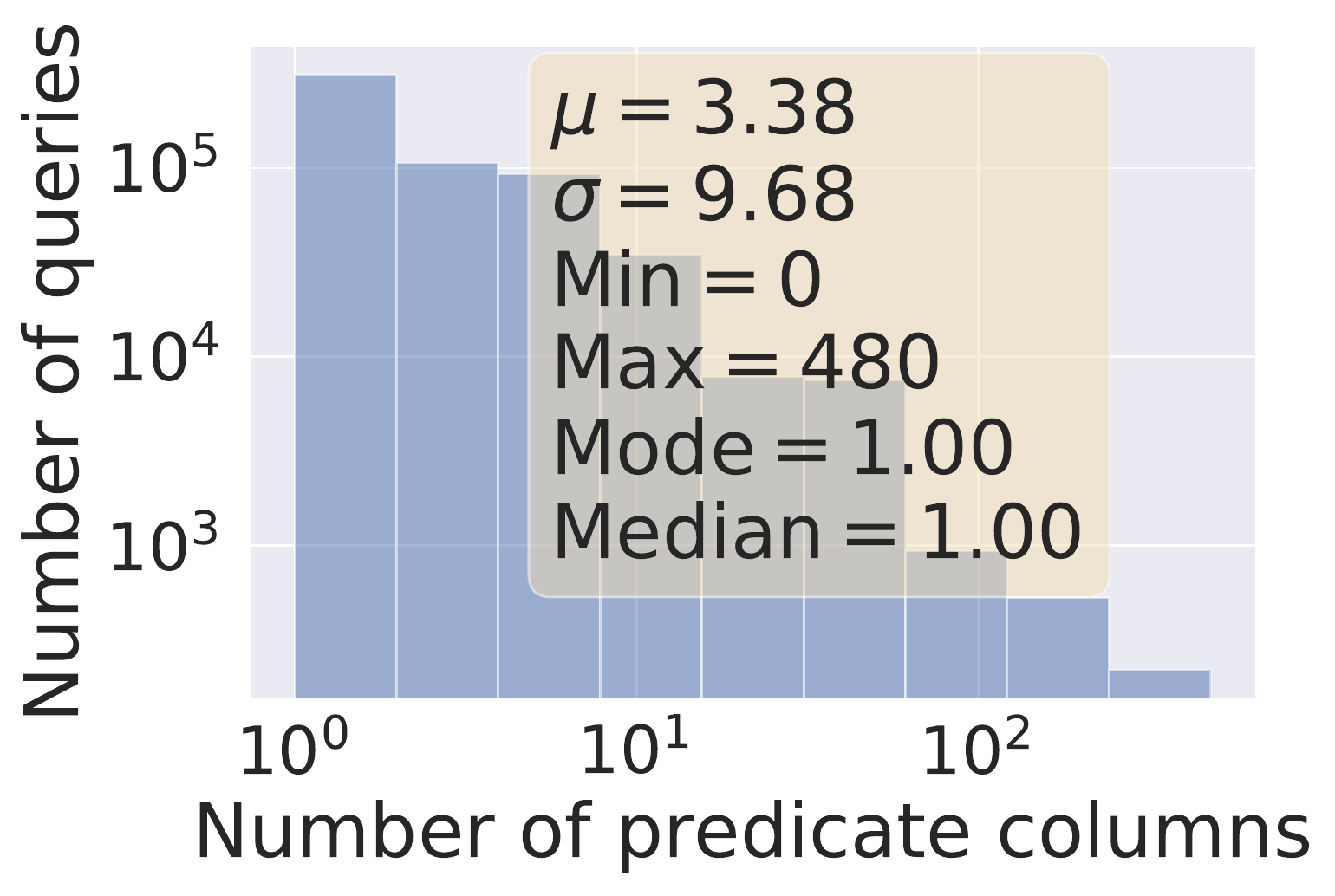} 
		\label{fig:sdss_predicate_column_count}
        }    
        \subfloat[]
        {
		\includegraphics[width=0.19\linewidth]{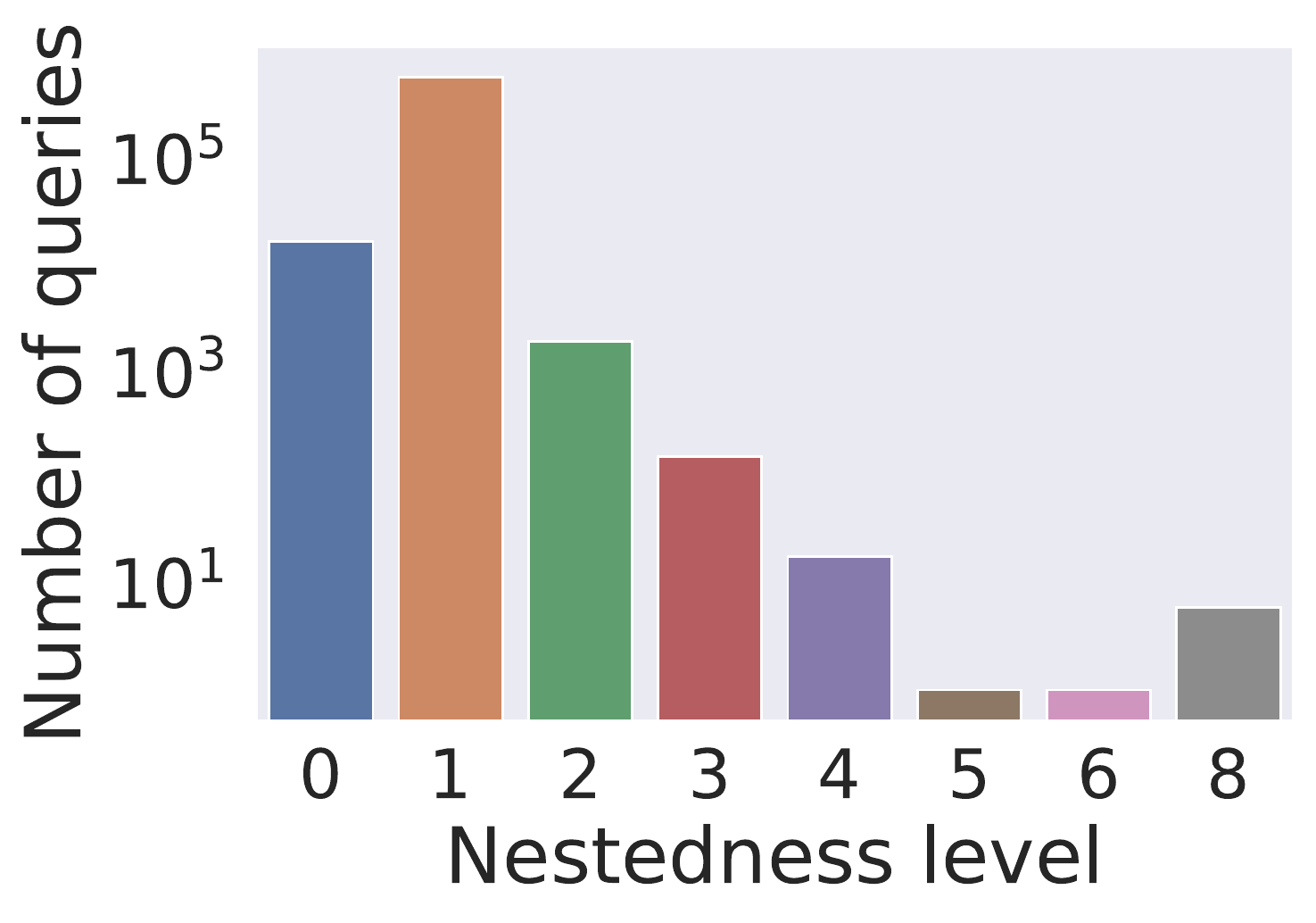} 
		\label{fig:sdss_nested_count}
        }
        \subfloat[]
        {
		\includegraphics[width=0.19\linewidth]{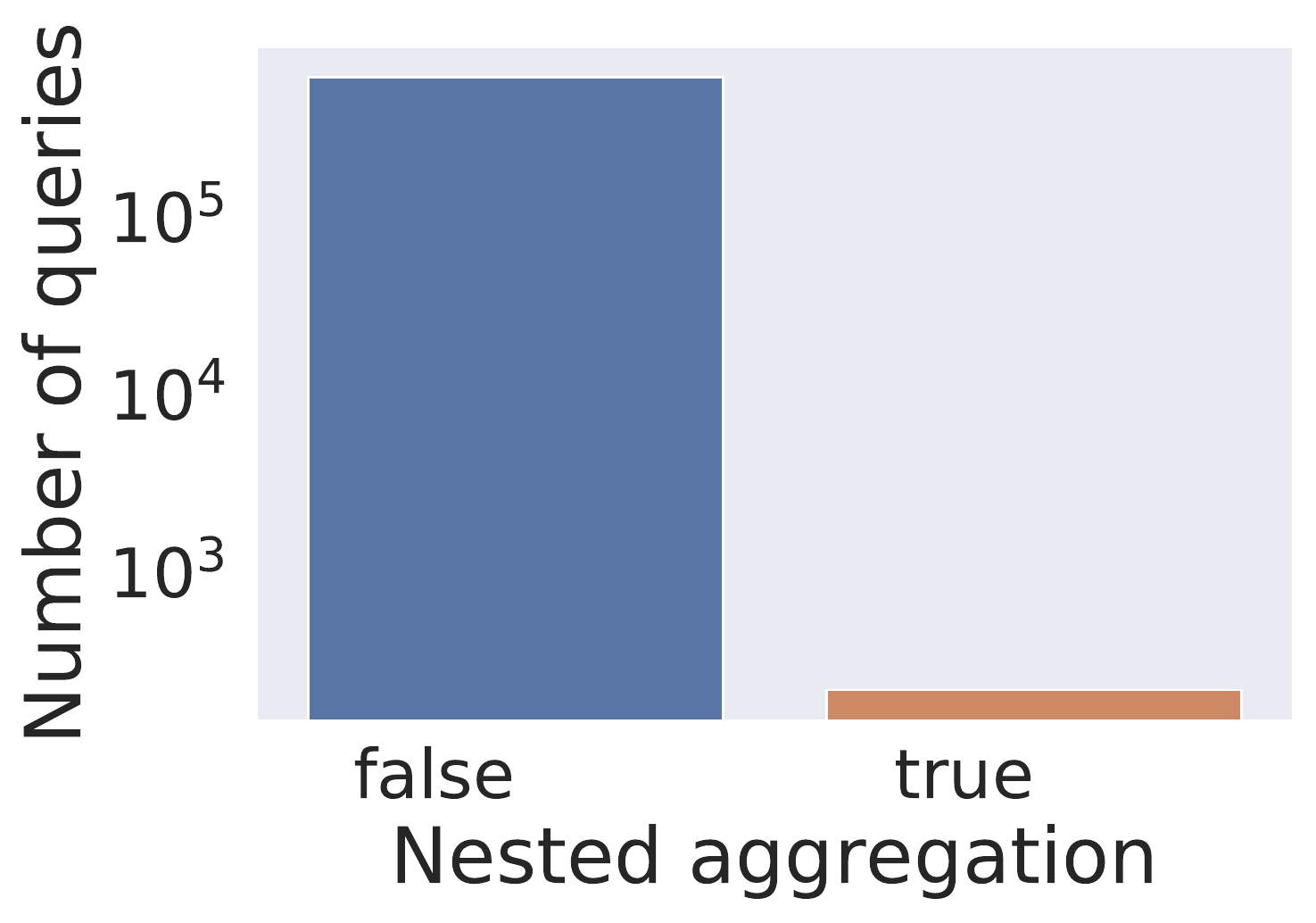} 
		\label{fig:sdss_nested_aggregate}
        }

\caption{Structural properties of SDSS query statements. 
}
\label{fig:sdss_query_statement_analysis}
\end{figure*}

\iffullpaperComposition

\begin{figure*}
\centering
        \subfloat[]
        {
		\includegraphics[width=0.19\linewidth]{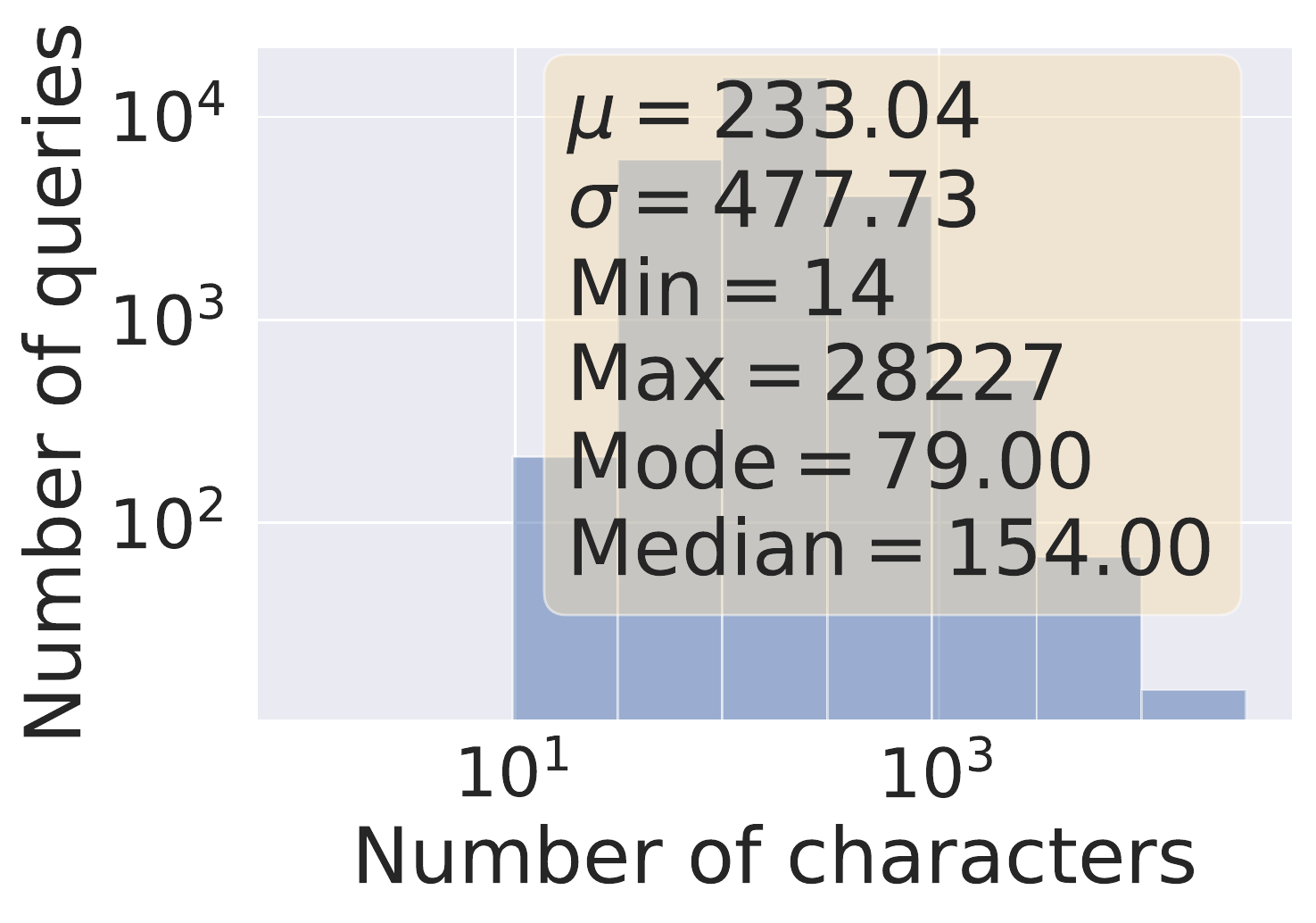} 
		\label{fig:sqlshare_analysis_char_level}
        }
		\subfloat[]
        {
		\includegraphics[width=0.19\linewidth]{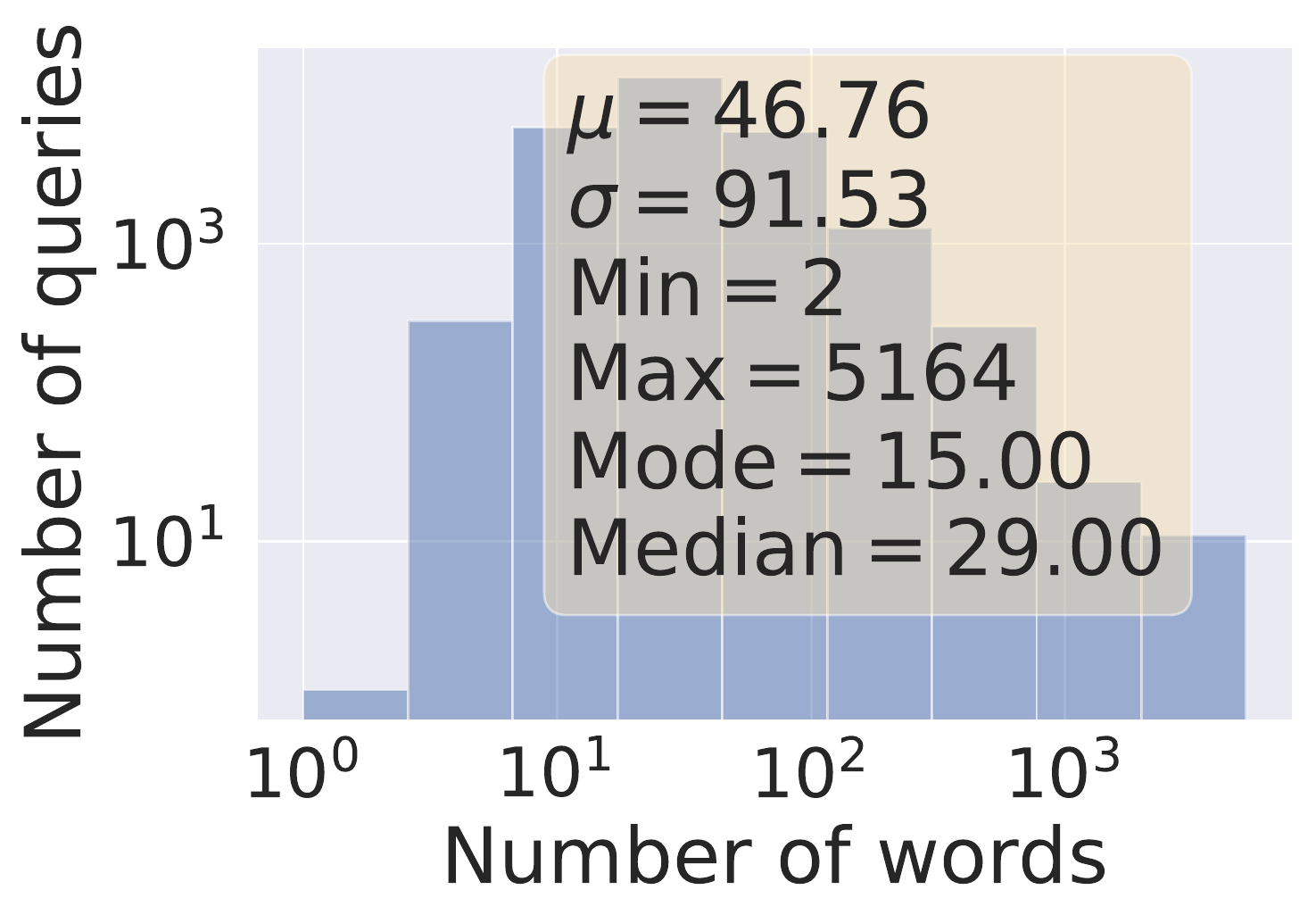}
		\label{fig:sqlshare_analysis_word_level}
        }
        \subfloat[]
        {
		\includegraphics[width=0.19\linewidth]{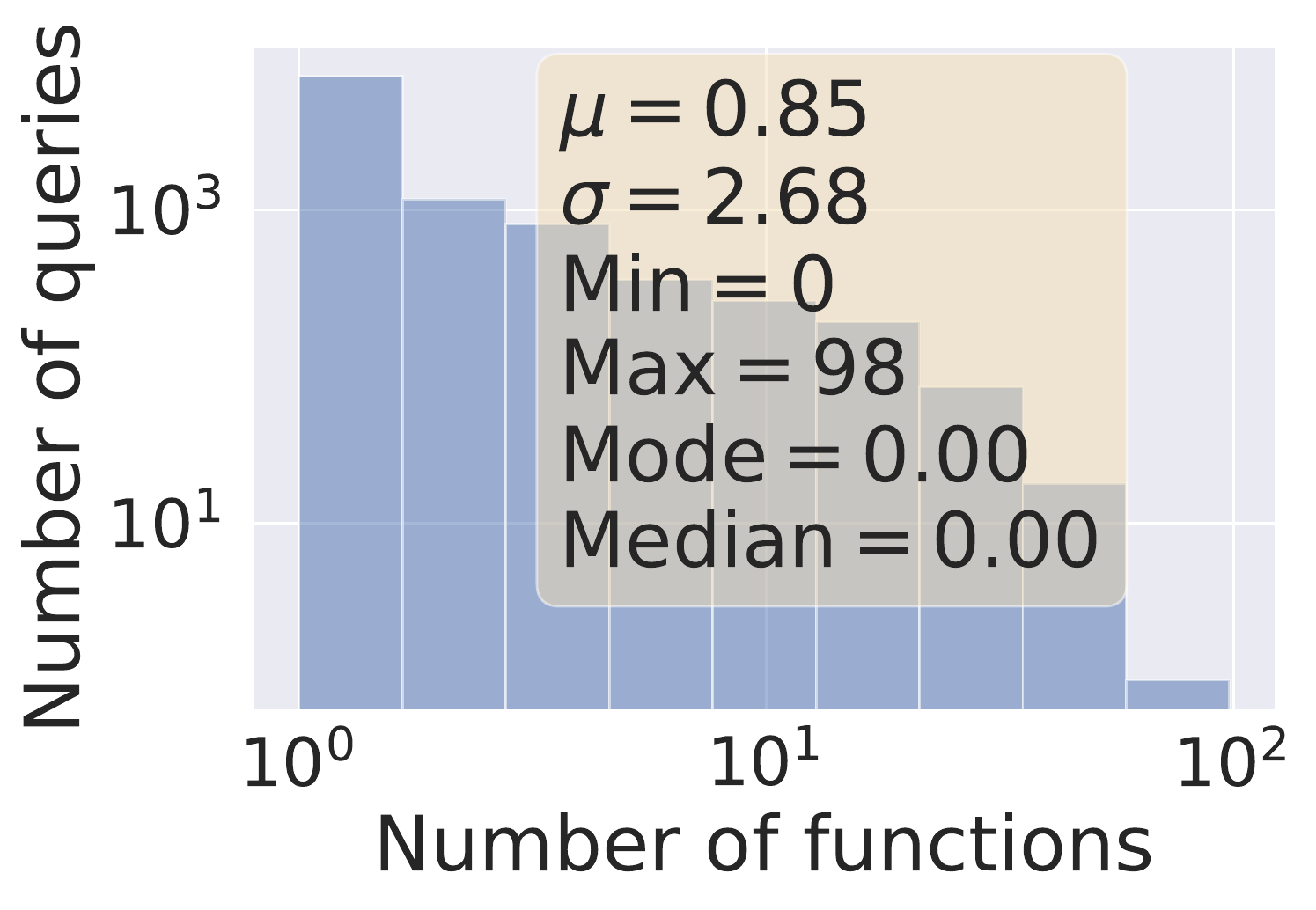} 
		\label{fig:sqlshare_function_count}
        }  
        \subfloat[]
        {
		\includegraphics[width=0.19\linewidth]{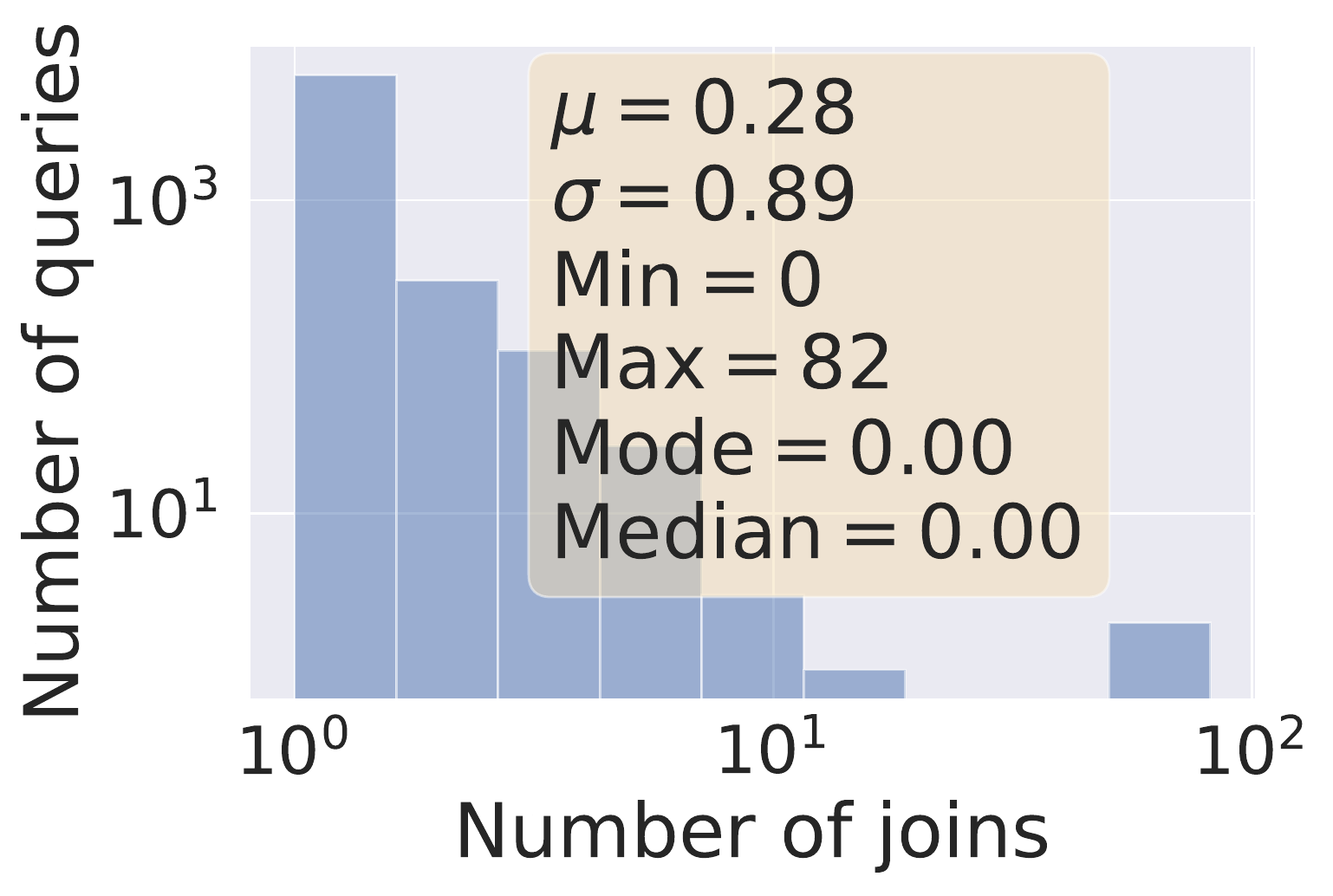} 
		\label{fig:sqlshare_join_count}
        } 
        \subfloat[]
        {
		\includegraphics[width=0.19\linewidth]{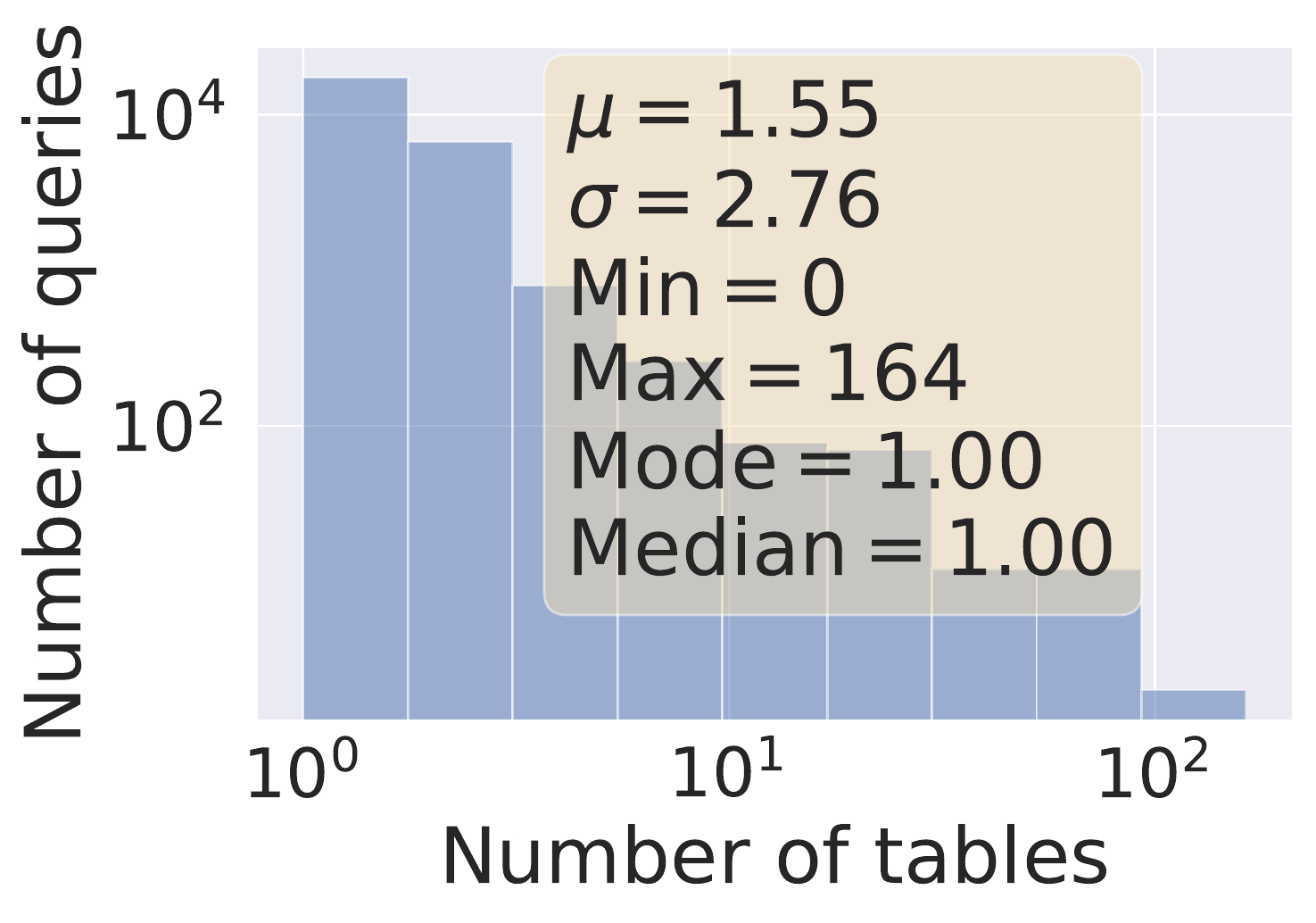} 
		\label{fig:sqlshare_table_count}
        } 
        
        \subfloat[]
        {
		\includegraphics[width=0.19\linewidth]{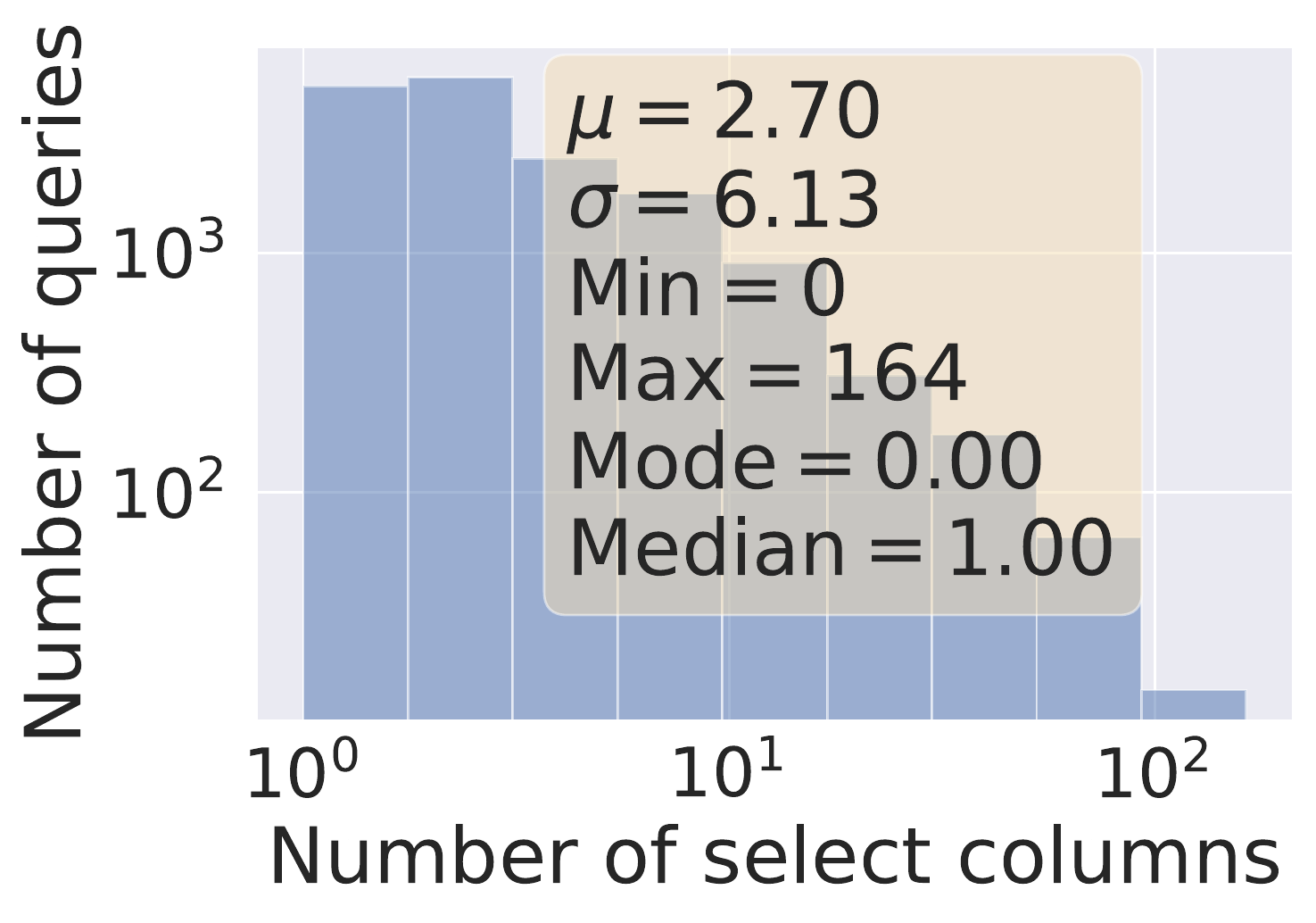} 
		\label{fig:sqlshare_select_columns}
        } 
         \subfloat[]
        {
		\includegraphics[width=0.19\linewidth]{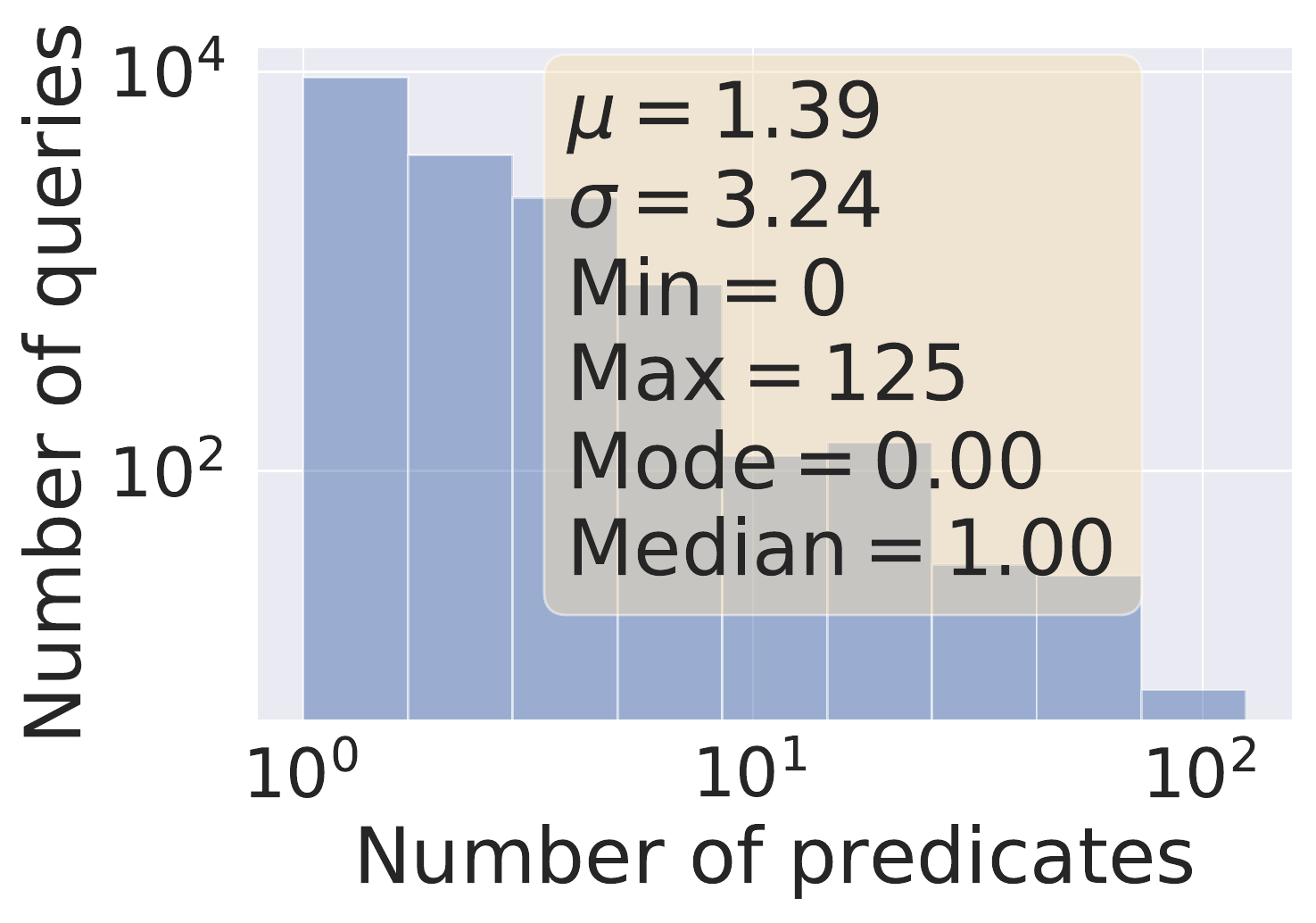} 
		\label{fig:sqlshare_predicate_count}
        } 
        \subfloat[]
        {
		\includegraphics[width=0.19\linewidth]{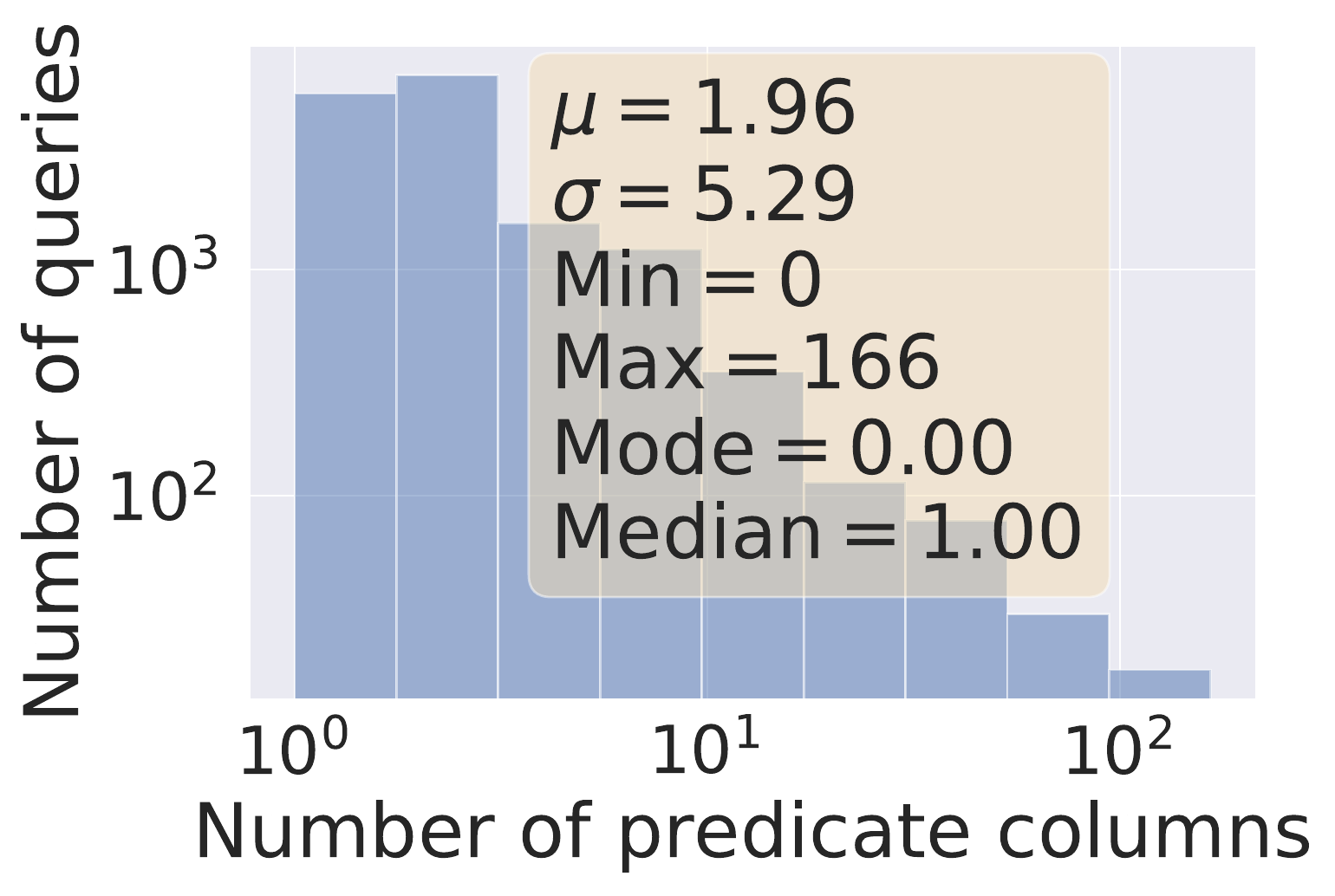} 
		\label{fig:sqlshare_predicate_column_count}
        }
        \subfloat[]
        {
		\includegraphics[width=0.19\linewidth]{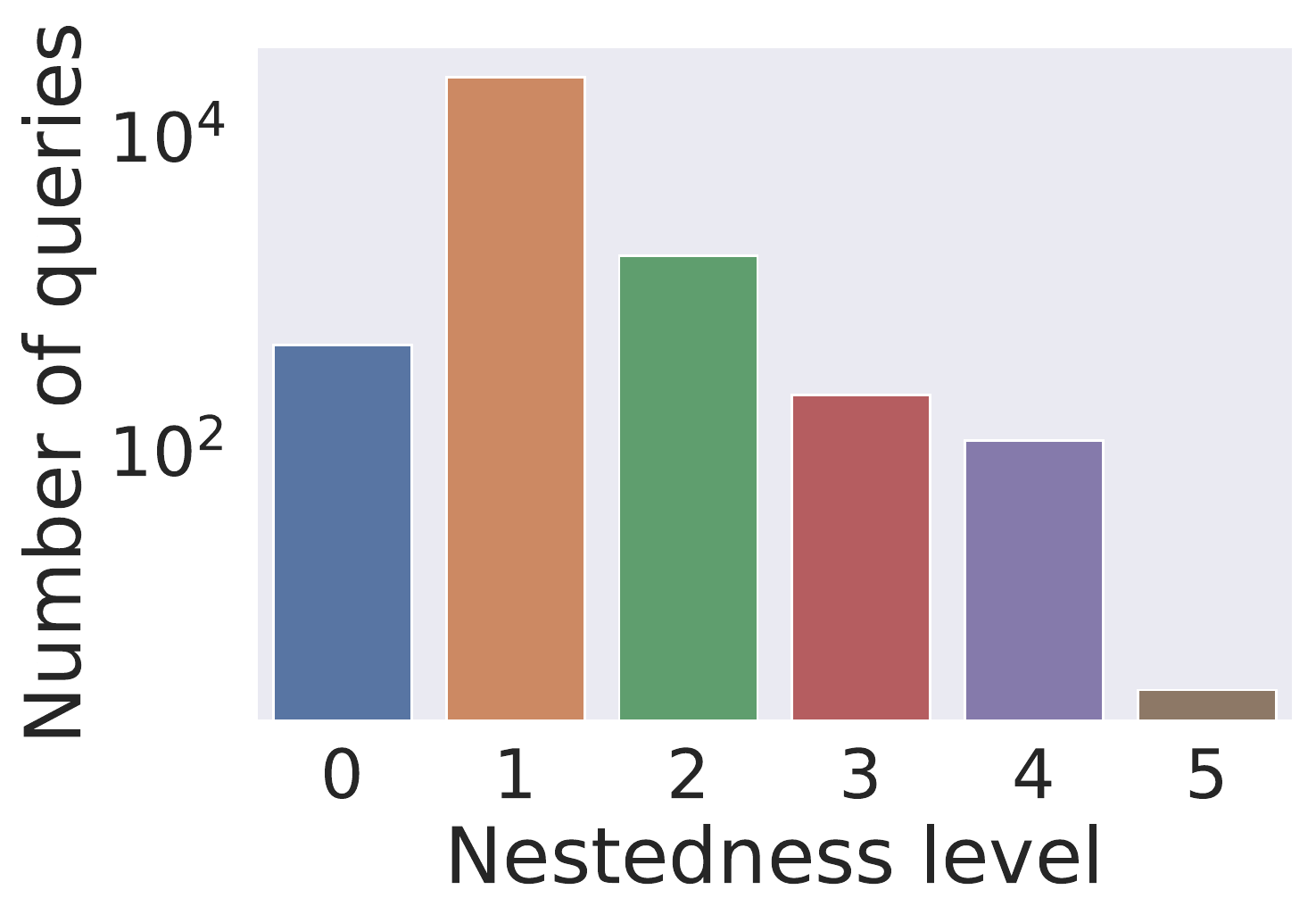} 
		\label{fig:sqlshare_nested_count}
        }
        \subfloat[]
        {
		\includegraphics[width=0.19\linewidth]{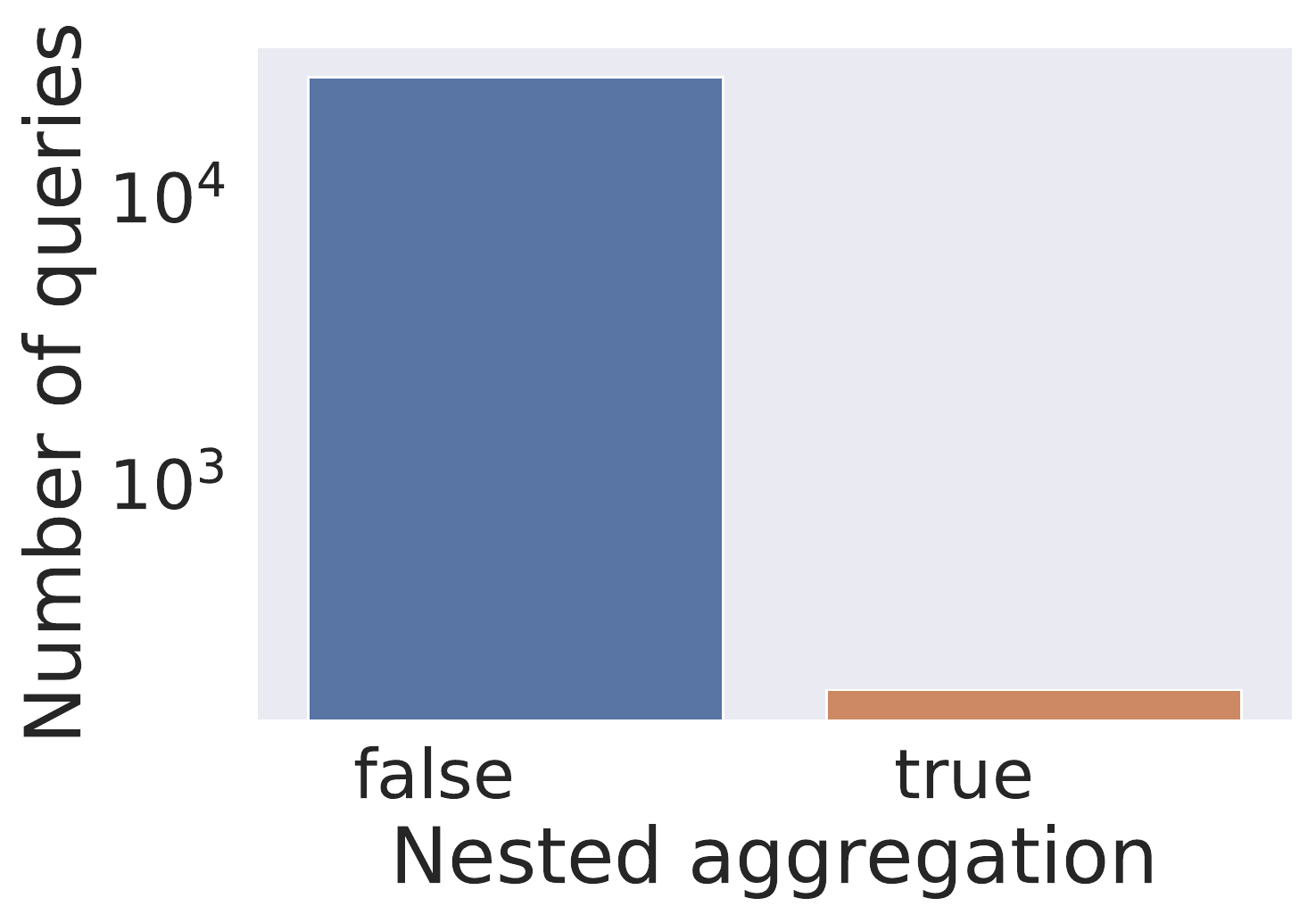} 
		\label{fig:sqlshare_nested_aggregate}
        }

\caption{Structural properties of SQLShare query statements. Most plots are on log-log scale, due to the wide data range. 
}
\label{fig:sqlshare_query_statement_analysis}
\end{figure*}

\fi

%% file: LabelDistribtionFiguresPhDThesis.tex
\begin{figure*}
    \centering
		\subfloat[SDSS]
		{
		\includegraphics[width=0.32\linewidth,valign=t]{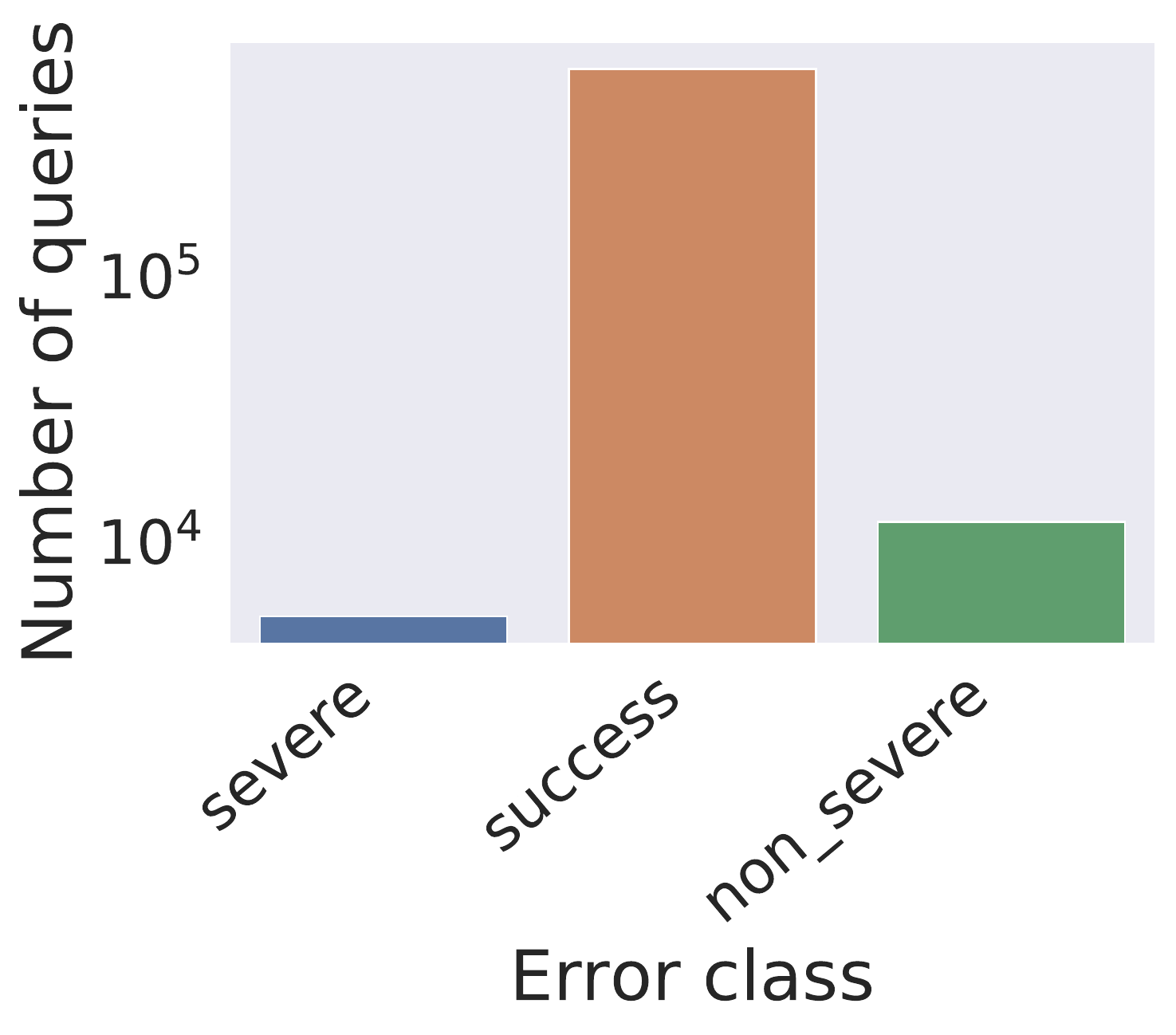}
		\label{fig:analysis_error}
		}
		\subfloat[SDSS]
		{
		\includegraphics[width=0.32\linewidth,valign=t]{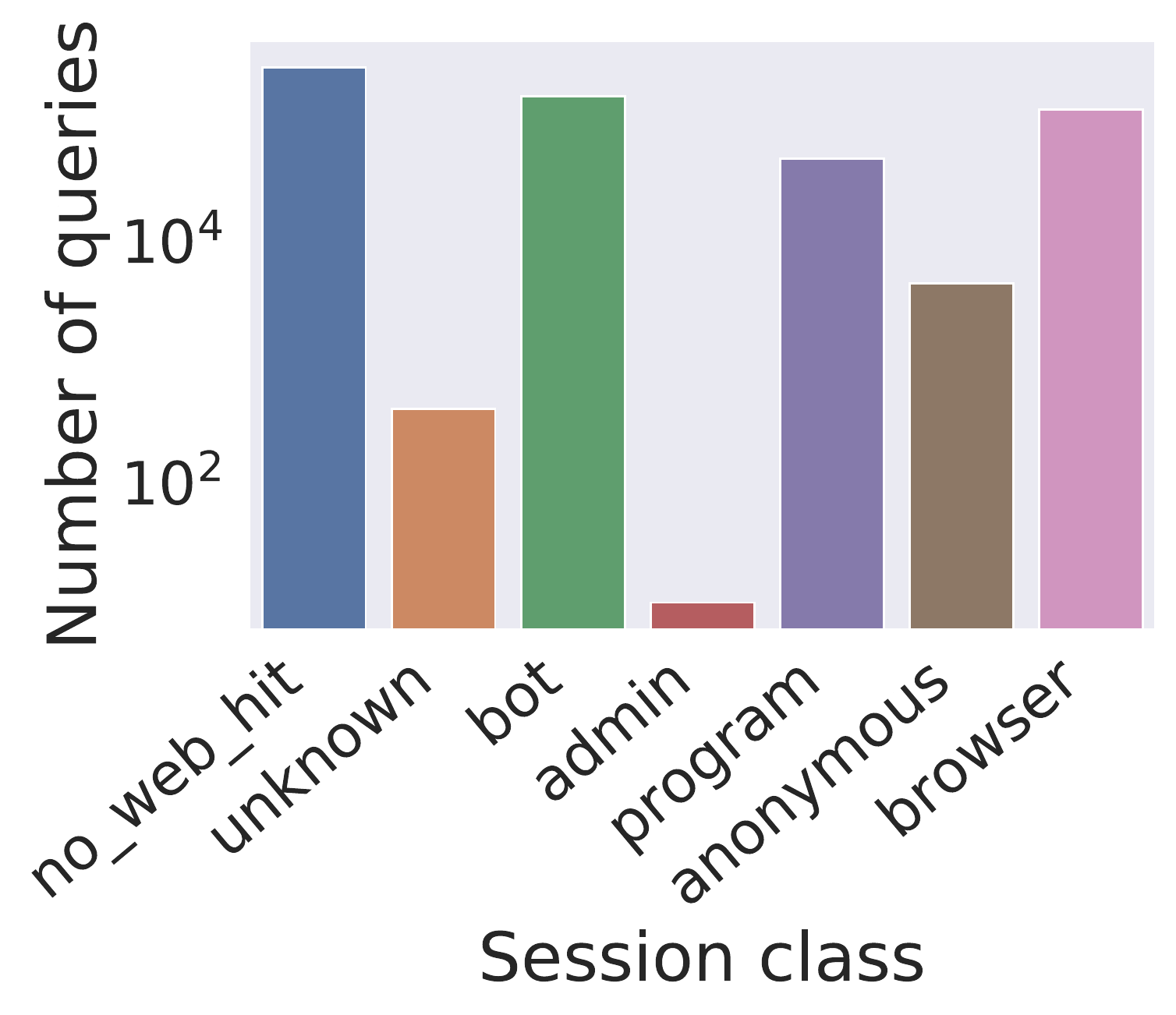}
		\label{fig:analysis_session_class}
		}
		
		\subfloat[SDSS]
		{
		\includegraphics[width=0.32\linewidth,valign=t]{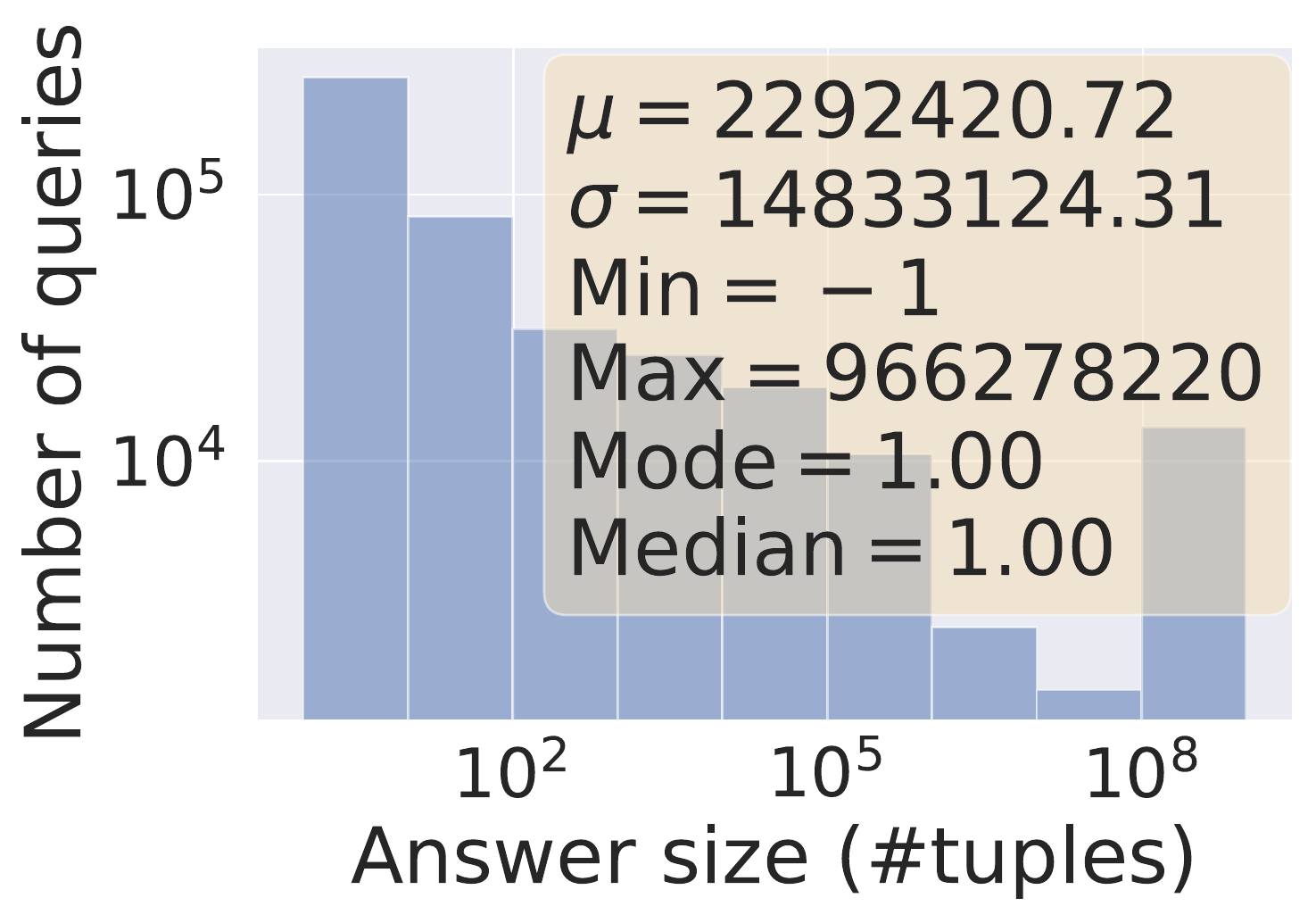}
		\label{fig:analysis_rows}
		}
		\subfloat[SDSS]
		{
		\includegraphics[width=0.32\linewidth,valign=t]{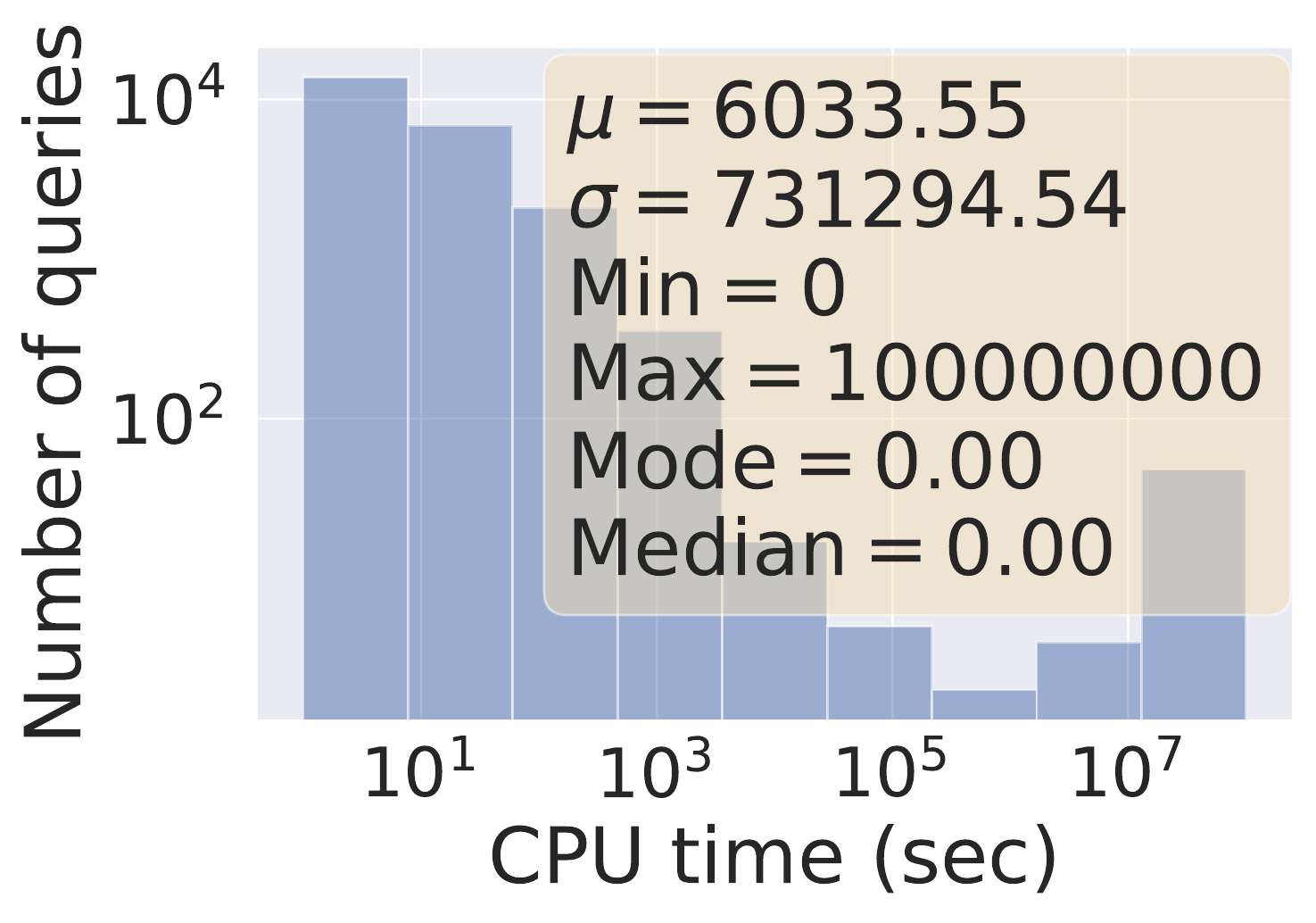}
		\label{fig:analysis_busy}
		}
		\subfloat[SQLShare]
		{
		\includegraphics[width=0.32\linewidth,valign=t]{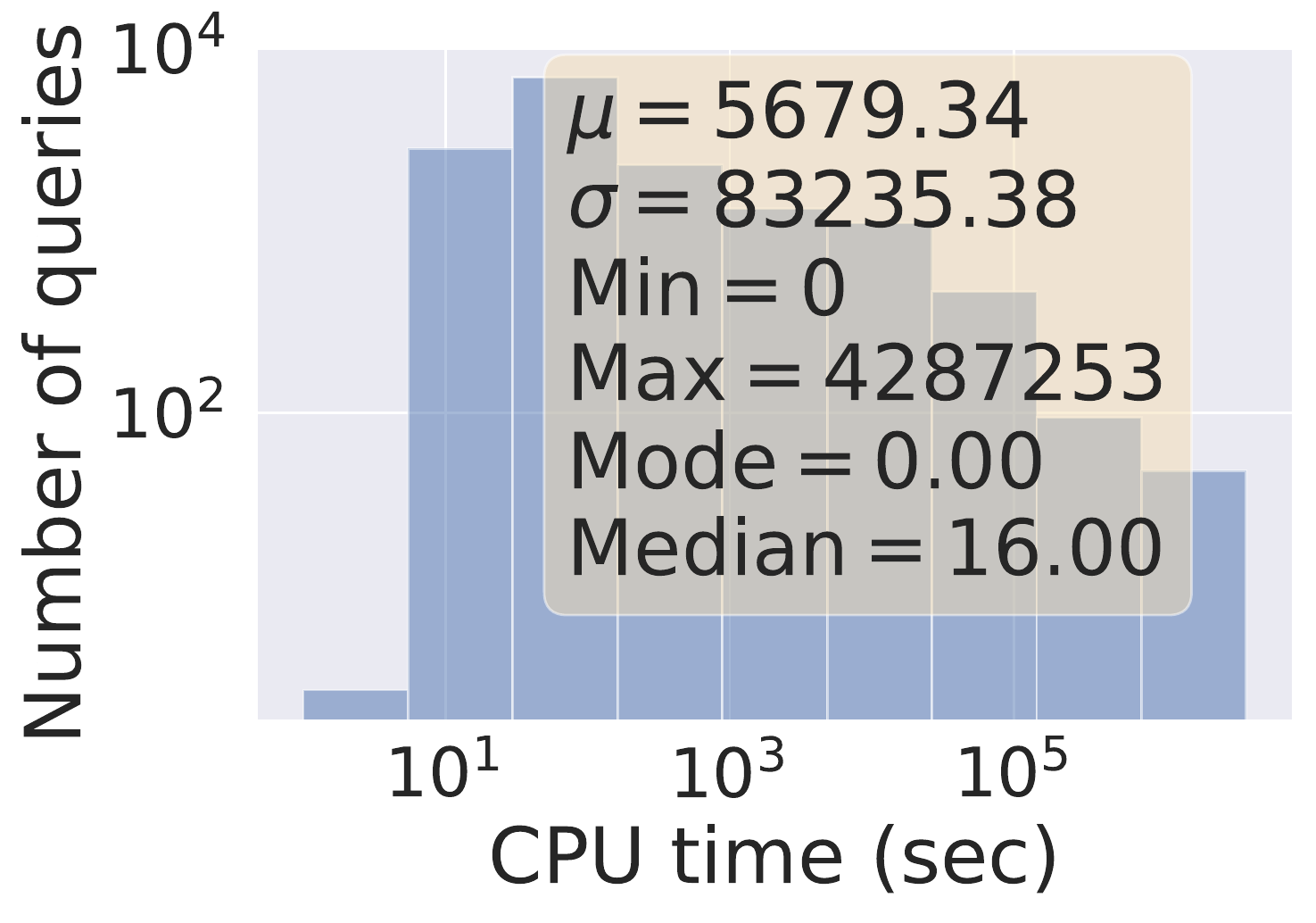}
		\label{fig:sqlshare_analysis_busy}
		}			
		\vspace*{-4mm}		
\caption{Label distributions for classification problems (Figures~\ref{fig:analysis_error} and \ref{fig:analysis_session_class}) and   regression problems (Figures~\ref{fig:analysis_rows}-\ref{fig:sqlshare_analysis_busy}).}  
\label{fig:analysis_of_labels_regression}
\end{figure*}

%% file: LabelDistribtionFigures.tex
\begin{figure*}
    \centering
		\subfloat[SDSS]
		{
		\includegraphics[width=0.19\linewidth,valign=t]{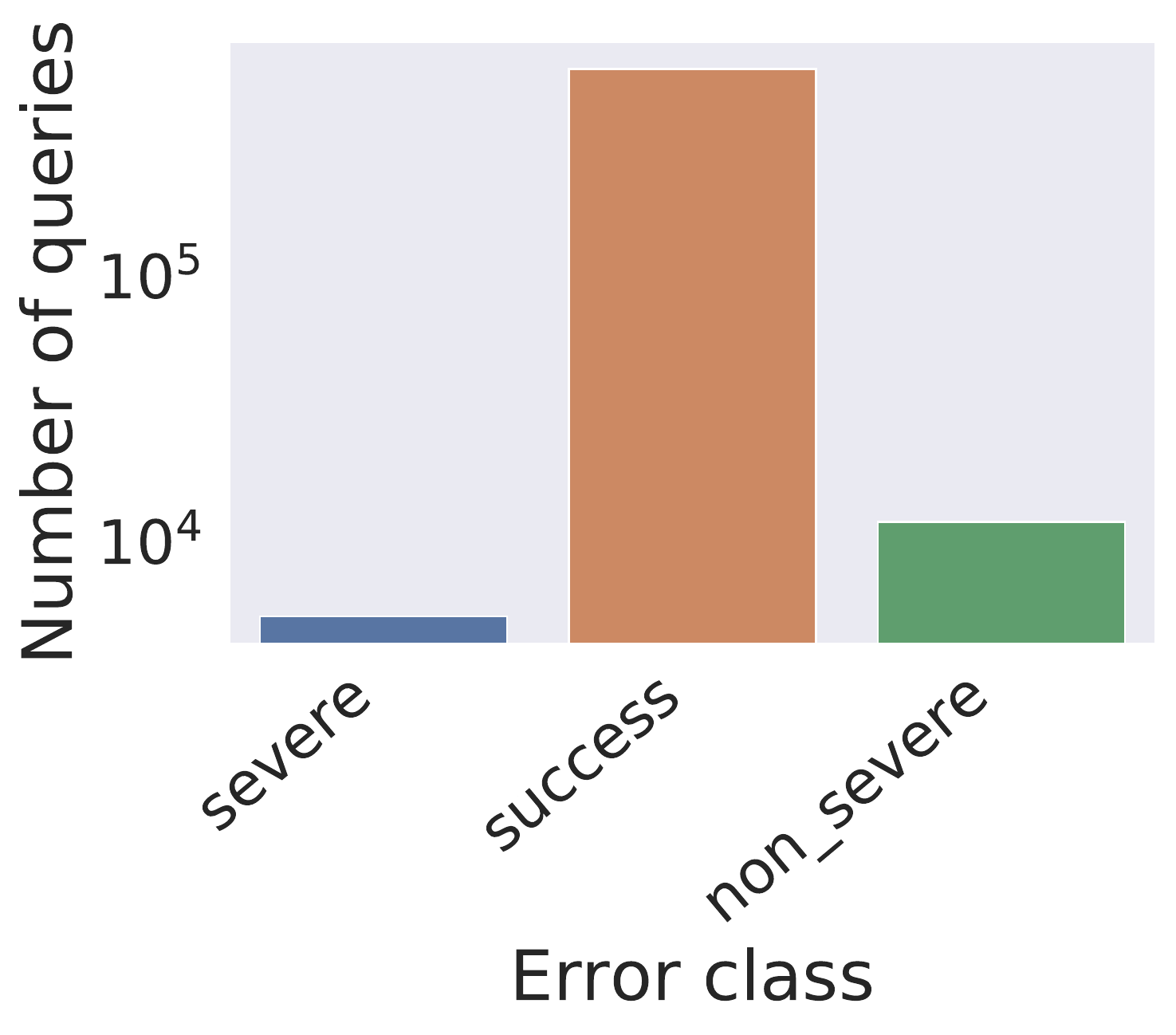}
		\label{fig:analysis_error}
		}
		\subfloat[SDSS]
		{
		\includegraphics[width=0.19\linewidth,valign=t]{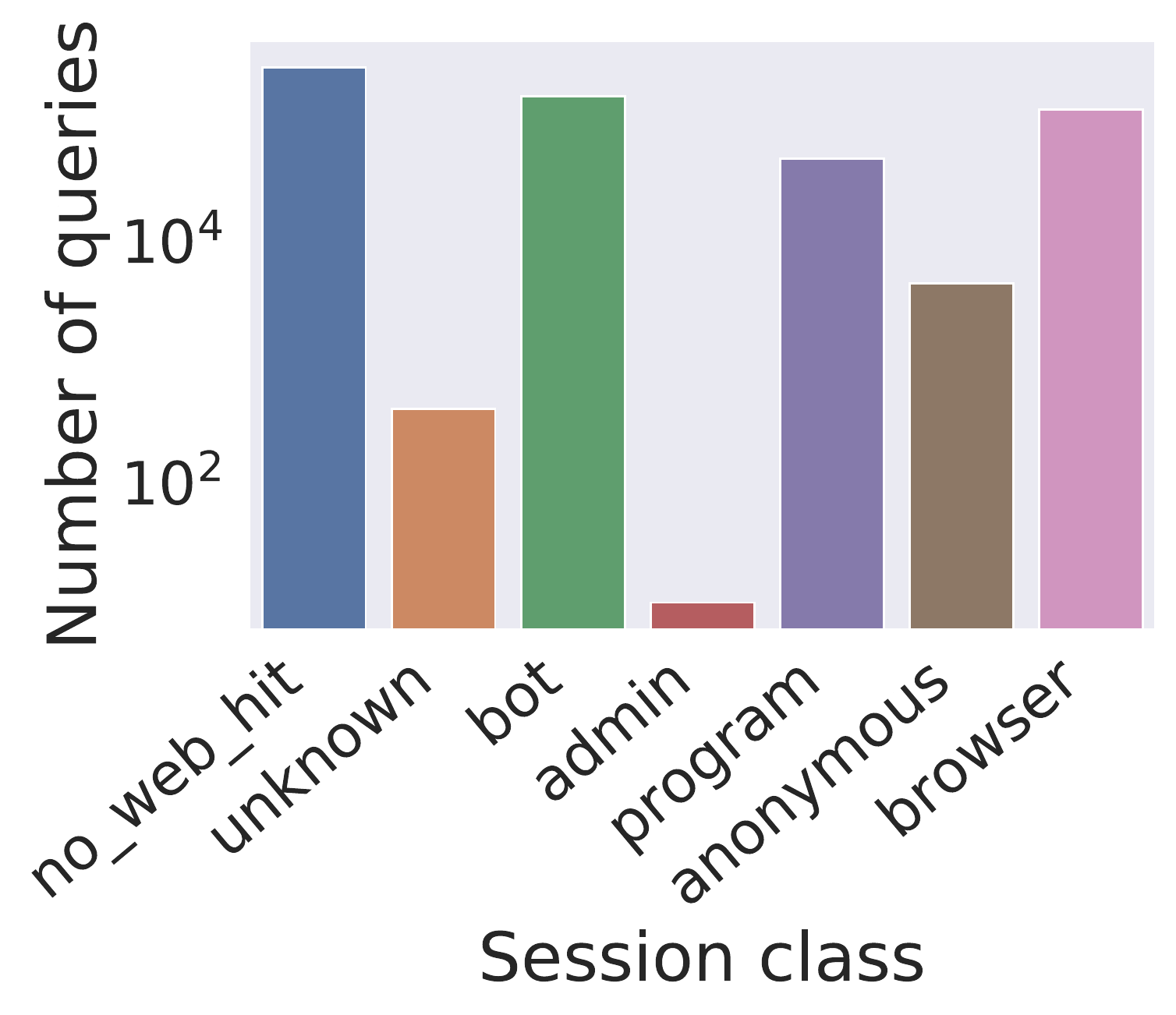}
		\label{fig:analysis_session_class}
		}
		\subfloat[SDSS]
		{
		\includegraphics[width=0.19\linewidth,valign=t]{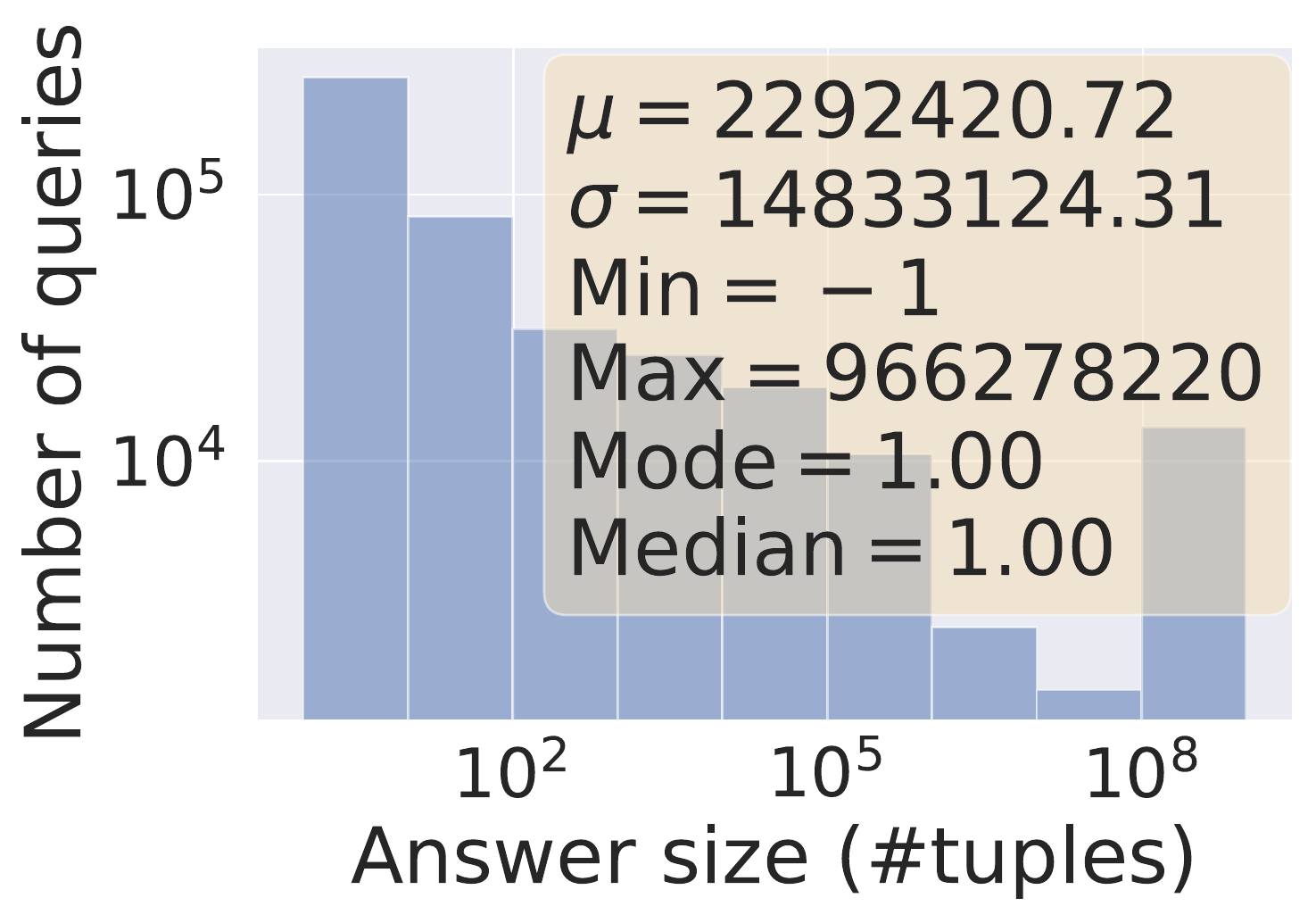}
		\label{fig:analysis_rows}
		}
		\subfloat[SDSS]
		{
		\includegraphics[width=0.19\linewidth,valign=t]{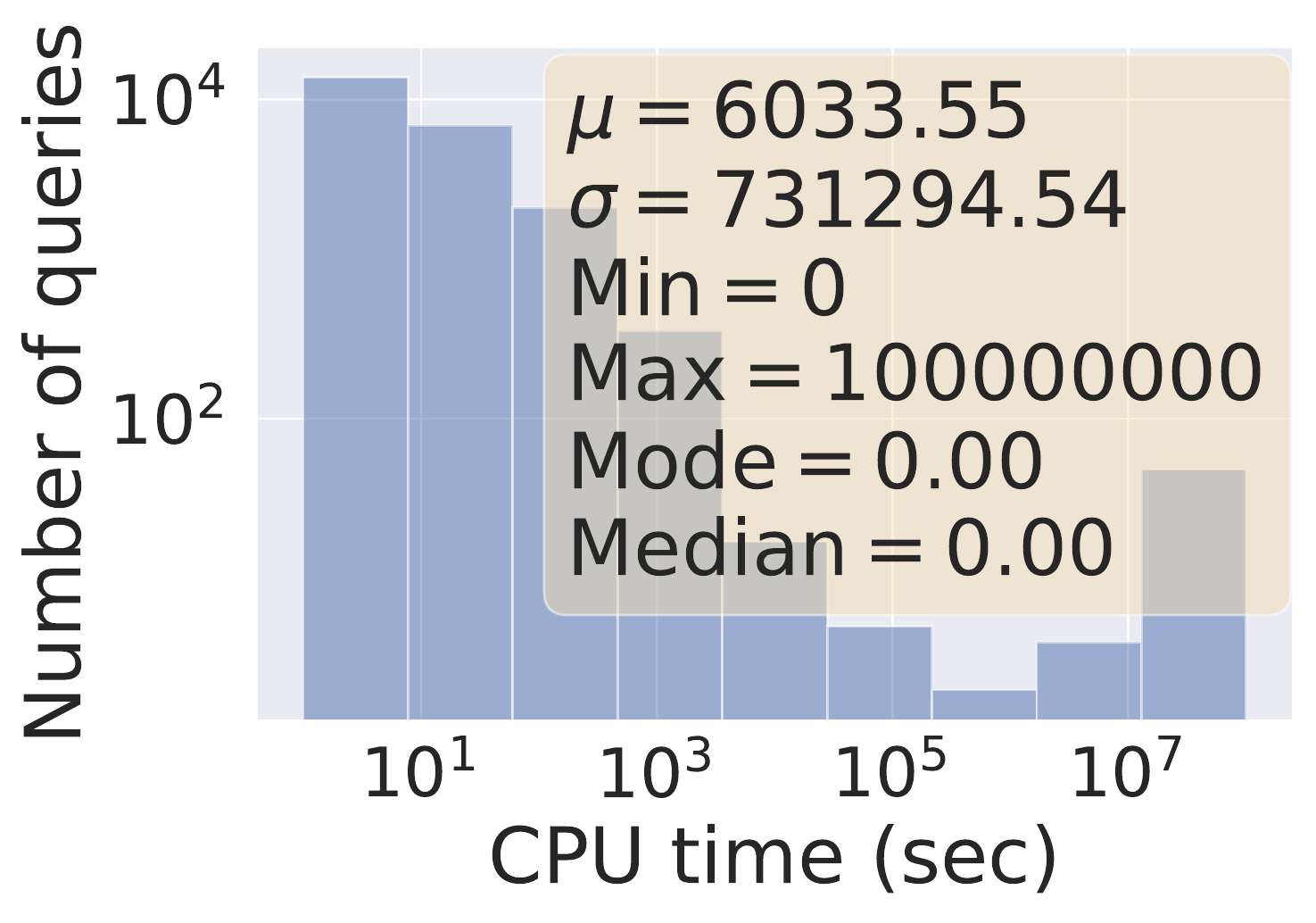}
		\label{fig:analysis_busy}
		}
		\subfloat[SQLShare]
		{
		\includegraphics[width=0.19\linewidth,valign=t]{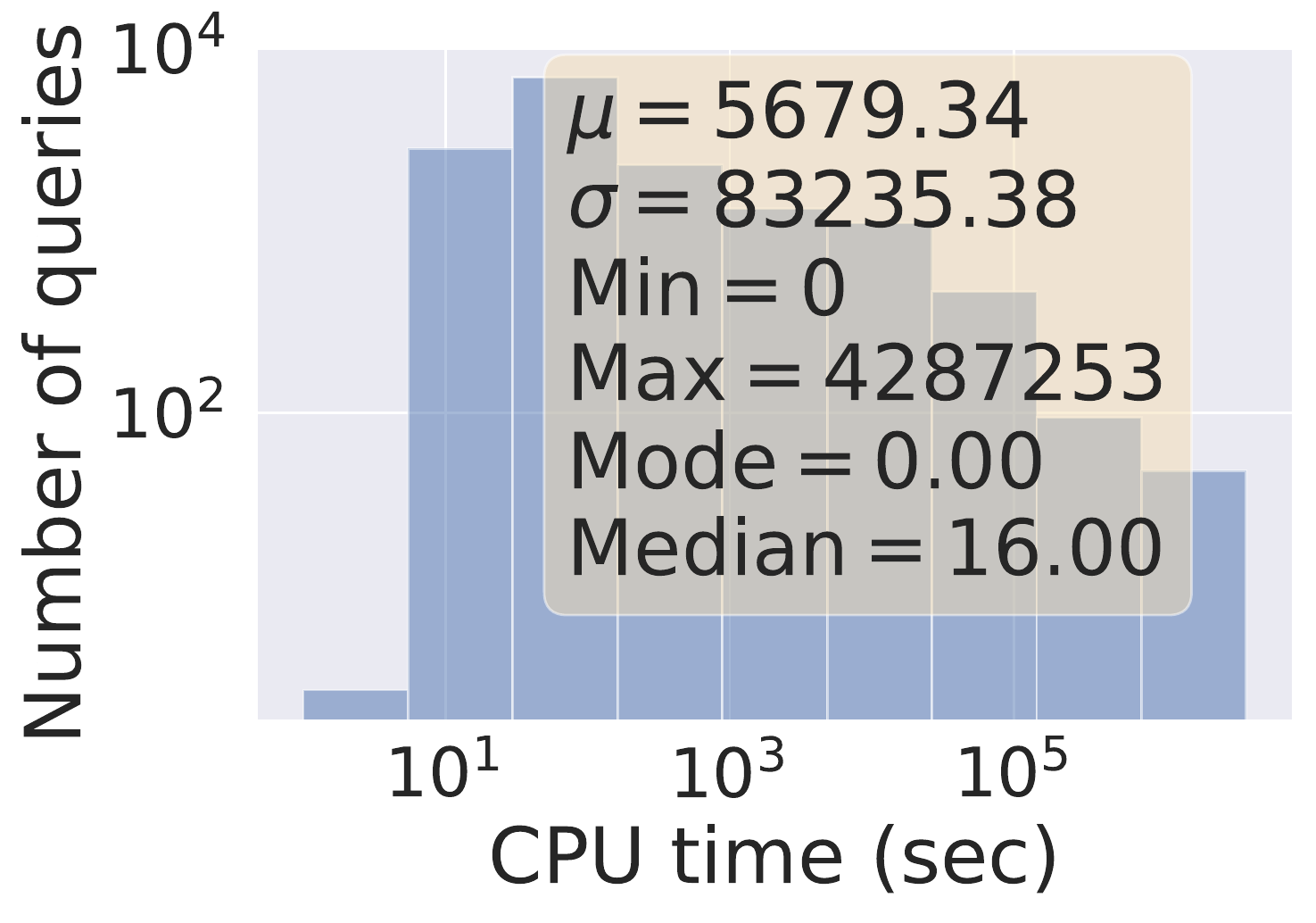}
		\label{fig:sqlshare_analysis_busy}
		}			
		\vspace*{-4mm}		
\caption{Label distributions for classification  (Figures~\ref{fig:analysis_error} and \ref{fig:analysis_session_class}) and   regression problems (Figures~\ref{fig:analysis_rows}-\ref{fig:sqlshare_analysis_busy}).}  
\label{fig:analysis_of_labels_regression}
\end{figure*}

%% file: SessionClassBreakDownFigure.tex
\begin{figure}
\centering
        \subfloat[Answer size]
        {
		\includegraphics[width=0.484\linewidth]{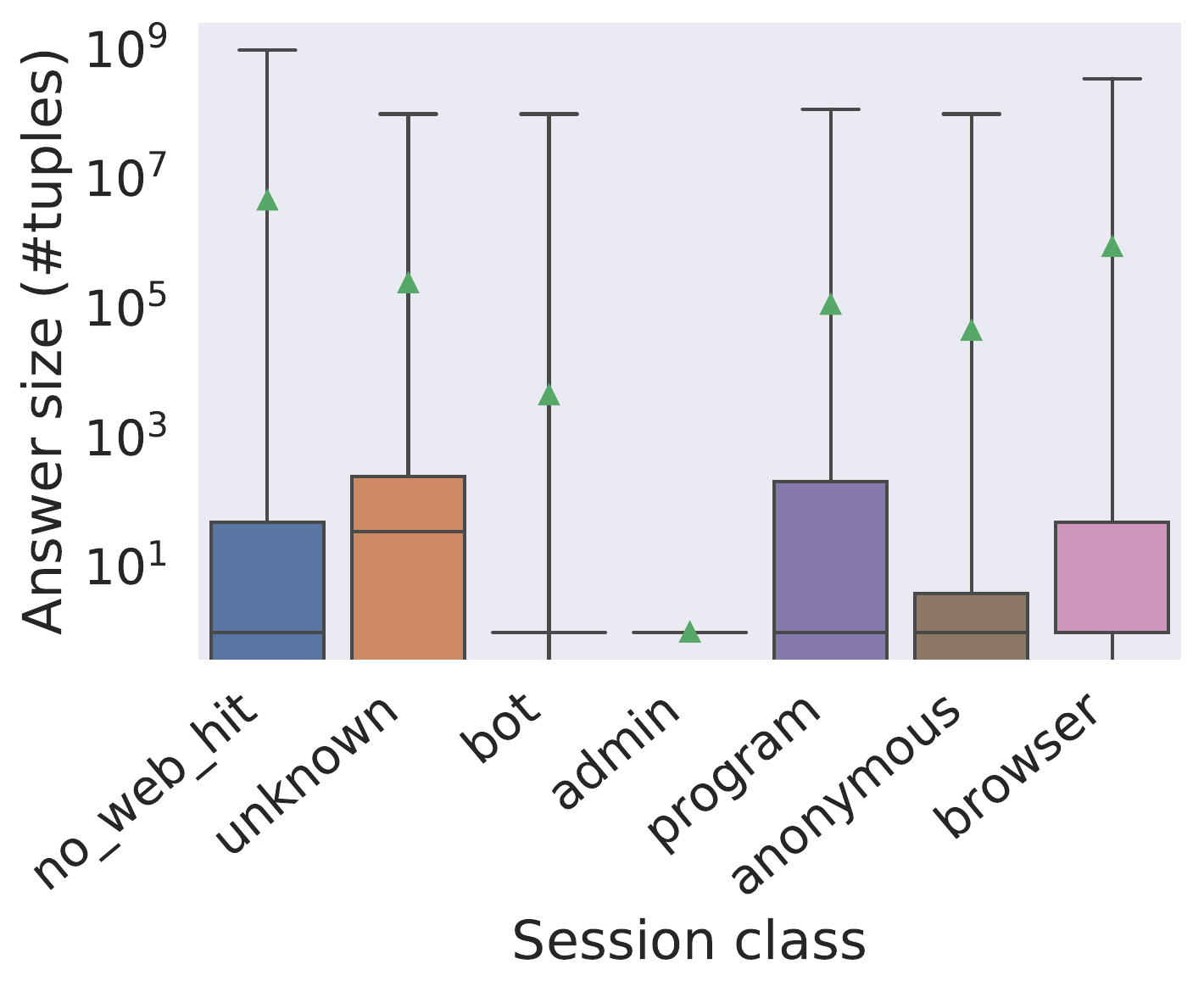} 
		\label{fig:bsa_row}
        }
		\subfloat[CPU time]
        {
		\includegraphics[width=0.484\linewidth]{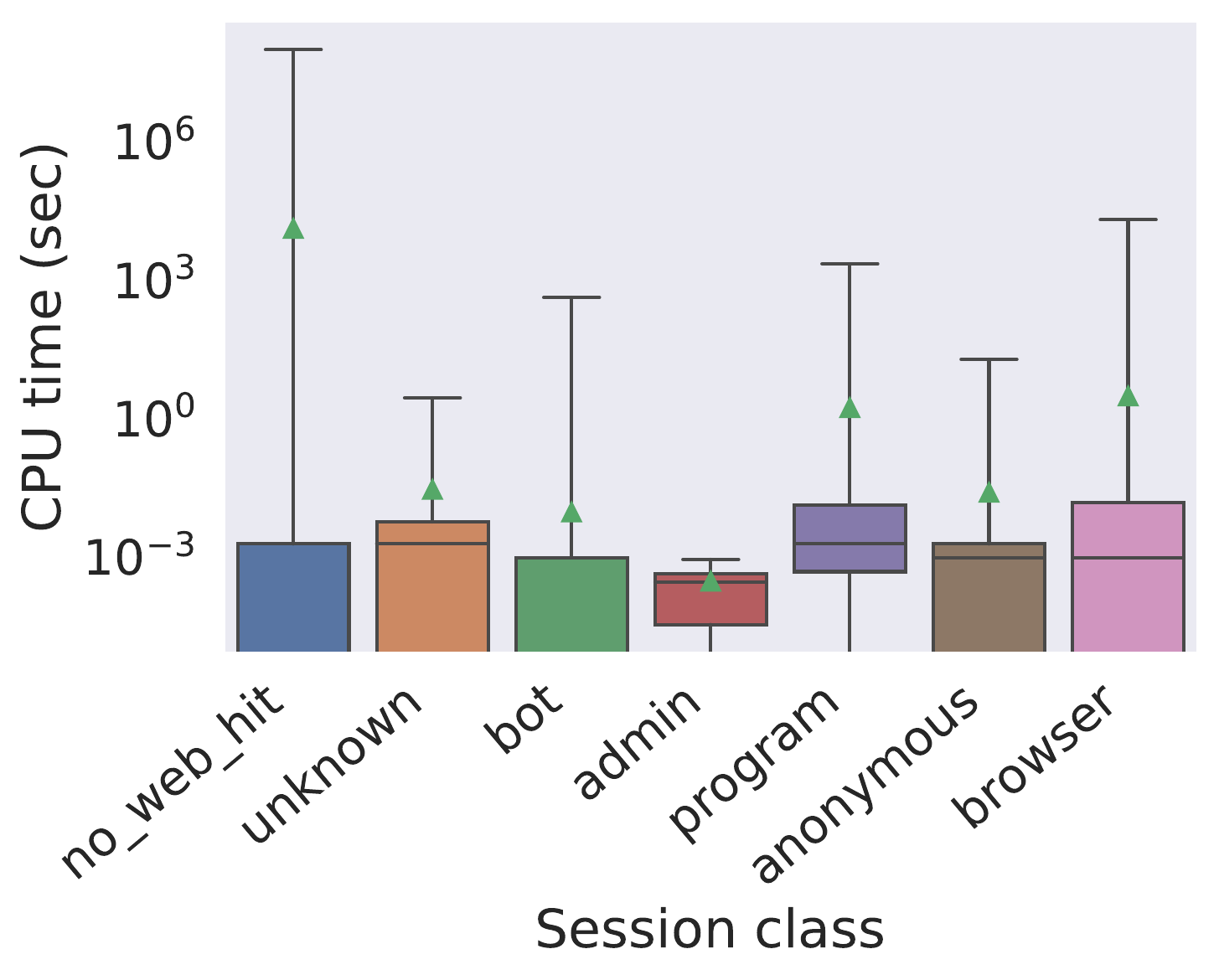} 
		\label{fig:bsa_busy}
        }
        
        \subfloat[Number of characters]
        {
		\includegraphics[width=0.484\linewidth]{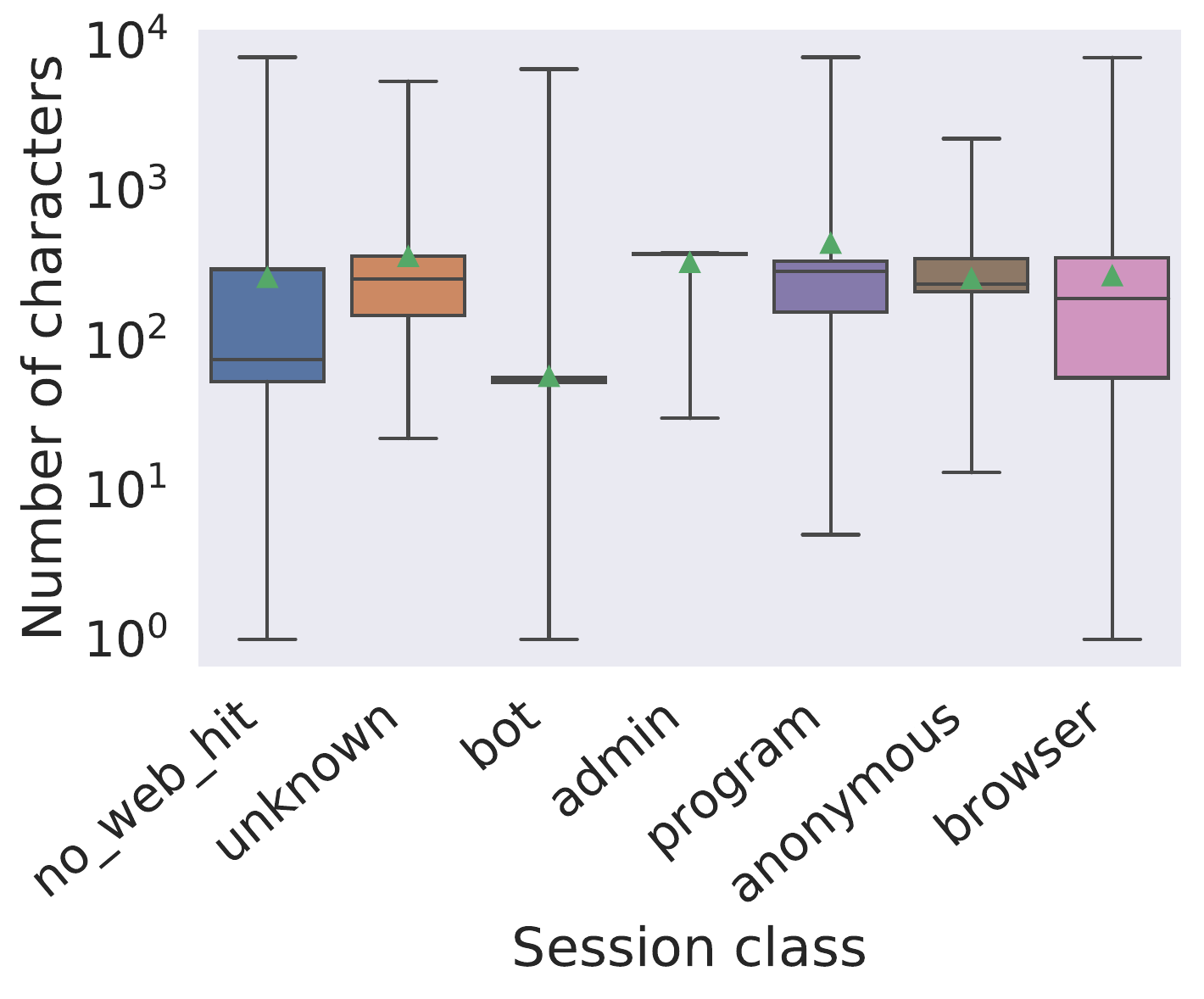} 
		\label{fig:full_bsa_length_char}
        }
        \subfloat[Number of words]
        {
		\includegraphics[width=0.484\linewidth]{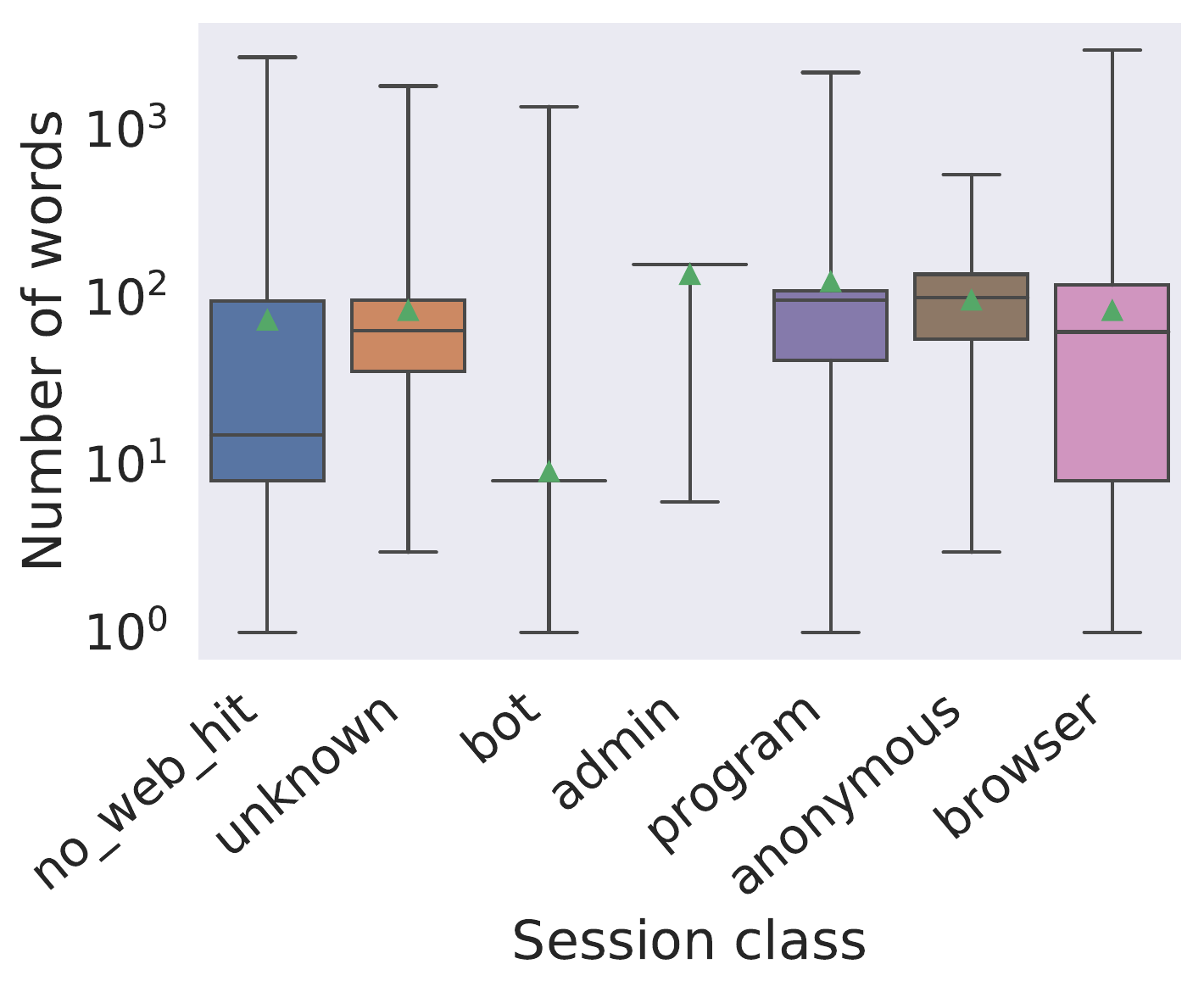} 
		\label{fig:full_bsa_length_word}
        }
        \vspace{-2mm}
\caption{SDSS analysis by session class. 
The top and bottom of each box represent the first and third quartiles of the class data distribution. The horizontal line in each box is the median, the mean is a green triangle. \ignore{The whiskers represent the  full range of the data.}}    
\label{fig:sdss_query_statement_bsa}    
\end{figure}

%% file: Methods.tex
\section{Methods}
\label{sec:dbrec_methods}
Based on our analysis in Section~\ref{subsec:dbrec_analysis_implications_model_selection}, we extend three models from the NLP  domain and benchmark their performance for our problems. In Section~\ref{subsec:dbrec-TFIDF} we describe a traditional model. In Section~\ref{subsec:dbrec-3layerlstm} we describe a three-layer LSTM model and in  Section~\ref{subsec:dbrec-shallowCNN} we describe a shallow CNN.  \iffullpaperComposition Details of the models are in  Appendix~\ref{appendix:dbrec-models}.
 \fi

\subsection{Traditional Model}
\label{subsec:dbrec-TFIDF}
  Traditional machine learning models work in two stages: a feature extraction phase and a prediction phase.
For the feature extraction phase,  Bag-of-ngrams and its TFIDF (term-frequency inverse-document-frequency) are commonly used in NLP applications.
For the Bag-of-ngrams,  we select the most frequent $n$-grams (up to 5-grams) from the training set. These features comprise the domain vocabulary $V$ with size $v$. Thus, this representation  maps each query to a  $v$-dimensional vector obtained by computing the sum of the one-hot representation of the $n$-grams that appear in the query.
Next, we compute the  TFIDF weight of each token $t_i$ in the $v$-dimensional representation of query $Q$ w.r.t. the collection of queries $\mathcal{Q}$. 
In particular, the weight of token  $t_i$ is computed using  
$\nit{TFIDF}(t_i, Q, \mathcal{Q} ) = \nit{TF}(t_i,Q) \times \nit{IDF}(t_i,\mathcal{Q})
$
. Here $TF(t_i,Q)$ is the normalized  frequency of $t_i$ in $Q$. The normalization prevents bias towards longer queries.
The $\nit{IDF}(t_i,\mathcal{Q})$ component describes the discriminative power of $t_i$ in $\mathcal{Q}$ and helps to control for the fact that some tokens are generally more common than other tokens. It can be computed by $\log{\frac{|\mathcal{Q}|}{ 1 +|\{Q \in \mathcal{Q}, t_i \in Q\}|}}$, where the denominator is the number of queries in $\mathcal{Q}$ that contain $t_i$. The TFIDF value increases proportionally to the frequency of a token in a query but is counterbalanced by the frequency of the term in the collection.  \iffullpaperComposition 
As a token appears in more queries, the ratio inside the logarithm approaches $1$, bringing the IDF and TFIDF closer to 0. \fi

We then  apply a prediction model given this fixed $v$-dimensional feature vector. For classification problems, we apply the multinomial logistic regression model. 
For regression problems, we use  Huber loss~\cite{huber1964robust}.
We optimize the parameters of the prediction model using scikit-learn~\cite{scikit-learn}.

\subsection{Three-layer LSTM}
\label{subsec:dbrec-3layerlstm}
Long-Short Term Memory (LSTM)  is a type of recurrent neural network (RNN)~\cite{zaremba2014learning}.  RNNs can process  sequential inputs of arbitrary length. \iffullpaperComposition Figure~\ref{fig:rnn-m}  shows a standard RNN unit. It works by reading the  input sequence one token at a time from left to right.\else
Standard RNN units work by reading the  input sequence one token at a time from left to right. \fi
At every step $i$, a hidden state  $\bm{h}_i \in \mathbb{R}^k$  is emitted, which   is a semantic representation of the sequence of tokens observed so far.   Specifically, $\bm{h}_i$ is produced  using the  recurrent equation $\bm{h}_i = f(\bm{W} \bm{x}_i  + \bm{U}  \bm{h}_{i-1} + \bm{b})$ where $\bm{x}_i \in \mathbb{R}^d$ is the distributed representation of the input token $q_i$, 
  and $\bm{h}_{i-1} \in \mathbb{R}^k$ is the hidden state at $i-1$. The parameters of this RNN unit include  weight matrices $\bm{W}$  and $\bm{U}$, and a bias vector $\bm{b}$.  $f(.)$ is a point-wise non-linear activation function, such as the Sigmoid or Rectified Linear unit (Relu)~\cite{Goodfellow-et-al-2016}. 

 \iffullpaperComposition
\begin{figure}
\centering
       \includegraphics[width=0.75\linewidth]{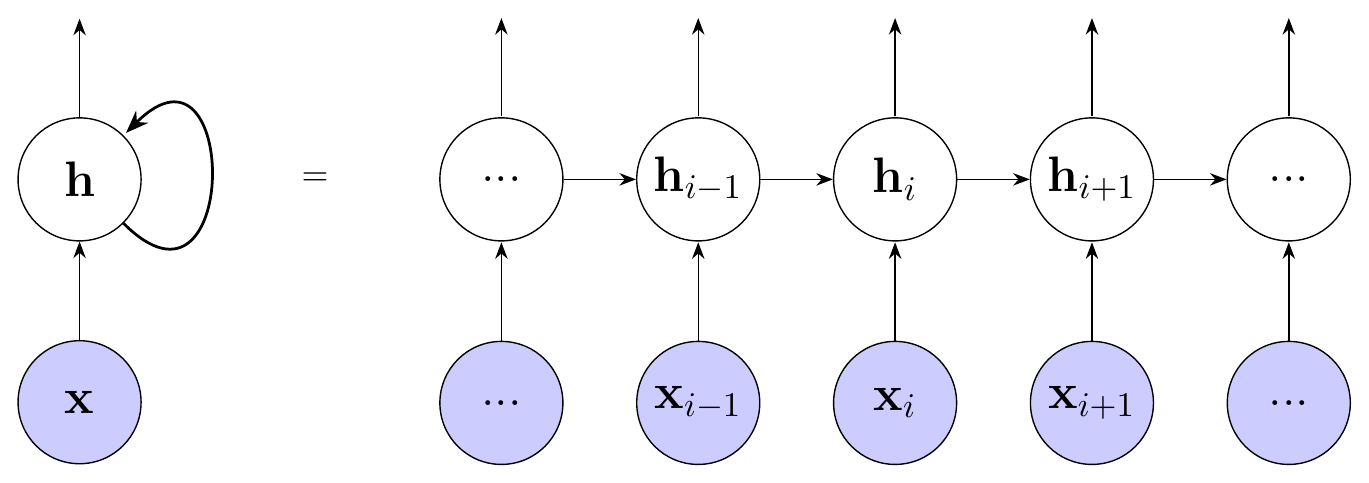}

\caption{An RNN network with hidden-to-hidden recurrent connections.}
\label{fig:rnn-m}
\end{figure}
\fi
Standard RNNs suffer from  the \textit{vanishing gradient} problem. In particular, during training, the gradient vector can grow or decay exponentially~\cite{tai2015improved,Goodfellow-et-al-2016}. LSTMs are a more effective variant of RNNs. They are equipped with a memory cell $\bm{c} \in \mathbb{R}^k$, which helps  preserve the long-term dependencies better than standard RNNs.
  The LSTM unit~\cite{zaremba2014learning} has a  hidden state $\bm{h}_i$ that is a partial view of the unit's memory cell. The unit is equipped with additional parameters and machinery to produce $\bm{h}_i$ from  $\bm{c}_{i-1}$  (memory cell at step $i$), $\bm{x}_i$, and $\bm{h}_{i-1}$\iffullpaperComposition (details in Appendix~\ref{subsec:dbrec-lstm}).\else (details in~\cite{t-report}). \fi

Since well-known RNN  architectures do not exist  for our problems,   we explore those used in similar domains.
 Deep architectures consisting of many layers  are often developed to learn hierarchical representations for the input 
and to learn non-linear functions of the input~\cite{conneau2016very}.
 However, increasing the number of layers and units increases the number of  parameters to learn, and training time increases substantially.

 A two-layer character-level LSTM architecture  was used to predict program execution in~\cite{zaremba2014learning}. Motivated by their success, we  use a three-layered LSTM model. We use the output of the last layer  as the query vector representation. 
For classification problems, we apply the softmax operation to generate the output probability distribution. Similar to the traditional models, we use the cross-entropy loss for classification problems.   For regression problems, we pass the the vector through a linear unit and use Huber loss.
To optimize the network, we examined both Adam and AdaMax~\cite{kingma2014adam}  which are gradient-based optimization techniques that are well suited for problems with large data and many parameters.  We found the latter performed better.

\subsection{Shallow CNN}
\label{subsec:dbrec-shallowCNN}
Convolutional Neural Networks (CNNs) 
are  feed-forward neural networks that   process data  with grid-like topology, e.g., a sequence of concatenated distributed representations of tokens in NLP. Their application in NLP enables the model to extract the most important n-gram features from the input and create a semantic representation.  As a result,  long-term dependencies may not be preserved and  token order information is preserved locally.  CNNs, however,  have comparable performance to RNNs,  they are easier to train, and  are also parallelizable~\cite{young2018recent,bai2018empirical,yin2017comparative}.

\begin{figure}[ht]
\centering
       \includegraphics[width=1\linewidth]{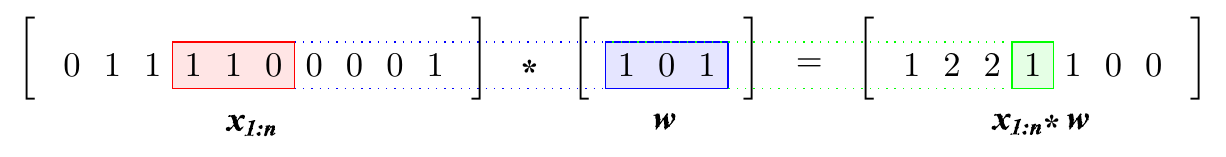}
       \vspace{-8mm}
\caption{1D convolution operation with input $\bm{x}_{1:n}$ and kernel $\bm{w}$. The output $\bm{p}=\bm{x}_{1:n}*\bm{w}$ is produced by sliding $\bm{w}$ over $\bm{x}_{1:n}$ and computing the dot product.}
\label{fig:cnn-op-m}
\end{figure}

Each layer in a CNN consists of  three stages~\cite{Goodfellow-et-al-2016}: a convolution stage, a detection stage, and a pooling stage. We explain each of stages based on a 1D  convolution operation, although higher dimensions are also possible.

The \textbf{convolution stage} applies several  convolution operations. A convolution operator has two operands: a multidimensional array of weights, called the \textit{kernel}, and a multidimensional array of input data. Convolving the input with the  kernel consists of sliding the kernel over all possible windows of the input. At every position $j$,  a linear activation $p_j$ is obtained by  computing  the dot product between the kernel entries and local regions of the input. Figure~\ref{fig:cnn-op-m} shows an example of a 1D convolution operation. Let $\oplus$ denote the concatenation operation, and $\bm{x}_{1:n} = \bm{x}_{1} \oplus \bm{x}_{2} \oplus \dots \oplus \bm{x}_{n}$ denote the concatenated distributed representations $\bm{x}_i$ in an input query (stacked length-wise as a long column).   Let  $\bm{x}_{j:j+m-1}$ represent a window of $m$ words and $\bm{w} \in \mathbb{R}^{m}$ denote a kernel. The dot product of $\bm{w}$ and each $m$-gram in the sequence is computed to obtain $p_j = \bm{w}^{T} \bm{x}_{j:j+m-1}+b$
where  $b \in \mathbb{R}$ is a bias term. By sliding the kernel over all possible windows of the input, we obtain a sequence $\bm{p} \in \mathbb{R}^ {n-m+1}$ ($b=0$ in Figure~\ref{fig:cnn-op-m}).  Note, in the convolution stage, several kernels with varying window sizes   may be convolved with the data to produce different linear sequences. In the \textbf{detector stage}, the linear sequence $\bm{p}$ is run through a non-linear activation function. This results  in a sequence of non-linear activations called  the  activation map $\bm{a} = f(\bm{p})$,
where $\bm{a} \in \mathbb{R}^{n-m+1}$ and $f(.)$ is a non-linear activation function, e.g., Relu. In the \textbf{pooling stage} a pooling operation is applied to the activation map to summarize its values which also enables the model to handle inputs of varying size. For example, the max-pooling function  returns the maximum, i.e.,~ $g = \max\{\bm{a}\}$. 

\begin{figure}
\centering
       \includegraphics[width=1\linewidth]{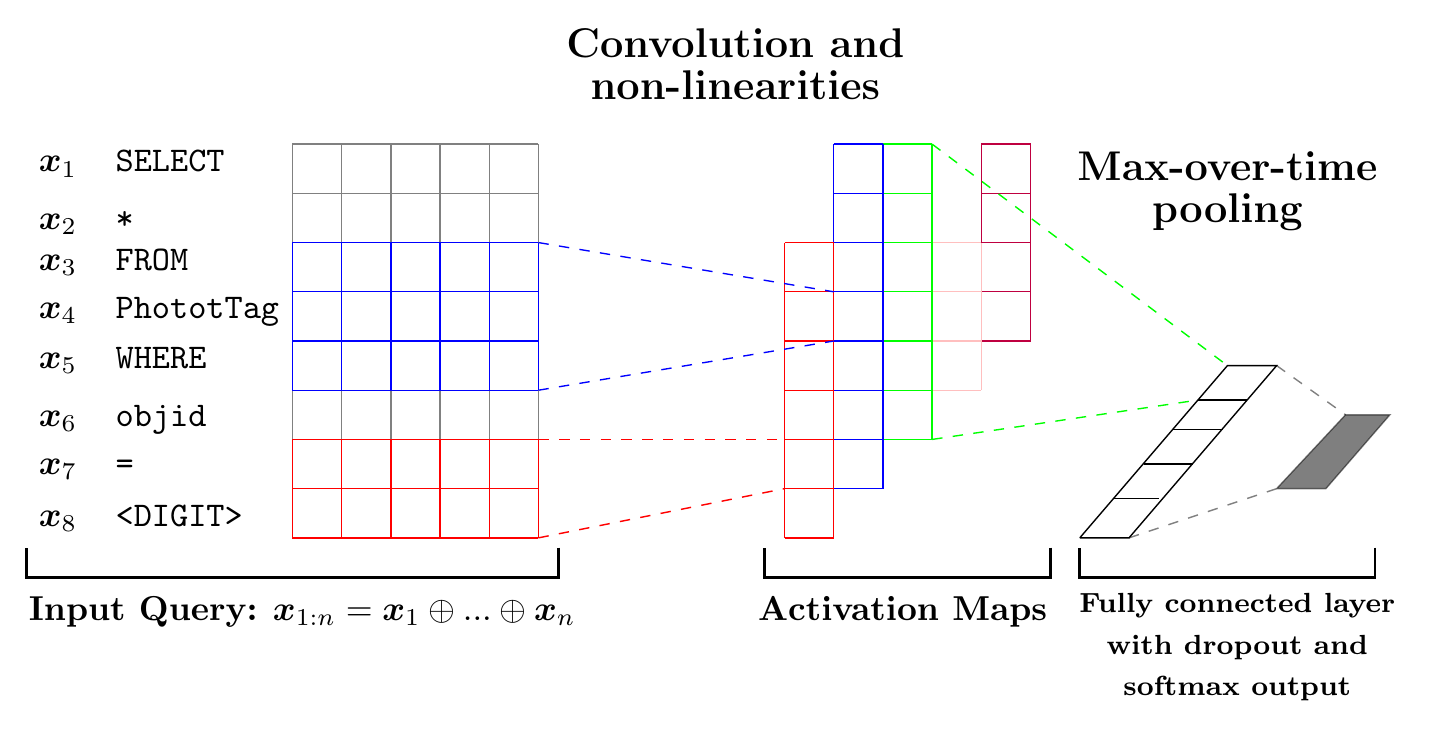}
       \vspace{-8mm}
\caption{1-layer CNN, adapted from~\protect\cite{kim2014convolutional}.}	
	\label{fig:traditional-convolutional-network}
\label{fig:cnn-m}
\end{figure}

Figure~\ref{fig:cnn-m} shows the shallow CNN architecture in~\cite{kim2014convolutional}, which we adapt for our application. 
The input query $\bm{x}_{1:n}$ and convolution operations are shown in 2D for easier presentation.  In the convolution stage, several filters of varying window size  $m \in \{3,4,5\}$ are applied, and the resulting sequence $\bm{p}_1,\bm{p}_2,...$ is passed through a Relu function to generate activations $\bm{a}_1,\bm{a}_2,...$ . Different size sequence inputs and kernels result in activation maps of different sizes. To deal with variable length of the input and also obtain the most important feature, a max pooling operation is applied to obtain a single feature  per kernel, $g = \max\{\bm{a}\}$. The resulting features for all kernels are concatenated to produce a fixed size vector $\bm{g} \in \mathbb{R}^{K}$, where $K$ is the number of kernels. This output is   used to create a fully connected layer, followed by a dropout layer.



We tried changing the architecture by increasing the number of kernels and the window size but did not obtain significant improvements. Similar to the other models,  for classification problems we apply the softmax operation to generate the output probability distribution. We use the cross-entropy loss for classification problems.  For regression problems, we pass the the vector through a linear unit and use Huber loss to learn the parameters.  We used AdaMax as the  optimizer.


%% file: ExperimentalSetup.tex
\section{Empirical Evaluation}
\label{sec:dbrec_emp_eval}
We evaluate the performance of models in Section~\ref{sec:dbrec_methods} on the four query facilitation problems, considering two aspects:
\begin{enumerate*}
\item the different query facilitation problem settings and
\item  the query statement complexity (described in Section~\ref{subsec:dbrec_statement_complexity}).
\end{enumerate*}

\subsection{Setup}
\label{sec:dbrec_exp_setup}
\noindent \textbf{Data split. }
For \settingOne, we used our extracted SDSS workload. For \settingTwo, we used SQLShare.  In both settings, we randomly split the queries. For \settingThree, we used SQLShare and randomly  split the data based on users, so as to  decrease the likelihood of  data sharing.
\iffullpaperComposition Table~\ref{tab:sdss_extracted data} summarizes our datasets. \else The SDSS workload has a total  of 618,053 queries. SQLShare has a total of 26,728 queries. In all three settings, we  used a  (80/10/10) split  for the train, validation, and test sets, respectively.  \fi

\ignore{\begin{table}
\centering
\small
\begin{tabular}{rrll}
\toprule
{\bf Setting} &{\bf Workload} & & {\bf \#SQL queries}\\
  \toprule
1& SDSS & Total &  618,053\\
&& Train & 494,443\\
&& Validation & 61,805\\
&& Test & 61,805\\
\midrule
2& SQLShare &Total & 26,728 \\
&&Train & 21,382 \\
&&Validation & 2,673 \\
&&Test & 2,673 \\
\midrule
3&  SQLShare&Total & 26,728 \\
&&Train & 22,068 \\
&&Validation & 1,893 \\
&&Test & 2,767 \\
\bottomrule
\end{tabular}
\caption{Workload and data split description.}
\label{tab:sdss_extracted data}
\end{table}}

\iffullpaperComposition
\begin{table}
\centering
\begin{tabular}{llll}

    & {\bf Homogeneous} & {\bf Homogeneous} & {\bf Heterogeneous}\\
& {\bf Instance} & {\bf Schema} & {\bf Schema}\\


\toprule

\hspace{-5pt}Total &  618,053 & 26,728 & 26,728 \\
\hspace{-5pt}Train & 494,443 & 21,382 & 22,068 \\
\hspace{-5pt}Valid. & 61,805 & 2,673 & 1,893  \\
\hspace{-5pt}Test & 61,805 & 2,673 & 2,767  \\
\bottomrule
\end{tabular}
\caption{The number of queries and data split in SDSS (\settingOne), and SQLShare (\settingTwo and \settingThree).}
\label{tab:sdss_extracted data}
\end{table}
\fi

\vspace{2mm}
\noindent \textbf{Methods compared. }
We compare the models in Section~\ref{sec:dbrec_methods}, where   character-level model names begin with c and word-levels models with w. The traditional models are \ctfidf  and \wtfidf. The 3-layer LSTM models are \clstm and \wlstm.  The CNN models are  \ccnn and \wcnn. For each prediction task, we also include a simple baseline. For classification problems,  \mfreq predicts the most frequent label, i.e., it predicts \scss for error classification, and \nowebh for session class prediction.
For regression problems,  \median predicts  the median of the corresponding train distribution, i.e., the median of train answer size distribution is 1.099, 
and the median of the  train CPU time distribution is 0. 
Following~\cite{li2012robust,ganapathi2009predicting,akdere2012learning} we report results for an \opt model which uses linear regression to predict  CPU time from the query optimizer estimates cost estimates.

\vspace{2mm}
\noindent \textbf{Hyper-parameter tuning. }
We tune the hyper-parameters based on  \settingOne (SDSS).
To keep this problem tractable, we restrict the set of hyper-parameters of each model 
 and  choose the best set of hyper-parameters based on performance on the validation set.  
  We fixed the  learning rate  1e-3, batch size to 16, \iffullpaperComposition token embedding size  to $100$, and weight decay set to 0
  \else 
  and token embedding size  to $100$.
  \fi For   \clstm and \wlstm, we tested the number of hidden dimensions in $\{150, 300\}$ and clipping rate in  $\{0.25,0\}$.
For   \ccnn and\ wcnn, we tested number of kernels in $\{100,250\}$,
 drop out  in $\{0.5,0\}$, and clipping rate in $\{0.25, 0\}$.
 We report results for the best performing model and  also use the model in \settingTwo and \settingThree settings.
 \ifPhDThesis
Full results are in Appendix~\ref{appendix:deb_rec_additional_results}.
\fi

\vspace{2mm}
\noindent \textbf{Performance metrics. }
For the classification problems we report the test average loss  computed according to cross entropy (loss)\iffullpaperComposition Eq.~\ref{eq:dbrec_class_loss}
\fi. We  also report Accuracy, which is the number of correct predictions divided by the total number of predictions.
Due to the class imbalance for both error and session classification, for every class $C$,  we  report the per class F-measure computed by  $\text{F}_C= \frac{2 . \text{Precision}_C.\text{Recall}_C}{ \text{Precision}_C + \text{Recall}_C}$.   $\text{Precision}_C$ is the number of correct predictions for class $C$ divided by the total number predictions for class $C$. $\text{Recall}_C$   is the number of correct predictions for class $C$ divided by the total number queries in  class $C$.  
For the regression problems we report the   test average loss computed according to Huber loss\iffullpaperComposition  (Eq.~\ref{eq:dbrec_reg_loss})
\fi
. We also  report Mean Square Error (MSE)  computed as $\text{MSE} = \frac{1}{m} \sum_{i}^m (y'_{i}  - \hat{y}_{i})^2$,
where $y'_{i} $ is the log-transformed label (CPU time or answer size) of query $i$, and  $\hat{y}_{i}$ is the predicted value.  We also use qerror which measures  the quality of  estimates~\cite{leis2015good}. The qerror of a query $Q_i$ is $\nit{q-error}_i=\max(\frac{y_i}{\hat{y_i}},\frac{\hat{y_i}}{y_i})$.

%% file: Results.tex
\subsection{Model Performance}
\label{sec:results}

\ifPhDThesis
\begin{table*}
\centering
\small
\begin{tabular}{llllllll}
\toprule
{\bf Model} & {\bf $v$} & {\bf \prm} &  {\bf Loss} & {\bf F$_\severe$} & {\bf F$_\scss$} & {\bf F$_\nsevere$} &  {\bf Accuracy}  \\ \hline
\input{results/results-final/sdss-results/tex/error-sel-short-v.tex}
\bottomrule
\end{tabular}
\caption{Query error classification in \settingOne (SDSS). 
Here  $v$ is \#tokens in the vocabulary, \prm is \#model parameters, Loss is the average test loss (lower loss is better). F$_{\tt C}$ is the F-measure of class {\tt C}. The \#samples in the test set for each class are:
\severe $=533$,
\scss $=60138$,
\nsevere $=1134$.
}
\label{tab:error_prediction_results_sel}
\end{table*}

\else
\begin{table*}
\centering
\small
\begin{tabular}{llllllll|ll|ll}
& & \multicolumn{6}{c}{{\bf Error Classification}} & \multicolumn{2}{c}{{\bf CPU Time}} & \multicolumn{2}{c}{{\bf Answer Size}} \\
\toprule
{\bf Model} & {\bf $v$} & {\bf \prm} &  {\bf Accuracy} & {\bf F$_\severe$} & {\bf F$_\scss$} & {\bf F$_\nsevere$} &  {\bf Loss} & {\bf $p$} &  {\bf Loss} & {\bf $p$} &  {\bf Loss}  \\ \hline
\input{modified-results/merge.tex}
\bottomrule
\end{tabular}
\caption{Query error classification (left), CPU time (middle), and answer size (right) prediction in \settingOne (SDSS). Here  $v$ is \#tokens in the vocabulary, \prm is \#model parameters, Loss is the average test loss (lower loss is better). {\tt baseline} is \median in the regression problems and \mfreq in the classification problem. In the classification problem, F$_{\tt C}$ is the F-measure of class {\tt C}. The \#samples of each class in the test set are:
\severe $=533$,
\scss $=60138$,
\nsevere $=1134$.
}
\label{tab:error_prediction_results_sel}
\end{table*}
\fi

\subsubsection{\settingOne}
\label{sec:results_setting1}
\ifPhDThesis
 Table~\ref{tab:error_prediction_results_sel}
 \else
 Table~\ref{tab:error_prediction_results_sel} (left)
 \fi  shows the error classification results. The \mfreq baseline achieves a high   $\text{F}_\scss$ but performs poorly w.r.t.~other classes. All other models  improve upon this baseline. The \ccnn model obtains a high $\text{F}_\severe= 0.7961$ and has the highest test accuracy.
\ifPhDThesis Table~\ref{tab:busy_prediction_results_sel}
 \else Table~\ref{tab:error_prediction_results_sel} (middle)
 \fi
 shows results for CPU time prediction. The  \wcnn model obtains the lowest test loss, followed closely by the \wlstm model.
\ifPhDThesis
  Table~\ref{tab:busy_prediction_results_sel} also
  \else
 Table~\ref{tab:error_prediction_results_sel} (right)
 \fi shows results for  answer size prediction. The \ccnn model obtains the lowest test loss, followed closely by the \clstm model. In Section~\ref{sec:dbrec_performance_by_session_class} we assess both problems w.r.t.~MSE values.
Table~\ref{tab:session_class_prediction_results_sel} shows results for session  classification. Again the \mfreq baseline achieves a high $\text{F}_{\nowebh}$ (the majority class) but  under-performs all other models w.r.t. other classes. The highest test accuracy is obtained by the \ctfidf model, which
outperforms other models in the F-measure of individual classes, except for $\text{F}_{\unknown}$ which is 0.
$\text{F}_{\admin}$ is 0 since \admin  only has 2 queries in  the test set.

Table~\ref{tab:qerror} shows the qerror of the answer size predictions of 62K test queries in SDSS. The percentage of  queries that have at most the reported qerror is shown, e.g.,~ the qerror of $75\%$ of the test queries is less than $2.38$ for \clstm. Intuitively, qerror for answer size is the factor by which an estimate differs
from the true answer size. We observe \ccnn and \clstm have lowest qerrors.  Note, all models perform well for $50\%$ of the queries and the main comparison is for the other $50\%$ of the queries for which prediction is more difficult.

\begin{table}
\centering
\small
\begin{tabular}{lllllll}
\toprule
{\bf Model} & {\bf $50\%$} & {\bf $75\%$} & {\bf $80\%$} & {\bf $85\%$} & {\bf $90\%$} & {\bf $95\%$} \\ \hline
\ifPhDThesis
\input{modified-results/qerror-rows-thesis.tex}
\else
\input{modified-results/qerror-rows.tex}
\fi
\bottomrule
\end{tabular}
\caption{Answer size prediction qerror (SDSS).}
\label{tab:qerror}
\end{table}

\ignore{\begin{table}
\centering
\begin{tabular}{lllll}
\toprule
Model & $v$ & \prm & Loss  \\ \hline
\input{results/results-final/sdss-results/tex/busy-sel-short-v.tex}
\bottomrule
\end{tabular}
\caption{SDSS Query CPU time prediction.  $v$ is the number of tokens in the vocabulary, \prm is the number of model parameters,   Loss is the average test loss.  Lower loss is better. }
\label{tab:busy_prediction_results_sel}
\end{table}

\begin{table}
\centering
\small
\begin{tabular}{lllll}
\toprule
Model & $v$ & \prm  & Loss  \\ \hline
\input{results/results-final/sdss-results/tex/rows-sel-short-v.tex}
\bottomrule
\end{tabular}
\caption{SDSS Query answer size prediction.  $v$ is the number of tokens in the vocabulary, \prm is the number of model parameters,   Loss is the average test loss.  Lower loss is better. }
\label{tab:rows_prediction_results_sel}
\end{table}}

\ifPhDThesis
\begin{table}
\centering
\small
\begin{tabular}{llll|ll}
 & & \multicolumn{2}{c}{{\bf CPU Time}} & \multicolumn{2}{c}{{\bf Answer Size}} \\
\toprule
{\bf Model} & {\bf $v$} & {\bf \prm} & {\bf Loss} & {\bf \prm} & {\bf Loss} \\ \hline
\input{modified-results/busy-rownumber.tex}
\bottomrule
\end{tabular}
\caption{Query CPU time (middle) and answer size (right) prediction, in \settingOne (SDSS)\iffullpaperComposition $v$ is the vocabulary size, \prm is \#model parameters, Loss is the average test loss.\fi}
\label{tab:busy_prediction_results_sel}
\end{table}
\fi

\ifPhDThesis
\begin{landscape}
\begin{table*}
\centering
\small
\begin{tabular}{llllllllllll}
\toprule
{\bf Model} & {\bf $v$} &  {\bf \prm} &  {\bf Loss} &  {\bf F$_\nowebh$} & {\bf F$_\unknown$} & {\bf F$_\bott$} & {\bf F$_\admin$} & {\bf F$_\browser$} & {\bf F$_\anonym$} &  {\bf F$_\browser$} &  {\bf Accuracy} \\ \hline
\input{results/results-final/sdss-results/tex/session_class-sel-short-v.tex}
\bottomrule
\end{tabular}
\caption{Query Session  classification in \settingOne (SDSS). Here $v$ is the number of tokens in the vocabulary, \prm is the number of model parameters,   Loss is the average test loss. Lower loss is better.
 F$_{\tt C}$ is the F-measure of class {\tt C}. High F-measure and accuracy are better. Note, the \#examples in the test set for each class is:
\nowebh $ = 27677$,
\unknown $ = 42$,
\bott $ = 16148$,
\admin $ = 2$,
\program $ = 4882$,
\anonym $ = 467$,
\browser $ = 12587$.
}
\label{tab:session_class_prediction_results_sel}
\end{table*}
\end{landscape}
\else

\begin{table*}
\centering
\small
\begin{tabular}{lllllllllll}
\toprule
{\bf Model} & {\bf $v$} &  {\bf \prm} &  {\bf Loss} &  {\bf F$_\nowebh$} & {\bf F$_\unknown$} & {\bf F$_\bott$} \ignore{& {\bf F$_\admin$}} & {\bf F$_\program$} & {\bf F$_\anonym$} &  {\bf F$_\browser$} &  {\bf Accuracy} \\ \hline
\input{modified-results/session_class-sel-short-v-modified.tex}
\bottomrule
\end{tabular}
\caption{Query Session  classification in \settingOne (SDSS). \ignore{Here $v$ is the number of tokens in the vocabulary, \prm is the number of model parameters,   Loss is the average test loss. Lower loss is better.
 F$_{\tt C}$ is the F-measure of class {\tt C}. High F-measure and accuracy are better. Note,}The \#examples in the test set for each class is:
\nowebh $ = 27677$,
\unknown $ = 42$,
\bott $ = 16148$,
\ignore{\admin $ = 2$,}
\program $ = 4882$,
\anonym $ = 467$,
\browser $ = 12587$.
}
\label{tab:session_class_prediction_results_sel}
\end{table*}
\fi

\subsubsection{\settingTwo}
\label{sec:results_setting2}

Table~\ref{tab:sqlshare_busy_prediction_results_sel} reports performance for CPU time prediction in \settingTwo. The \ccnn model outperforms other models. Compared to \settingOne, the  overall loss value is higher for all models in \settingTwo. This is because the latter  poses an additional challenge where the distribution of the queries in individual database instances is different, and to get accuracy compared to \settingOne, we need to increase model capacity (e.g., add more layers in the architecture).  Moreover, observe that the \opt model, that is based on the query optimizer cost model,  is closer to median in it's error. Our qerror analysis for 2,674 test queries in SQLShare shows \ccnn performs better across different percentiles. For $50\%$ and $75\%$ of the queries, qerror is less than $1.94$ and $27$,
\iffullpaperComposition
resp (Table~\ref{tab:qerror-cputime-settingtwo}).
\else
resp(see~\cite{t-report}).
\fi

\begin{table}
\centering
\small
\begin{tabular}{llll|ll}
& & \multicolumn{2}{l}{\shortstack{{\bf Homogeneous}\\ {\bf Schema}}\ignore{\blue{{\bf Hom. Sch.}}}\ignore{\settingTwo}} & \multicolumn{2}{l}{\shortstack{{\bf Heterogeneous}\\ {\bf Schema}}\ignore{\blue{Het. Sch.}}\ignore{\settingThree}} \\
\toprule
{\bf Model} & {\bf $v$} & {\bf \prm} & {\bf Loss} & {\bf \prm} & {\bf Loss}  \\ \hline
\ifPhDThesis
\input{modified-results/busy-setting-2-3.tex}
\else
\input{modified-results/busy-setting-2-3-modified.tex}
\fi
\bottomrule
\end{tabular}
\caption{Query CPU time prediction (SQLShare)\ignore{ in \settingTwo (middle) and \settingThree (right).  $v$ is the number of tokens in the vocabulary, \prm is the number of model parameters,   Loss is the average test loss.  Lower loss is better.} }
\label{tab:sqlshare_busy_prediction_results_sel}
\end{table}

\ignore{\begin{table}
\centering
\begin{tabular}{lllll}
\toprule
Model & $v$ & \prm & Loss  \\ \hline
\input{results/results-final/sqlshare_dataset-results/tex/busy-sel-short-v.tex}
\bottomrule
\end{tabular}
\caption{Query CPU time prediction (SQLShare).  $v$ is the number of tokens in the vocabulary, \prm is the number of model parameters,   Loss is the average test loss.  Lower loss is better. }
\label{tab:sqlshare_busy_prediction_results_sel}
\end{table}

\begin{table}
\centering
\begin{tabular}{lllll}
\toprule
Model & $v$ & \prm & Loss  \\ \hline
\input{results/results-final/sqlshare_dataset_by_users-results/tex/busy-sel-short-v.tex}
\bottomrule
\end{tabular}
\caption{SQLShare Query CPU time prediction different split.  $v$ is the number of tokens in the vocabulary, \prm is the number of model parameters,   Loss is the average test loss.  Lower loss is better. }
\label{tab:sqlshare_byusers_busy_prediction_results_sel}
\end{table}}

\subsubsection{\settingThree}
\label{sec:results_setting3}
Table~\ref{tab:sqlshare_busy_prediction_results_sel} reports  performance for CPU time prediction in \settingThree. Similar to \settingTwo, the \ccnn model outperforms others. However, compared to \settingOne and \settingTwo,  the loss value achieved by all models is higher. This is expected since the data is extracted from databases with different schemata, which makes it more challenging for the models to predict, i.e.,~the  train and test sample distributions are different.   For \opt,  prediction is more difficult, too. As explained in~\cite{akdere2012learning},  the query optimizer cost model assumes I/O is most time consuming,  even though certain  computations (e.g., nested aggregates over numeric types) are performed in memory. Moreover, a non-linear regression model may improve performance of  \opt. Our qerror analysis
\iffullpaperComposition
 shown in Table~\ref{tab:qerror-cputime-settingthree},
\else
 (see~\cite{t-report})
 \fi
  shows \ccnn performs better across different percentiles in \settingThree. For $30\%$ of the queries, qerror is less than $34$. The substantial qerror increase means prediction is harder in \settingThree.

\iffullpaperComposition

\begin{table}
\centering
\small
\begin{tabular}{lllllll}
\toprule
{\bf Model} & {\bf $40\%$} & {\bf $50\%$} & {\bf $60\%$} & {\bf $70\%$} & {\bf $75\%$} & {\bf $80\%$} \\ \hline

\input{modified-results/qerror-cputime-settingtwo.tex}

\bottomrule
\end{tabular}
\caption{CPU time prediction qerror (SQLShare, \settingTwo).}
\label{tab:qerror-cputime-settingtwo}
\end{table}

\begin{table}
\centering
\small
\begin{tabular}{lllllll}
\toprule
{\bf Model} & {\bf $10\%$} & {\bf $20\%$} & {\bf $30\%$} & {\bf $40\%$} & {\bf $50\%$} & {\bf $60\%$} \\ \hline
\input{modified-results/qerror-cputime-settingthree.tex}
\bottomrule
\end{tabular}
\caption{CPU time prediction qerror (SQLShare, \settingThree).}
\label{tab:qerror-cputime-settingthree}
\end{table}

\fi

\subsubsection{Discussion}\label{sec:discussion_model}
We found the following:
\begin{enumerate*}

\item Character-level models (\ccnn and \ctfidf) obtain the best performance for all  problems except CPU time prediction in \settingOne, where word-level models (\wcnn  and \wlstm) obtained the lowest test loss and MSE.  Intuitively, as the problem heterogeneity increases, the number of rare words increases, making it difficult to learn word-level patterns. In \settingOne setting, however, queries have more words in common (e.g., table names and SQL keywords), and the models can learn the underlying distributions better.  

\item Overall, CNN and LSTM architectures outperform others on all problems except session classification, where \ctfidf obtains better results  in predicting several classes. The frequency of the classes (see Table~\ref{tab:session_class_prediction_results_sel}) shows that \ctfidf performs better for majority classes (e.g., \nowebh and \bott); and CNN and LSTM beat \ctfidf in non-frequent classes (e.g., \unknown and \program) where prediction is more difficult. In addition, \ccnn achieves almost the same overall accuracy with much fewer parameters ($16801$ vs $500000$). The neural networks  learn features w.r.t. task, but \ctfidf and \wtfidf are limited to pre-determined features.

\item Regarding generalization of a single model under various settings, \ccnn  identifies local sequential character patterns which help it learn the underlying data distribution  better.
\end{enumerate*}

\subsection{Detailed Qualitative Analysis}

\subsubsection{Performance by Session Class}
\label{sec:dbrec_performance_by_session_class}
  \iffullpaperComposition
  We perform  a  finer-level of  analysis and  use the session class information as a proxy for complexity under the \settingOne setting. We analyze CPU time prediction performance in Figure~\ref{fig:mse_by_session_class_busy} and answer size prediction in Figure~\ref{fig:mse_by_session_class_rows}. The figures show  MSE of prediction  by session class for each model. The MSE trends show that predicting CPU time for \nowebh, \program, and \browser is more difficult. Moreover, the simple baseline \median under-performs all models across all sessions.  Interestingly, \ctfidf and \wtfidf perform similarly to \median  for CPU time prediction, and under-perform all other models for Answer size prediction.  This shows the neural network models perform better on complex session classes.
  \else
  We refine the analysis and use session class  as a proxy for complexity in \settingOne. We analyze answer size  prediction  in Figure~\ref{fig:mse_by_session_class_rows}, which shows  MSE  by session class. The MSE trends show that predicting answer size for \nowebh, \program, and \browser is more difficult. Moreover, the simple baseline \median under-performs all models across all sessions.   Interestingly, \ctfidf and \wtfidf perform similarly to \median  for CPU time prediction (see~\cite{t-report}) and under-perform all other models for answer size prediction. This shows the neural network models perform better on complex session classes.
  \fi

\iffullpaperComposition
\begin{figure}
\centering

        \subfloat[CPU time prediction by session class. ]
        {
		\includegraphics[width=0.8\linewidth]{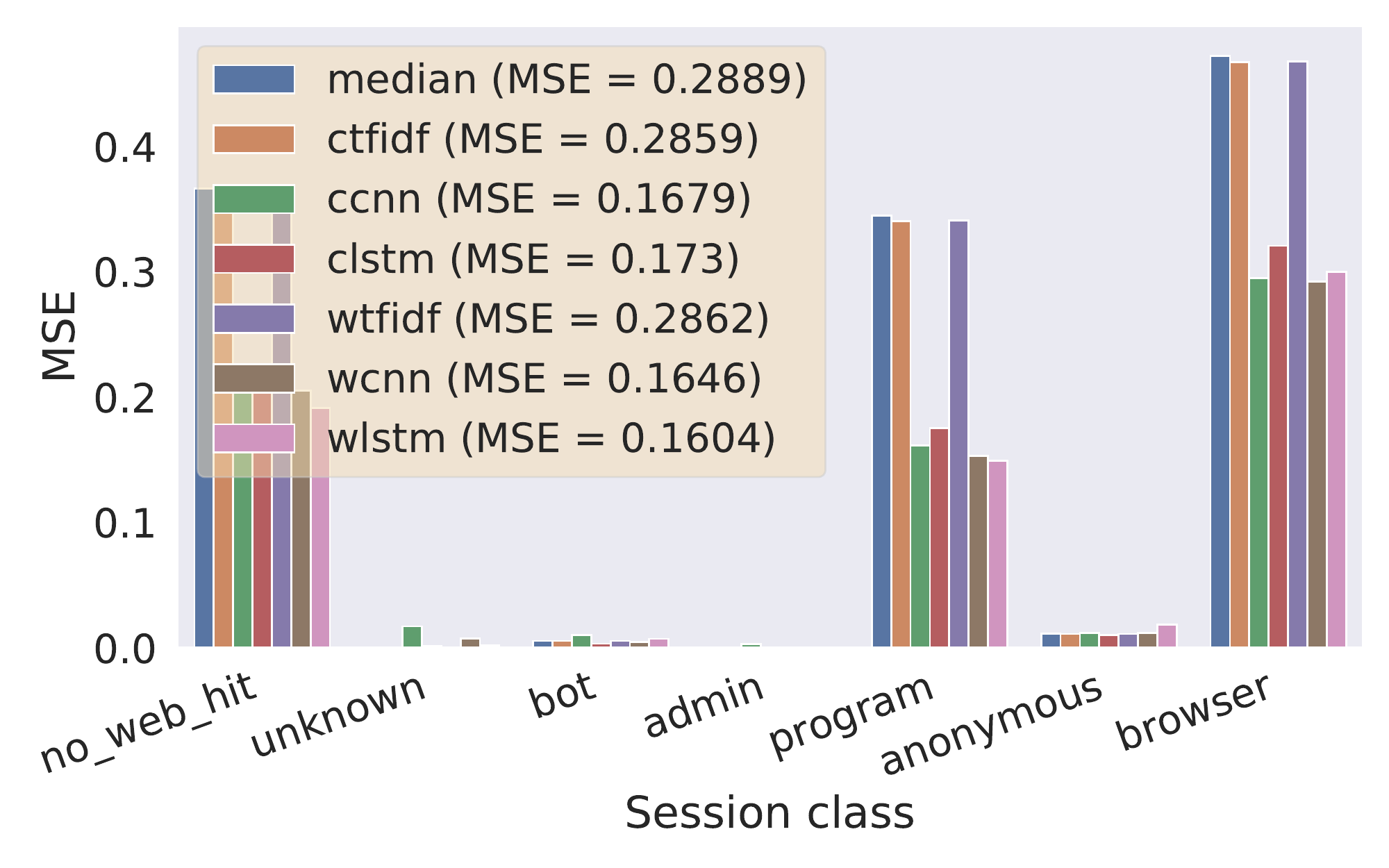}
		\label{fig:mse_by_session_class_busy}
        }

        \subfloat[Answer size prediction by session class.  ]
        {
	\includegraphics[width=0.8\linewidth]{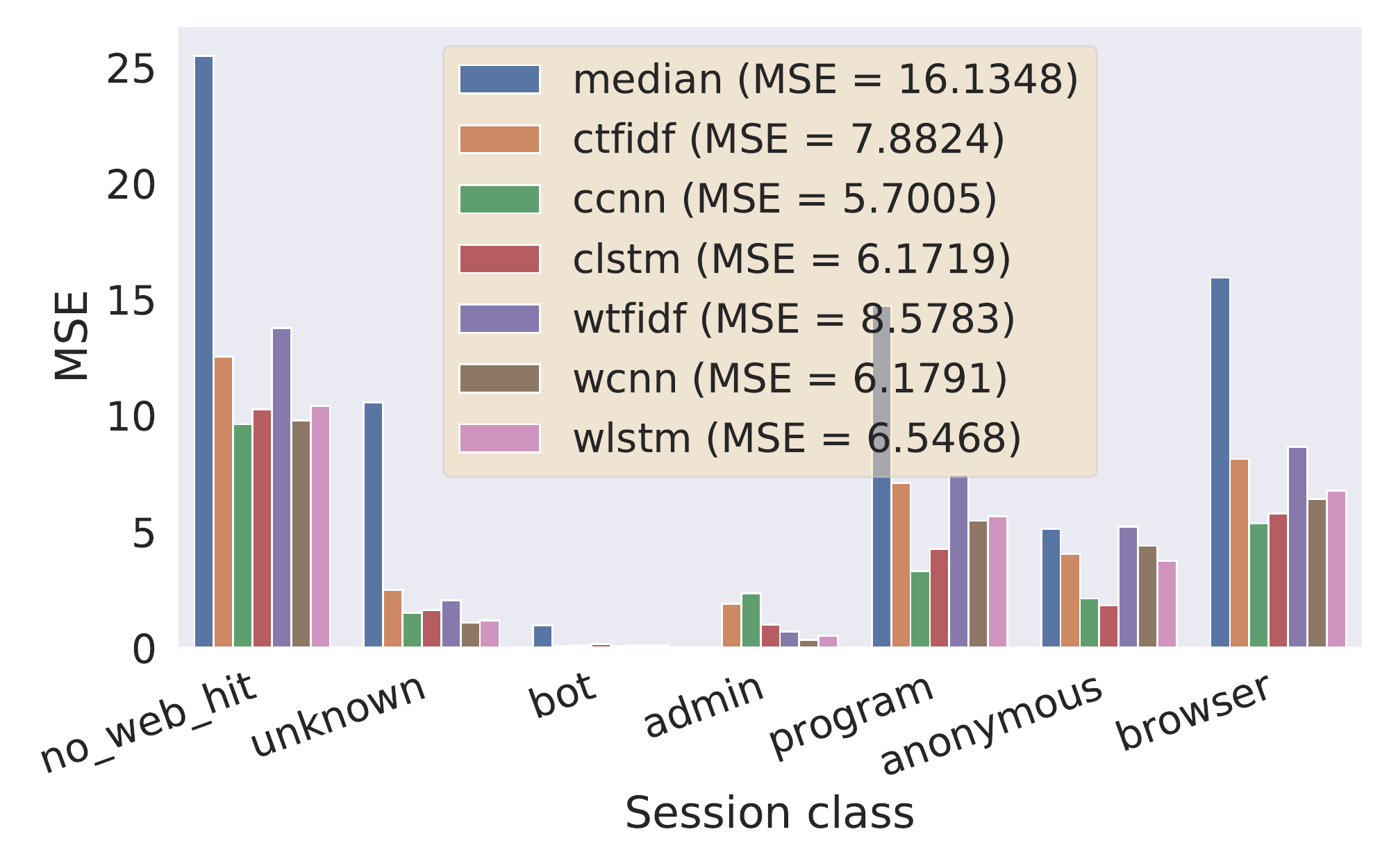}
		\label{fig:mse_by_session_class_rows}
        }
\caption{MSE of regression problems by session class in \settingOne (SDSS).  In~\ref{fig:mse_by_session_class_busy},   \ctfidf, \wtfidf, are similar to \median baseline while they outperform it  in~\ref{fig:mse_by_session_class_rows}. The neural network models outperform all baselines. }
\label{fig:mse_by_session_class}
\end{figure}
\fi

\subsubsection{Performance by Structural Properties}
\label{sec:perfstr}
Figures~\ref{fig:mse_clevel_rows_set1}-\ref{fig:mse_na_rows_set1} analyse error of answer size prediction  for varying structural properties under \settingOne. As  expected, error increases for more complex queries (with larger \clevel, \fcount, \jcount and \ncount). The decrease of error  in the middle and   end of the graphs in Figures~\ref{fig:mse_clevel_rows_set1}-\ref{fig:mse_jcount_rows_set1} is due to fewer answers for the corresponding queries, which makes prediction easier for all the models (including \median, which supports this claim).

\begin{figure}
\centering

        \subfloat[\settingOne  ]
        {
	    \includegraphics[width=0.5\linewidth]{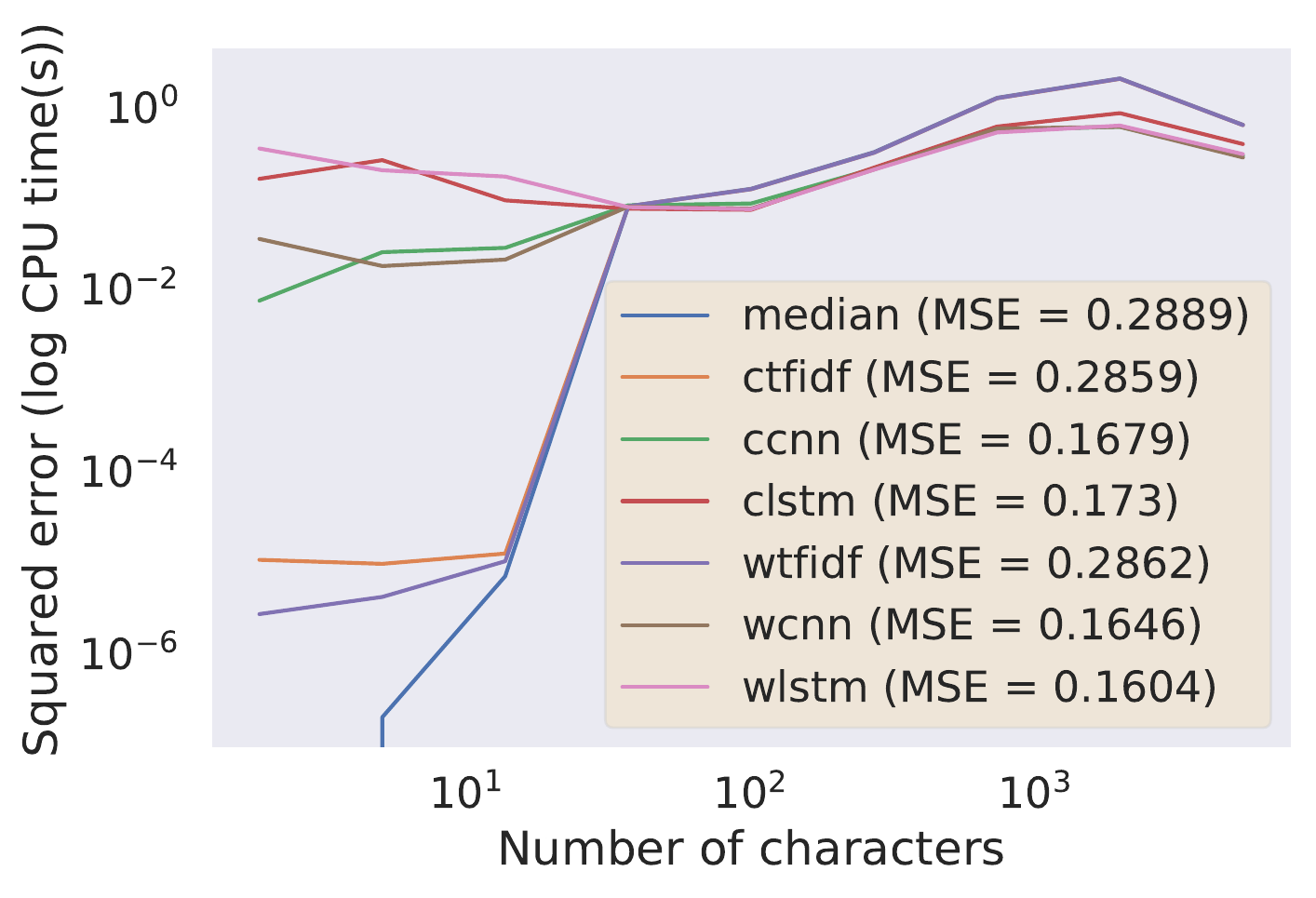}
		\label{fig:mse_by_original_char_busy_1}
        }
        \subfloat[\settingOne  ]
        {
	    \includegraphics[width=0.5\linewidth]{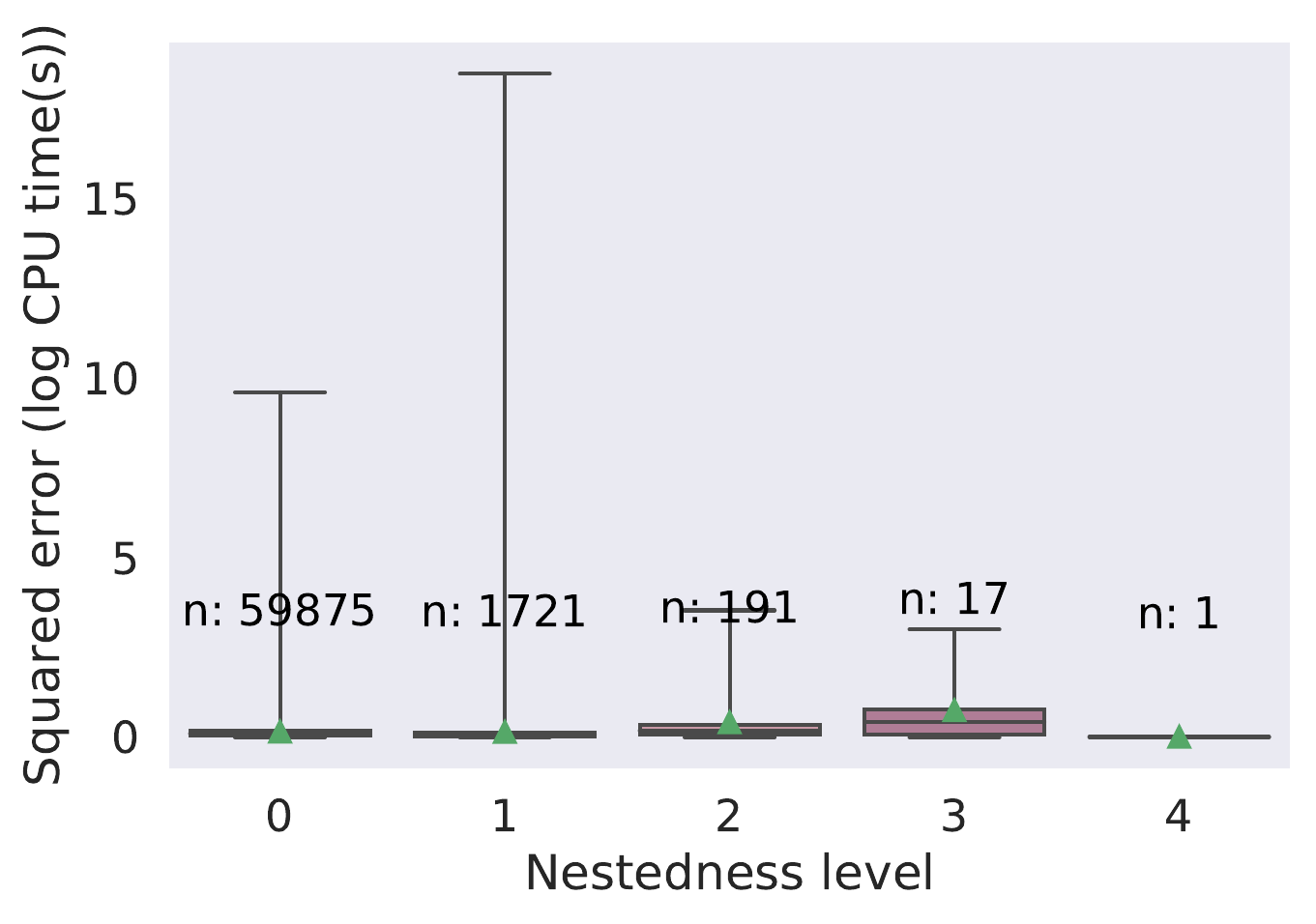}
		\label{fig:mse_cnested_time_set1}
        }

        \subfloat[\settingTwo  ]
        {
	    \includegraphics[width=0.5\linewidth]{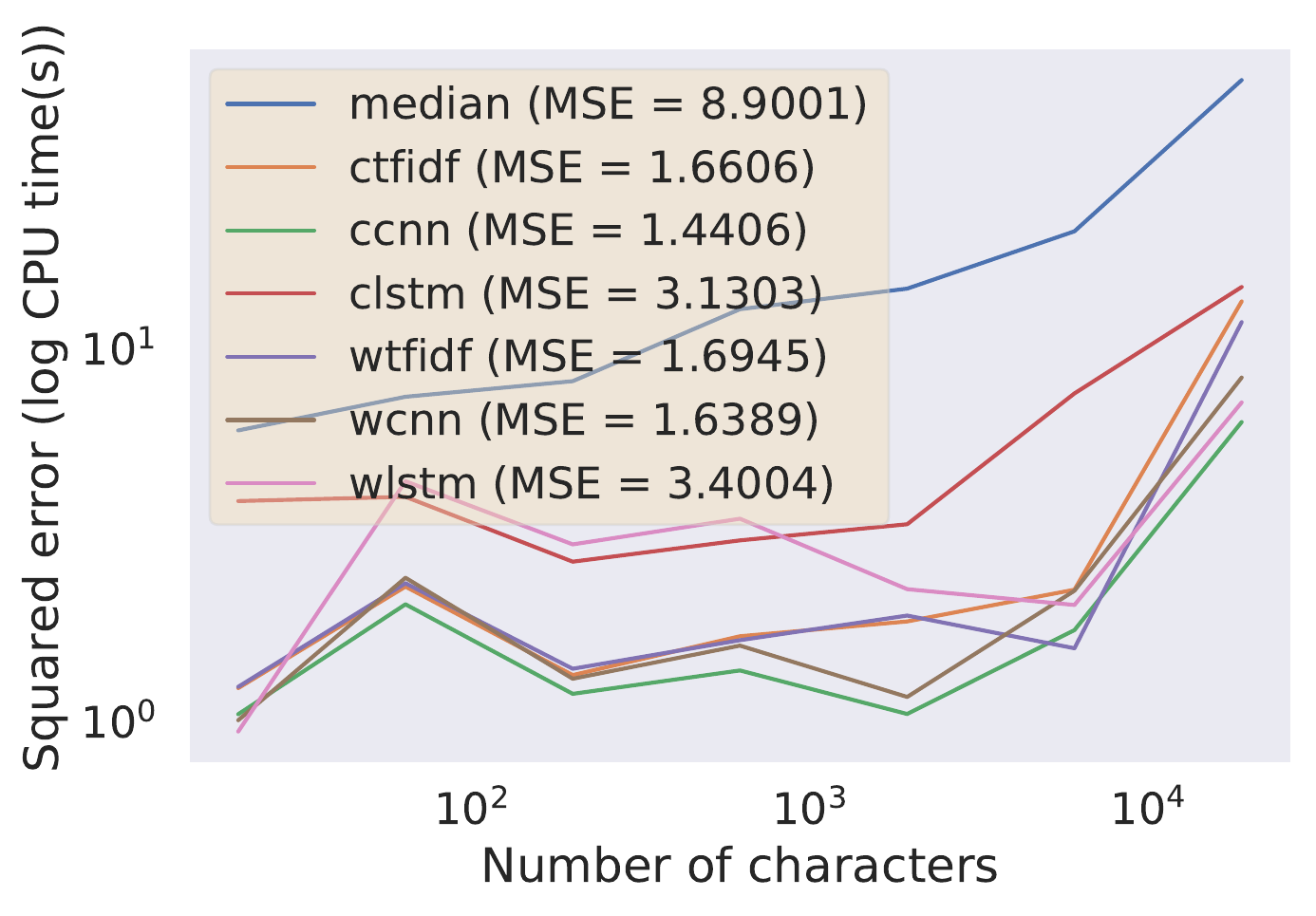}
		\label{fig:mse_by_original_char_busy_2}
        }
        \subfloat[\settingTwo  ]
        {
	    \includegraphics[width=0.5\linewidth]{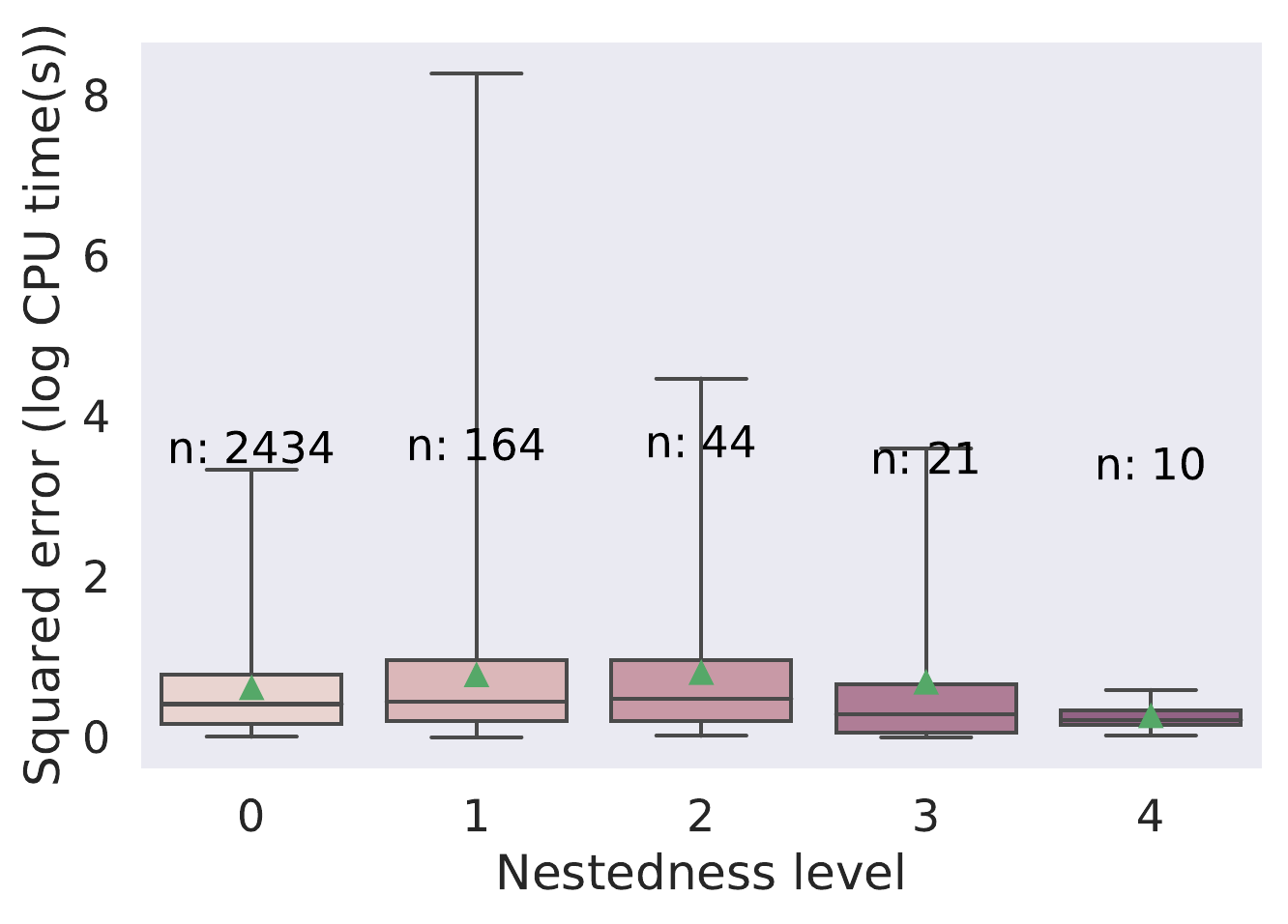}
		\label{fig:mse_cnested_time_set2}
        }

        \subfloat[\settingThree  ]
        {
	    \includegraphics[width=0.5\linewidth]{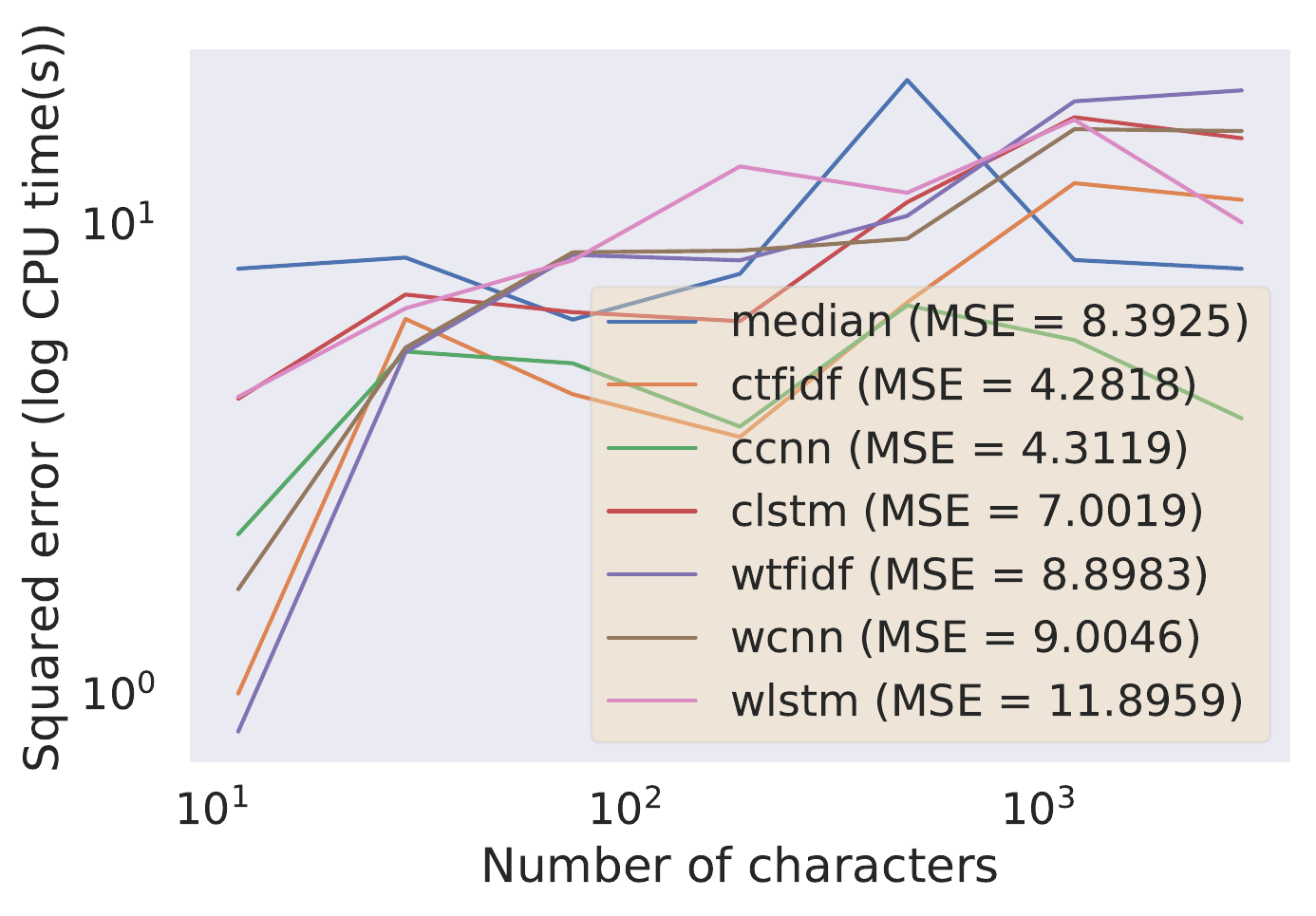}
		\label{fig:mse_by_original_char_busy_3}
        }
        \subfloat[\settingThree  ]
        {
	    \includegraphics[width=0.5\linewidth]{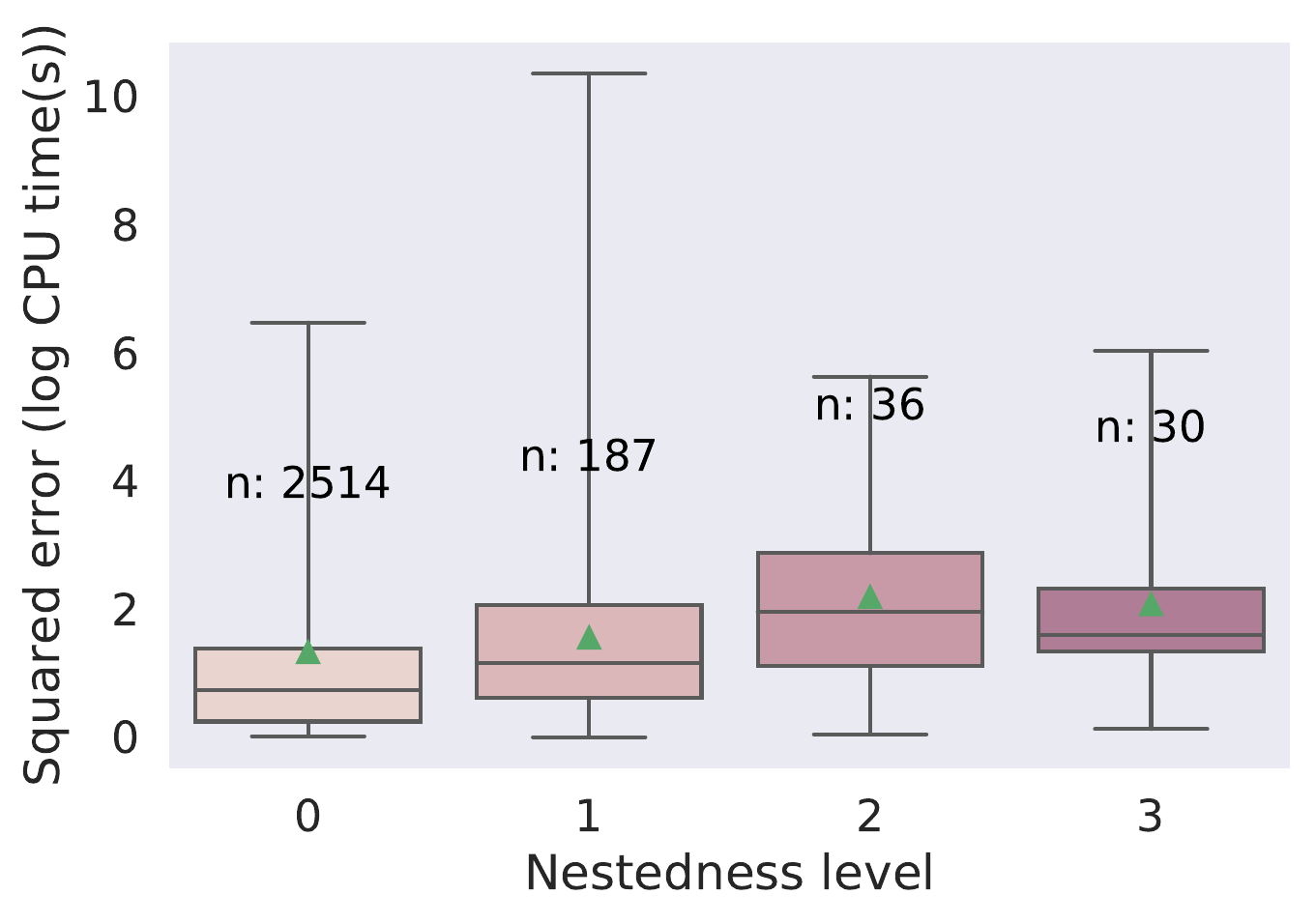}
		\label{fig:mse_cnested_time_set3}
        }

\caption{ Error analysis of CPU time prediction  by \clevel (left). Error analysis of CPU time prediction for   \ccnn by \ncount (right).}
\label{fig:mse_by_properties_busy}
\end{figure}

Figure~\ref{fig:mse_by_properties_busy} shows MSE for CPU time predication. For all models, error increases from \settingOne to \settingTwo to \settingThree, because the problem setting heterogeneity makes  prediction more difficult. Figure~\ref{fig:mse_by_properties_busy} also shows the MSE of \ccnn  increases for more complex queries. Again, the unexpected decrease in MSE of  queries with high nestedness (see \ncount\!=$3,4$ in Figures~\ref{fig:mse_cnested_time_set1},~\ref{fig:mse_cnested_time_set2} and~\ref{fig:mse_cnested_time_set3}) is due to  better prediction for the few queries with small CPU time.

\ifPhDThesis
\input{QualAnalysisFiguresPhDThesis}
\else
\input{QualAnalysisFigures}
\fi

\subsubsection{Case Study}
\label{sec:dbrec_qualatative_analysis}

We study performance for two sample queries with different structural properties. $Q_1$ in Figure~\ref{fig:q1} is a large query (\clevel=$1,247$ and \wlevel=$376$) that joins three large tables (e.g., {\tt Specobj} and {\tt Photoobj} contain $4,311,571$ and $794,328,715$ rows, respectively), selects 49 columns in the answer, and calls 3 functions. The query is from the \browser  and ran successfully (error class: \scss) with CPU time of $105.37$ sec and returned $304$ answers. Comparing \ccnn and \clstm, the former predicts $116.40$ sec for CPU time while \clstm's estimation is $980$ sec. The query length makes it hard for \clstm to capture the long-term dependencies,  whereas \ccnn detects local patterns and combines them globally to make a better prediction.

 $Q_2$ in Figure~\ref{fig:q2}, is shorter than $Q_1$ (\clevel\!=$645$ and  \wlevel\!=$181$), but it is more complex (\ncount\!=$3$, \fcount\!=$5$, and \pcount\!=$11$). The query runs instantly since it accesses  tables ({\tt Jobs}, {\tt Users}, {\tt Status} and {\tt Servers})  with fewer rows. Its answer size is  $27$ rows. The CPU time prediction of \ccnn, \wcnn and \clstm are $1.00$ sec, $1.28$ sec and $1.01$ sec, respectively. Their answer size predictions are $45$, $46$ and $49$. For $Q_2$ all 3  models perform fairly well. The small CPU time and the answer size of $Q_2$ compared to $Q_1$, contributes to more accurate predictions (due to the logarithmic label transformation and Huber loss (cf.~Section~\ref{subsec:dbrec_analysis_implications_model_selection})). $Q_2$ is shorter in length compared to $Q_1$, which makes it easier for \clstm to make predictions.

\subsubsection{Discussion} \label{sec:discussion_qualitative}

Using session information, CPU time and answer size prediction were more difficult for \nowebh, \program, and \browser\ sessions, for  all models. This is because queries in these classes are more complex compared to other classes (see Figure~\ref{fig:sdss_query_statement_bsa}) and are likely issued by humans.  Our evaluation  by structural properties showed that predicting labels is more difficult for complex queries (e.g. with large \clevel, \fcount, \jcount), and in settings where data is from heterogeneous sources. Word-level models suffer from many rare tokens in heterogeneous settings,  and do not generalize well.
Among the character-level models, \clstm is   sensitive to the query length and is outperformed by \ccnn as statement complexity increases.


\begin{figure}
    \centering
    \begin{lstlisting}
SELECT q.name AS qname,
       dbo.fDistanceArcMinEq(q.ra,q.dec,p.ra,p.dec), ...
FROM SpecObj AS s,
     SDSSSQL010.MYDB_670681563.test.QSOQuery1_DR5 AS q, PhotoObj AS p
WHERE ((s.bestobjid=p.objid) AND (s.ra BETWEEN 185 AND 190) AND ...) ORDER BY q.ra
\end{lstlisting}
\caption{Sample query $Q_1$}
\label{fig:q1}
\end{figure}

\begin{figure}
    \centering
    \begin{lstlisting}
SELECT  j.target,cast(j.estimate AS varchar) AS queue,...
FROM Jobs j,Users u,Status s,
     (SELECT DISTINCT target,queue FROM Servers s1
      WHERE s1.name NOT IN
           (SELECT name FROM Servers s,
                 (SELECT target,min(queue) AS queue
                  FROM Servers GROUP BY target) AS a
            WHERE a.target=s.target)) b
WHERE j.outputtype LIKE '%QUERY%' AND ...

\end{lstlisting}
\caption{Sample query $Q_2$}
\label{fig:q2}
\end{figure}

%% file: modified-results/merge.tex
{\tt baseline}&-&-&0.9730&0.0000&0.9863&0.0000&0.5951&-&0.0675&-&1.6357\\
\ctfidf&500000&1500000&0.9778&0.7131&0.9888&0.0053&0.5860&500000&0.0668&500000&1.0400\\
\ccnn&159&17403&\HL{0.9797}&\HL{0.7961}&\HL{0.9897}&0.1669&0.1106&16801&0.0471&16801&\HL{0.7517}\\
\clstm&159&1944003&0.9786&0.6922&0.9893&\HL{0.2206}&0.0830&1943401&0.0452&529651&0.7678\\
\wtfidf&500000&1500000&0.9773&0.6546&0.9885&0.0620&0.5836&500000&0.0668&500000&1.0922\\
\wcnn&85942&8597953&0.9790&0.7441&0.9894&0.2006&0.1006&8595101&\HL{0.0441}&8595101&0.8472\\
\wlstm&85942&10522303&0.9776&0.6971&0.9887&0.0018&\HL{0.0691}&10521701&0.0443&9107951&0.8256\\

%% file: modified-results/qerror-rows-thesis.tex
\median&1&36&50&144&1885&50000\\
\ctfidf & 1.13 & 4.86 &10& 25 & 88 & 727\\
\ccnn& 1.36 & 2.60 &3.75& \textbf{6.79} &	\textbf{18} & 174\\
\clstm& 1.07 & \textbf{2.38} &\textbf{3.50}& \textbf{6.79} & 19 & \textbf{172}\\
\wtfidf& 1.00 & 5.37 &11.04& 31.98 & 100 & 879\\
\wcnn& 1.33 & 3.42 &5.14& 10.93 & 36 & 295\\
\wlstm& 1.12 & 2.62 &4.27& 10.43 & 30 & 292\\

%% file: modified-results/qerror-rows.tex
\median&1&36&50&144&1885&50000\\
\ctfidf & 1.13 & 4.86 &10& 25 & 88 & 727\\
\ccnn& 1.36 & 2.60 &3.75& \HL{6.79} &	\HL{18} & 174\\
\clstm& 1.07 & \HL{2.38} &\HL{3.50}& \HL{6.79} & 19 & \HL{172}\\
\wtfidf& 1.00 & 5.37 &11.04& 31.98 & 100 & 879\\
\wcnn& 1.33 & 3.42 &5.14& 10.93 & 36 & 295\\
\wlstm& 1.12 & 2.62 &4.27& 10.43 & 30 & 292\\

%% file: modified-results/busy-rownumber.tex
\median&-&-&0.0675&-&1.6357\\
\ctfidf&500000&500000&0.0668&500000&1.0400\\
\ccnn&159&16801&0.0471&16801&\textbf{0.7517}\\
\clstm&159&1943401&0.0452&529651&0.7678\\
\wtfidf&500000&500000&0.0668&500000&1.0922\\
\wcnn&85942&8595101&\textbf{0.0441}&8595101&0.8472\\
\wlstm&85942&10521701&0.0443&9107951&0.8256\\

%% file: modified-results/session_class-sel-short-v-modified.tex
\mfreq&-&-&1.7848&0.6186&0.0000&0.0000&0.0000&0.0000&0.0000\ignore{&0.0000}&0.4478\\
\ctfidf&500000&3500000&1.5786&\HL{0.6235}&0.0000&\HL{0.7272}\ignore{&0.0000}&0.6128&\HL{0.6176}&\HL{0.5618}&\HL{0.6421}\\
\ccnn&159&18607&\HL{0.7960}&0.5921&0.2373&0.6940\ignore{&0.0000}&0.6076&0.5441&0.5389&0.6152\\
\clstm&159&1945207&0.8600&0.5774&0.0455&0.6817\ignore{&0.0000}&\HL{0.6366}&0.4868&0.5451&0.6102\\
\wtfidf&500000&3500000&1.6077&0.5487&0.2903&0.6958\ignore{&0.0000}&0.6241&0.6042&0.5416&0.6068\\
\wcnn&85942&8596907&0.8373&0.5411&\HL{0.3778}&0.6955\ignore{&0.0000}&0.5344&0.5970&0.5615&0.6004\\
\wlstm&85942&10523507&0.8452&0.5558&0.0000&0.6628\ignore{&0.0000}&0.6278&0.4482&0.5303&0.5911\\

%% file: modified-results/busy-setting-2-3.tex
\median&-&-&2.0049&-&2.1616\\
\opt&-&-&1.8909&-&2.2841\\
\ctfidf&500000&500000&0.4742&500000&1.2360\\
\ccnn&101&11001&\textbf{0.4625}&10901&\textbf{1.1547}\\
\clstm&101&1937601&0.7935&1937501&1.6046\\
\wtfidf&500000&500000&0.4898&500000&1.9702\\
\wcnn&14843&1485201&0.5230&1395901&2.0416\\
\wlstm&14843&3411801&0.8081&3322501&1.8546\\

%% file: modified-results/busy-setting-2-3-modified.tex
\median&-&-&2.0049&-&2.1616\\
\opt&-&-&1.8909&-&2.2841\\
\ctfidf&500000&500000&0.4742&500000&1.2360\\
\ccnn&101&11001&\HL{0.4625}&10901&\HL{1.1547}\\
\clstm&101&1937601&0.7935&1937501&1.6046\\
\wtfidf&500000&500000&0.4898&500000&1.9702\\
\wcnn&14843&1485201&0.5230&1395901&2.0416\\
\wlstm&14843&3411801&0.8081&3322501&1.8546\\

%% file: modified-results/qerror-cputime-settingtwo.tex
\median& 16.875 & 332.06 & - &  - & - & - \\
\ctfidf & 1.69 & 2.35 & 4.29 & 33.93 & - & -\\
\ccnn& \HL{1.49} & \HL{1.94} & \HL{3.03} & \HL{7.43} & \HL{27.08} & -\\
\clstm& 2.66 & 5.07 & 16.61 & - & - & -\\
\wtfidf& 1.53 & 2.10 & 3.33 & 9.42 & 46.61 & - \\
\wcnn& 1.72 & 2.48 & 4.33 & 16.51 & - & -\\
\wlstm& 4.93 & 29.75 & - & - & - & - \\

%% file: modified-results/qerror-cputime-settingthree.tex
\median& 2.06 &	12.06 & 4194.56 & - & - & - \\
\ctfidf & 2.03 & 6.57 & 375.46 & - & - & -\\
\ccnn& \HL{1.32} & \HL{3.85} & \HL{33.21} & - & - & -\\
\clstm& 2.19 & 20.75 &	2984.67 & - & - & -\\
\wtfidf& 2.03 &	8.29 & 441.51 & - & - & - \\
\wcnn& 2.13 & 6.19 & 181.49 & - & - & -\\

%% file: QualAnalysisFiguresPhDThesis.tex
\begin{figure*}
\centering
        \subfloat[Error Analysis for all models]
        {
	    \includegraphics[width=0.5\linewidth]{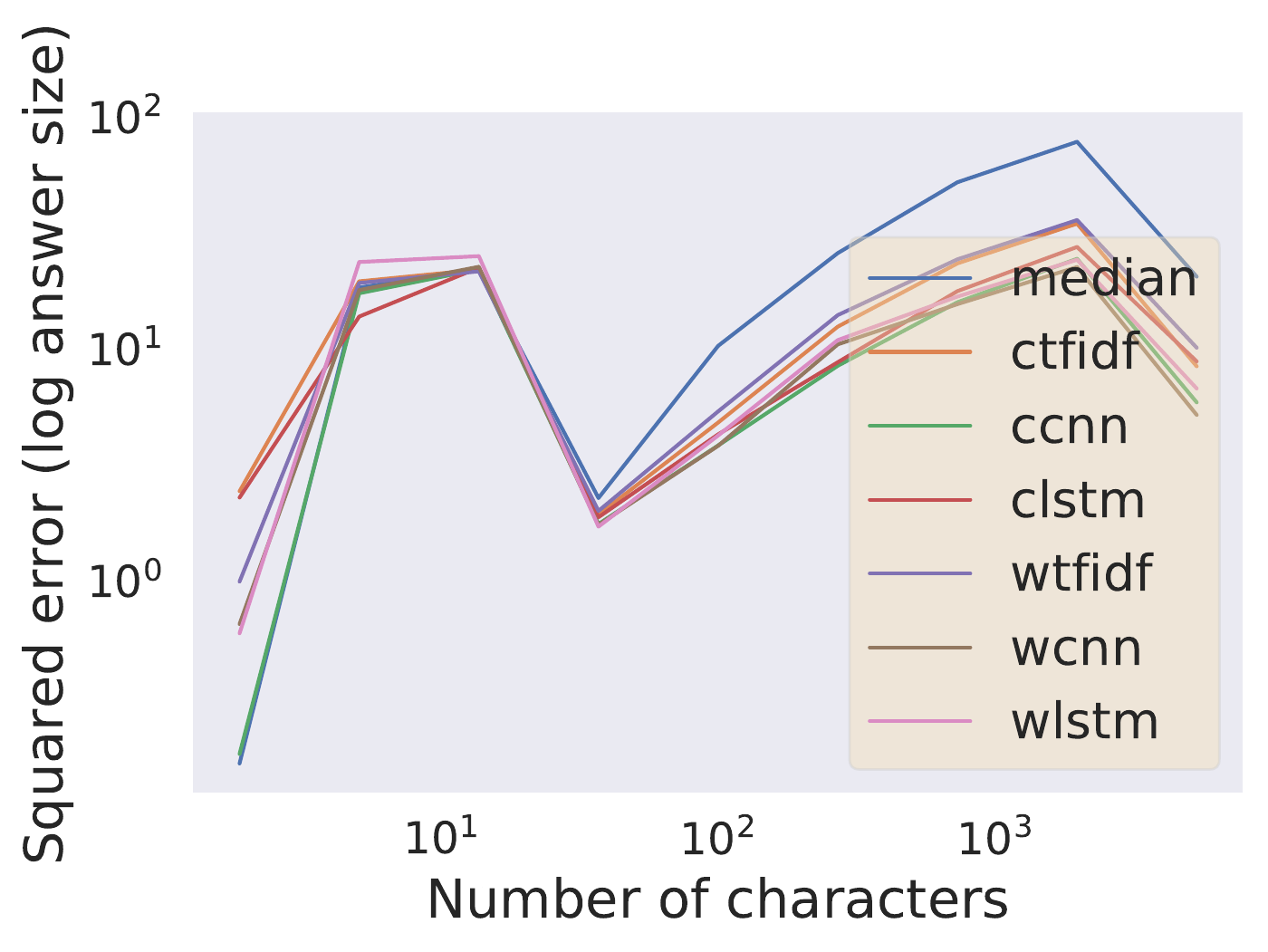}
		\label{fig:mse_clevel_rows_set1}
        }
        \subfloat[Error Analysis for all models]
        {
	    \includegraphics[width=0.5\linewidth]{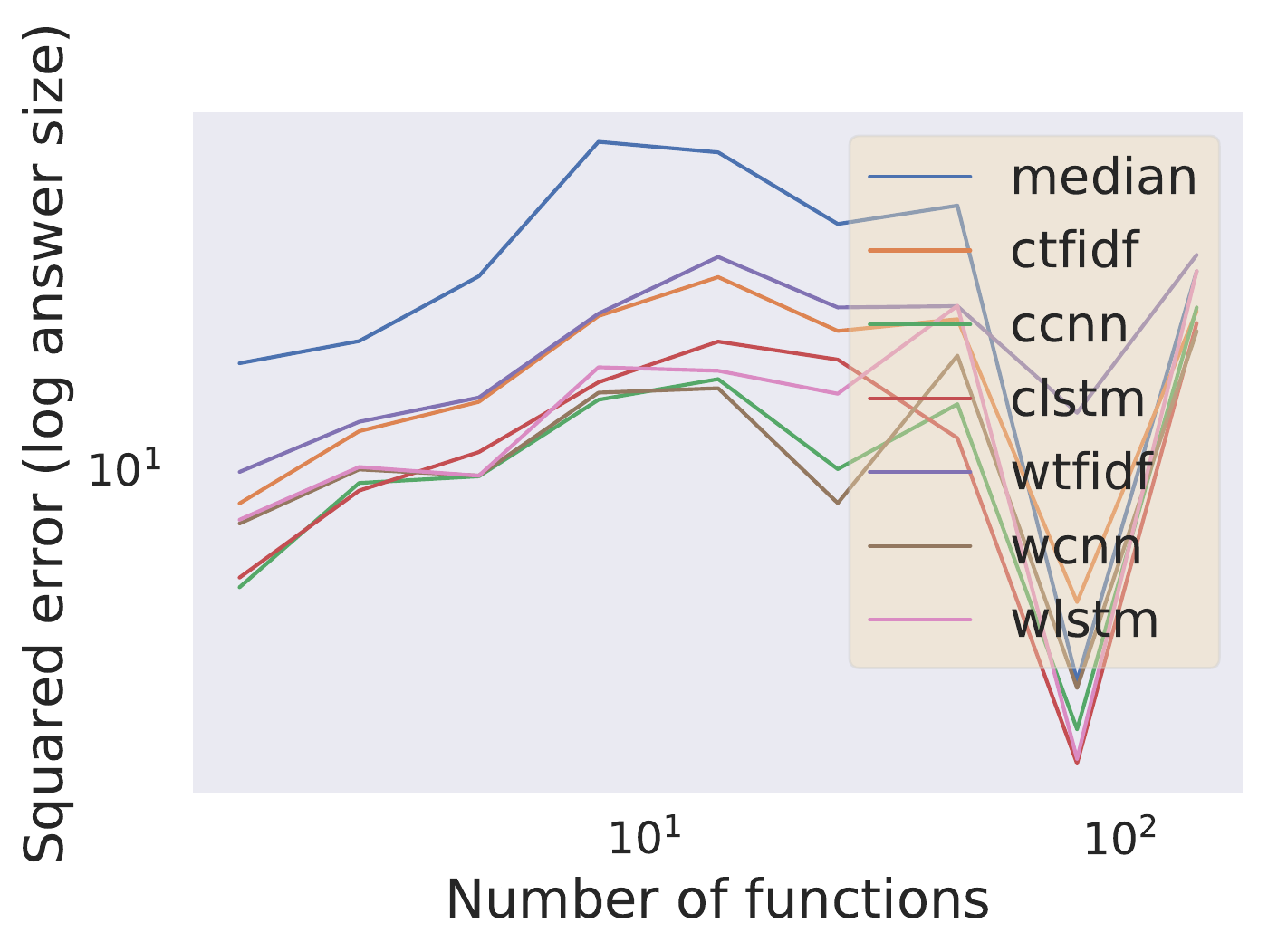}
		\label{fig:mse_fcount_rows_set1}
        }
        
        \subfloat[Error Analysis for all models]
        {
	    \includegraphics[width=0.5\linewidth]{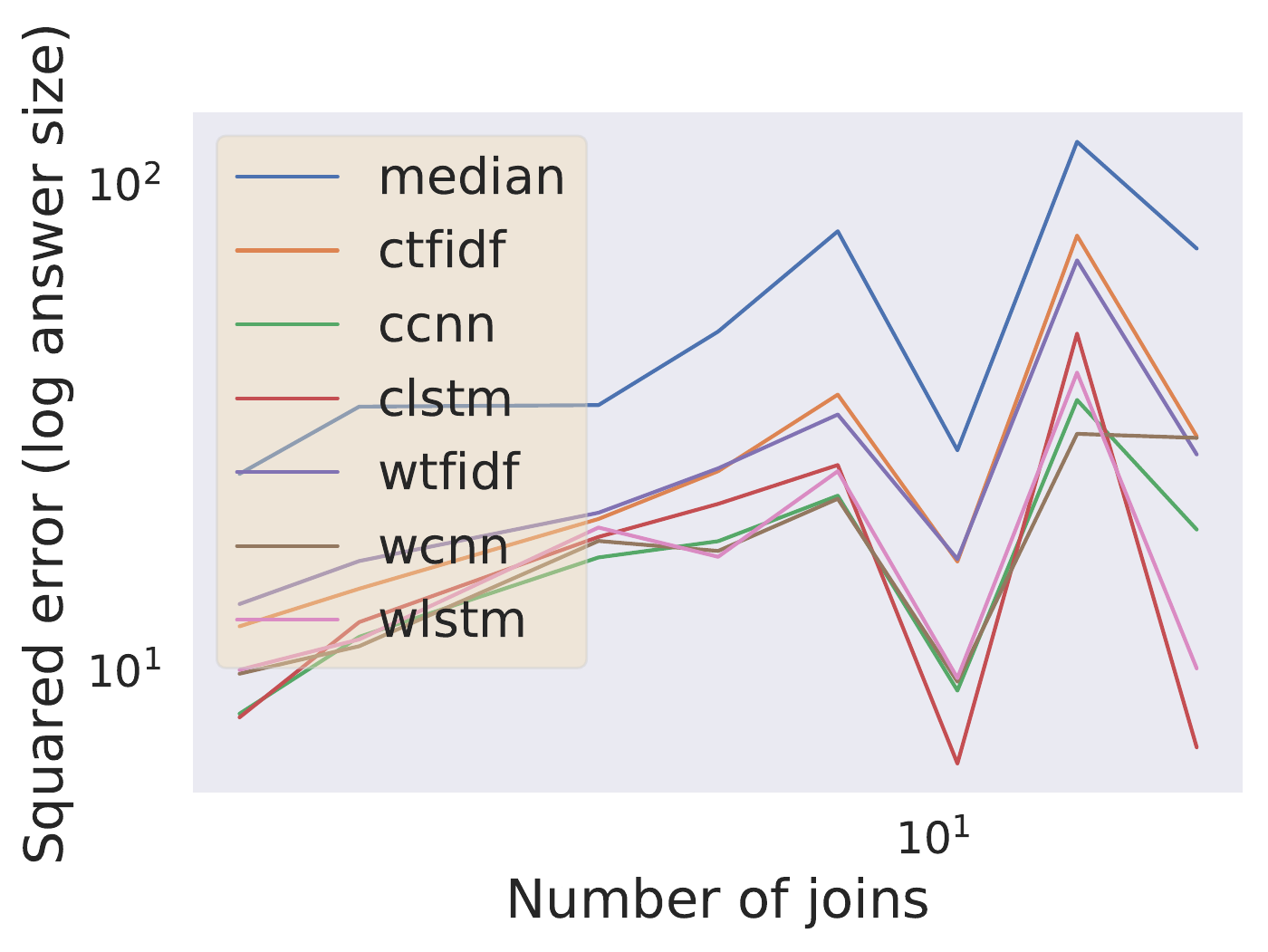}
		\label{fig:mse_jcount_rows_set1}
        }
        
        \subfloat[Error Analysis for \ccnn]
        {
	    \includegraphics[width=0.5\linewidth]{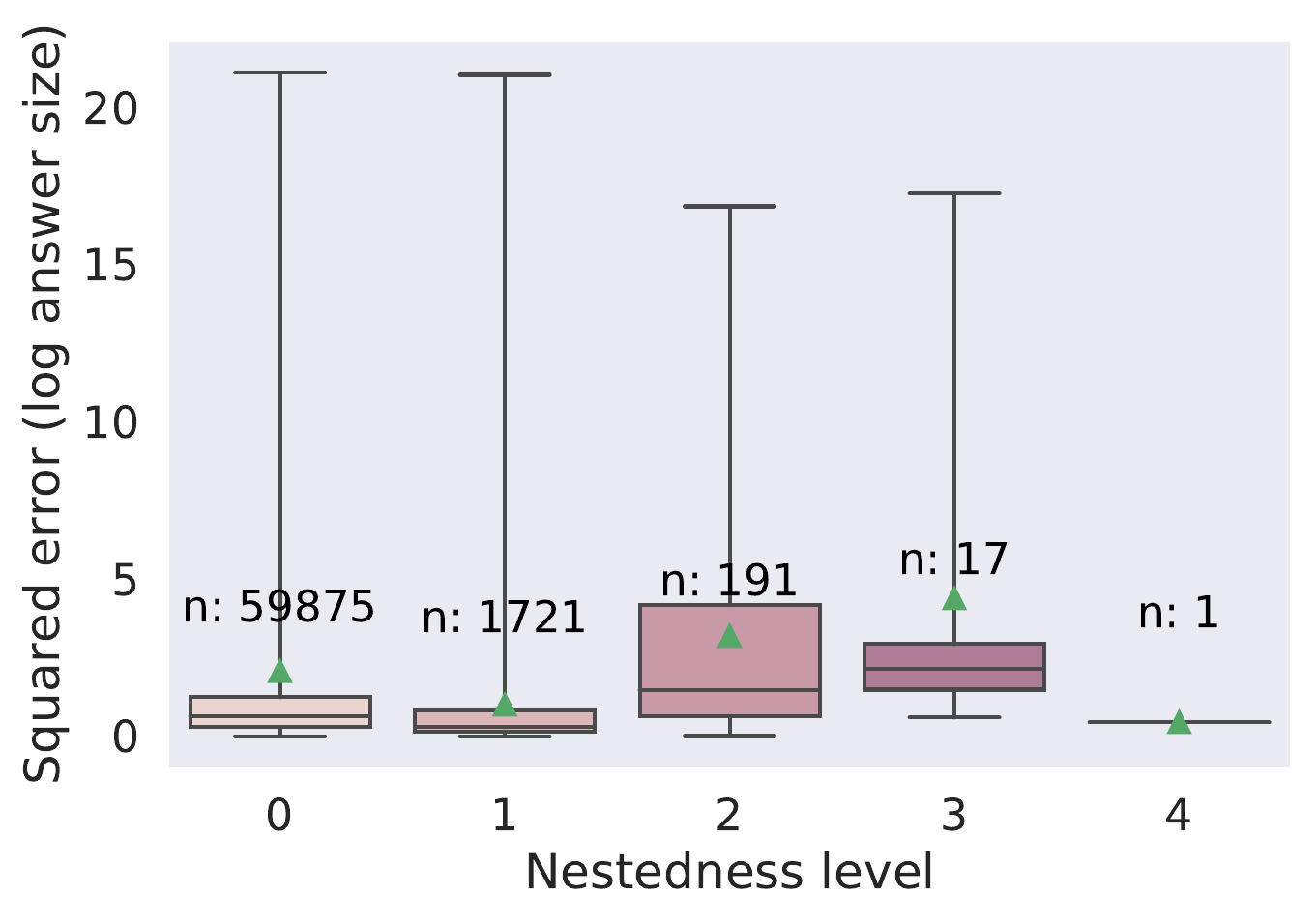}
		\label{fig:mse_ncount_rows_set1}
        }
        \subfloat[Error Analysis for \ccnn]
        {
	    \includegraphics[width=0.5\linewidth]{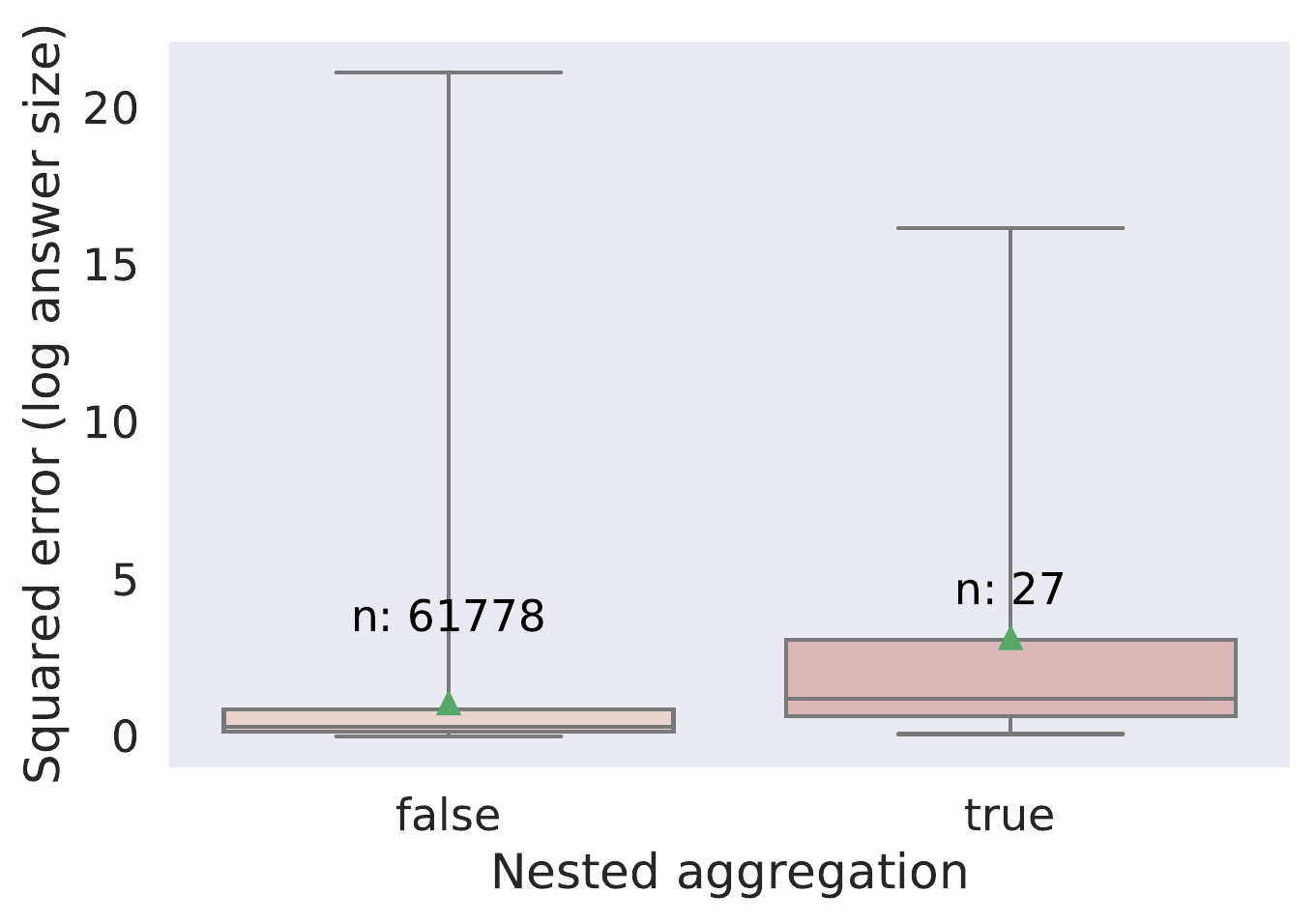}
		\label{fig:mse_na_rows_set1}
        }
\caption{Error analysis of answer size prediction  by structural properties on SDSS (\settingOne). } 
\label{fig:mse_by_properties_rows}
\end{figure*}
 

%% file: QualAnalysisFigures.tex
\iffullpaperComposition
\begin{figure*}
\centering
        \subfloat[]
        {
	    \includegraphics[width=0.2\linewidth]{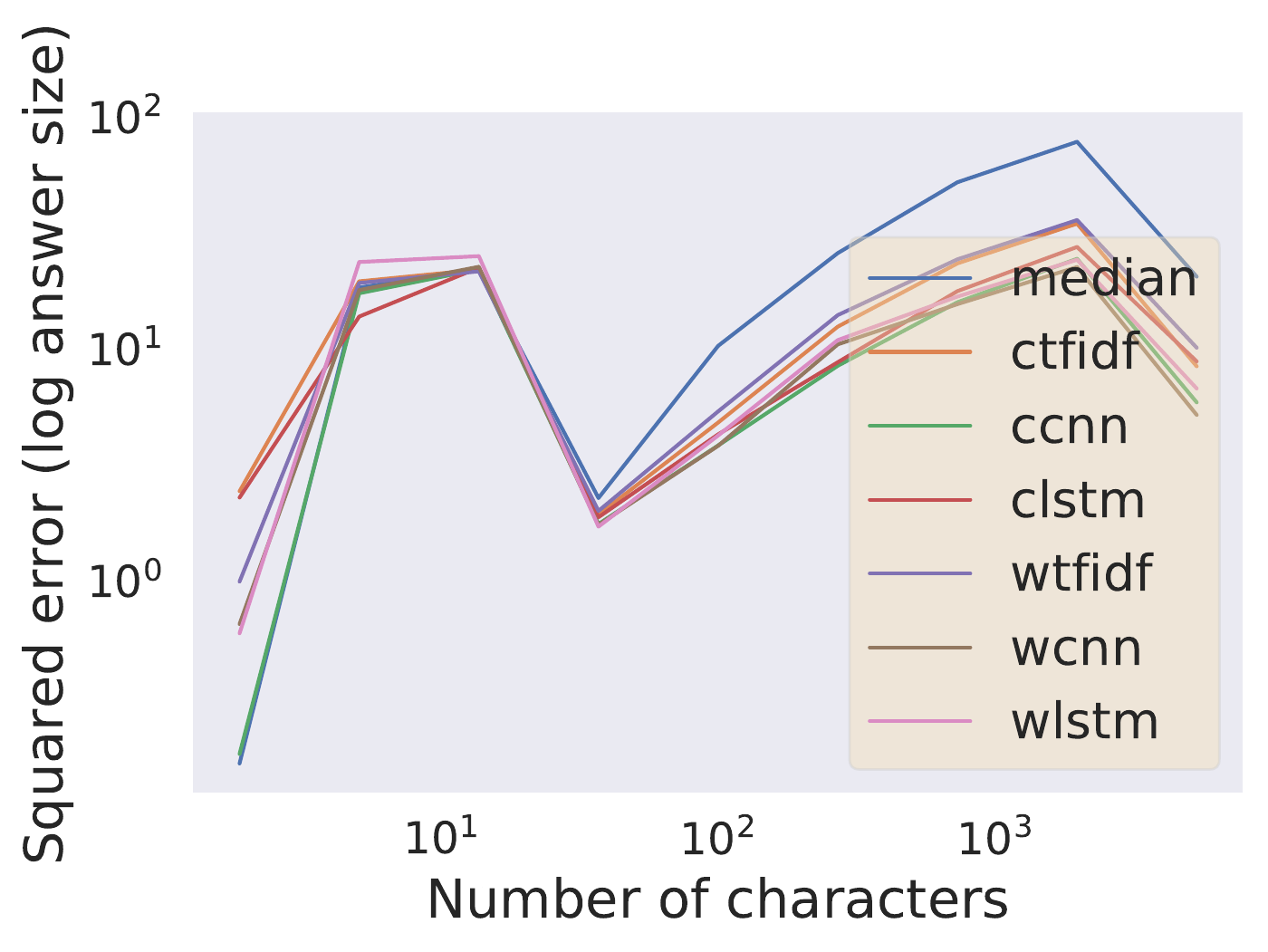}
		\label{fig:mse_clevel_rows_set1}
        }          
        \subfloat[]
        {
	    \includegraphics[width=0.2\linewidth]{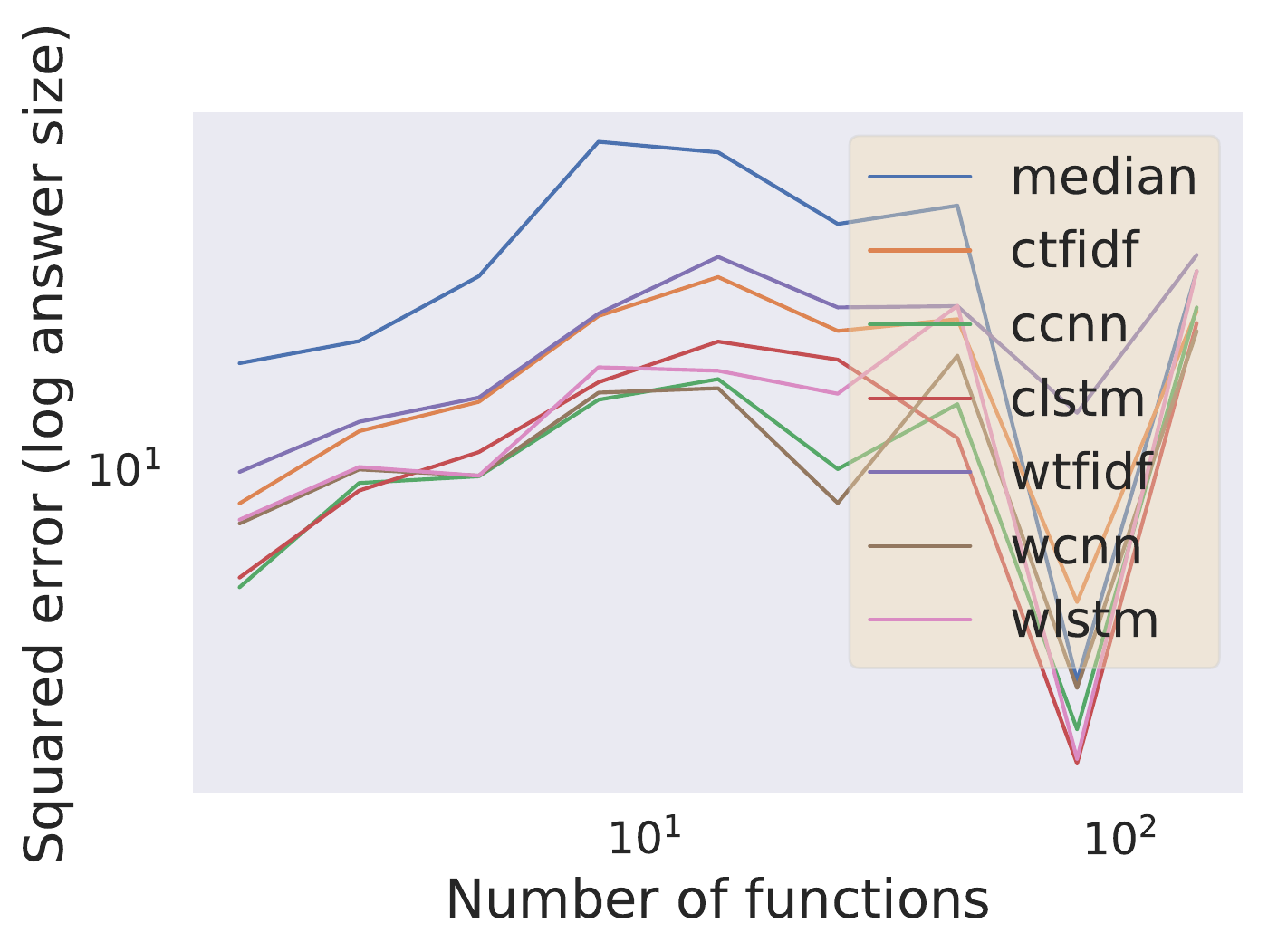}
		\label{fig:mse_fcount_rows_set1}
        }
        \subfloat[]
        {
	    \includegraphics[width=0.2\linewidth]{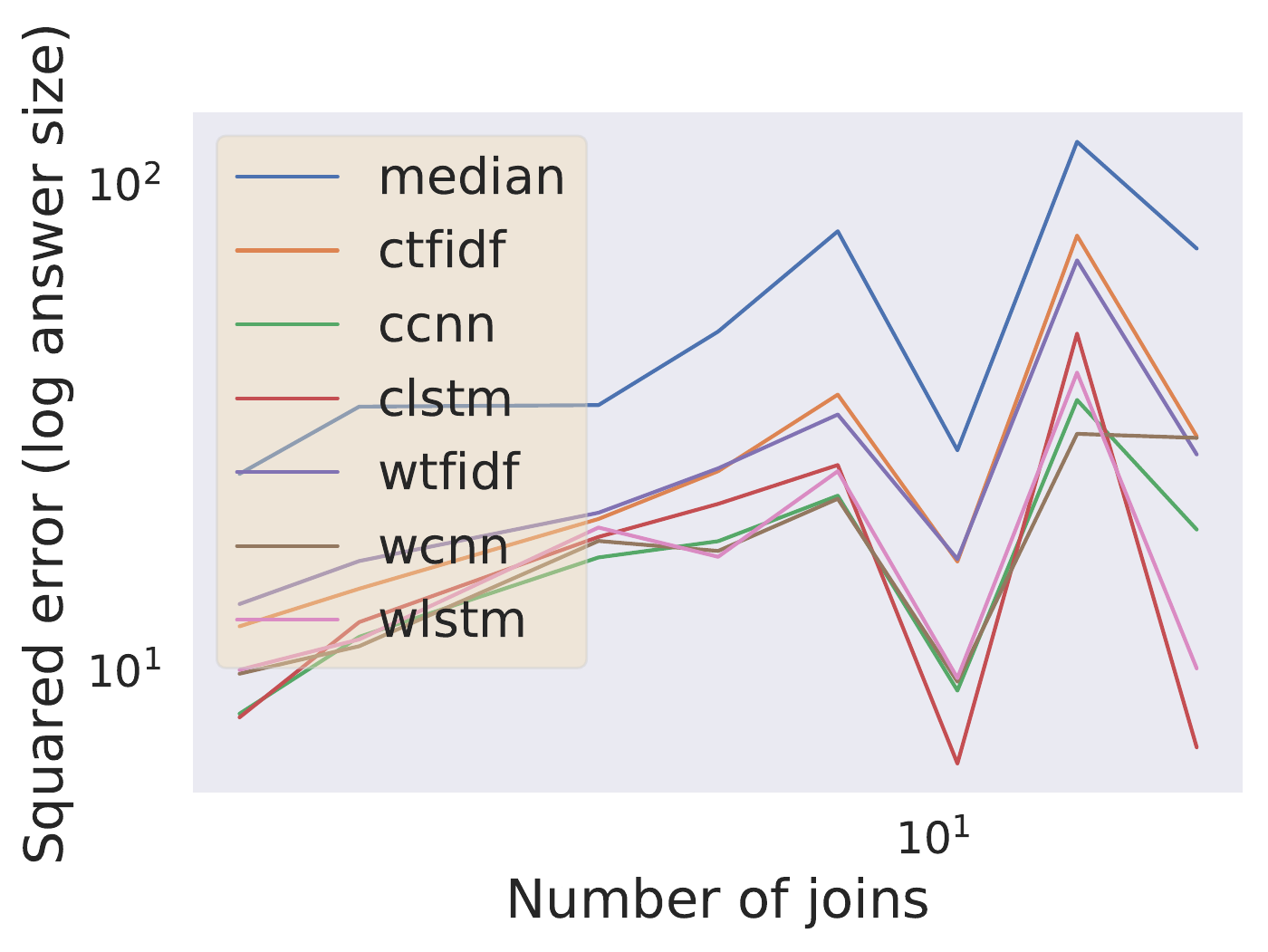}
		\label{fig:mse_jcount_rows_set1}
        }
        \subfloat[error analysis for \ccnn]
        {
	    \includegraphics[width=0.2\linewidth]{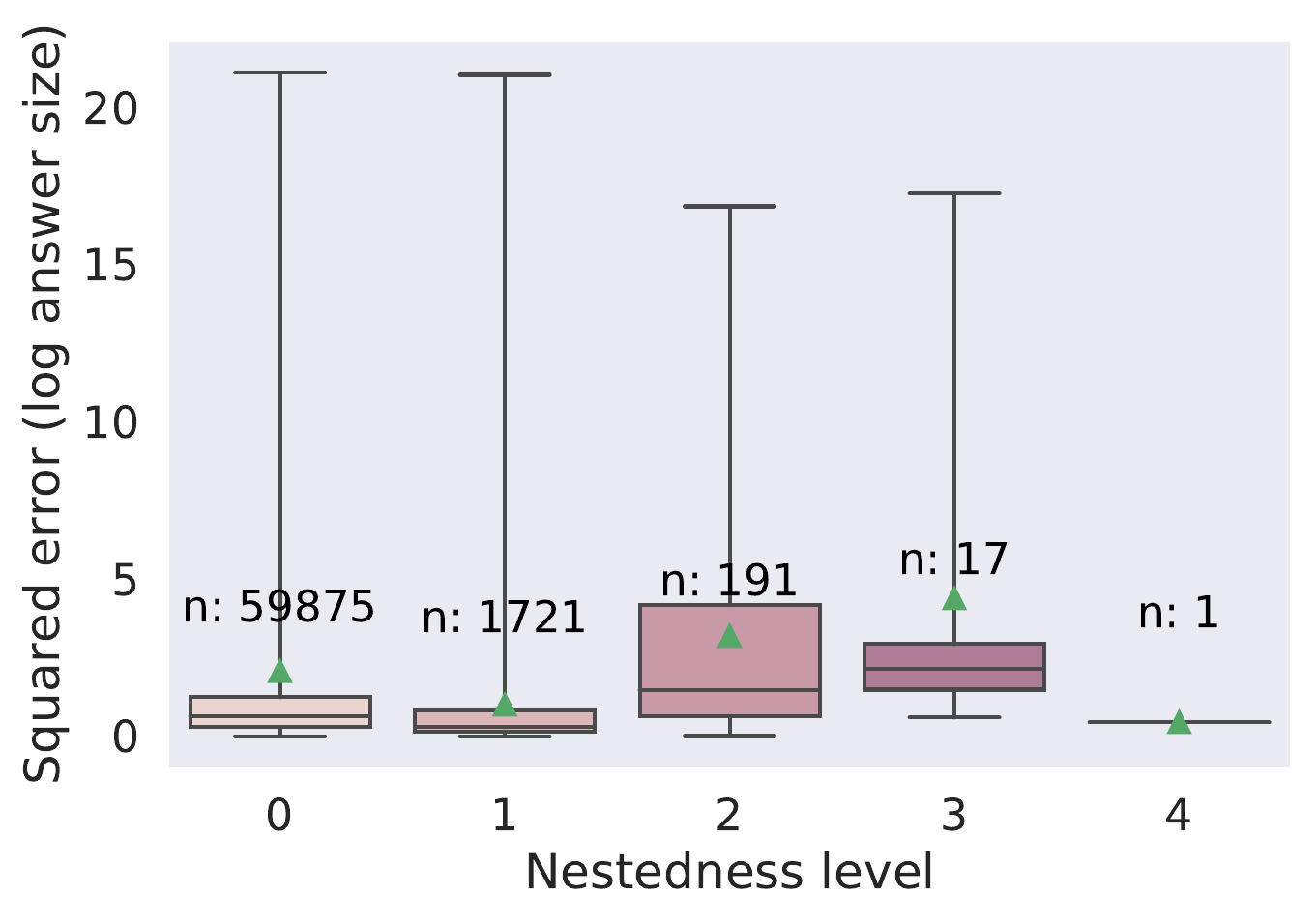}
		\label{fig:mse_ncount_rows_set1}
        }
        \subfloat[error analysis for \ccnn]
        {
	    \includegraphics[width=0.2\linewidth]{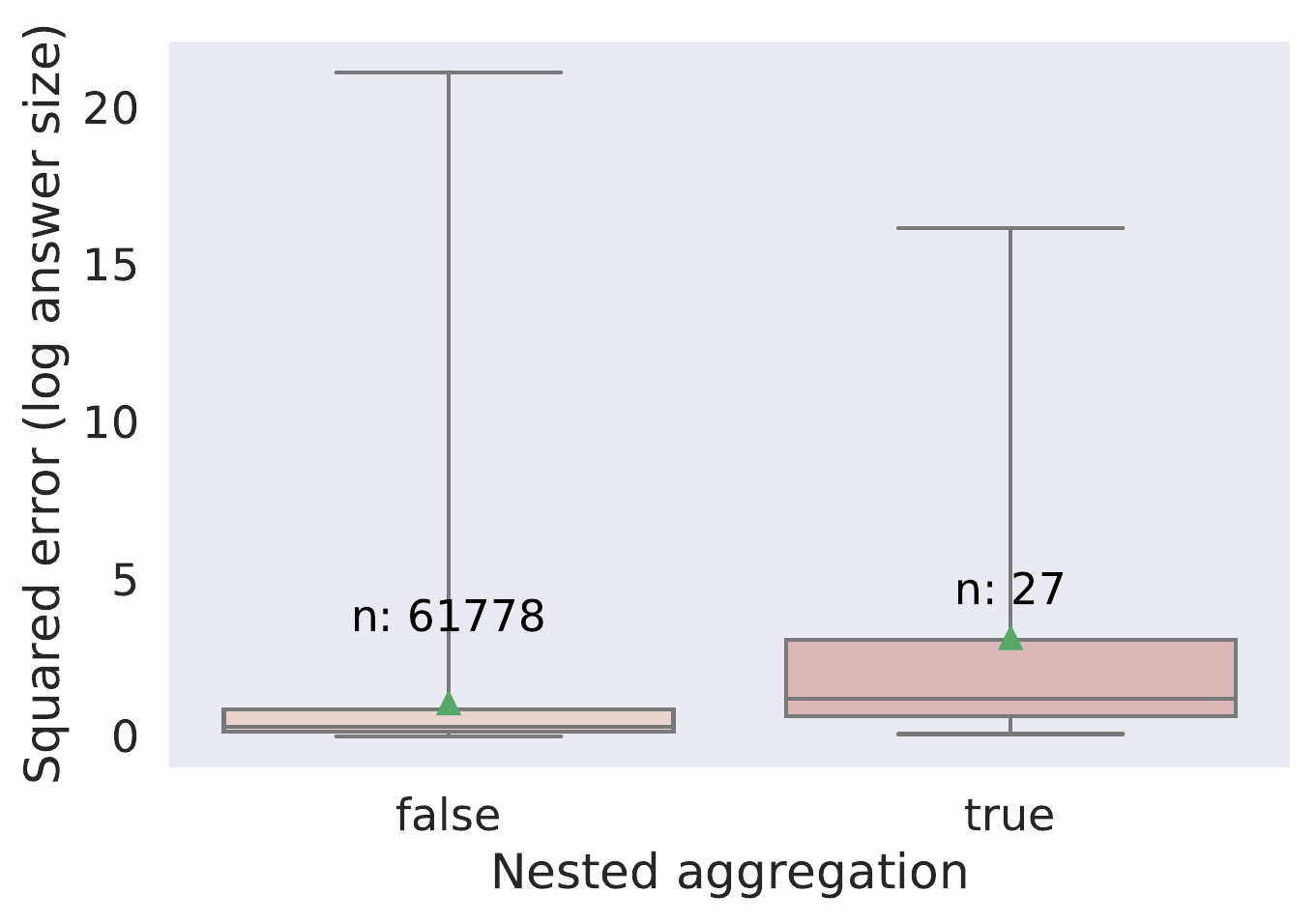}
		\label{fig:mse_na_rows_set1}
        }
\caption{Error analysis of answer size prediction on SDSS (\settingOne). Considering session class 9.a, and  considering structural properties 9.b-9.e. MSE of each model is reported in legend of 9.a. } 
\label{fig:mse_by_properties_rows}
\end{figure*}

\else 
\begin{figure}
\centering
        \subfloat[]
        {
		\includegraphics[width=0.8\linewidth]{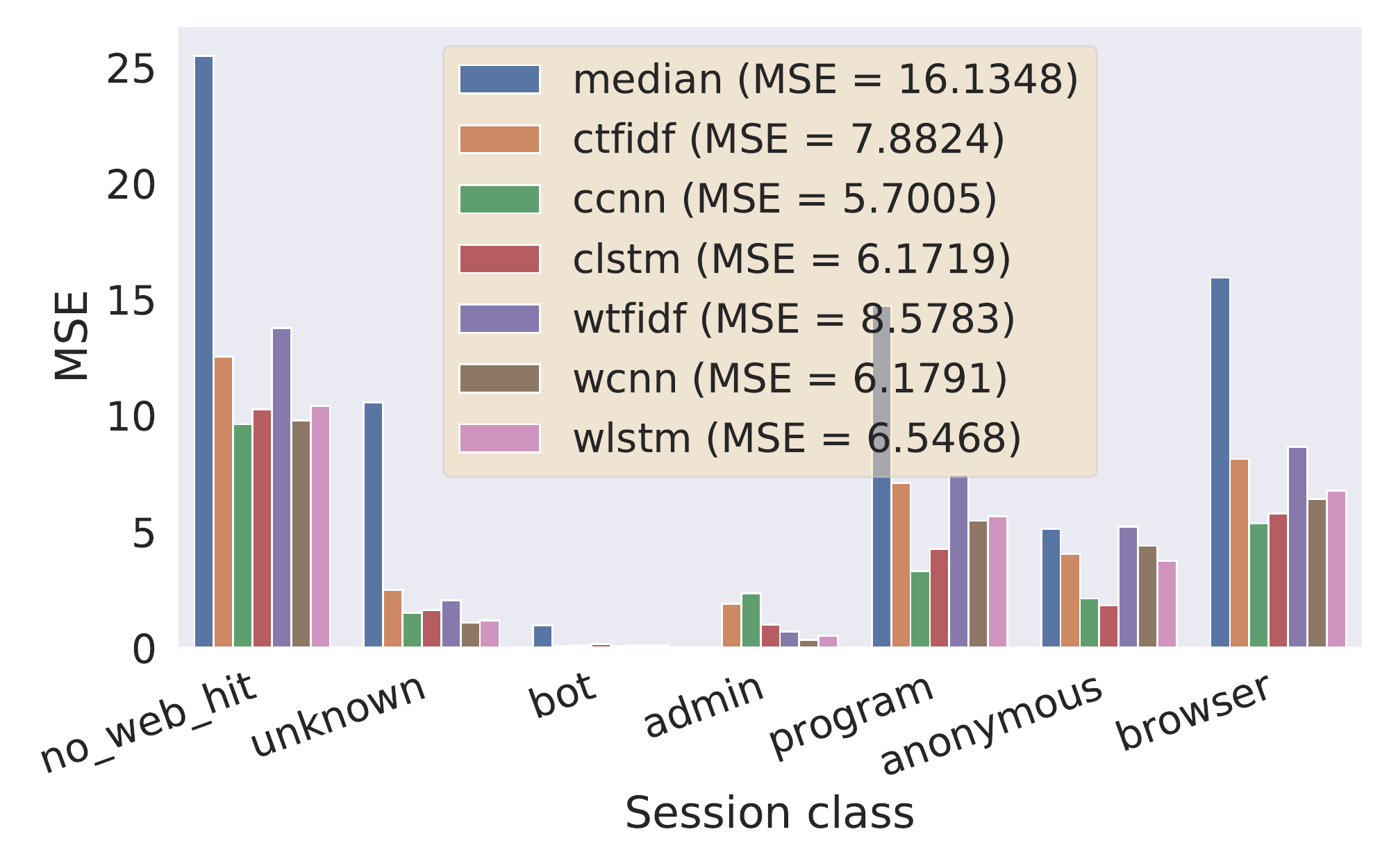} 
		\label{fig:mse_by_session_class_rows}
		}
		
        \subfloat[]
        {
	    \includegraphics[width=0.5\linewidth]{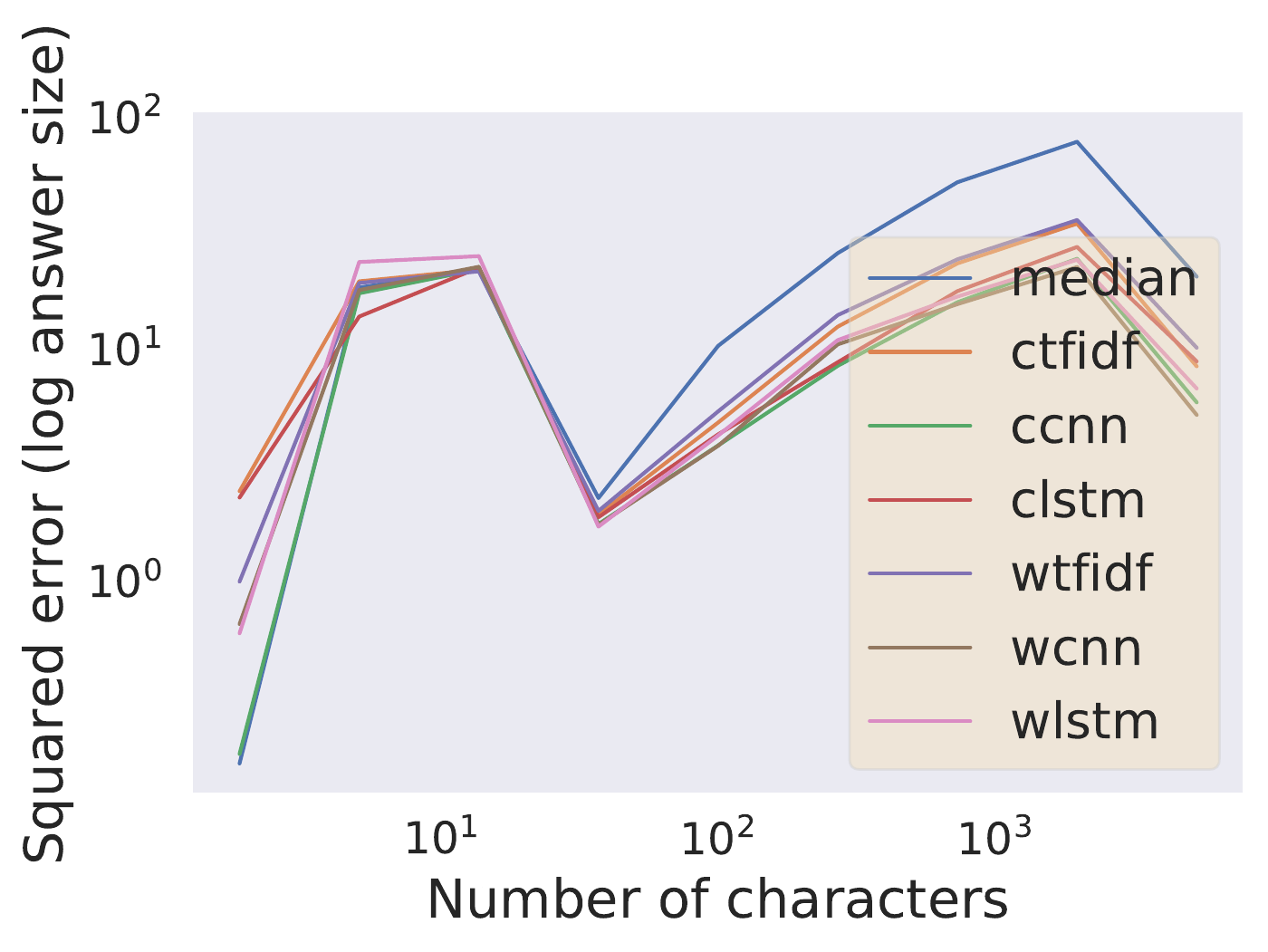}
		\label{fig:mse_clevel_rows_set1}
        }      
        \subfloat[]
        {
	    \includegraphics[width=0.5\linewidth]{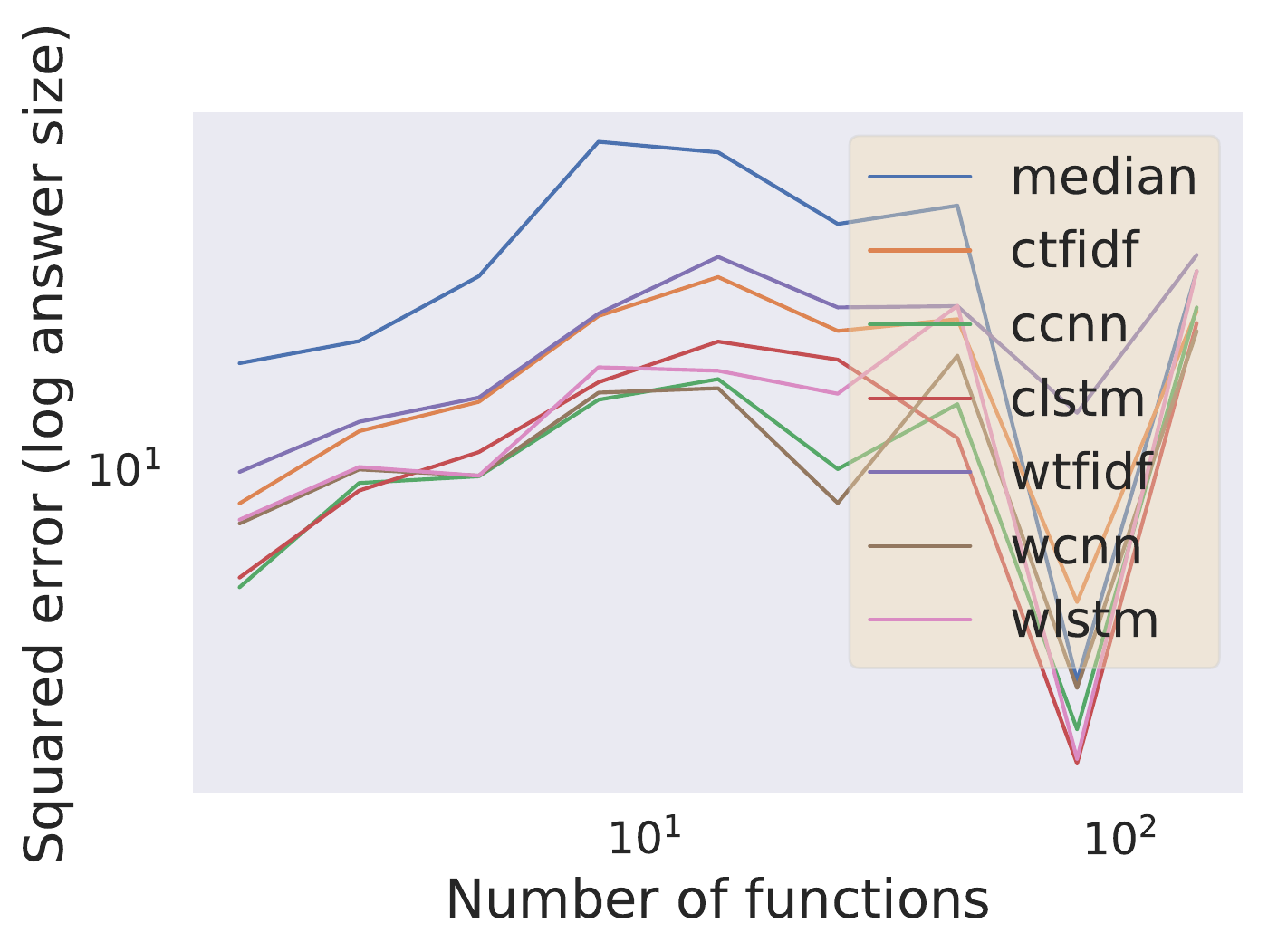}
		\label{fig:mse_fcount_rows_set1}
        }
        
        \subfloat[]
        {
	    \includegraphics[width=0.5\linewidth]{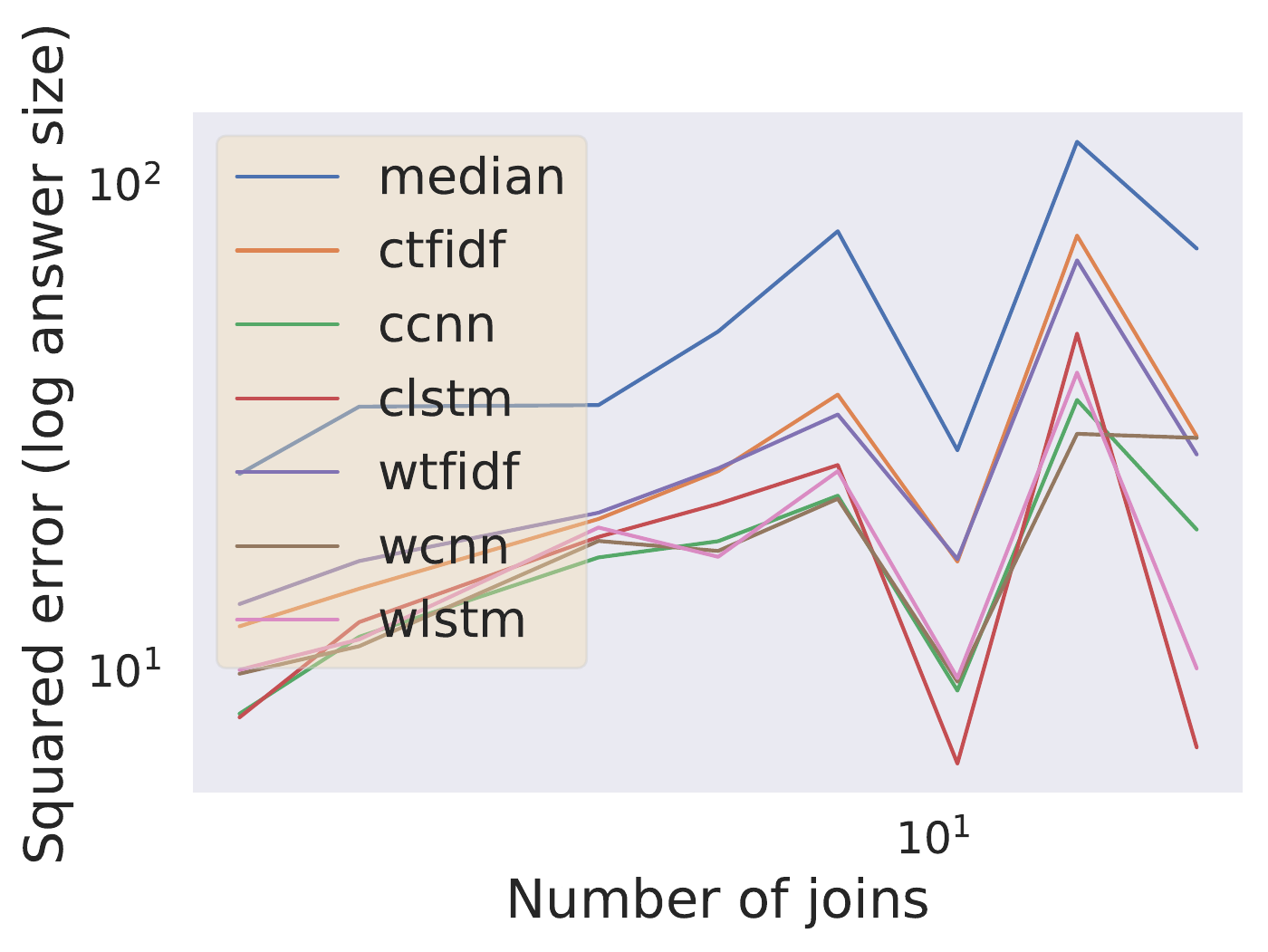}
		\label{fig:mse_jcount_rows_set1}
        }
        \subfloat[error analysis for \ccnn]
        {
	    \includegraphics[width=0.5\linewidth]{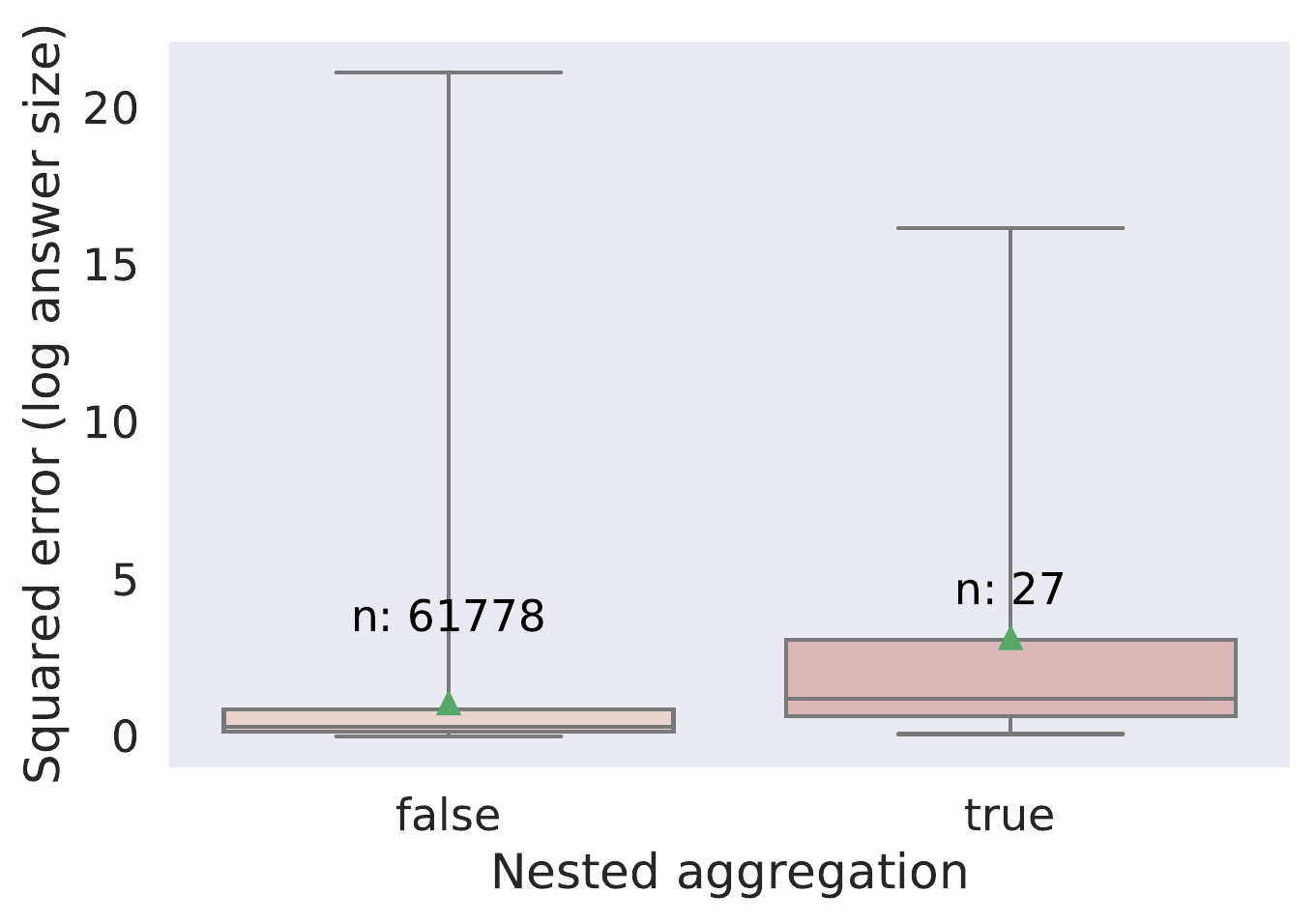}
		\label{fig:mse_na_rows_set1}
        }
	
\caption{Error analysis of answer size prediction  in \settingOne (SDSS). Performance breakdown across different session classes (9.a) and w.r.t structural properties (9.b-9.d). The total MSE of each model is reported in the legend of 9.a. } 
\label{fig:mse_by_properties_rows}
\end{figure}
\fi

%% file: RelatedWork.tex
\section{Related Work}
\label{sec:dbrec-related-work}

\vspace{2mm}
\noindent \textbf{Deep learning, Machine Learning, and NLP. }
RNNs and CNNs are dominant  in many text applications~\cite{conneau2016very}. Character-level LSTMs were used  for program execution in~\cite{zaremba2014learning}.
In~\cite{kim2014convolutional}, a one-layer word-level CNN model  outperformed tree-structured models that use syntactic parse trees as their input, for text categorization.
 Deep character-level CNN models~\cite{zhang2015character,conneau2016very,johnson2016convolutional} outperformed shallow word-level CNNs~\cite{johnson2016convolutional}. Although shallow word-level models have more parameters and need more storage,  their computations are faster. Subsequently, deep word-level CNNs have been applied in~\cite{johnson2017deep}.
LSTMs and CNNs are compared in~\cite{bai2018empirical,yin2017comparative}. \iffullpaperComposition In language modeling and other domains, CNNs can obtain comparable or better performance compared to RNNs for sequence modeling tasks. They are also parallelizable, which leads to speeds up in their execution~\cite{yin2017comparative}. We examine both LSTM and CNN models at the character and word levels. Our work is also related to machine learning for Big Code and naturalness~\cite{allamanis2018survey}, however we leave a more detailed analysis of those approaches for future work.
\else
  CNNs outperform RNNs, for sequence modeling tasks. They are also parallelizable, which leads to speeds up in their execution~\cite{yin2017comparative}. We examined  both character-level and word-level LSTMs and CNNs. Our work is related to machine learning for Big Code and naturalness~\cite{allamanis2018survey}; we leave that analysis  for future work.
\fi

\vspace{2mm}
\noindent \textbf{Deep learning in databases. } Research  problems at the intersection of deep learning and databases are introduced in~\cite{Wang:2016:DMD:3003665.3003669}.
Examples include  query optimization and natural language query interfaces~\cite{Wang:2016:DMD:3003665.3003669}. A feed-forward neural network (with 1 hidden layer) for cardinality estimation of simple range queries (without joins) is proposed and  evaluated  on a synthetic dataset~\cite{liu2015cardinality}.  Recently,~\cite{zhong2017seq2sql} developed  a natural language interface for  database systems using deep neural networks. 
In~\cite{jain2018query2vec},  an LSTM autoencoder and  a paragraph2vec model were applied  for the tasks of query workload summarization and error prediction, with experiments on Snowflake, a private query workload, and TPC-H~\cite{TPC-H1}. 
Compared to the datasets in~\cite{jain2018query2vec,zhong2017seq2sql}, SDSS and SQLShare are publicly-available and real-world. 

\vspace{2mm}
\noindent \textbf{Modeling SQL query performance. }
Estimates of SQL query properties and performance are  used in  admission control, scheduling, and costing during query optimization. Commonly, these estimates are based on manually constructed cost  models  in the query  \iffullpaperComposition optimizer. However, the cost model may not be precise and requires access to the database instance.
\else optimizer, which may be imprecise.
\fi Prior work has used machine learning to accurately estimate SQL query properties~\cite{leis2015good,li2012robust,liu2015cardinality,ganapathi2009predicting,akdere2012learning}. Most works use relatively small synthetic workloads, like TPC-H and TPC-DS, along with traditional two-stage machine learning models.  Their results are better with query execution plans as  input.  Similar to us, the  database-agnostic approach  in~\cite{jain2018database} automatically learns  features  from large query workloads rather than devising task-specific heuristics and feature engineering for pre-determined conditions.   However, they focus on  index selection and security audits. \iffullpaperComposition Note, devising robust  prediction models  that generalize well to unseen queries and changes in workloads, is studied in~\cite{li2012robust}. The approach is based on operator-level query execution plan feature engineering,  focuses on  CPU time and logical I/O for a query execution plan,  and is evaluated on small-scale query workloads.  We extend~\cite{li2012robust} by considering large-scale query workloads, and using data-driven machine learning models which learn features and their compositions.\fi

\vspace{2mm}
\noindent \textbf{Facilitating SQL query composition. }
\iffullpaperComposition
Earlier methods  provided forms for querying over databases~\cite{jayapandian2009automating}. 
But forms are restrictive.  Keyword queries are an alternative~\cite{bergamaschi2011keyword,zeng2014expressq},  but it is difficult to identify  user intention from a flat list of keywords. Both~\cite{bergamaschi2011keyword,zeng2014expressq}  tackle this problem by considering contextual dependencies between keywords, and the database structure.
\else
Form-based~\cite{jayapandian2009automating}  
and keyword queries~\cite{bergamaschi2011keyword,zeng2014expressq}  can help users write queries.
\fi
Natural language interfaces,  like NaLIR~\cite{li2014constructing}, allow complex query intents to be expressed. 
\iffullpaperComposition  Initially, the system communicates its query interpretation to the user  via a Query Tree structure.  The user can then verify, or select the likely interpretations. Next,  the system  translates the  verified or corrected query tree to the correct SQL statement.
\fi
Query recommendation  by mining query logs~\cite{khoussainova2010snipsuggest, chatzopoulou2009query,eirinaki2014querie} is another approach.
\iffullpaperComposition
QuerIE~\cite{chatzopoulou2009query,eirinaki2014querie}  assumes access to database tuples and a SQL query log. It recommends queries by identifying data tuples that are related to the interests (past query tuples) of the users.
\else
\fi
Given the schema, tuples,  and some keywords, the approach in~\cite{fan2011interactive} suggests SQL queries from templates. 
\iffullpaperComposition The evaluation is in the form of a user study with 10 experts. \fi Additional query results are recommended  for  each query  in~\cite{stefanidis2009you}. However, other than~\cite{khoussainova2010snipsuggest},  these works  access tuples. 
Other work assume the user is familiar with  samples in the query answer.  \iffullpaperComposition AIDE~\cite{dimitriadou2014explore} 
 helps the user refine their query and iteratively guides them toward interesting data areas . It is limited to linear queries, and predicts queries using decision tree classifiers. \else
  AIDE~\cite{dimitriadou2014explore} 
 helps the user refine linear queries using decision tree classifiers.
 \fi
Finding minimal project join queries based on a sample table of tuples contained in the  query answer, is studied in~\cite{Shen:2014:DQB:2588555.2593664}. \iffullpaperComposition ~\cite{cumin2017data} re-write alternate forms for the queries w.r.t.~their answer  tuples. \fi
These works  are  complementary to ours.

\vspace{2mm}
\noindent \textbf{Mining SQL query workloads. }
Several usability works use the TPC-H benchmark dataset~\cite{TPC-H1}. TPC-H has 8 tables, contains  (22) ad-hoc queries, and data content modifications. A synthetic workload can be simulated  from the ad-hoc queries.
\iffullpaperComposition WikiSQL~\cite{zhong2017seq2sql}, is a recent public query workload that  contains natural language  descriptions for  SQL queries over small datasets collected from the Wikipedia, but it does not contain the meta-information we require. We use  two publicly available and real-world  query workloads, SDSS  and SQLShare~\cite{szalay2002sdss,raddick2014ten1,raddick2014ten2,jain2016sqlshare}
\fi
 Query workloads are also used for tasks like index  selection~\cite{jain2018query2vec},   improving query optimization~\cite{li2012robust}, and workload compression~\cite{chaudhuri2002compressing}. \iffullpaperComposition The motivation  in workload compression  is that large-scale SQL query workloads can create practical problems for tasks like index selection~\cite{chaudhuri2002compressing}. While  data-driven machine learning models  rely on data abundance to  train models with many parameters, data redundancies and size can pose computational challenges. Therefore, workload compression techniques can provide an orthogonal extension for data extraction part of our work.\else
Workload compression techniques can provide an orthogonal extension for data extraction part of our work.
\fi  SDSS  has been  used to identify  user interests and access areas within the data space~\cite{nguyen2015identifying}. Ettu~\cite{kul2016ettu}, is a system that identifies insider attacks, by  clustering SQL queries in a query workload.  We focus on different problems. 

%% file: Conclusion.tex
\section{Conclusion}
\label{sec:dbrec-conclusion}

We address facilitating user interaction with the database by  providing insights  about SQL queries ---  prior to query execution.  We  leverage   (only)   the abundant information in large-scale query workloads. We conduct  an empirical study on SDSS  and SQLShare query workloads and  adapt  various  data-driven machine learning models. We found the neural networks (character-level CNNs in particular) outperformed  other models, for query error classification, answer size prediction, and  CPU time prediction.

There are several avenues for future work.  We intend to apply transfer-learning ideas to improve \ccnn under heterogeneous settings~\cite{Bengio11,Yosinski14}. More sophisticated  models, e.g.,~deep character CNNs~\cite{conneau2016very} or tree-structured architectures~\cite{tai2015improved} may lead to performance gains. Query workload extraction is another  direction. The SDSS dataset is large and noisy.  To understand the challenges, we extracted a sample  and analyzed our problems.  However, more adequate query workloads can be extracted, separately, for various problems. Another direction is to  use multi-task models that learn correlations between the query labels, although our models are applicable in broader settings where workloads have only one label.  Incorporating other types of  meta-data, e.g.,~the  database version that was queried, may increase accuracy.
\iffullpaperComposition
While our work offers a preliminary study of the challenges  in using  large-scale query workloads  for improving database usability,  the techniques are generalizable. Similar methods can be used for predict the elapsed time of queries, or  general workload analytics and management problems   such as  workload compression.  We leave addressing these challenges for future work.
\else
We leave addressing these challenges for future work.
\fi


%% file: appendix.tex
\input{Background-v3}

\section{SDSS Query Log Information}
\label{appendix:sdss_schema}

\subsection{Query Session Class Assignment}
\label{appendix:sdss_data_preprocessing}
\iffullpaperComposition
\begin{figure*}
\centering
       \includegraphics[width=1\linewidth]{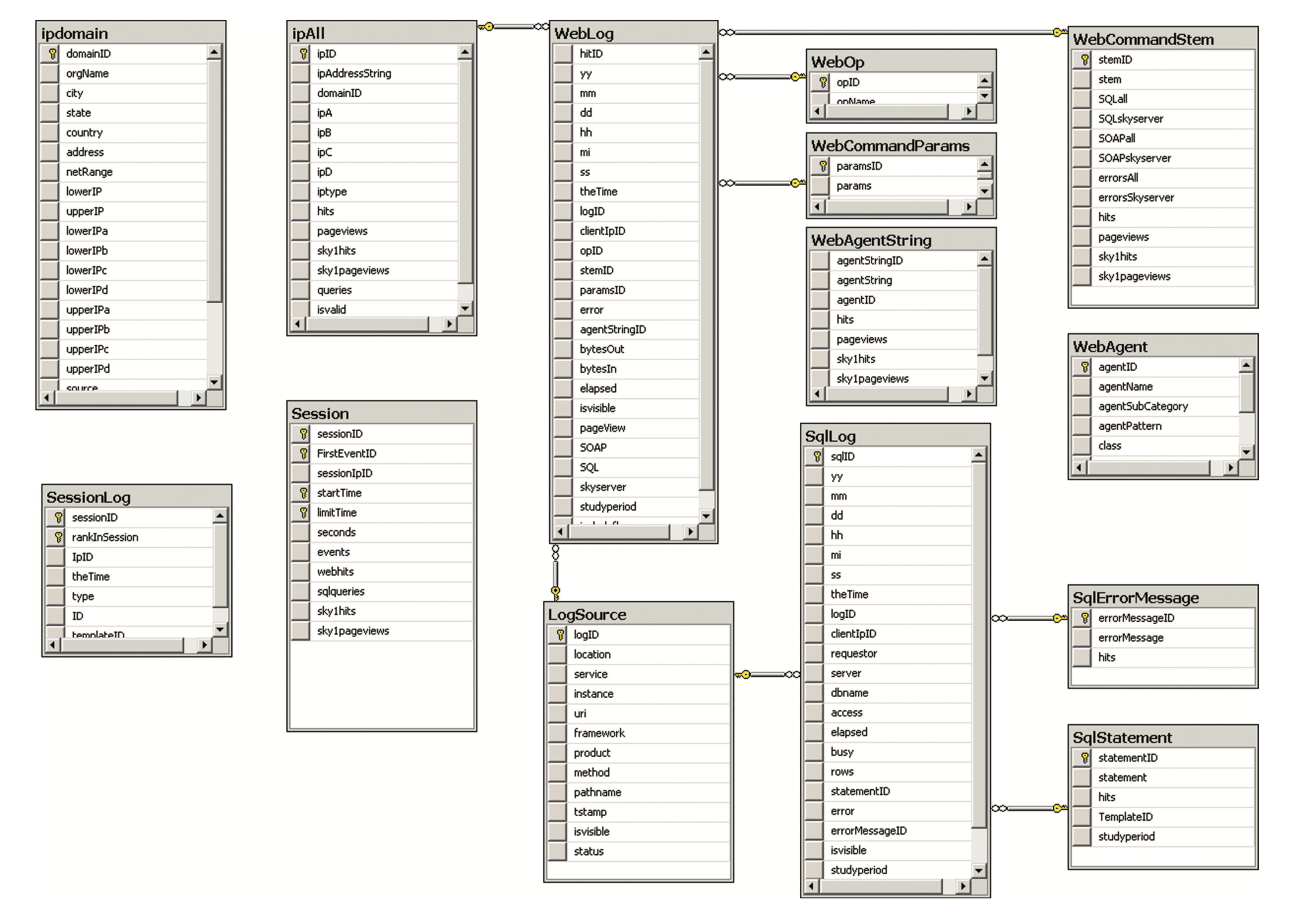}
\caption{SDSS Query Logs Schema (from~\protect\cite{raddick2014ten1})}
\label{fig:db-rec-sdss_querylogs_schema} 
\end{figure*}

\begin{table}
\centering
\scriptsize
\begin{tabular}{p{1.5cm}p{1.5cm}p{4.2cm}}
\toprule

Name&	Type&	Description\\
\toprule
yy&	smallint&	the year of the event\\
mm&	tinyint&	the month of the event\\
dd&	tinyint&	the day of the event\\
hh&	tinyint&	the hour of the event\\
mi&	tinyint&	the minute of the event\\
ss&	tinyint&	the second of the event\\
logID&	int&	the log that this came from, foreign key: LogSource.logID\\
seq&	bigint&	sequence number\\
clientIP&	char(256)&	the IP address of the client\\
op&	char(8)&	the operation (GET,POST,...)\\
command&	varchar(7000)&	the command executed\\
error&	int&	the error code if any\\
browser&	varchar(2000)&	the browser type\\
location&	varchar(32)&	the location of the site (FNAL, JHU,..\\
service&	varchar(32)&	type of service (SKYSERVER, SKYSERVICE, SKYQUERY,...)\\
instance&	varchar(32)&	The log underneath the service (V1, V2,.. )\\
uri&	varchar(32)&	The url or other ID for this service.\\
framework&	varchar(32)&	the calling framework (ASP,ASPX,HTML,QA,SOAP,...)\\
product&	varchar(32)&	the type of product acessed (EDR, DR1, DR2,...\\
PRIMARY KEY && (yy desc ,mm desc,dd desc,hh desc,mi desc,ss desc,seq desc,logID)\\

\bottomrule
\end{tabular}
\caption{WebLog Schema  (from \url{http://skyserver.sdss.org/log/en/traffic/sql.asp},  Access date: May 2018.) }
\label{dbrec:tab-weblog-detail}
\end{table}

\begin{table}
\centering
\scriptsize
\begin{tabular}{p{1.5cm}p{1.5cm}p{4.2cm}}
\toprule
Name&	Type&	Description\\
\toprule
theTime &	datetime&	the timestamp\\
webserver&	varchar(64)&	the url\\
winname&	varchar(64)&	the windows name of the server\\
clientIP&	varchar(16)&	client IP address\\
seq&	int&	sequence number to guarantee uniqueness of PK\\
server&	varchar(32)&	the name of the database server\\
dbname&	varchar(32)&	the name of the database\\
access&	varchar(32)&	The website DR1, collab,...\\
sql&	varchar(7800)&	the SQL statement\\
elapsed&	real&	the lapse time of the query\\
busy&	real&	the total CPU time of the query\\
rows&	bigint&	the number of rows generated\\
error&	int&	0 if ok, otherwise the sql error \#; negative numbers are generated by the procedure\\
errorMessage&	varchar(2000)&	the error message.\\
\bottomrule
\end{tabular}
\caption{SqlLog Schema (from \url{http://skyserver.sdss.org/log/en/traffic/sql.asp}, Access date: May 2018.)}
\label{dbrec:tab-sqllog-detail}
\end{table}

Figure~\ref{fig:db-rec-sdss_querylogs_schema} shows the SDSS schema from~\cite{raddick2014ten1}. Tables~\ref{dbrec:tab-weblog-detail} and\ref{dbrec:tab-weblog-detail} show a detailed description of the WebLog and SqlLog tables.  
\else
The SDSS schema is presented in~\cite{raddick2014ten1}.
\fi
Each session in SDSS can have a mix of SQL query entries and webhit entries. The SQL query entries are in the  ``SQLLog'' table with a unique SQLLog.sqlID, while webhit entries are in the ``WebLog'' table with a unique WebLog.hitID. The  ``SqlLog'' table contains around 194M entries which we refer to SQL query logs.  
 The session information is in the ``Session'' table, which contains the number of SQL entries (Session.sqlqueries)   and the number of webhits (Session.webhits). The ``SessionLog'' table contains more detailed information for the sessions. In particular, 
if SessionLog.type  is 0, then SessionLog.ID is a foreign key pointing to  WebLog.hitID. 
if SessionLog.type is 1, then SessionLog.ID is a foreign key pointing to SqlLog.sqlID.

Sessions with webhits have entries in the WebLog table, and include a WebLog.agentStringIDs. Based on~\cite{singh2007skyserver}, we submitted the following SQL query on CasJobs: 

\begin{lstlisting}
SELECT dbo.webAgentString.agentStringID, dbo.WebAgent.class into mydb.agentStringIDClass 
FROM dbo.webAgentString, dbo.WebAgent 
WHERE  dbo.webAgentString.agentID = dbo.WebAgent.agentID 
ORDER BY dbo.webAgentString.agentStringID
\end{lstlisting}

It retrieved a table ``agentStringIDClass'' with 578,627 entries. In this table, the WebAgent.class attribute has 6 unique values: 
UNKNOWN,
BOT,
ADMIN,
PROGRAM,
ANONYMOUS,
BROWSER. 
Therefore, sessions with webhit entries are assigned at least one of these 6 values. 
To assign a ``Session Class'' to these sessions, 
and their corresponding SQL queries,  we do a majority vote among the webhits of the session. However, even if one webhit entry with ``BOT'' exists for a session, then we assign the class ``BOT'' to the entire session, regardless of the outcome of the vote. 
For sessions  that  do not have  have a webhit entry,  an explicit class is not considered in the SDSS dataset.  We assigned the ``noWebhit'' class for SQL queries that belong to these sessions. 

\iffullpaperComposition
\subsection{Exception Handling} In our processing, we had to handle  two exceptions:

 \begin{itemize}
 
\item There were entries  in SqlLog  with errorMessageID that did not exist in corresponding SqlErrorMessage table. But SqlErrorMessage.errorMessageID is a primary key. Instead of eliminating these entries we handled this exception by creating a temporary entry for them in the final dataset output, with an appropriate description.  

\item Around  4870578 entries  in SqlLog  had SqlLog.statementID equal to zero. But SqlStatement.statementID   is a primary key in the SqlStatment table, and  does not have a zero value. We eliminated these SqlLog  entries in our full  dataset. So the difference of  194113641 - 4870578 = 189243063 is due to this step. Therefore, although SqlLog  had  194113641 entries, our final full dataset has  189243063 entries.

\end{itemize}

Another problem,  is that several entries in SqlStatement  had an empty statement but had different statementIDs and TemplateIDs. We handle this problem when we sample from the full data. Specifically, for these SqlLog entries we modify the statement to 'Empty' string.

\fi
\subsection{Repetition of Query Statements}
\label{appendix:repetition_query_statement}
\begin{figure}
\centering
       \includegraphics[width=0.7\linewidth]{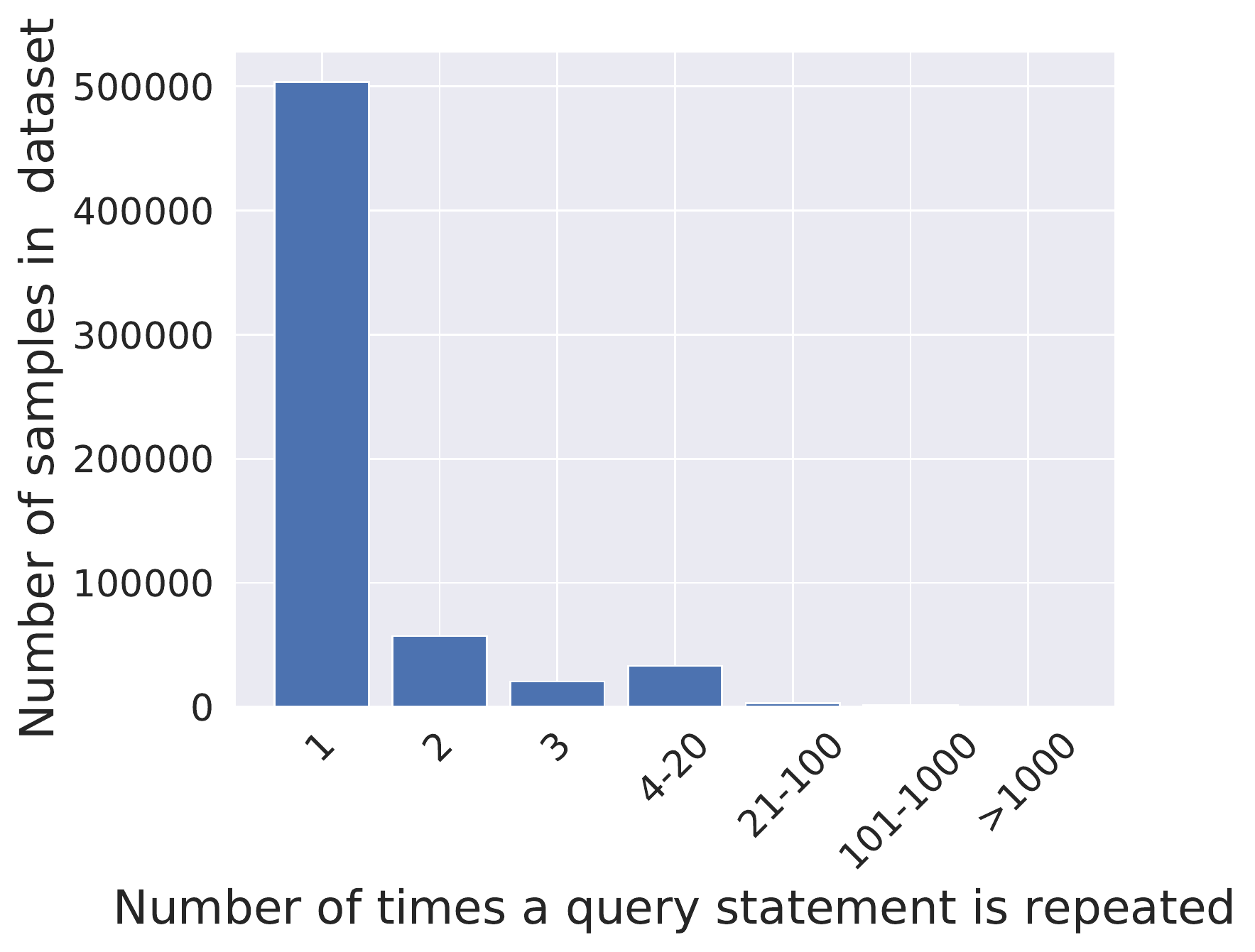}
\caption{Histogram of the number of times a query statement is repeated in a dataset consisting of randomly sampled queries from each session. }
\label{fig:statement_rep_info} 
\end{figure}

\iffullpaperComposition
We found that some SQL query logs have the same query statement, albeit varying values for properties such as session class, error class, answer size, and CPU time.  This is because the same query statement may be submitted in different sessions,   via different access interfaces, and against different versions of the database. 

To collect a dataset we proceed as follows: First,   we randomly sampled a SQL query log from each session in the pre-processed query workload.  This resulted in a total of 1,563,386 SQL query logs. 
Next, we  grouped SQL query logs  with the same query statement. This results in 618,053 groups or unique statements. We found  approximately 81.5\% of these unique statements only appear in one SQL query log. Figure~\ref{fig:statement_rep_info} shows the histogram of the number of times a query statement is repeated in the dataset.  
For the remaining 18.5\% of the query statements, we aggregated their  meta-data labels. In particular,  for answer size, and CPU time we use the average of value  as the label of the statement. For session class, and error class, we considered the majority class and used it as the label of the statement. Our final dataset consists of  618,053  statements and corresponding meta-data. 
\else 
After randomly sampling a SQL query log from each session, we obtained  a total of 1,563,386 SQL query logs.  We  grouped SQL query logs  with the same query statement, which  results in 618,053   unique statements. We found  approximately 81.5\% of these unique statements only appear in one SQL query log. Figure~\ref{fig:statement_rep_info} shows the histogram of the number of times a query statement is repeated in the dataset.  
For the remaining 18.5\% of the query statements, we aggregated their  meta-data labels.
\fi

\ifPhDThesis

\input{additionalResults}

\fi

%% file: Background-v3.tex
\section{Background: Models}
\label{appendix:dbrec-models}

\subsection{Overview}
\label{subsec:dbrec_vector_rep}

Similar to   natural language, SQL queries  have a compositional structure where smaller units are combined to create larger units. In particular characters are combined into  tokens, which are  combined into longer constructs, like  lines and blocks of code. 
This compositional structure imposes  the following considerations for generating vector representations:

\vspace{4mm}\noindent \textbf{Model granularity level. } The  level of granularity of the input data that the model considers  determines the smallest unit of input, which we refer to as \textit{tokens}. Common granularity levels are   character-level and word-level.
 For example, the query  in Figure~\ref{dbrec:fig_Q1} has 48 tokens at  the character level (excluding spaces), and  8 tokens at the word-level. 
 In text applications,  n-gram granularity level can also be considered.  An n-gram is a sequence of $n$ tokens  that appear in the document. For example a 3-gram considers sequences of length 3 as tokens. The domain \textit{vocabulary}  is the set of all possible tokens.
Word-level models are commonly used in NLP tasks.   To prevent the vocabulary from getting too large,   rare tokens and digits are  removed,    and  out-of-vocabulary (OOV) words  are replaced with an unknown token,  $<$UNK$>$~\cite{zolaktaf2011modeling}. 
 When there are many  rare or OOV tokens in a domain, character-level models can instead be used. Although they result in longer input sequences,   they can  provide a boost in accuracy~\cite{kim2016character,zhang2015character}.


\vspace{4mm}\noindent \textbf{Token representation. }  Two alternatives are \textit{symbolic} and \textit{distributed} representations. 
In  symbolic representations, each token  is represented by one symbol or feature. 
Let $V$ denote the size of the domain vocabulary. In vector space terms, each   token  is associated with an index $j$ in a  $V$-dimensional binary vector $\bm{e} \in \{0,1\}^V$,  where element $j$ is equal to 1, and the remaining elements are zero.    This is also called the  \textit{one-hot} encoding scheme.
But this  representation  does not define an inherent notion of  meaning  for  tokens. 

In  distributed representations,  each token is represented by many features. In particular, the meaning of each token  is encoded by  a real-valued d-dimensional vector $\bm{x} \in \mathbb{R}^d$.  These vectors can be based on co-occurrence statistics in the corpora. Deep learning frameworks can use large-scale corpora to learn the  semantic and syntactic aspects of the corpora in these representations~\cite{mikolov2013distributed}.

\vspace{4mm}\noindent \textbf{Combining tokens  into global vectors. }  Traditional models used the bag-of-words (BOW) model to combine   symbolic representations into a global vector that represents longer constructs, like sentences.  In BOW, token order is not considered. For example,  a simple BOW representation is the sum  of  one-hot encoding vectors of words. 
For distributed representations, in addition to  BOW, sequence and tree-structured models are commonly used~\cite{tai2015improved,li2015tree}. 
Sequence models consider token order. They process and combine the tokens sequentially to obtain the global vector. 
Examples include standard Recurrent Neural Networks (RNN), which we explore in Section~\ref{subsec:dbrec-3layerlstm}.
Tree-structured models  process and combine the words according to a tree order, bottom-up until they reach the root.  
Examples include standard recursive models 
or Tree-LSTM models~\cite{tai2015improved}.  RNNs are also regarded as a simple recursive model that processes words from left to right~\cite{conneau2016very}.

\subsection{LSTM}
\label{subsec:dbrec-lstm}

\begin{figure}
\centering
	 \includegraphics[width=0.7\linewidth]{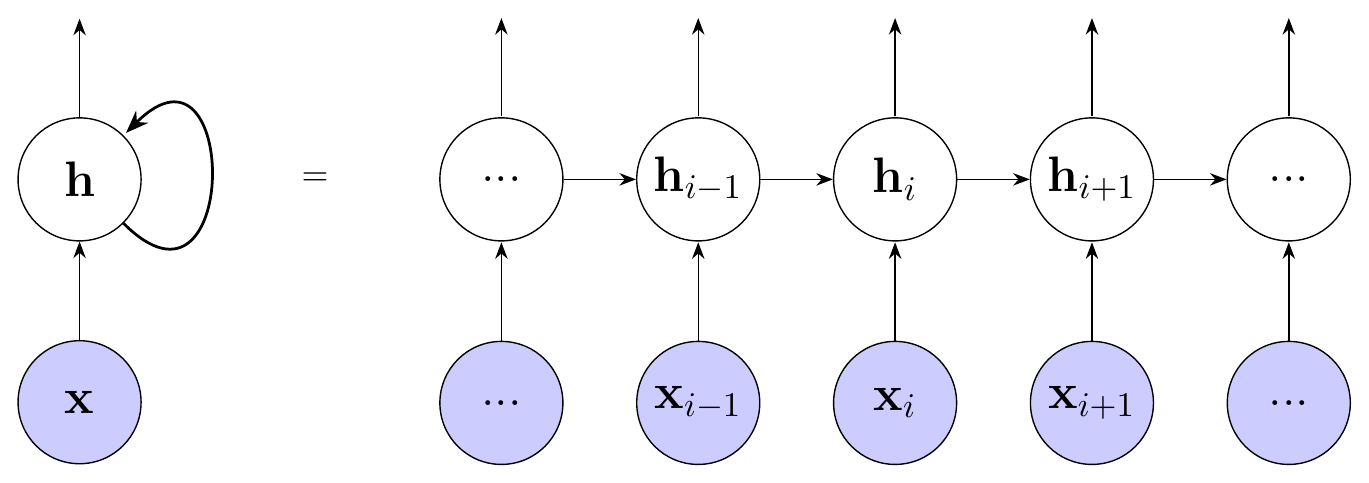}
\caption{An example RNN network with hidden-to-hidden recurrent connections.  The RNN has a repetitive structure. The left hand side,   the recurrent graph is shown, while it is  unfolded on the right. At every step $i$, the embedding $\bm{x}_i$ of a token in the sequence is  fed into the RNN. 
} 
\label{fig:rnn} 
\end{figure}

\begin{figure}
\centering
\includegraphics[width=0.6\linewidth]{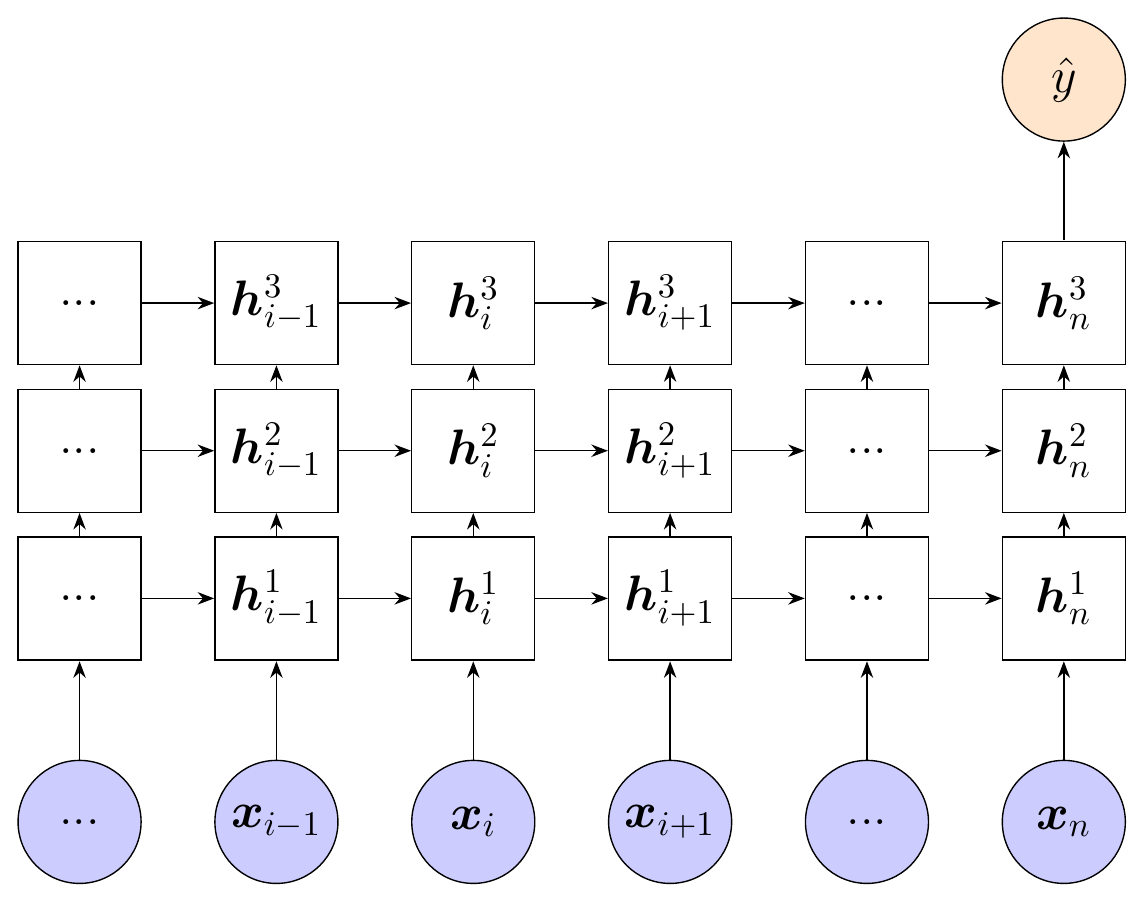}
\caption{3-layer LSTM model. Here $\bm{h}_{i}^{l}$ denotes the hidden state of an LSTM in layer $l$ at step $i$.  At each step,  $\bm{h}_{i}^{l}$  is fed as input to layer $l+1$. This allows the higher layers to capture more abstract concepts and longer term-dependencies. }
\label{fig:threelayerlstm} 
\end{figure}

Recurrent Neural Networks (RNN) can process varying length input sequences such as  characters  or word tokens in a SQL query.
Figure~\ref{fig:rnn} shows a standard  RNN network. %
Standard RNNs suffer from  the \textit{vanishing gradient} problem.
In particular, during training, the gradient vector can grow or decay exponentially~\cite{tai2015improved,Goodfellow-et-al-2016}.
LSTMs are a more effective variant of RNNs. They are equipped with a memory cell, which helps  preserve the long-term dependencies better than standard RNNs. We describe the LSTM version from~\cite{zaremba2014learning}. 
The mathematical formulation  for the LSTM is as follows: 
\begin{align*}
\bm{\tilde{c}}_i &= \tanh(\bm{W}_c \bm{x}_i + \bm{U}_c \bm{h}_{i-1} + \bm{b}_c) \\
\bm{\Gamma}_u &= \sigma (\bm{W}_u \bm{x}_i +  \bm{U}_u \bm{h}_{i-1} + \bm{b}_u) \\
\bm{\Gamma}_f &= \sigma (\bm{W}_f \bm{x}_i +  \bm{U}_f \bm{h}_{i-1} + \bm{b}_f) \\
\bm{\Gamma}_o &= \sigma (\bm{W}_o \bm{x}_i +  \bm{U}_o \bm{h}_{i-1} + \bm{b}_o) \\
\bm{c}_i &= \bm{\Gamma}_u \odot \bm{\tilde{c}}_i + \bm{\Gamma}_f \odot \bm{c}_{i-1}\\
\bm{h}_i &= \bm{\Gamma}_o \odot \tanh(\bm{c}_i) 
\end{align*}
where $\odot$ is element-wise multiplication. The LSTM parameters are $\bm{W}_c$, $\bm{W}_u$, $\bm{W}_f$, $\bm{W}_o, \bm{U}_c$, $\bm{U}_u$, $\bm{U}_f$, $\bm{U}_o$ that are weight matrices, and   $\bm{b}_c$, $\bm{b}_u$, $\bm{b}_f$, $\bm{b}_o$ that are biases.   
The memory cell $\bm{c}_i \in \mathbb{R}^k$  helps  model the long-term dependencies.  The gates $\bm{\Gamma}_i, \bm{\Gamma}_f,\bm{\Gamma}_o,\bm{\tilde{c}}_i \in \mathbb{R}^k$ help control the flow of information~\cite{Goodfellow-et-al-2016}.
We  use a three-layered LSTM model (shown in Figure~\ref{fig:threelayerlstm}).

\subsection{Prediction and training}
Having obtained a vector representation for each query, we need to learn the mapping from the vector to the query property, e.g.,~ answer size.  In the prediction stage,  for the regression problems we  pass this vector through a linear unit 
\begin{align*}
\hat{y} &=  \bm{W}_{r} \bm{h} + \bm{b}_r   
\end{align*}
where   $\bm{W}_{r}$  and   $\bm{b}_r$ are the prediction model weight and bias parameters  that should be learned.
In classification problems, we pass $\bm{h}$ through a softmax layer, to produce the predicted target value 
\begin{align*}
\hat{\bm{y}} &=  \textup{softmax} (\bm{W}_{s} \bm{h} + \bm{b}_s ) 
\end{align*} 
where $\bm{W}_{s}$  and   $\bm{b}_s$ are     parameters   that the  model  must learn. The softmax function is $\textup{softmax} (\bm{z})_i = \frac{ \exp(z_i)}{\sum_j^n \exp{z_j} } $. 

Note, for three-layered LSTM $\bm{h} = \bm{h}_{n}^3$. For the CNN architecture, we also apply dropout and use $\bm{h} = (\bm{v} \circ \bm{g} )$. Here,  $\bm{v} \in \mathbb{R}^K$ is  a masking vector  of $K$ Bernoulli random variables with probability $p$ of being 1, and $\circ$ is element-wise   multiplication. At train time,  $\bm{v}$   randomly masks  $\bm{g}$. Dropout  helps prevent co-adaptation of the feature detectors~\cite{hinton2012improving}. 

Next, we discuss the  objective functions  we used to train the   models and learn  the parameters.  In the regression problems, the goal to predict a real number for each query, which corresponds to either the answer size or CPU time. In Section~\ref{subsec:dbrec_label_analysis} we observed the answer time and CPU time distributions had  a large number of outliers. Therefore, for the regression problems we use the Huber loss function~\cite{huber1964robust}, that is a hybrid between $l_2$ for small residuals and $l_1$ for large residuals 
\begin{align}
J(\theta) & = \sum_{i=1}^m h (\hat{y}(\bm{x}_i,\bm{\theta}) - y_i) \label{eq:dbrec_reg_loss} \\
h(r) &=  
\begin{cases}
0.5 r^2, &  |r| \leq \epsilon \\
|r| - 0.5, & |r| > \epsilon.
\end{cases} 
\end{align}
where $m$ is the number of instances (queries) in the train set,  $\bm{\theta}$ is the set of all parameters that the model must learn, and $\hat{y}(\bm{x}_i,\bm{\theta})$ is the  predicted value, i.e., predicted CPU time for query $i$.
 
 For error classification and session type classification, we use the 
cross entropy objective function
\begin{align}
J(\bm{\theta}) & = \frac{1}{m} \sum_{i=1}^m -  \log \hat{y}_{y_i}(\bm{x}_i,\bm{\theta}) 
\label{eq:dbrec_class_loss}
\end{align}
here $y_i$ is the label of the correct class for instance $i$, and $\hat{y}_{y_i}(\bm{x}_i,\bm{\theta})$ is the probability of class $y_i$.

Note that  while  the traditional models fix the query vector representation (e.g.,~using the model explained in Section~\ref{subsec:dbrec-TFIDF}), and only optimize the weights of the prediction model, the neural network  models jointly train the representation and the weights of the prediction model. This is captured in the set of parameters $\bm{\theta}$ that are learned for each model.

%% file: additionalResults.tex
\section{Additional Results}
\label{appendix:deb_rec_additional_results}
In Tables~\ref{tab:error_prediction_results_full},~\ref{tab:session_class_prediction_results_full},~\ref{tab:busy_prediction_results_full},~\ref{tab:rows_prediction_results_full}, we show the results for all the experiments we ran (additional experiments modifying batch size and optimizer are not included). 
For optimizing the neural network models, we examined Adam and  Adamax~\cite{kingma2014adam}, and found the latter obtained better results.   We fixed the  learning rate  1e-3, batch size to 16, with weight decay set to 0. 
For the clstm and wlstm, we tested number of hidden dimensions in $\{150, 300\}$, clipping rate in  $\{0.25,0\}$, 
For the ccnn and wcnn, we tested number of kernels in $\{100,250\}$, 
 and drop out  in $\{0.5,0\}$, and clipping rate in $\{0.25, 0\}$.  
 We  also tried  two set of  Kernel Sizes:  $k \in \{3,4,5\}$ and $k \in \{3,4,5,9\}$. We   did not find better results for the latter and do not report the results.   

In Tables~\ref{tab:error_prediction_results_full},~\ref{tab:session_class_prediction_results_full},~\ref{tab:busy_prediction_results_full},~\ref{tab:rows_prediction_results_full},  we report  the value of the hyper-parameters that were modified, next to the model name. 
Note,  in  Section~\ref{sec:dbrec_emp_eval} we only report the models which have obtained the best results from these tables:

\begin{landscape}
\begin{table*}
\centering
\small
\begin{tabular}{llllllll}
\toprule
{\bf Model} & {\bf $v$} & {\bf \prm} &  {\bf Loss} & {\bf F$_\nsevere$} & {\bf F$_\scss$} & {\bf F$_\severe$} &  {\bf Accuracy}  \\ \hline
\input{results/results-final/sdss-results/tex/error_baseline_params.tex}
\midrule
\input{results/results-final/sdss-results/tex/error_ccnn_params.tex}
\midrule
\input{results/results-final/sdss-results/tex/error_wcnn_params.tex}
\midrule
\input{results/results-final/sdss-results/tex/error_clstm_params.tex}
\midrule
\input{results/results-final/sdss-results/tex/error_wlstm_params.tex}
\bottomrule
\end{tabular}
\caption{Error classification. 
Here  $v$ is the number of tokens in the vocabulary, \prm is the number of model parameters,   Loss is the average test loss. Lower loss is better. 
 F$_{\tt C}$ is the F-measure of class {\tt C}. High F-measure and accuracy are better. The number of samples in the test set for each class is as follows: 
\severe $=533$,
\scss $=60138$,
\nsevere $=1134$.
}
\label{tab:error_prediction_results_full}

\end{table*}
\end{landscape}

\begin{landscape}
\begin{table*}
\footnotesize
\begin{tabular}{llllllllllll}
\toprule
{\bf Model} & {\bf $v$} & {\bf \prm} &  {\bf Loss} &  {\bf F$_\nowebh$} & {\bf F$_\unknown$} & {\bf F$_\bott$} & {\bf F$_\admin$} & {\bf F$_\program$} & {\bf F$_\anonym$} & {\bf F$_\browser$} &  {\bf Accuracy}  \\ \hline
\input{results/results-final/sdss-results/tex/session_class_baseline_params.tex}
\midrule
\input{results/results-final/sdss-results/tex/session_class_ccnn_params.tex}
\midrule
\input{results/results-final/sdss-results/tex/session_class_wcnn_params.tex}
\midrule
\input{results/results-final/sdss-results/tex/session_class_clstm_params.tex}
\midrule
\input{results/results-final/sdss-results/tex/session_class_wlstm_params.tex}
\bottomrule
\end{tabular}


\caption{Session class classification. Here $v$ is the number of tokens in the vocabulary, \prm is the number of model parameters,   Loss is the average test loss. Lower loss is better. 
 F$_{\tt C}$ is the F-measure of class {\tt C}. High F-measure and accuracy are better. Note, the number of examples in the test set for each class is as follows:  
\nowebh $ = 27677$,
\unknown $ = 42$,
\bott $ = 16148$,
\admin $ = 2$,
\program $ = 4882$,
\anonym $ = 467$,
\browser $ = 12587$.
}

\label{tab:session_class_prediction_results_full}
\end{table*}
\end{landscape}

\begin{table}
\centering
\small
\begin{tabular}{lllll}
\toprule
{\bf Model} & {\bf $v$} & {\bf \prm} & {\bf Loss}  \\ \hline
\input{results/results-final/sdss-results/tex/busy_baseline_params.tex}
\midrule
\input{results/results-final/sdss-results/tex/busy_ccnn_params.tex}
\midrule
\input{results/results-final/sdss-results/tex/busy_wcnn_params.tex}
\midrule
\input{results/results-final/sdss-results/tex/busy_clstm_params.tex}
\midrule
\input{results/results-final/sdss-results/tex/busy_wlstm_params.tex}
\bottomrule
\end{tabular}
\caption{CPU time prediction.  $v$ is the number of tokens in the vocabulary, \prm is the number of model parameters,   Loss is the average test loss.  Lower loss is better.}
\label{tab:busy_prediction_results_full}
\end{table}

\begin{table}
\centering
\small
\begin{tabular}{lllll}
\toprule
{\bf Model} & {\bf $v$} & {\bf \prm} & {\bf Loss}  \\ \hline
\input{results/results-final/sdss-results/tex/rows_baseline_params.tex}
\midrule
\input{results/results-final/sdss-results/tex/rows_ccnn_params.tex}
\midrule
\input{results/results-final/sdss-results/tex/rows_wcnn_params.tex}
\midrule
\input{results/results-final/sdss-results/tex/rows_clstm_params.tex}
\midrule
\input{results/results-final/sdss-results/tex/rows_wlstm_params.tex}
\bottomrule
\end{tabular}
\caption{Answer size prediction.  $v$ is the number of tokens in the vocabulary, \prm is the number of model parameters,   Loss is the average test loss.  Lower loss is better.}
\label{tab:rows_prediction_results_full}
\end{table}